\documentclass{aa}  

\usepackage{graphicx}
\usepackage{xcolor}
\usepackage{txfonts}
\usepackage[colorlinks=true,linkcolor=bluea,allcolors=blue]{hyperref}

\begin{document}

   \title{The role of grain size in AGN torus dust models}


\author{
Omaira Gonz\'alez-Mart\'in\inst{1}\thanks{e-mail: o.gonzalez@irya.unam.mx} 
\and
Cristina Ramos Almeida\inst{2,3}
\and
Jacopo Fritz\inst{1}
\and
Almudena Alonso-Herrero\inst{4}
\and
Sebastian F. H\"{o}nig\inst{5}
\and
Patrick F. Roche\inst{6}
\and
Donaji Esparza-Arredondo\inst{2,3}
\and
Ismael Garc\'ia-Bernete\inst{6}
\and
Santiago Garc\'ia-Burillo\inst{7}
\and
Natalia Osorio-Clavijo\inst{1}
\and
Ulises Reyes-Amador\inst{1}
\and
Marko Stalevski\inst{8,9}
\and 
C\'esar Victoria-Ceballos\inst{1}
}
\institute{Instituto de Radioastronom\'ia y Astrof\'isica (IRyA), Universidad Nacional Aut\'onoma de M\'exico, Antigua Carretera a P\'atzcuaro \#8701, Ex-Hda. San Jos\'e de la Huerta, Morelia, Michoac\'an, M\'exico C.P. 58089 
\and
Instituto de Astrof\'isica de Canarias, Calle V\'ia L\'actea, s/n, E-38205 La Laguna, Tenerife, Spain
\and
Departamento de Astrof\'isica, Universidad de La Laguna, E-38206 La Laguna, Tenerife, Spain
\and
Centro de Astrobiolog\'ia (CAB), CSIC-INTA, Camino Bajo del Castillo s/n, 28692 Villanueva de la Cañada, Madrid, Spain
\and
School of Physics \& Astronomy, University of Southampton, Southampton SO17 1BJ, UK
\and
Astrophysics, University of Oxford, DWB, Keble Road, Oxford, OX1 3RH, UK
\and
Observatorio de Madrid, OAN-IGN, Alfonso XII, 3, 28014, Madrid, Spain
\and
Astronomical Observatory, Volgina 7, 11060 Belgrade, Serbia
\and
Sterrenkundig Observatorium, Universiteit Gent, Krijgslaan 281-S9, Gent B-9000, Belgium
    }

   \date{Received December 20, 2022; accepted ??}

 
  \abstract
   {Active galactic nuclei (AGN) are surrounded by dust within the central parsecs. The dusty circumnuclear structures, referred to as the torus, are mainly heated by radiation from the AGN and emitted at infrared wavelengths, producing the emergent dust continuum and silicate features. Fits to the infrared spectra from the nuclear regions of AGN can place constraints on the dust properties, distribution, and geometry by comparison with models. However, none of the currently available models fully describe the observations of AGN currently available.}
   {Among the aspects least explored, here we focus on the role of dust grain size. We offer the community a new spectral energy distribution (SED) library, hereinafter [GoMar23] model, which is based on the two-phase torus model developed before with the inclusion of the grain size as a model parameter, parameterized by the maximum grain size $\rm{P_{size}}$ or equivalently the mass-weighted average grain size $\rm{<P>}$.}
   {We created 691,200 SEDs using the SKIRT code, where the maximum grain size can vary within the range $\rm{P_{size}= 0.01-10.0\mu m}$ ($\rm{ <P>=0.007-3.41\mu m}$). We fit this new and several existing libraries to a sample of 68 nearby and luminous AGN with \emph{Spitzer}/IRS spectra dominated by AGN-heated dust.}
   {We find that the [GoMar23] model can adequately reproduce up to $\rm{\sim}$85-88\% of the spectra. The dust grain size parameter significantly improves the final fit in up to 90\% of these spectra. Statistical tests indicate that the grain size is the third most important parameter in the fitting procedure (after the size and half opening angle of the torus). The requirement of a foreground extinction by our model is lower compared to purely clumpy models. We find that $\rm{\sim 41\%}$ of our sample requires that the maximum dust grain size is as large as $\rm{P_{size}\sim 10 \mu m}$ ($\rm{ <P>\sim 3.41\mu m}$). Nonetheless, we also remark that disk+wind and clumpy torus models are still required to reproduce the spectra of a non-negligible fraction of objects, suggesting the need of several dust geometries to explain the infrared continuum of AGN. 
   }
   {This work provides tentative evidence for dust grain growth in the proximity of the AGN.}


   \maketitle
%

\section{Introduction} \label{sec:intro}

Active galactic nuclei (AGN) are powered by supermassive black holes (SMBHs) at the center of galaxies with accretion disks emitting $\rm{\sim 10^{40-47}erg/s}$ from a region as small as the solar system. AGN were originally classified based on their spectroscopic signatures, ranging from type-1 AGN, showing both broad and narrow emission lines, to type-2 AGN, showing only narrow lines. \citet{Antonucci85} suggested that the continuum and broad-line emission regions are located inside an optically and geometrically thick `disk', which was later on referred as the torus \citep{Krolik86}. Continuum and broad-line photons are scattered into the line of sight by free electrons above and below the torus. This orientation bias, as \citet{Urry95} wrote, is what makes the ``AGN depend so strongly on the orientation that our current classification schemes are dominated by random pointing directions instead of more interesting physical properties''.

However, the torus does not only act as a bridge for unifying the AGN phenomenon but it is also expected to explain the bulk of the infrared dust emission of AGN \citep{Ramos-Almeida17}. Indeed, a significant proportion of the optical/UV photons produced by the AGN accretion disk are reprocessed by dust located beyond the sublimation radius and re-emitted in the infrared. However, one of the challenges faced by such unified theories is to match the observed infrared emission of AGN to that predicted for such a dust structure \citep{Pier92,Granato94,Efstathiou95}. 

Until the arrival of extremely large single-aperture telescopes (e.g. the Extreme Large Telescope), the dust emission morphology will not be unambiguously resolved in the near or mid infrared to be compared with models \citep[][]{Nikutta21A, Nikutta21B}, except for the brightest AGN with interferometric observations with VLTI/MIDI \citep[][]{Hoenig12, Tristram14, Lopez-Gonzaga16, Hoenig17, Leftley18} and MATISSE \citep[][]{Isbell22, Gamez-Rosas22}. Although unresolved point-like ($\rm{<}$1-2\,pc) emission at 800$\rm{\mu m}$ imaged by ALMA in AGN can be contaminated by contribution from synchrotron or free-free emission \citep[][]{Pasetto19}, we rely on dust continuum-dominated sub-mm images for constraining the torus properties \citep{Garcia-Burillo21}. However, sub-mm observations are more sensitive to colder dust, which is located further away from the accretion disk, as compared to mid-infrared ones. This explains the large and geometrically thin disks with radial sizes up to 50\,pc found with ALMA \citep{Garcia-Burillo21} while the bulk of the mid-infrared emission remains confined to the central parsecs \citep[][and references therein]{Alonso-Herrero21} except for a few cases \citep[][]{Bock00, Radomski03, Packham05, Asmus16, Asmus19}.

In this work, we fit the infrared spectral energy distribution (SED) of AGN to constrain the parameters of the torus models. For this purpose, we use the so-called geometrical/phenomenological models \citep[][for a review]{Ramos-Almeida17}. According to \citet{Efstathiou95}, a successful AGN torus model must fulfill the following three minimum requirements, dictated by the shape of the infrared dust continuum \citep[see also the updated description of this shape given by][]{Hao05}: (1) the predicted spectrum should be quite broad in terms of wavelength range, at least for face-on views, and peak in the mid infrared; and (2) the model should also predict (moderate) absorption features at 10$\rm{\mu m}$ for edge-on views, and very weak (or featureless) spectra for face-on views. Indeed, early observations showed that silicate emission features were weak in AGN, which was unexpected if the dust is heated by the central AGN \citep[][]{Roche91,Laor93}. Furthermore, for those objects in which the silicate dust is seen in emission, it often shows an anomalous spectral profile where the peak is shifted to longer wavelengths \citep{Sturm05,Mason13} and with a broad profile \citep{Li08,Smith10}; and (3) the infrared SEDs of some type-1 Seyferts and quasars show a $\rm{\sim 3 \mu m}$ bump \citep{Edelson86} that cannot be easily reproduced \citep{Mor09,Garcia-Bernete17}. We now know that this feature can be seen up to $\rm{7-8\mu m}$ for most QSOs and bright AGN \citep{Garcia-Bernete19,Martinez-Paredes20,Martinez-Paredes21}. 

Several radiative transfer models in the past couple of decades have focused on the study of infrared emission and absorption from this putative dusty torus at several viewing angles, including pioneering works such as those by \citet{Pier92} and \citet{Rowan-Robinson95}. Initially, the debate was focused on the distribution of the dust. Torus models with a homogeneous dust distribution (smooth torus), predicted strong 9.7 $\mu$m silicate features, both in absorption and in emission. However, this does not match the weak absorption features generally observed for type-2 AGN or the almost absent features for type-1 AGN \citep[][]{Roche91,Laor93,Nenkova02,Nikutta09}. The clumpy torus model can naturally explain the weakness of the silicate feature but cannot produce sufficient near-infrared emission from hot dust to explain the near-infrared excess of some AGN \citep[][]{Mor09,Ramos-Almeida09,Garcia-Bernete19,Garcia-Bernete22}. This near-infrared excess might be accounted for when including a two-phase dust distribution \citep[i.e. smooth+clumpy,][]{Stalevski12,Stalevski16}. However, this two-phase model has only moderate success at explaining the overall observed infrared spectral shape of AGN \citep[failing at describing the silicate features and the mid-infrared slopes of bright AGN,][]{Gonzalez-Martin19B}.

Geometry is indeed key in these models. The near-infrared excess could also be reproduced with a disk (i.e. torus/disk equatorial dust component) and a wind (conical/polar dust component) that produces the bulk of the mid-infrared emission \citep{Hoenig17}, although other aspects as chemistry might play a significant role \citep{Garcia-Bernete22}. The AGN in the Circinus galaxy is a clear example of polar/wind dust component present from scales going from the central parsecs up to hundreds of parsecs  \citep[][]{Tristram14,Packham05,Roche06,Stalevski17,Stalevski19,Isbell22}. However, although successful at explaining the brightest AGN \citep{Martinez-Paredes21}, the disk+wind model has only moderate success at describing low and intermediate luminosity (type-2) AGN \citep{Gonzalez-Martin19B, Garcia-Bernete22} which show striking differences in the mid-infrared spectral shape compared to luminous AGN \citep{Gonzalez-Martin15, Gonzalez-Martin17}. The reason is that this wind model fails to reproduce moderately deep absorption silicate features observed in some type-2 and low-luminosity AGN due to the enhancement of graphite grains for the wind.

One of the least explored aspects of these torus models is the grain size distribution. Grain size should have a strong impact in the resulting infrared spectral shape. In particular, the size of graphite grains could contribute to the enhancement of the near-infrared continuum because grain temperature increases with decreasing dust grain size. In particular graphite grains with sizes of $\rm{0.005\, \mu m}$ have their emission peak at around 1$\rm{\mu m}$ while grain sizes of $\rm{0.25\, \mu m}$ more efficiently contribute to the 3-5$\rm{\mu m}$ flux \citep[see Fig.\,9 by][]{Laor93}. At the same time, small grains are preferentially destroyed by the AGN radiation/shocks, while large grains are more resilient \citep[][]{Schartmann08,Almeyda17}. Moreover, the silicate feature is broader with a lower strength if the grain size is micron-sized, up to its disappearance when the grain size exceeds 10$\rm{\mu m}$ \citep{Laor93}. Indeed, micron-sized grains have also been suggested to explain the anomalous silicate feature observed in the low-luminosity AGN in M81 \citep{Smith10}. \citet{Lyu14} also suggested that the low $\rm{A_{V}/\tau_{9.7\mu m}}$ ratio found in their analysis could be due to the predominance of large grains in the AGN torus \citep[see also][]{Shao17} while \citet{Xie17} found that micron-sizes particles can help to reproduce up to 90\% of the type-1 AGN explored in their analysis \citep[see also][]{Smith10}, although other aspects as geometry (e.g. wind, disk or torus) were not included in their analysis. Near-infrared interferometry suggests, throughout the normalization of the relation between inner radius and luminosity, an emission region that is more compact than expected, which might be explained by the presence of large graphite dust grains \citep{Kishimoto07,Kishimoto11,Gravity20}. \citet{Lyu21} also found that the hot dust reverberation signals for NGC\,4151 come from two distinct dust populations at separate radii, consistent with the expected properties of sublimating graphite and silicate dust grains.

Most of the current geometrical models assume the distribution of dust grains reported for the ISM in our Galaxy \citep{Mathis77,Laor93,Weingartner01,Li01}. The clumpy \citep{Nenkova08B,Hoenig10B}, smooth \citep{Fritz06} and two-phase \citep{Stalevski16} torus models assume a maximum grain size of 0.25$\rm{\mu m}$ while the minimum grain size is in the range between 0.005--0.025$\rm{\mu m}$. \citet{Hoenig10B} investigated an alternative grain size distribution called ``ISM large grains" containing grains between 0.1--1$\rm{\mu m}$. In the ISM large grain configuration, the disk/torus area of hot dust is smaller than in the standard ISM dust grain distribution (due to a steeper temperature profile), resulting in a shift of the mid-infrared emission peak towards longer wavelengths \citep[see Fig. 5 in][]{Hoenig10B}. For the disk+wind geometry, the hot dust is assumed to be primarily composed by large graphite grains, which is theoretically and observationally motivated \citep{Kishimoto07,Hoenig17,Hoenig19} where these grains produce the near-infrared excess due to the high temperatures they can survive \citep[][]{Garcia-Gonzalez17}. Indeed, this model is better at reproducing spectra below $\rm{\sim 7\mu m}$ when compared with observations \citep[see also][]{Gonzalez-Martin19B,Martinez-Paredes21,Garcia-Bernete22}.  

Although some works have investigated the role of grain size for a given geometrical model \citep[e.g.][]{Hoenig10B} or it has been used to explain a particular spectrum \citep[e.g. N- and Q-band spectra of NGC\,1068][]{Victoria-Ceballos22}, maximum grain size has not been included as a free parameter of the model. This manuscript presents a new SED library that includes variations of the maximum grain size into a geometrical model. This paper will show that the inclusion of this parameter significantly improves the adequacy of the model to AGN mid-infrared spectra observed with \emph{Spitzer}. The paper is organized as follows: Section\,\ref{sec:Data} shows the AGN sample and data used to confront currently available models (described in Section\,\ref{sec:PreviousModels}) and our new model (described in Section\,\ref{sec:NewModel}) with mid-infrared spectra. The methodology applied to the data includes an analysis of the spectral shape compared with models and spectral fits (see Section\,\ref{sec:Methodology}). Results are included in Section\,\ref{sec:Results} and discussed in Section\,\ref{sec:Discussion}. Finally, a summary of our main findings is included in Section\,\ref{sec:Summary}.

\begin{figure*}[!t]
\begin{center}
\includegraphics[width=2.\columnwidth, clip]{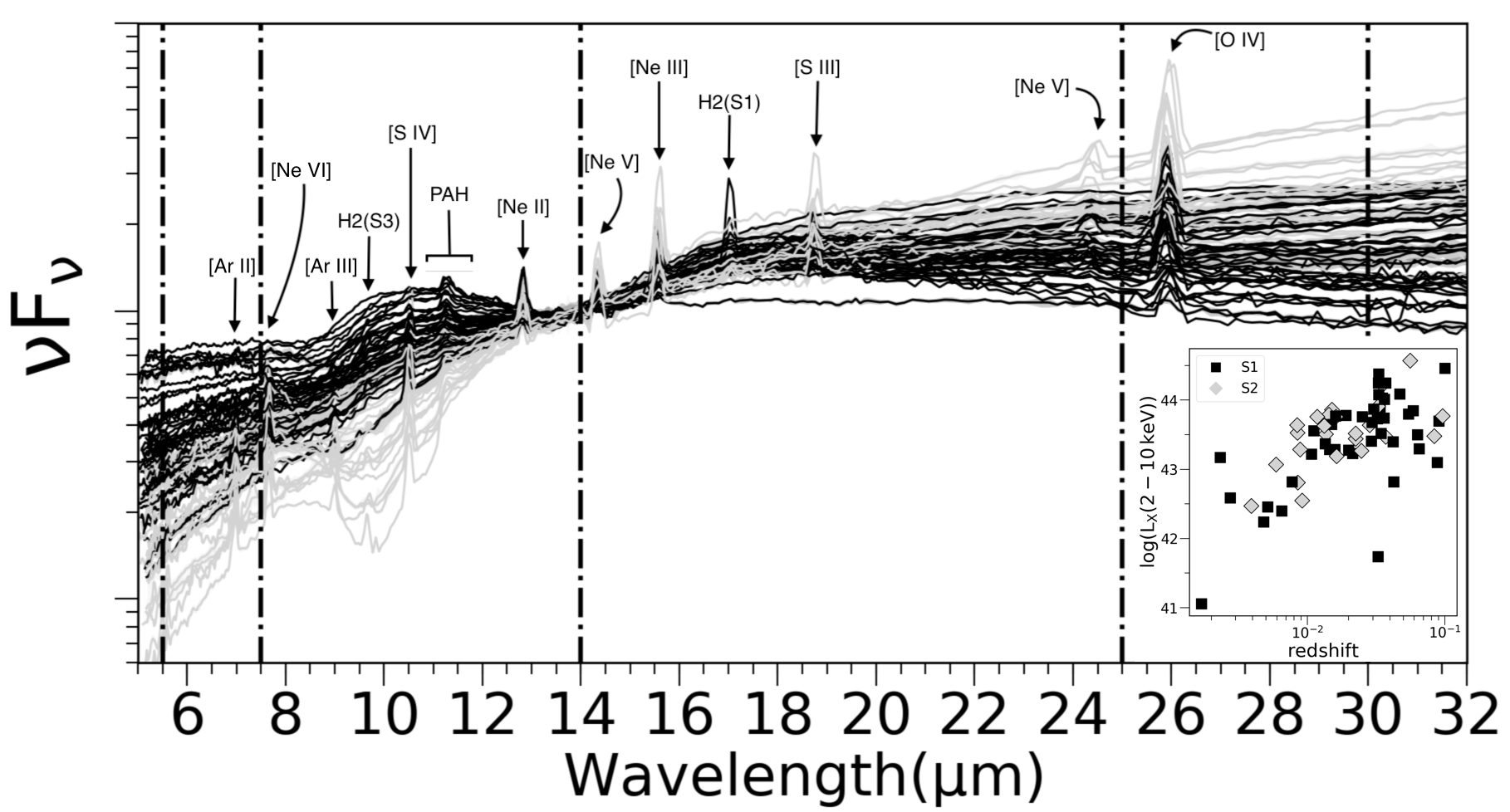}
\label{fig:spectra}
\caption{Spectra of our sample of 68 AGN normalized to the flux at 14$\rm{\mu m}$. Type-1 and type-2 AGN are shown in black and gray, respectively. Vertical dot-dashed lines indicate the locus of the wavelengths chosen to compute the spectral slopes (see text). The bottom-right inset shows the distribution of X-ray luminosities versus redshift in our sample.}
\end{center}
\end{figure*}

\section{Sample and data} \label{sec:Data}

We compile a sample of AGN with \emph{Spitzer}/IRS spectra representative of a variety of silicate features and continuum shapes. Our parent AGN samples are: (1) 53 local AGN with available \emph{NuSTAR} spectra reported by \citet{Esparza-Arredondo20}, (2) 110 nearby AGN drawn from the 105-months \emph{Swift}/BAT survey \citep{Oh18} reported by \citet{Gonzalez-Martin19B}, and (3) the 19 QSOs (one is excluded due to the lack of \emph{Spitzer} spectrum) reported by \citet{Martinez-Paredes17}. All of them have been chosen due to the availability of \emph{Spitzer}/IRS spectra needed for this analysis. Note that the first two parent samples are restricted to the nearby Universe to mitigate the low spatial resolution of \emph{Spitzer}/IRS spectra. The sample reported by \citet{Martinez-Paredes17} is not restricted to the nearby universe but they ensure the AGN dominates the mid-infrared \emph{Spitzer}/IRS spectrum from the comparison with ground-based N-band spectra. We downloaded all the spectra from Combined Atlas of Sources with \emph{Spitzer} IRS Spectra CASSIS\footnote{https://cassis.sirtf.com}. Only objects with full coverage 5--35$\rm{\mu m}$ spectra and good S/N ratio (SNR$\rm{>20}$) were selected. 

To ensure that the mid-infrared spectrum is dominated by AGN-heated dust, we impose the following criteria:

\begin{itemize}
    \item[-]We exclude \emph{Spitzer} spectra marked as extended by CASSIS because these sources are expected to have a significant contribution from circumnuclear emission.
    \item[-]We remove objects with strong PAH features (i.e. $\rm{EW(PAH)>0.233 \mu m}$) and very steep slopes above 20$\rm{\mu m}$ (i.e. slopes between 20 and 30$\mu m$ such that $\rm{log(\nu L_{\nu} (20\mu m)/\nu L_{\nu} (30\mu m))>-0.24}$), which excludes sources containing significant circumnuclear starbursts \citep[][]{Gonzalez-Martin15}.
    \item[-]The AGN dust component accounts at least for 80\% of the flux in the 5--30$\rm{\mu m}$ wavelength range inferred from spectral decomposition. Note that spectral decomposition was already done for the two nearby AGN parent samples reported above; i.e. fitting mid-infrared spectra with ISM, stellar, and AGN-heated dust templates \citep[see][and references there in]{Gonzalez-Martin19B}. However, we repeat the analysis here for homogeneity. 
    In the QSO sample, the spectra are AGN dominated because they show the same flux level and spectral shape as ground-based spectra \citep[see][]{Martinez-Paredes17}. 
\end{itemize}

With all these restrictions we selected 68 AGN with available \emph{Spitzer}/IRS spectra, good SNR and dominated by the AGN-heated dust within the \emph{Spitzer}/IRS aperture (3.6-7 arcsec). Table\,\ref{tab:sample} includes the redshift, AGN classification, X-ray 2-10\,keV intrinsic luminosity, and black-hole masses for our sample. Fig.\,\ref{fig:spectra} shows the mid-infrared spectra (type-1 and type-2 AGN in black and gray, respectively). A residual contamination of the spectra from star-forming processes is still visible, although the impact in the resulting spectral fit should be negligible \citep[see][]{Gonzalez-Martin19A}. Among them, 43 are type-1 and 25 are type-2 AGN according to the NASA extragalactic database (NED). The sample includes low- (e.g. M\,106), intermediate- (e.g. IC\,5063), and high-luminosity AGN (e.g. PG\,0804+761). We also show in Fig.\,\ref{fig:spectra} X-ray luminosities versus redshifts in our sample. No particular bias is observed when type-1 and type-2 AGN are compared. However, it is obviously biased toward high-luminosity sources and type-1 AGN because these are the objects less affected by circumnuclear star formation within the \emph{Spitzer}/IRS aperture. In this sense, our results are more representative of relatively luminous nearby AGN rather than low-luminosity AGN. Indeed, only two objects show X-ray luminosities below $\rm{L_{X}<10^{42}erg/s}$. \emph{JWST} observations with higher angular resolution would be required to explore AGN with lower luminosities. 

\section{Previous SED models} \label{sec:PreviousModels}

We aim to provide a new SED library of models that better reproduce the mid-infrared continuum emission of AGN. Thus a proper comparison with the results presented with previous SED libraries is mandatory. In the following, we provide a brief description of each model with particular emphasis on the dust grain size used for each model. Full description of the models can be found in the references below and in several recent works using these models \citep[][]{Gonzalez-Martin19B,Martinez-Paredes21,Garcia-Bernete22}. 

$\rm{\bullet}$ \underline{Smooth torus model} by \cite{Fritz06} (hereinafter {\textit{smooth [Fritz06] torus models}}): They use a simple toroidal geometry, consisting of a flared disk with the inner radius corresponding to the sublimation distance. The silicate and graphite grains have radii in the range of 0.025--0.25 and 0.005--0.25\,$\rm{\mu m}$, respectively. 

$\rm{\bullet}$ \underline{Clumpy torus model} by \cite{Nenkova08A,Nenkova08B} (hereinafter {\textit{clumpy [Nenkova08] torus models}}): They used a formalism that accounts for the concentration of dust in clouds, forming a torus-like structure. The silicate and graphite grains have both radii in the range of 0.005--0.25\,$\rm{\mu m}$. 

$\bullet$ \underline{Clumpy torus model} by \citet{Hoenig10B} \citep[see also][]{Hoenig06,Hoenig10A} (hereinafter {\textit{clumpy [Hoenig10] torus models}}): Radiative transfer model of three-dimensional clumpy dust tori using optically thick dust clouds and a low torus volume filling factor. The size of silicate and graphite grain is in the range of 0.025--0.25\,$\mu m$. 

$\rm{\bullet}$ \underline{Two-phase torus model} by \cite{Stalevski16} (hereinafter {\textit{two-phase [Stalev16] torus models}}): They model the dust in a torus geometry with a two-phase medium, consisting of high-density clumps embedded in a smooth dusty component of low density. The silicate and graphite grains have both radii in the range of 0.005--0.25\,$\rm{\mu m}$. 

$\rm{\bullet}$ \underline{Clumpy disk and outflow model} by \cite{Hoenig17} (hereinafter {\textit{clumpy disc$+$wind [Hoenig17] models}}): The model consists of a clumpy disk plus a polar outflow. This model uses a sublimation model that accounts for different sublimation conditions, with grain sizes of 0.025--0.25\,$\mu m$ for silicate, and 0.075--1.0\,$\mu$m for graphite.

Note that we also explore the clumpy disk model by \citet{Hoenig17} (i.e. removing the wind component in the disk+wind model), the anisotropically illuminated version of the clumpy torus model presented by \citet{Garcia-Gonzalez17}, and the two-phase torus model presented by \citet{Siebenmorgen15}. However, after performing spectral fitting to our sample, these models were not able to fully describe any of our objects (and provide acceptable fits for less than 10\% of the sample). Thus, we exclude them from our analysis for simplicity.

\section{New SED model GoMar23} \label{sec:NewModel}

We use the radiative transfer code SKIRT\footnote{www.skirt.ugent.be} \citep{Baes03,Baes15,Camps15,Camps20} in its version 9 to produce synthetic spectra with different dust and geometrical properties. We divide this section into the details of the overall setup of the simulations in common with the two-phase model created in \citet{Stalevski16} (Section\,\ref{sec:NewModel:setup}), the particularities of the grain size distribution (Section\,\ref{sec:NewModel:grainsize}), and a summary of the grid of parameter values for [GoMar23] model.

\subsection{Setup of the simulations}\label{sec:NewModel:setup}

\begin{figure}[!t]
\begin{center}
\includegraphics[width=1.\columnwidth]{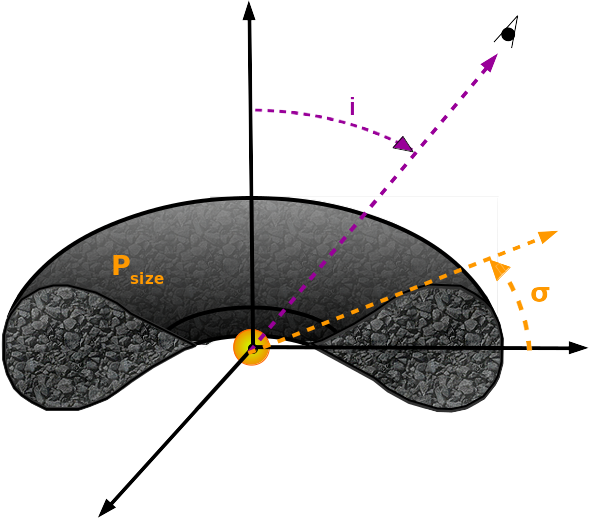}
\caption{Schematic view of [GoMar23] model (see text).} \label{fig:sketch}
\end{center}
\end{figure}

SKIRT has already been used by \citet{Stalevski16} to produce a two-phase torus model ([Stalev16], see Section\,\ref{sec:PreviousModels}). We set our initial geometry and parameter space to that of [Stalev16] model \citep[see Fig.\,1 by][]{Stalevski12}. This allows us to set a benchmark for the model because at least we should be able to provide good fits for the same objects that are already reproduced by [Stalev16] model. Furthermore, the two-phase medium (clumps within a smooth distribution) seems to be the most plausible scenario for the dust within the torus \citep{Wada15}. [Stalev16] model parameters are the viewing angle toward the observer, $i$, the ratio between the outer and the inner radius of the torus, $\rm{Y = R_{out}/R_{dust}}$ (with the inner radius anchored to the dust sublimation radius $\rm{R_{dust}}$), the half opening angle of the torus, $\sigma$, the indices that set dust density gradient with the radial, $p$, and polar, $q$, distribution of dust, and the $9.7 \mu m$ average edge-on optical depth, $\tau_{9.7\mu m}$. Fig.\,\ref{fig:sketch} shows a schematic view of the model. 

The geometry of the model consists of a flared disk ({\sc torusgeometry} within SKIRT). The inner radius is fixed to the dust sublimation radius $\rm{R_{dust}=0.14\,pc}$, which is obtained from \citet{Barvainis87}:
\begin{equation}
    R_{dust} = 1.3 (L_{bol}/10^{46})^{1/2}(T_{sub}/1500)^{-2.8} (pc)
\end{equation}
\noindent for an AGN with a bolometric luminosity of $\rm{L_{bol}=10^{11} L_{\odot}}$ and a sublimation temperature $\rm{T_{sub}}$ to 1250\,K. Note that, unlike \citet{Stalevski16} we assume that the accretion disk emits isotropically. 

To fully define the torus geometry, two parameters are required: the half-opening angle, $\rm{\sigma}$, and the outer radius of the torus, $\rm{R_{out}}$. Following several previous prescriptions \citep[e.g.][]{Fritz06,Nenkova08B, Stalevski16}, we report the values of the outer radius of the torus as a multiplier of the inner radius, i.e. $\rm{Y = R_{out}/R_{in}}$. In the [Stalev16] model, this parameter takes only three values at $\rm{Y=[10, 20, 30]}$. However, we find significant improvements by evaluating this parameter in a wider range and with a slightly finer step. In fact, this is the parameter with the lower number of values in [Stalev16] model. In particular, we choose the following set of values: $\rm{Y=[2,5,10,15,20,25,30,35,40]}$. For the half opening angle, we use the same values as in [Stalev16] model, i.e. $\rm{\sigma = 10-80^{\circ}}$ with $\rm{\Delta \sigma = 10^{\circ}}$. 

The model is indeed two-phase (i.e. smooth+clumpy) thanks to the `decorator' module {\sc clumpygeometrydecorator} within SKIRT that creates a clumpy-distributed medium from a continuously-distributed one. The geometry used is called flared disk, which consists of a sphere where two cones are removed from the polar direction. Therefore, the width of the torus grows with the radius, keeping the angular width constant. Following the parameters used by \citet{Stalevski16}, the percentage of the mass locked up in clumps is set to 25\%, the radius of each cloud is linked to the outer radius of the system as $\rm{R_{cl}=R_{out}/12.5}$, and the total number of clumps is set to 3000. Within the flared disk geometry, the dust is distributed following a variable radial power law and polar exponential density distribution of dust: 
\begin{equation}
    \rho (r,\theta) \propto r^{-p} e^{-q|cos (\theta)|}
\end{equation}

\noindent where $r$ and $\theta$ are the coordinates in polar coordinate system. The parameters associated with the radial and polar dust-density gradients are called \emph{p} and \emph{q}, respectively. Following the work by \citet{Stalevski16}, we use the same range for both \emph{p} and \emph{q} parameters including 0--1.5 (four values with $\rm{\Delta p = \Delta q = 0.5}$). The model also requires the edge-on optical depth at 9.7$\rm{\mu m}$, defined at the equator of the system, $\rm{\tau_{9.7}}$. The range explored by \citet{Stalevski16} for this parameter is $\rm{\tau_{9.7}=3-11}$ with $\rm{\Delta \tau_{9.7}=2}$. We set our model to the same range but extend the grid to $\rm{\tau_{9.7}=13}$. Note that in our library of models, $\rm{\tau_{9.7}}$ is defined along the x-axis of the system. 

\subsection{Grain size distribution}\label{sec:NewModel:grainsize}

Particles are assumed to obey a power-law distribution of sizes of the form \citep{Mathis77}:
\begin{equation}
\rm{dn=A n_{H}P^{-a}dP} \label{eq:sizedistribution}
\end{equation}

\noindent where $\rm{P}$ is the dust particle size, ``a" is the slope of the particle dust distribution (assumed to be $\rm{a = 3.5}$), $\rm{n_{H}}$ is the number density of H nuclei, and the constant A gives the normalization with respect to hydrogen abundance for each dust component (in units of particles per H atom per micron). These constants would only affect the normalization with respect to the hydrogen abundance, and not the SED scaling/normalization (which is dominated by the mass of dust). We use the normalization factors assumed in the two-phase torus model presented by \citet{Stalevski16} is log(A) = $\rm{-24.82}$ for silicate grains. Also following the prescription given by \citet{Stalevski16}, we choose a percentage of 51\% for silicate grains and 49\% for graphite grains. Note that we also ran some tests using a percentage of silicate grains in the range used in other works \citep[i.e. 47-53\%, see][]{Mathis77,Draine84,Weingartner01}, finding no significant impact on the resulting SEDs.

This new SED library explores the role of dust grain size. The grain size distribution (see Equation\,\ref{eq:sizedistribution}) also requires the minimum and maximum grain size for each population. Note that the effect of grain size can be investigated in various ways: changing the maximum/minimum of the grain size distribution, the slope of the grain size distribution, or combining these possibilities. \citet{Victoria-Ceballos22} recently explored the effect of both the minimum and maximum grain sizes in the shape of the mid-infrared spectra of NGC\,1068, finding no significant impact on the resulting SED for the minimum grain size, while changing the maximum grain size turned out to strongly affect the SEDs. Furthermore, the fluctuations induced by absorption of a single X-ray photon eliminate grains with sizes $\rm{P_{size} < 0.005\,\mu m}$ up to large distances from the center \citep[within tens of parsecs][]{Laor93}. \citet{Voit91} showed that grains with sizes below $\rm{0.001\, \mu m}$ are easily destroyed by the radiation field of the AGN while grains with sizes in the range of $\rm{0.001-0.005\, \mu m}$ have little effect on the total grain emission. Thus, we decided to focus on the maximum grain size, hereinafter $\rm{P_{size}}$, while the minimum grain size is fixed to $\rm{0.005 \mu m}$, consistent with the model presented by \citet{Stalevski16}. For that purpose, we explore SEDs with maximum dust grain sizes in the range $\rm{P_{size}=0.01-10\,\mu m}$ with $\rm{\Delta log(P_{size})\simeq 0.3}$ (i.e. 10 values). We set the maximum grain size to $\rm{10\mu m}$ because silicate features completely disappear when the grain size exceeds that value \citep{Laor93}. Finally, we remark that we choose to focus on the maximum grain size, but keep in mind that flatter grain size distributions compared to the usual $a=3.5$ have the same impact, i.e. it changes the small to large grain abundance \citep{Maiolino01}.  

Another way to report the grain size is throughout the weighted average grain size $\rm{<P>}$ \citep{Mishra17}:
\begin{equation}
\rm{<P>=\frac {\int_{P_{min}}^{P_{max}} P w(P) \frac{dn}{dP} dP} {\int_{P_{min}}^{P_{max}}  w(P) \frac{dn}{dP} dP }}
\label{eq:averagesize}
\end{equation}

\noindent where $\rm{w(P)}$ is the weight, which takes the form of $\rm{P^{\beta}}$ with $\rm{\beta=3}$ for the mass-weighted average. This equation is integrated between the minimum grain size $\rm{P_{min}=0.005\mu m}$ and the maximum grain size in the range $\rm{P_{size}=0.01-10\,\mu m}$. This results in a mass-weighted average grain size in the range $\rm{<P>=0.007-3.41\,\mu m}$. We use both definitions of $\rm{P_{size}}$ and $\rm{<P>}$ throughout the text.

\begin{table}[t!]
\begin{center}
\begin{tabular}{lcc}\hline \hline
        &  [Stalev16] & [GoMar23] \\
        & range / $\rm{\Delta par}$ / $\rm{N}$ & range / $\rm{\Delta par}$ / $\rm{N}$ \\ \hline
    $L_{bol}$ &  $\rm{10^{11}L_{\odot}}$   & $\rm{10^{11}L_{\odot}}$    \\
    $T_{sub}$ &  1250K   &  1250K  \\
    $a$ &   3.5  &  3.5  \\
    $P_{min}$ &   0.005$\rm{\mu m}$  &  0.005$\rm{\mu m}$  \\ \hline
    $i$    &  [0-90]/10$\rm{^{\circ}}$/10   &  [0-90]/10$\rm{^{\circ}}$/10 \\
    $\sigma$ & [10-80]/10$\rm{^{\circ}}$/8  & [10-80]/10$\rm{^{\circ}}$/8 \\
    $Y$ & [10,30]/10/3  & [2,40]/5/9 \\
    $p$ &  [0,1.5]/0.5/4 & [0,1.5]/0.5/4 \\
    $q$ & [0,1.5]/0.5/4  & [0,1.5]/0.5/4 \\
    $\tau_{9.7}$ &  [3,11]/2/5 & [3,13]/2/6 \\
    $P_{size}$ & 0.25$\rm{\mu m}$  & [0.01,10]$\rm{\mu m}$/$\rm{\Delta log\sim 0.3}$/10 \\
    $<P>$ & 0.097$\rm{\mu m}$  & [0.007,3.41]$\rm{\mu m}$/$\rm{\Delta log\sim 0.3}$/10 \\
    \# SEDs & 16,800 & 691,200  \\
\hline\hline
\end{tabular}
\end{center}
\caption{Parameter space for the new [GoMar23] model. We also provide the values of [Stalev16] model for comparison purposes. From top to the bottom row, the bolometric luminosity, $\rm{L_{bol}}$, the sublimation temperature, $\rm{T_{sub}}$, the slope of the size distribution, $a$, the minimum grain size, $\rm{P_{min}}$, the viewing angle toward the system, $i$, the half opening angle of the torus, $\rm{\sigma}$, the ratio between the outer and the inner radius of the torus, $Y$, the slope of the radial and polar dust density distributions, $p$ and $q$, the opacity at 9.7$\rm{\mu m}$ measured at the equator of the system, $\rm{\tau_{9.7}}$, and the maximum grain size, $\rm{P_{size}}$. The ratio between the outer and the inner radius, $\rm{Y}$, goes from 5 up to 40 in steps of 5 with the inclusion of $\rm{Y=2}$. Finally, the maximum grain size is set to the following values: 0.01, 0.025, 0.05, 0.1, 0.25, 0.5, 1.0, 2.5, 5.0, and 10$\rm{\mu m}$ ($\rm{<P>=}$ 0.07, 0.014, 0.024, 0.042, 0.097, 0.18, 0.36, 0.87, 1.72, and 3.41 $\rm{\mu m}$). The last row of this table includes the total number of SEDs performed.}
\label{tab:model}
\end{table}

\subsection{Overall grid}\label{sec:NewModel:grid}

Each SKIRT simulation can provide as many SEDs with different viewing angles as requested. Given the symmetry of the system, we requested viewing angles in the range $\rm{i = 0-90^\circ}$ with $\rm{\Delta i = 10^\circ}$, where the viewing angle is measured from the pole toward the equator of the system (see Fig.\,\ref{fig:sketch}). The final model has seven parameters (eight when including the flux-normalization of the final fit). A summary of the parameters involved in the model is included in Table\,\ref{tab:model}. For comparison purposes, we also include the original parameter space of [Stalev16] model. 

Each simulation was computed using $\rm{10^{6}}$ photon packages \citep[consistent with other torus simulations, see][]{Stalevski17} and a spatial hierarchical octree grid with a variable size of the cell named {\sc treespatialgrid}. We initially set the number of random density samples for determining spatial cell mass to 100. The cells are then recursively partitioned into sub-cells until the optical depth for each of them is lower than one or the simulation reaches the maximum number of cell subdivisions, $\rm{N_{cell}^{max}}$, that we set to $\rm{N_{cell}^{max}=11}$. Note that this maximum number, when reached, already requires at least 30GB of RAM, depending on the geometrical configuration of the simulation. Thus, further subdivisions are not allowed in order to limit computational times due to the maximum RAM available for these simulations. However, we run a test to quantify if the resulting SEDs depend on this maximum number of subdivisions. In particular, we simulated 7290 SEDs using the minimum, medium, and maximum values for each parameter with $\rm{N_{cell}^{max}=12}$. The difference between these SEDs and those using $\rm{N_{cell}^{max}=11}$ is always lower than 5\%. This ensures that $\rm{N_{cell}^{max}}$ does not have a strong impact on the resulting SEDs because the cells are optically thin for all of our simulations.

All together we produce 691,200 SEDs in the range of 0.001-5000\,$\rm{\mu m}$ with all the parameters evaluated at least at four grid points. We use a nested logarithmic wavelength grid, with 81 steps in the range 0.001-5000\,$\rm{\mu m}$ and 101 additional steps in the range 1-100\,$\rm{\mu m}$. This SED library has among the largest number of SEDs performed; below the $\rm{\sim}$1.2 million SEDs produced for the clumpy torus model by \citet{Nenkova08B} but well above the $\rm{\sim}$130 thousand SEDs produced for the disk+wind model by \citet{Hoenig17}, which is the second largest library. The computing time for this grid was around 200 days on a cluster\footnote{These simulations are produced thanks to Calzozin and Mouruka clusters at IRyA/UNAM} able to produce 16-20 SEDs per hour thanks to the parallel processing capabilities. Following our own nomenclature in previous works \citep{Gonzalez-Martin19A,Gonzalez-Martin19B,Martinez-Paredes21,Garcia-Bernete22}, we name this model according to the initials of the first author of this manuscript as [GoMar23] model. 

\section{Methodology}\label{sec:Methodology}

\subsection{Spectral shape characterization}\label{sec:SpectralShape}

The continuum shape of the mid-infrared spectra of many AGN is distinctive with a monotonic increase in flux density with wavelength and reaching a maximum at  $\rm{\sim18-20\mu m}$ \citep[see spectral slopes reported by][]{Garcia-Gonzalez17} and a flattening at longer wavelengths \citep{Gonzalez-Martin19B} (see Fig.\,\ref{fig:spectra}). Furthermore, silicate features have been reported in absorption as well as in emission \citep[e.g.][]{Roche91,Shi06,Alonso-Herrero16}. Two silicate features can be seen in the \emph{Spitzer}/IRS spectral window around $\rm{10}$ and $\rm{18\mu m}$. The $\rm{10\mu m}$ feature is easier to characterize because it is more prominent than the $\rm{18\mu m}$ feature in our sample. Although the vast majority of type-2 (type-1) AGN show silicate features in absorption (emission), the opposite behavior is not uncommon \citep[see Fig.\,\ref{fig:spectra} and also][]{Feltre12}. 

On top of the AGN dust continuum, observed AGN spectra also show emission lines associated with the ionization gas in the nuclei (e.g. [OIV]) or to some level of circumnuclear star formation (e.g. [NeII]). These lines must be carefully removed to properly measure the AGN infrared continuum. We fit each spectrum to a 6th-order polynomial and then smooth it to remove the effect of emission lines in the fitted continuum. We then divided the original spectrum by this fitted continuum to produce a clean spectrum of emission lines. Where prominent lines are detected in the latter, we replace the spectral window with the smooth polynomial fit to the continuum.

In order to characterize the spectral shape of the objects in our sample, we define three spectral slopes (denoted with $\rm{\alpha}$) within the \emph{Spitzer}/IRS wavelength range. These slopes are computed as:
\begin{equation}
    \alpha_{\lambda_1 - \lambda_2} = -log(F_{\lambda_1}/F_{\lambda_2})/ log(\lambda_1/\lambda_2)
\end{equation}
\noindent where $\rm{\lambda_1}$ and $\rm{\lambda_2}$ are the two wavelengths where the slope is computed and $\rm{\lambda_1< \lambda_2}$. Negative (positive) values are found when the flux increases (decreases) with wavelength. These three slopes are called $\rm{\alpha_{5.5-7.5\mu m}}$, $\rm{\alpha_{7.5-14\mu m}}$, and  $\rm{\alpha_{25-30\mu m}}$ and correspond to $\rm{[\lambda_1,\lambda_2]}$ equal to [5.5, 7.5], [7.5, 14], and [25, 30]$\rm{\mu m}$, respectively. These slopes are selected as in \citet{Gonzalez-Martin19A}.

\begin{figure*}[!t]
\begin{center}
\includegraphics[width=0.67\columnwidth]{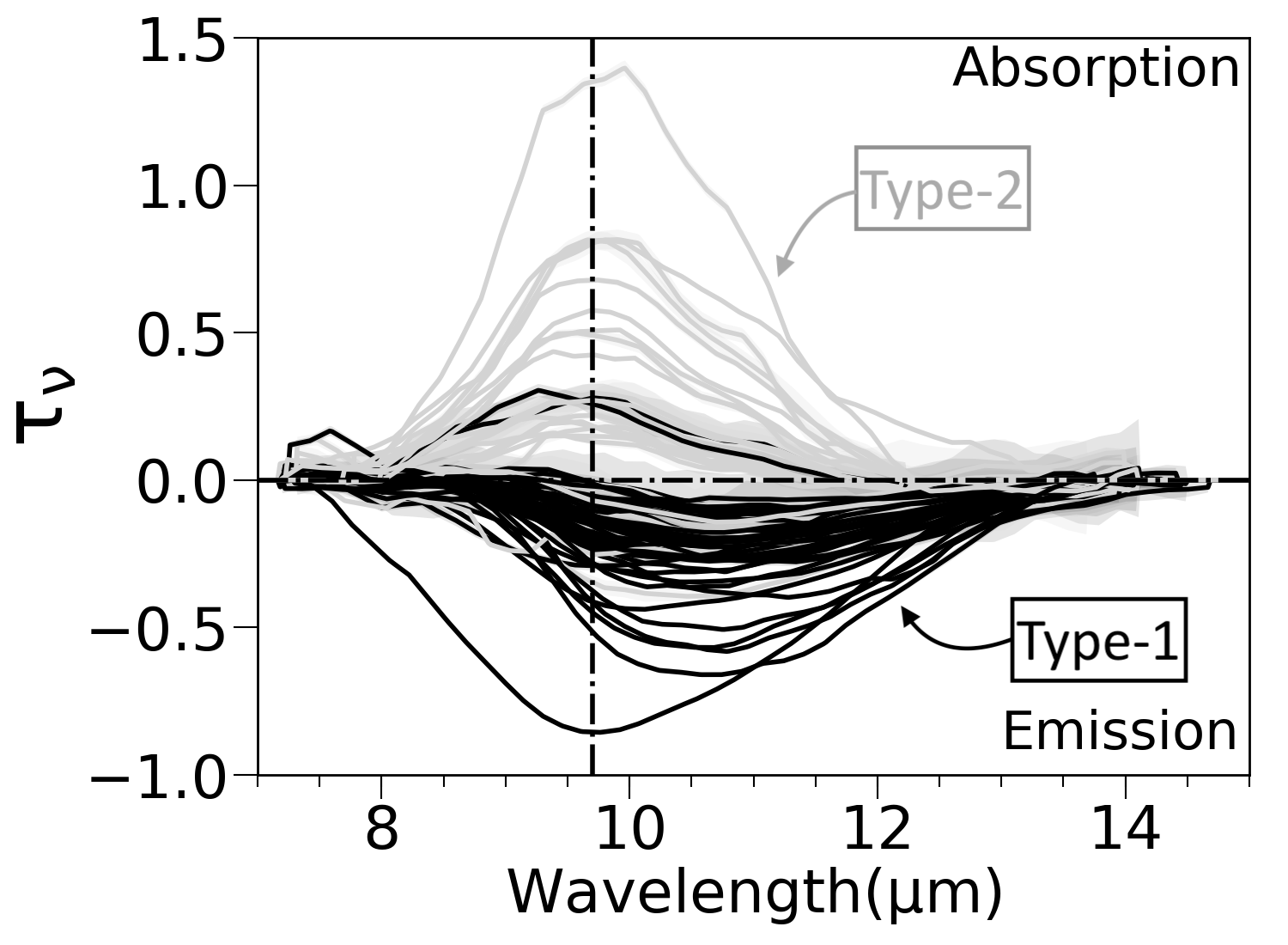} 
\includegraphics[width=0.67\columnwidth]{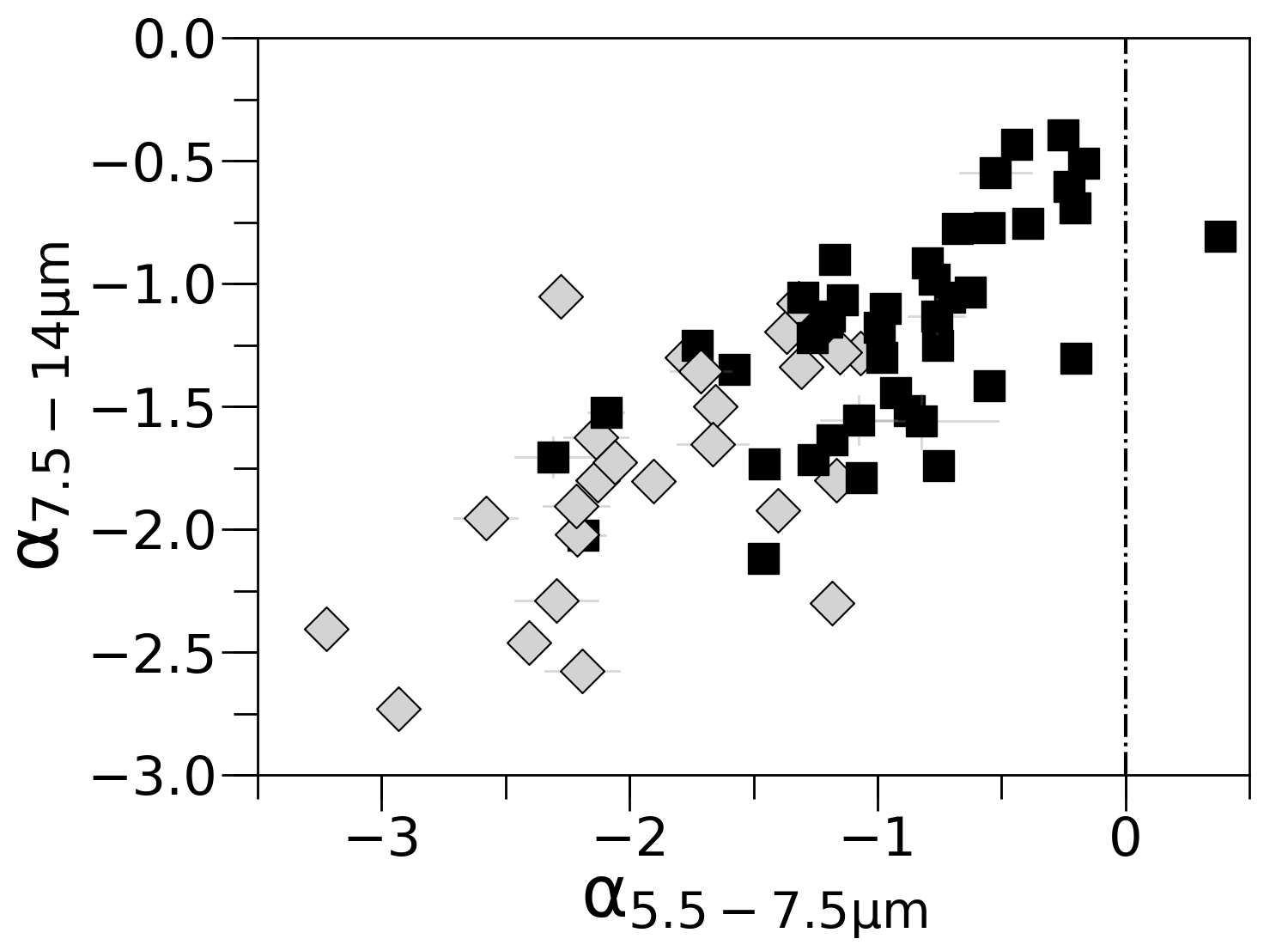}
\includegraphics[width=0.67\columnwidth]{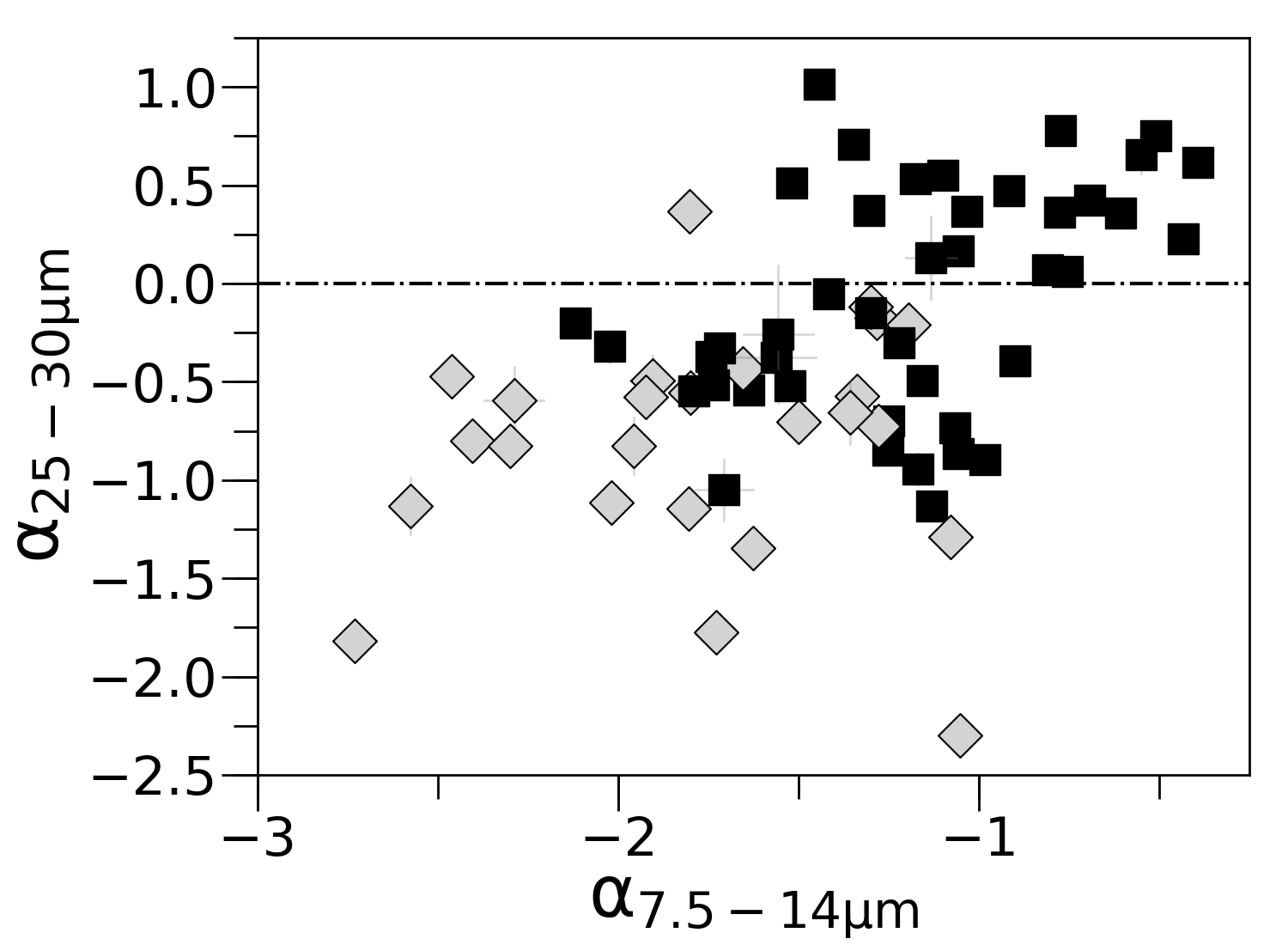}
\includegraphics[width=0.67\columnwidth]{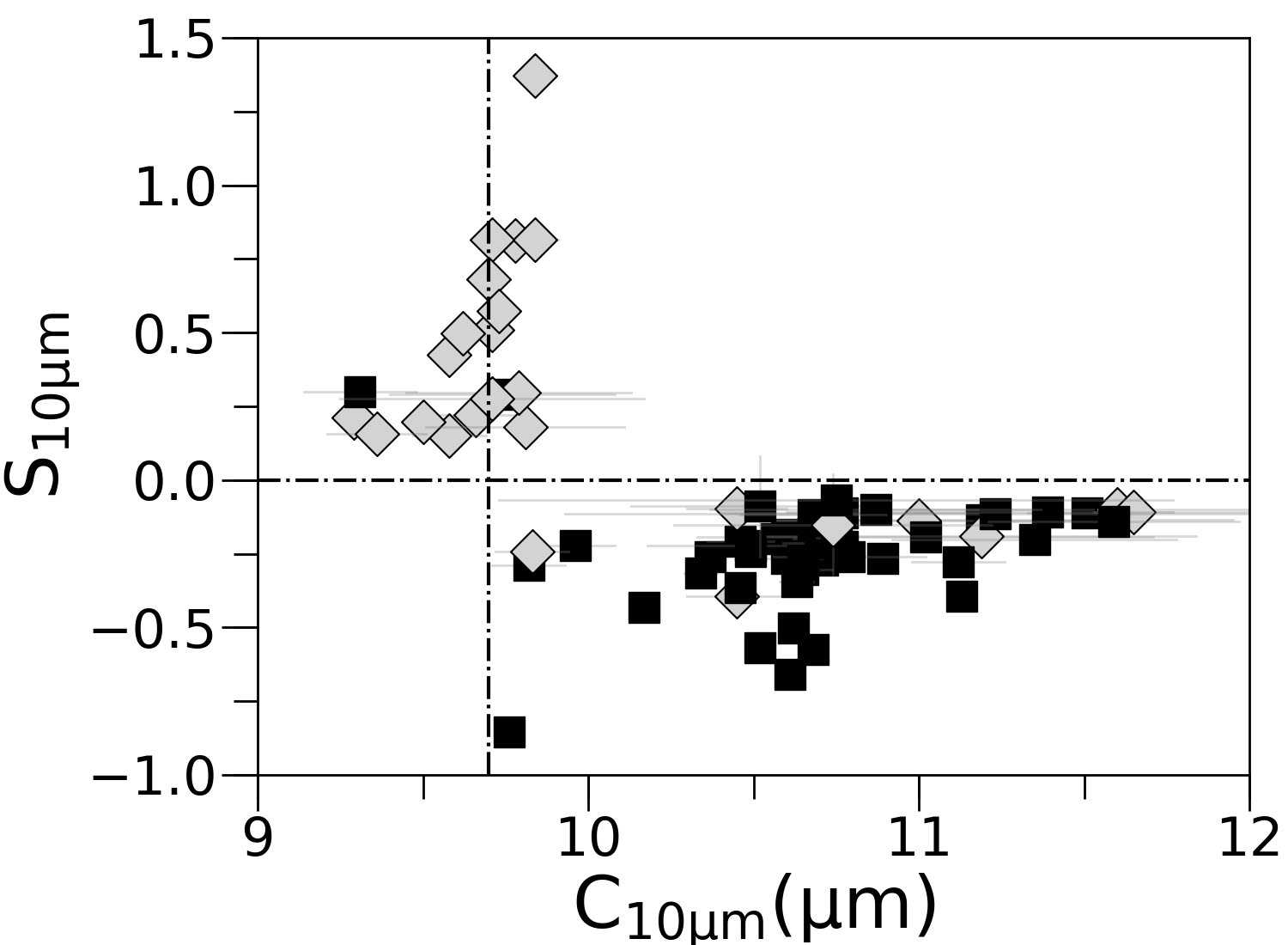}
\includegraphics[width=0.67\columnwidth]{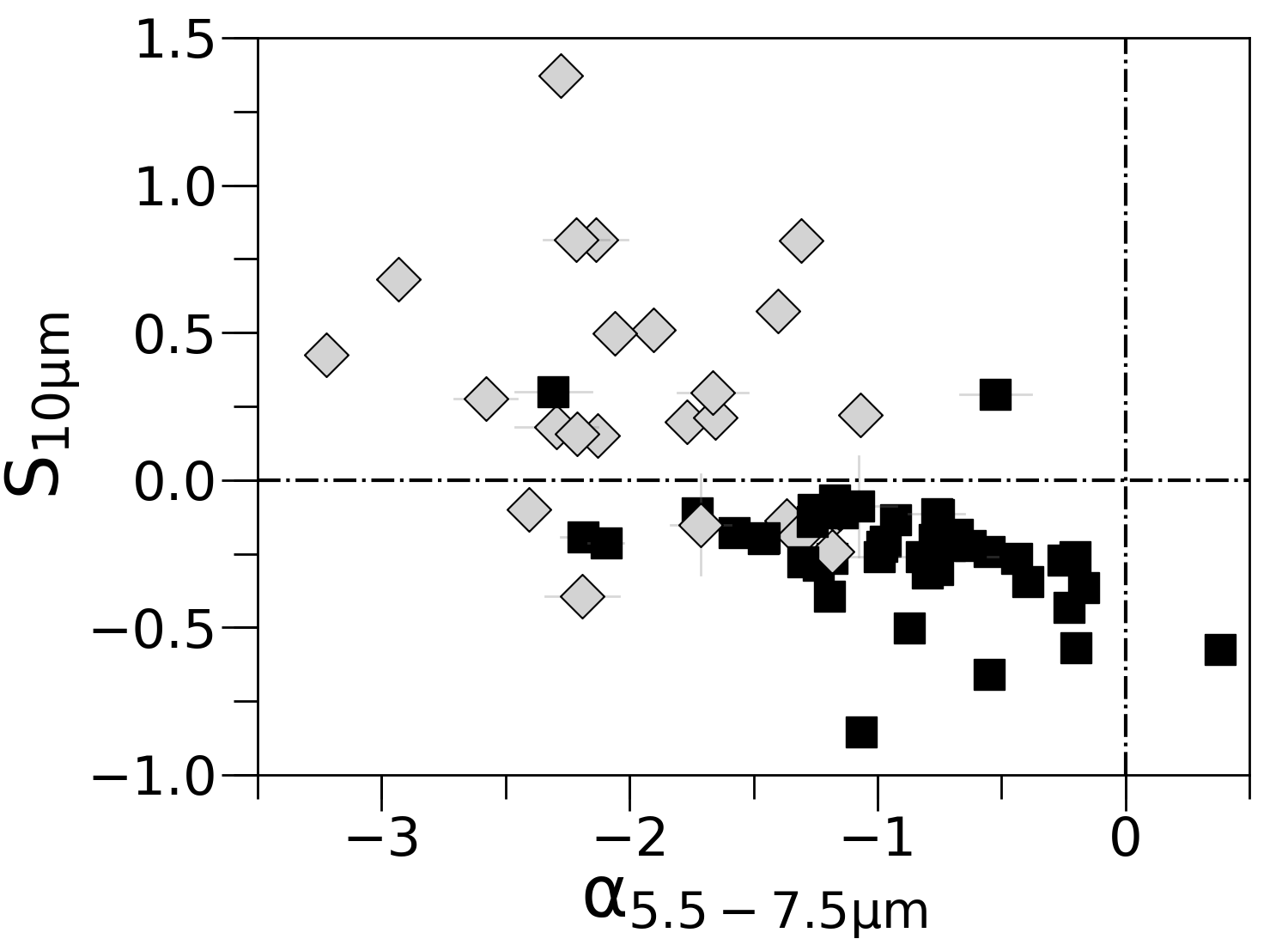}
\caption{Shape of the 9.7$\rm{\mu m}$ silicate feature in optical depth units for our AGN sample (see text, top-left). Black and gray lines show type-1 and type-2 AGN, respectively. Observational diagrams for the AGN sample: 7.5-14$\rm{\mu m}$ versus 5.5-7.5$\rm{\mu m}$ slopes (top-middle); 25-30$\rm{\mu m}$ versus 7.5-14$\rm{\mu m}$ slopes (top-right); $\rm{S_{10\mu m}}$ versus $\rm{C_{10\mu m}}$ (bottom-left), and $\rm{S_{10\mu m}}$ versus 5.5-7.5$\rm{\mu m}$ slope (bottom-right). Black squares and gray diamonds show the measurements in the \emph{Spitzer} spectra for type-1 and type-2 AGN, respectively.}
\label{fig:observations}
\end{center}
\end{figure*}

Following the analysis done by \citet{Kemper04}, we characterize the optical depth at $\rm{10\mu m}$ as:  
\begin{equation}
    \tau_{10\mu m} = -ln(F_{obs}/F_{cont})
\end{equation}
\noindent where $\rm{F_{cont}}$ is the underlying continuum, and it is obtained by fitting the continuum to a 6th order polynomial excluding the silicate feature. For the observed AGN spectra, the two spectral windows are at $\rm{\sim}$7-8.5$\rm{\mu m}$ (bluewards of the silicate feature) and at $\rm{\sim}$12-14.5$\rm{\mu m}$ (redwards of the silicate feature). However, note that this is particularized for each object to optimize the process. For SED models we fitted the 7-32\,$\mu m$ spectral window, excluding both the 10\,$\rm{\mu m}$ (8-13\,$\rm{\mu m}$) and the 18\,$\rm{\mu m}$ (15.5-24\,$\rm{\mu m}$) silicate features.

We define two quantities to quantify the 10$\rm{\mu m}$ silicate emission/absorption features: (1) the silicate feature central peak wavelength (denoted as $\rm{C_{10\mu m}}$); and (2) the silicate feature strength (denoted as $\rm{S_{10\mu m}}$). These two quantities are widely used in the literature to study the shape of the silicate feature \citep[see for instance Fig. 16 in][]{Nenkova08B}. Note that we use this sign convention so absorption (emission) features show positive (negative) values for $\rm{S_{10\mu m}}$. The central peak wavelength and the silicate strength were obtained by fitting the core\footnote{We defined the core of the emission/absorption silicate feature as the wavelength range where the feature is above 90\% of its minimum/maximum.} of the emission/absorption feature to a Gaussian profile to determine the locus of the minimum/maximum. In order to obtain the errors of the fits for the AGN spectra in our sample, we perform 100 Monte Carlo simulations including the error bars of the spectra. 

We construct diagrams combining the three slopes, and the silicate strength, and the central wavelength to compare the spectral shape of the AGN sample with models. Results are shown in Section\,\ref{sec:Results}. 

\begin{figure*}[!t]
\begin{center}
\includegraphics[width=0.49\columnwidth, clip, trim=45 120 20 15]{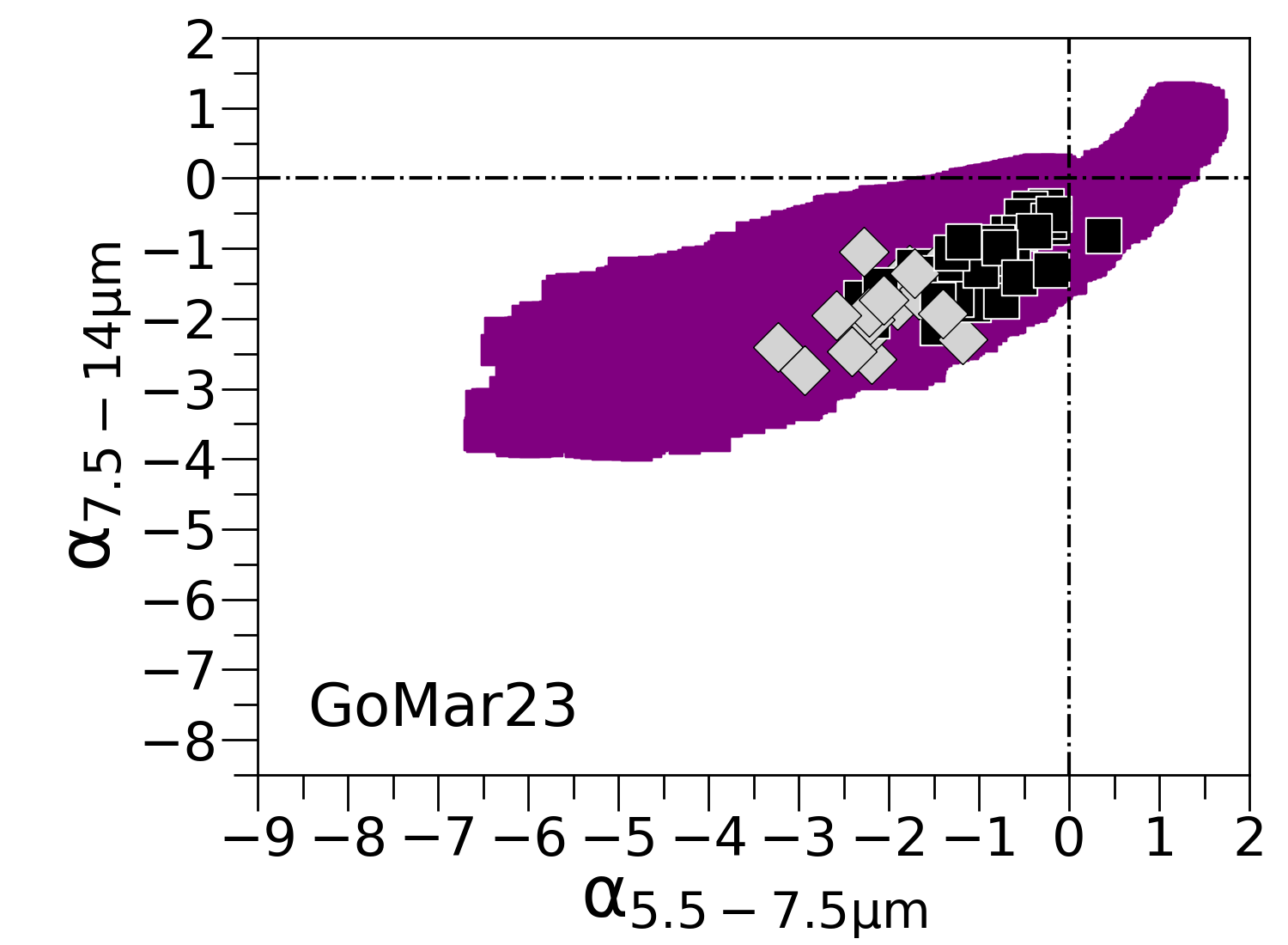}
\includegraphics[width=0.49\columnwidth, clip, trim=45 120 20 15]{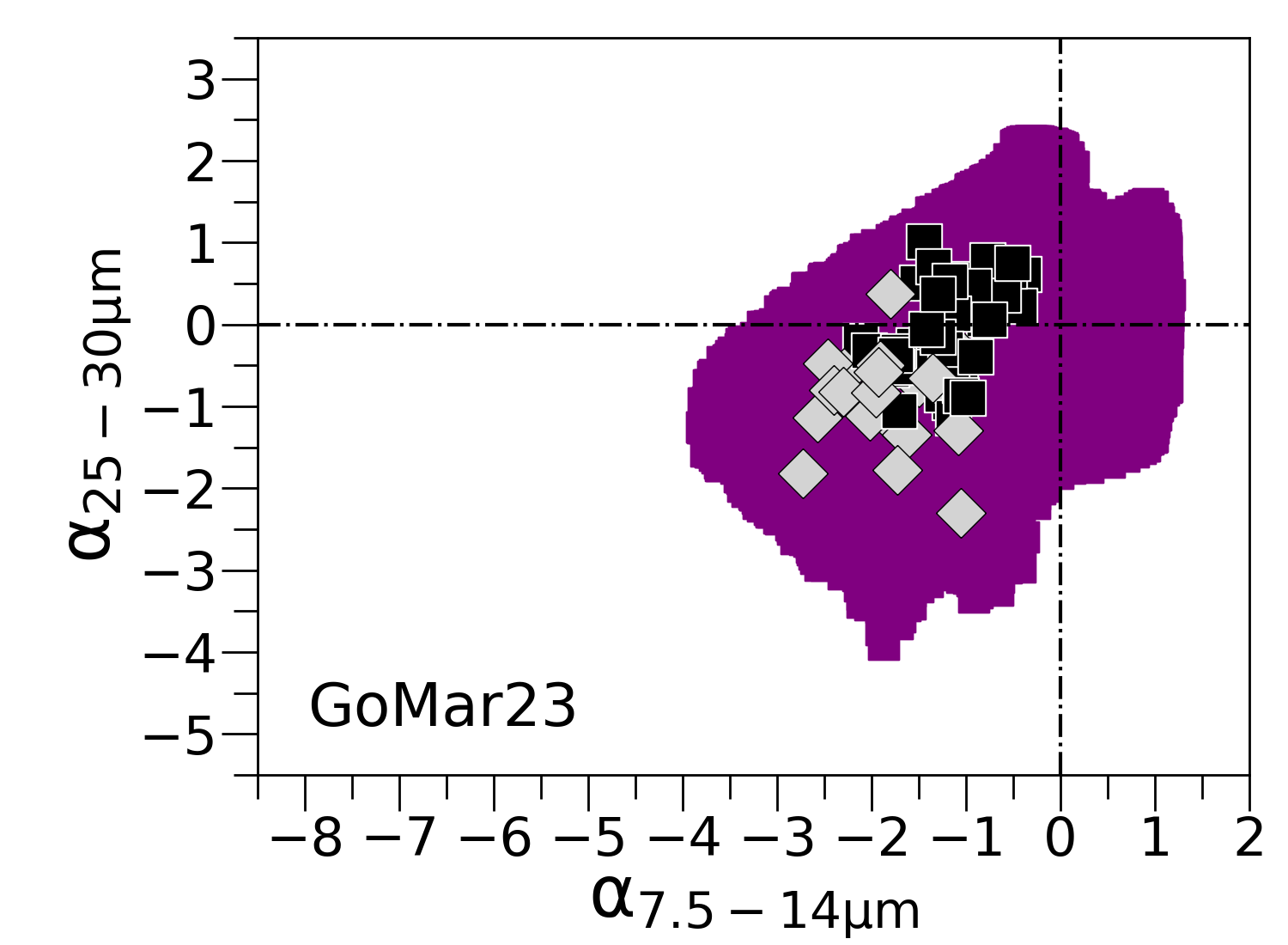}
\includegraphics[width=0.49\columnwidth, clip, trim=45 120 20 15]{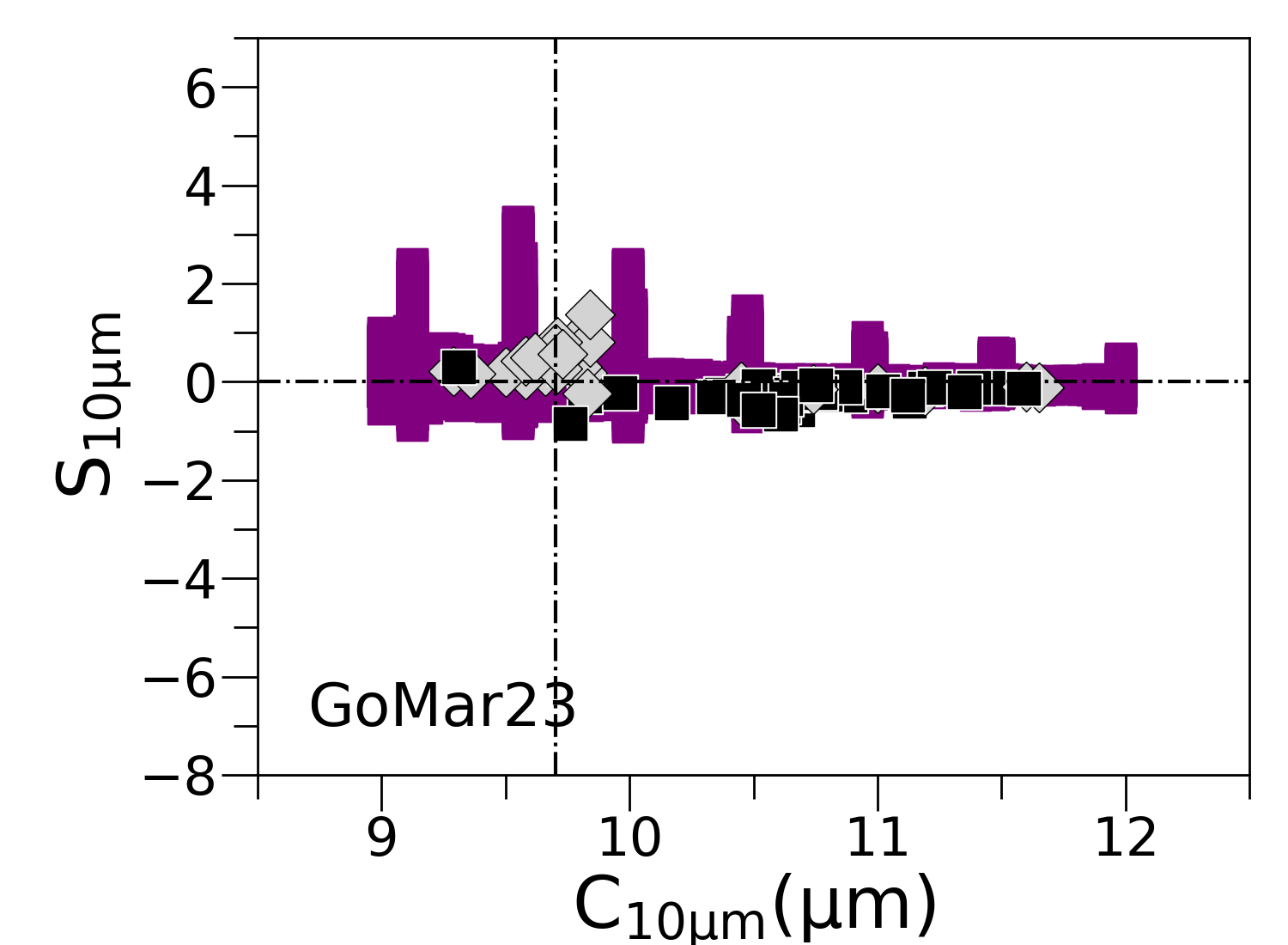}
\includegraphics[width=0.49\columnwidth, clip, trim=45 120 20 15]{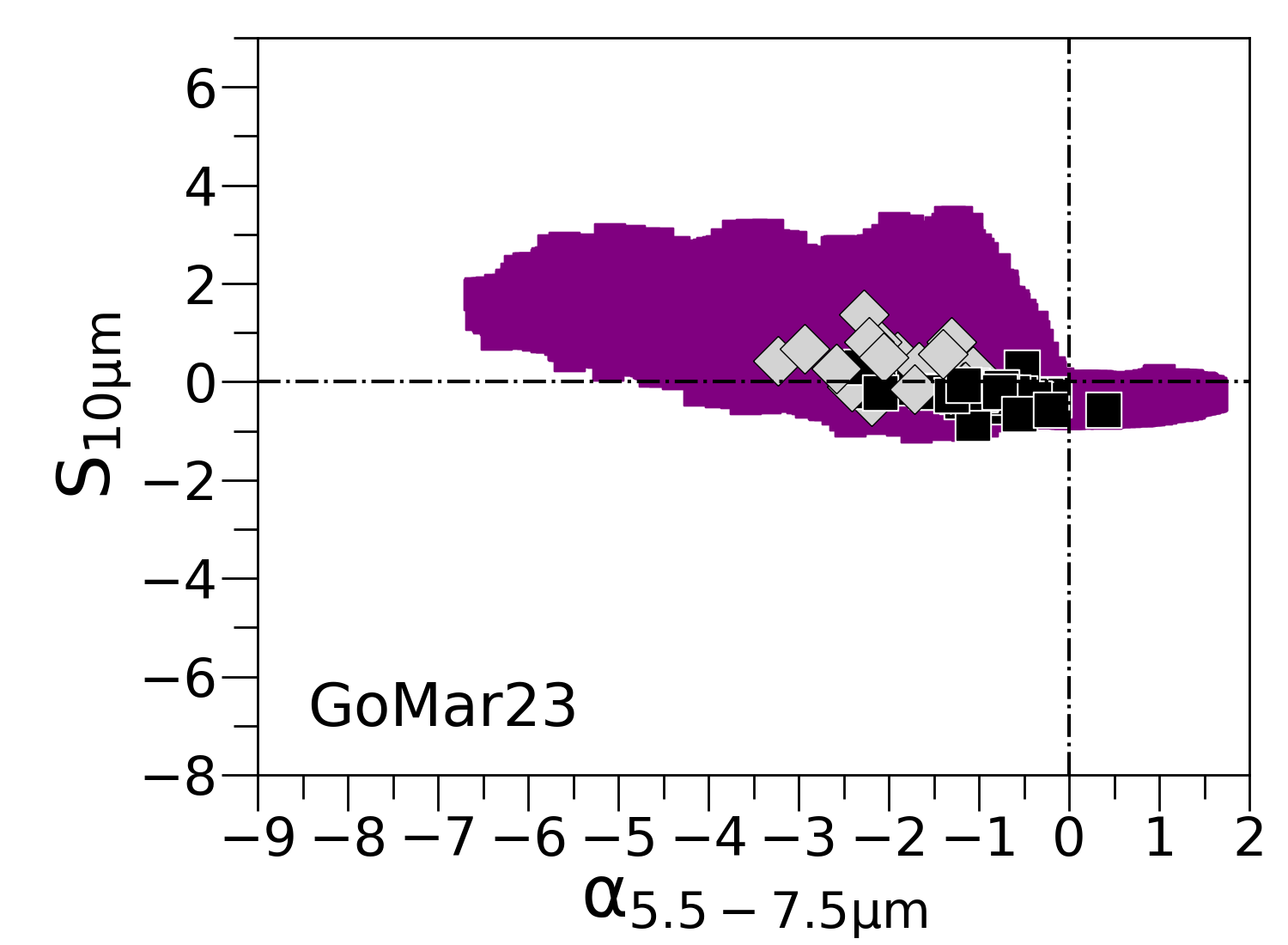}
\includegraphics[width=0.49\columnwidth, clip, trim=45 120 20 15]{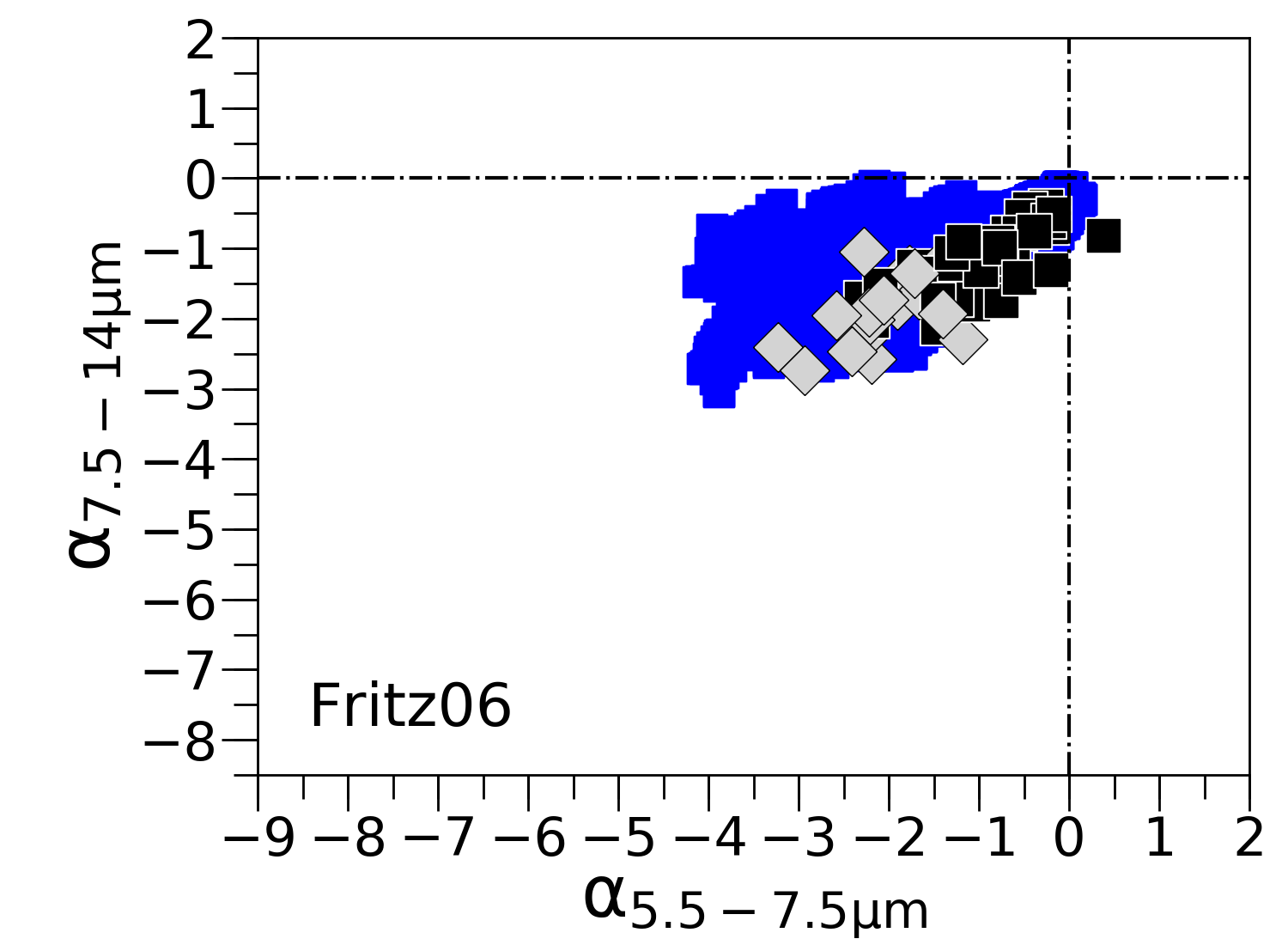}
\includegraphics[width=0.49\columnwidth, clip, trim=45 120 20 15]{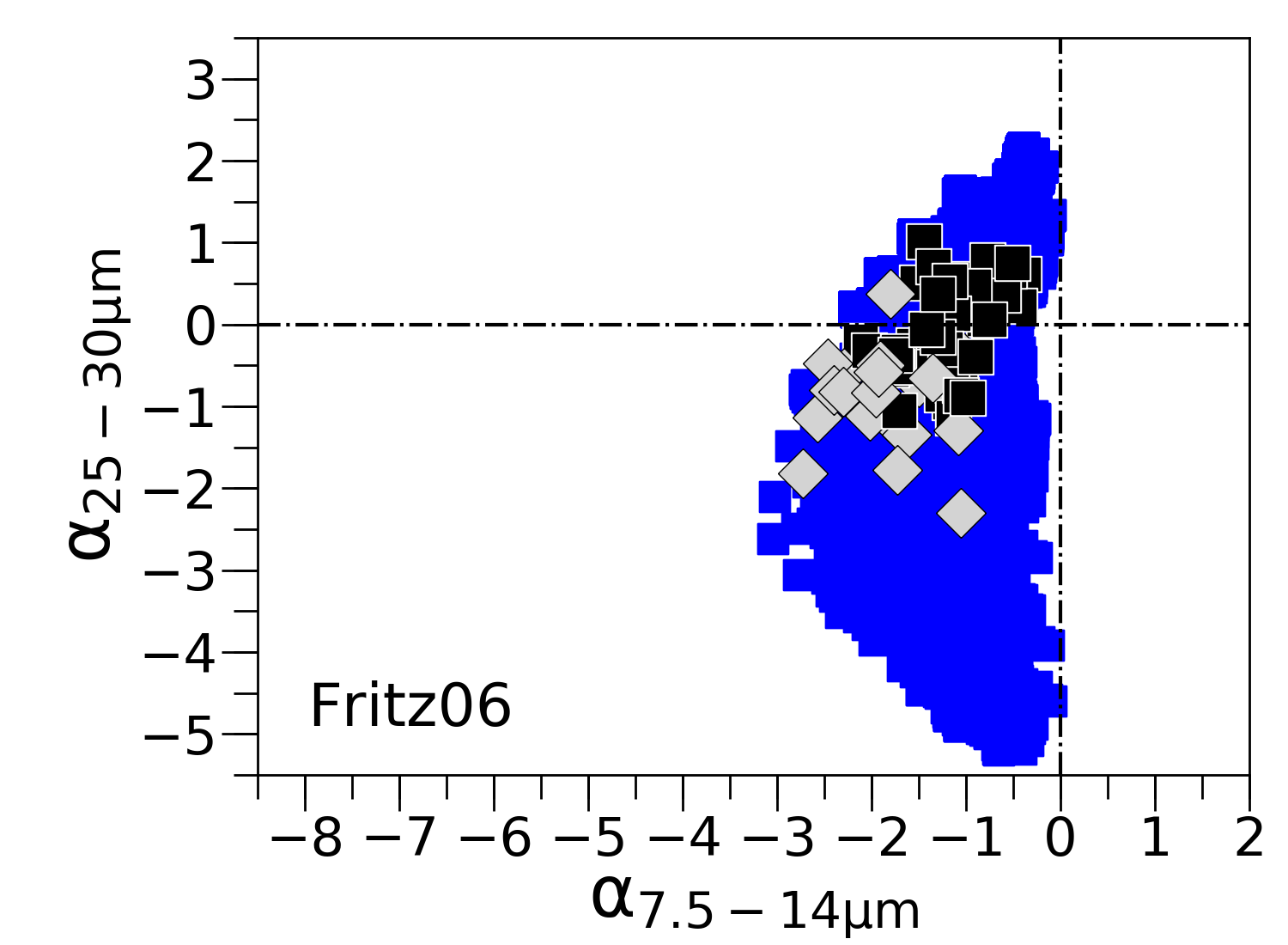}
\includegraphics[width=0.49\columnwidth, clip, trim=45 120 20 15]{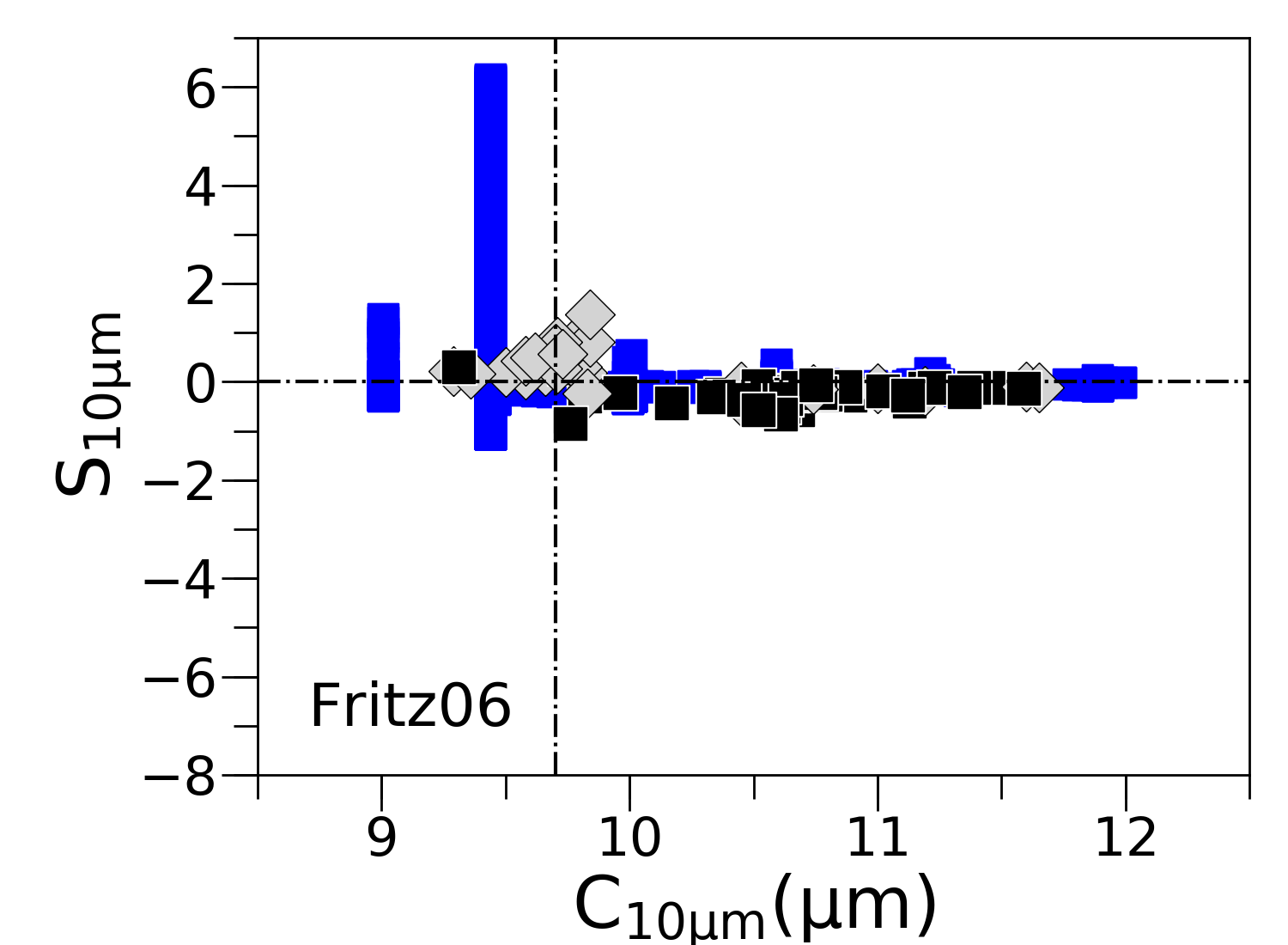}
\includegraphics[width=0.49\columnwidth, clip, trim=45 120 20 15]{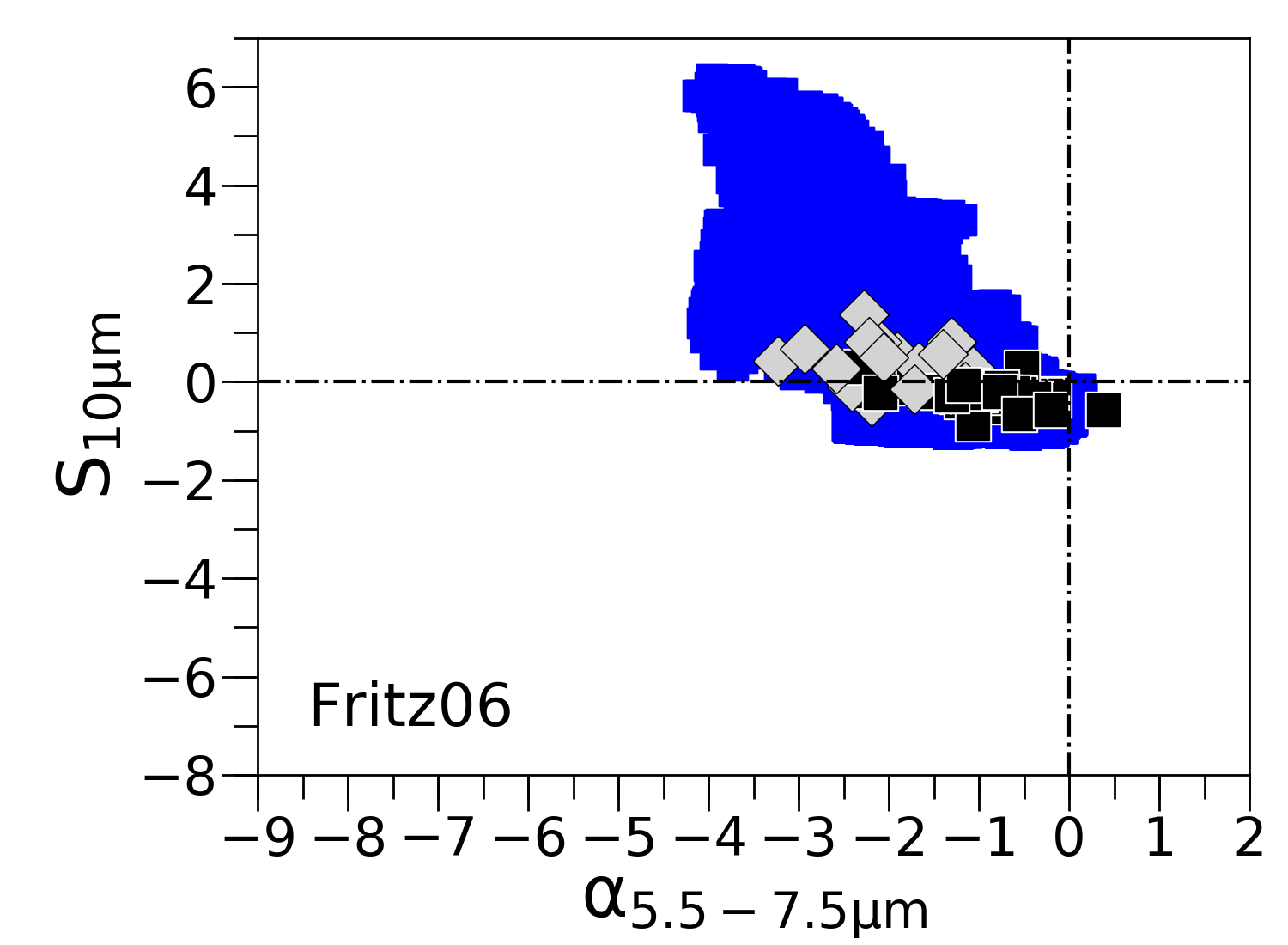}
\includegraphics[width=0.49\columnwidth, clip, trim=45 120 20 15]{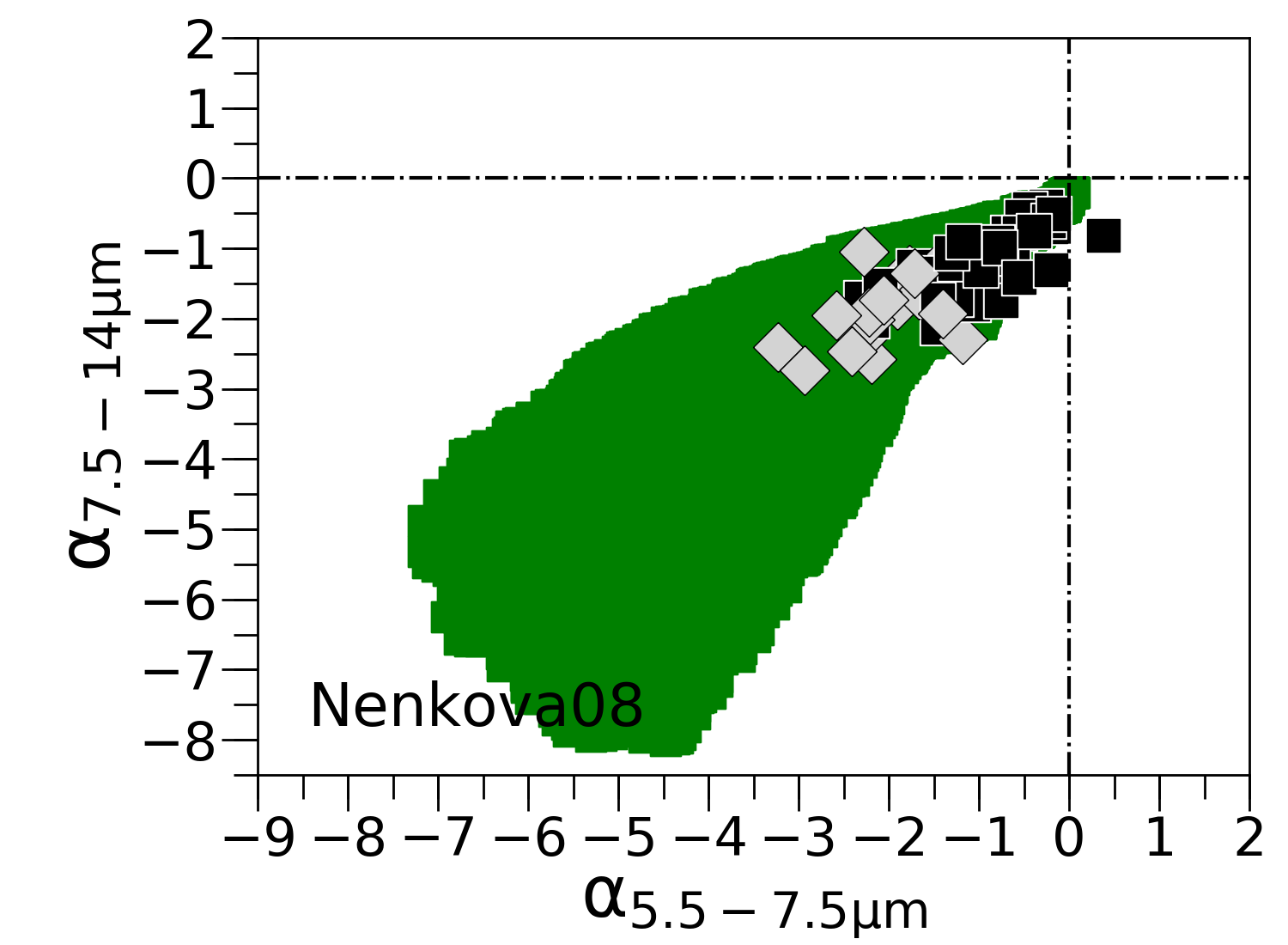}
\includegraphics[width=0.49\columnwidth, clip, trim=45 120 20 15]{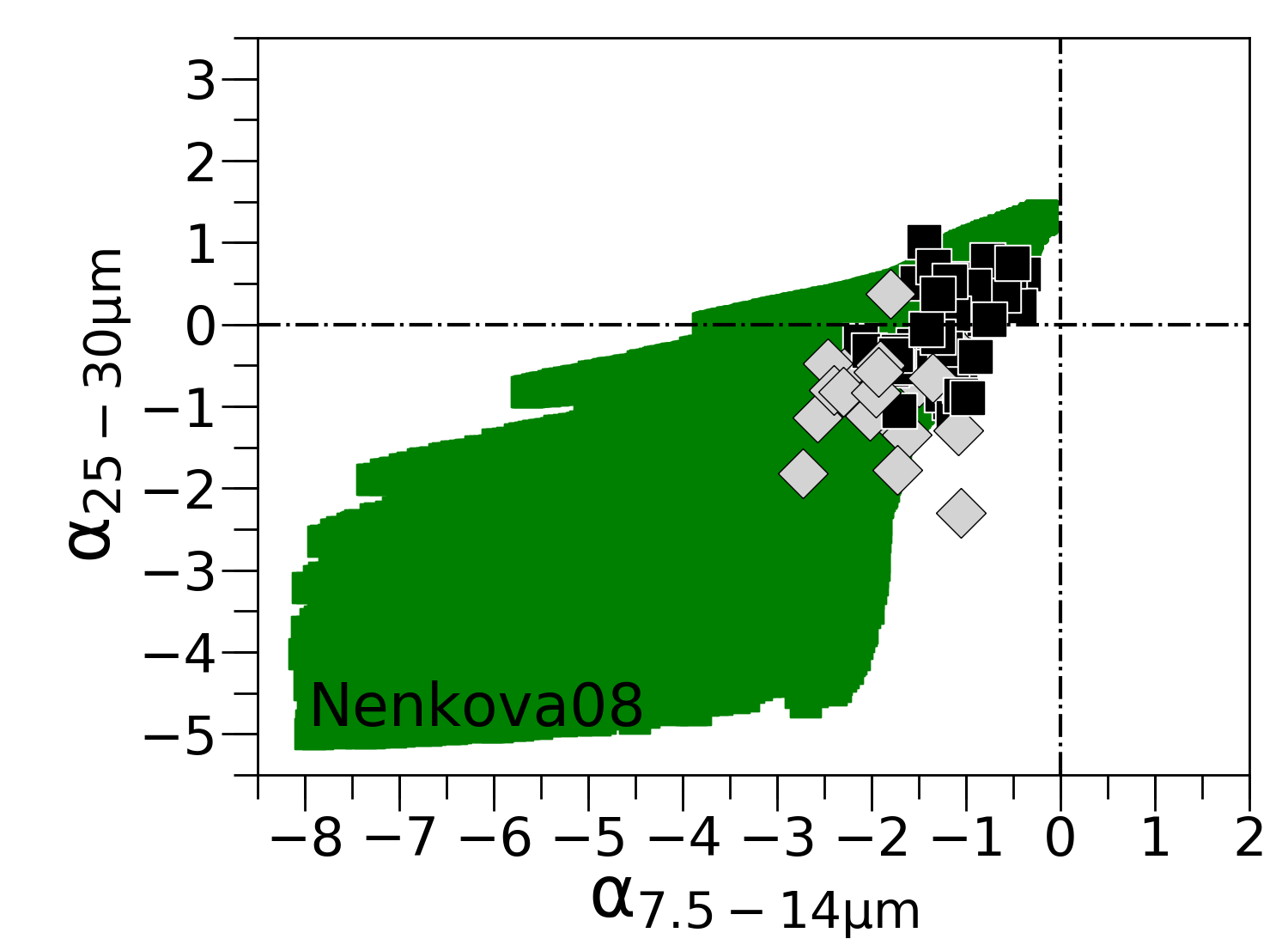}
\includegraphics[width=0.49\columnwidth, clip, trim=45 120 20 15]{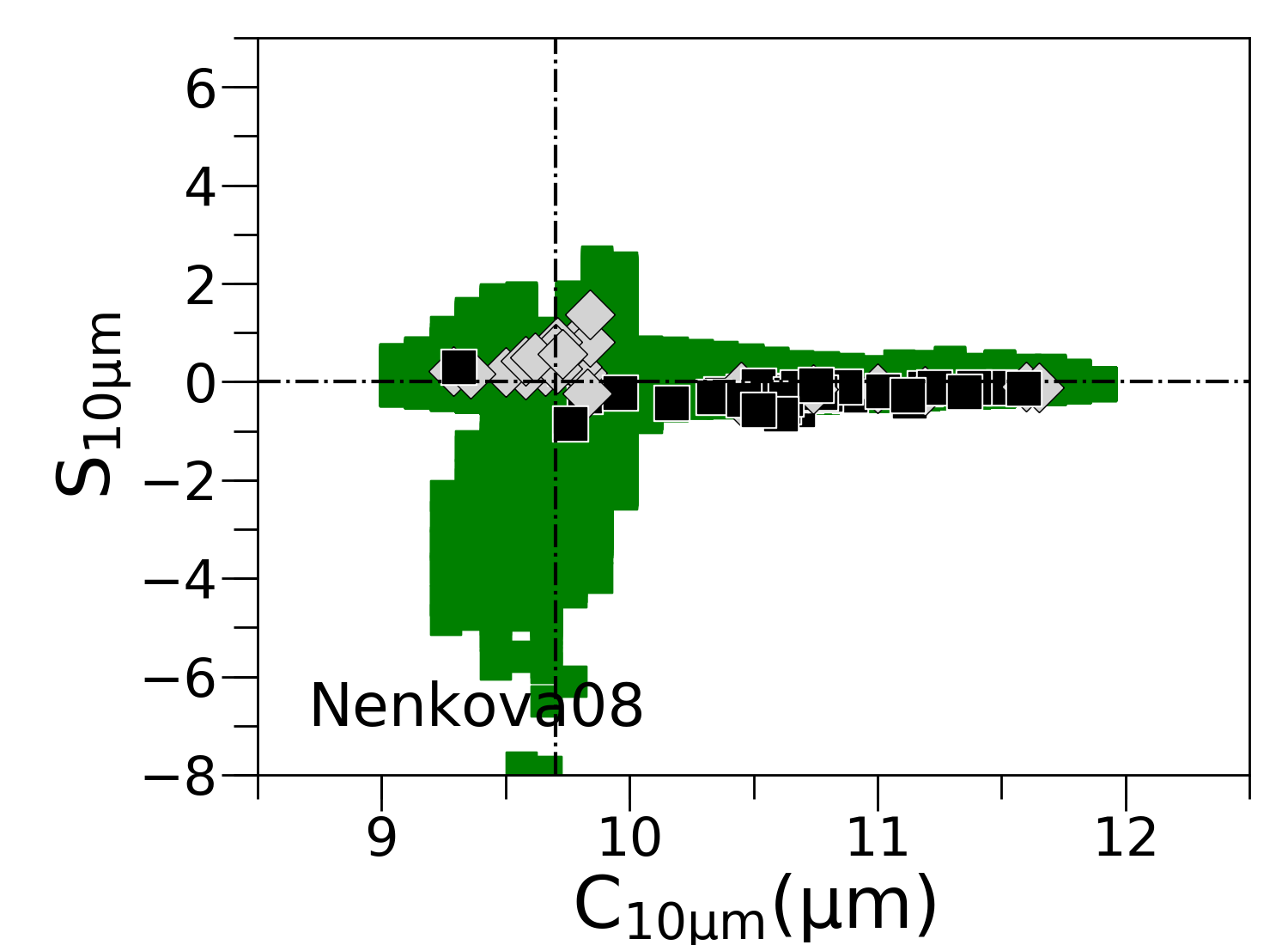}
\includegraphics[width=0.49\columnwidth, clip, trim=45 120 20 15]{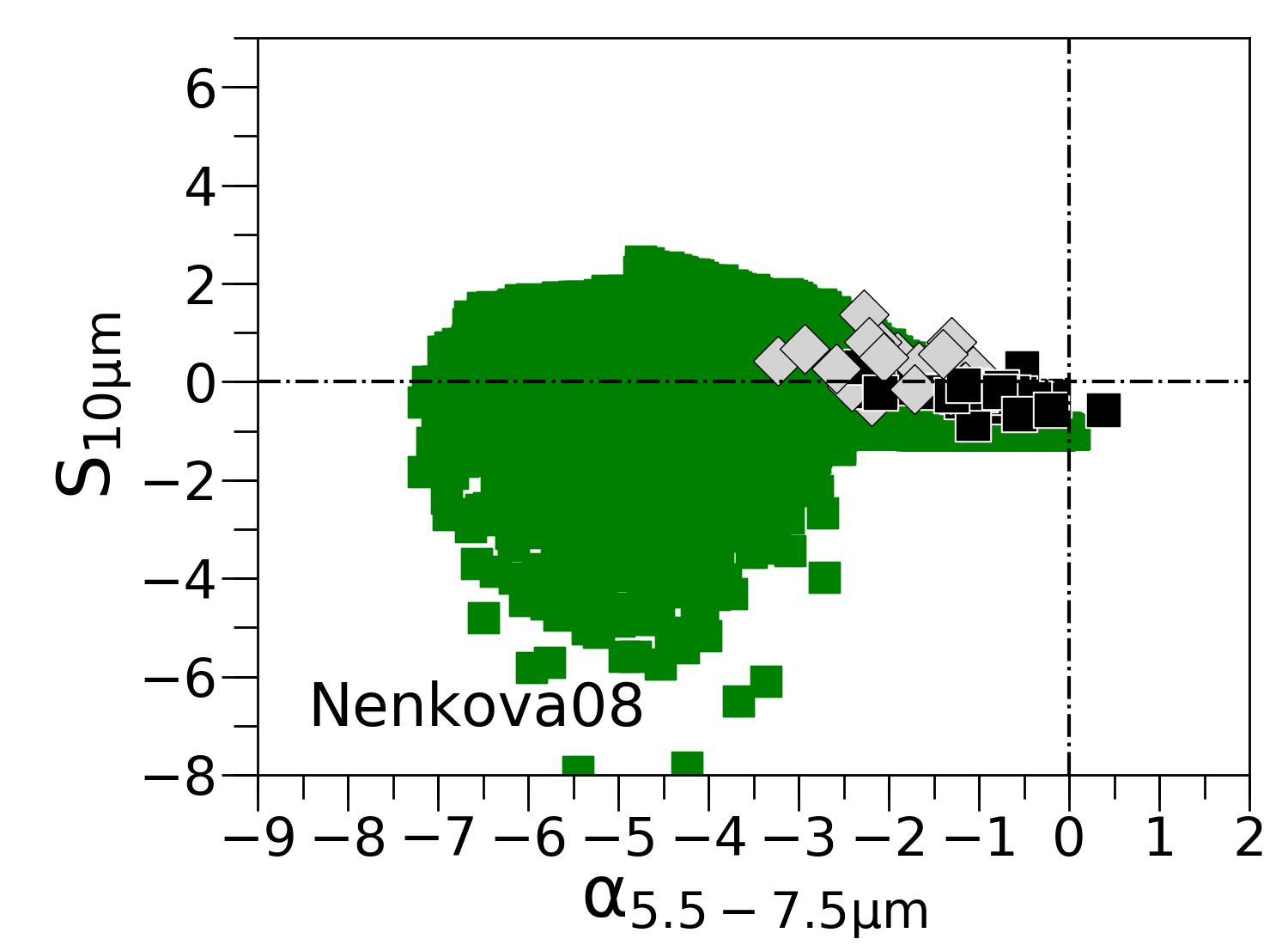}
\includegraphics[width=0.49\columnwidth, clip, trim=45 120 20 15]{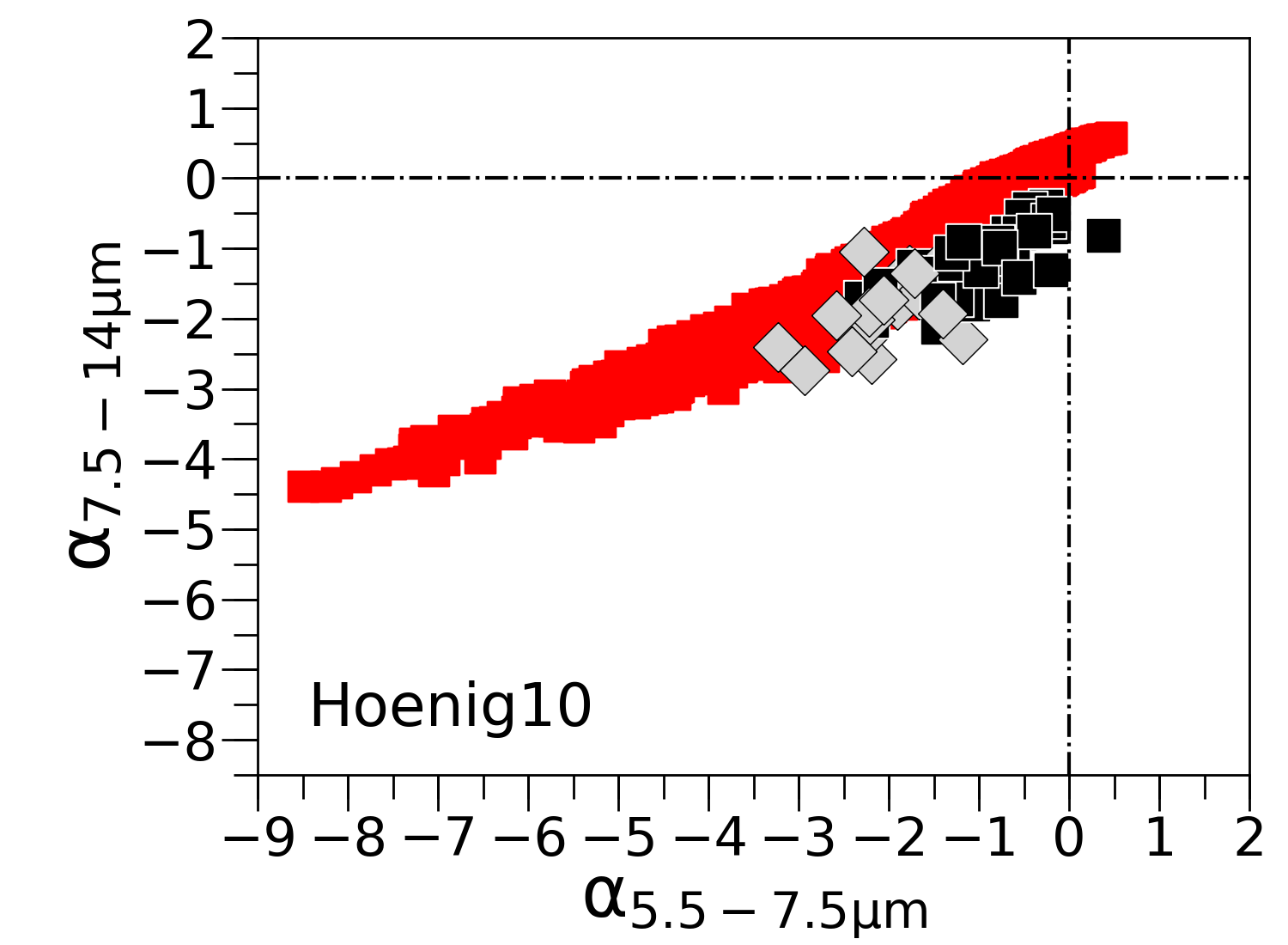}
\includegraphics[width=0.49\columnwidth, clip, trim=45 120 20 15]{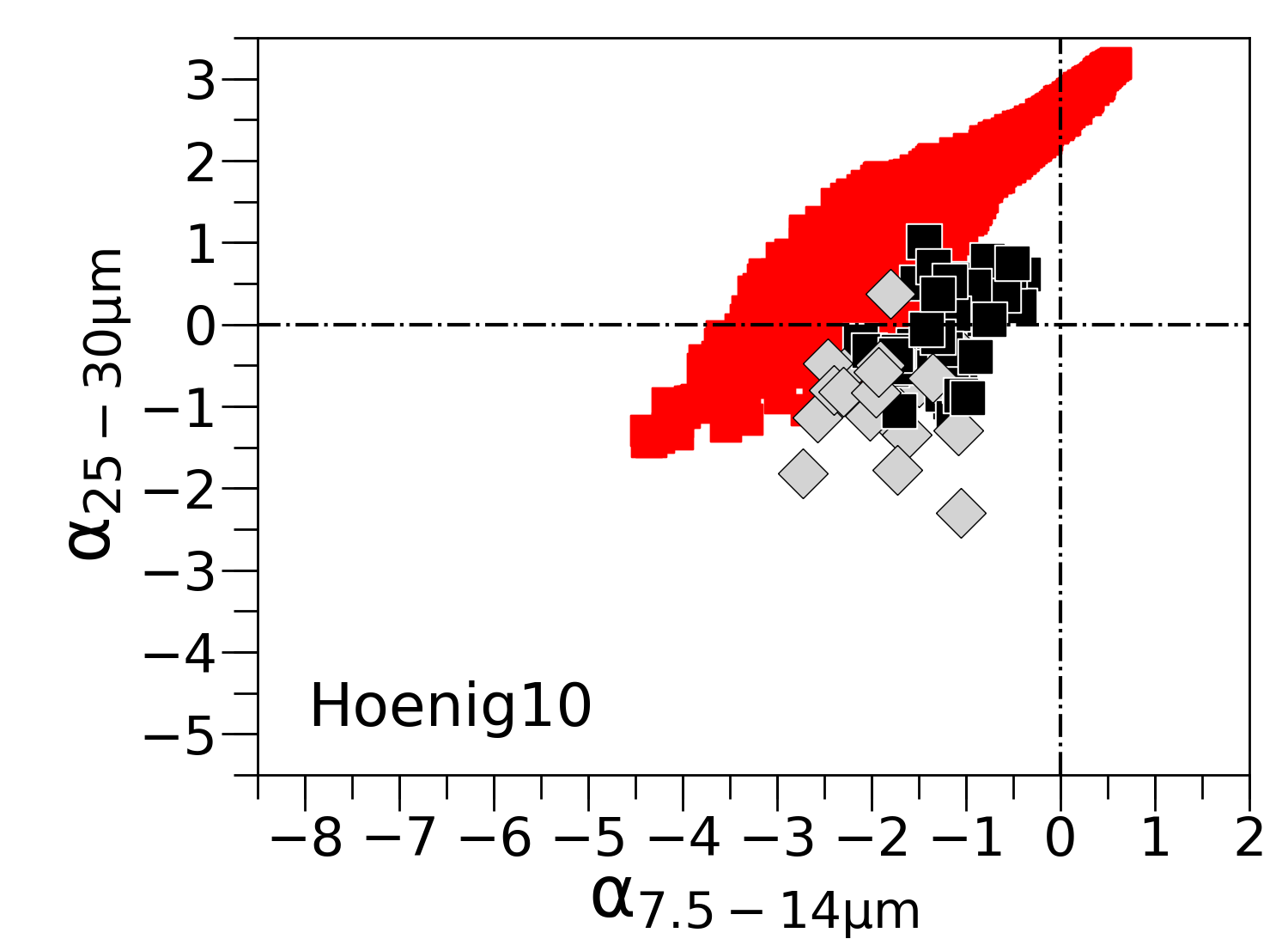}
\includegraphics[width=0.49\columnwidth, clip, trim=45 120 20 15]{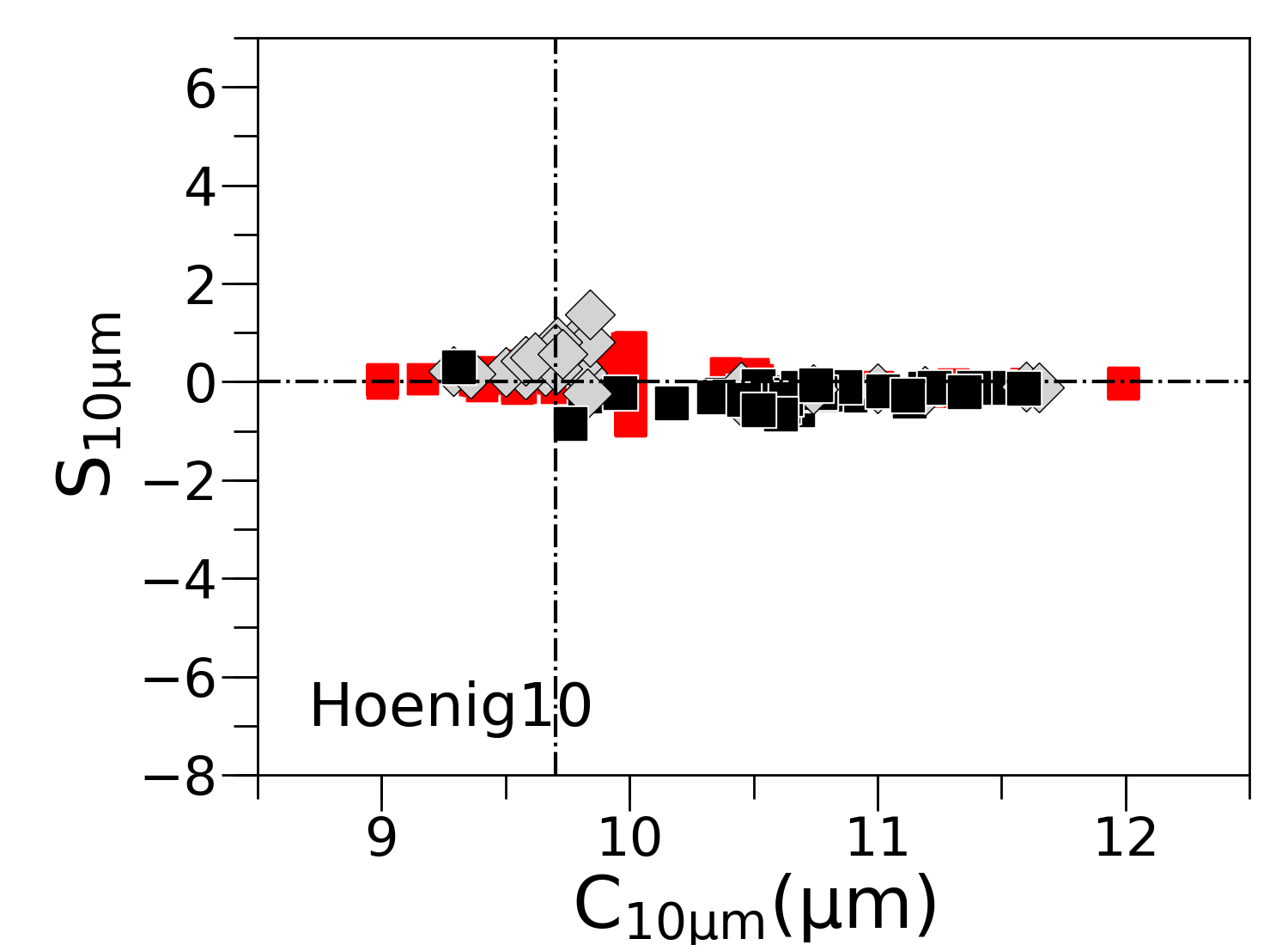}
\includegraphics[width=0.49\columnwidth, clip, trim=45 120 20 15]{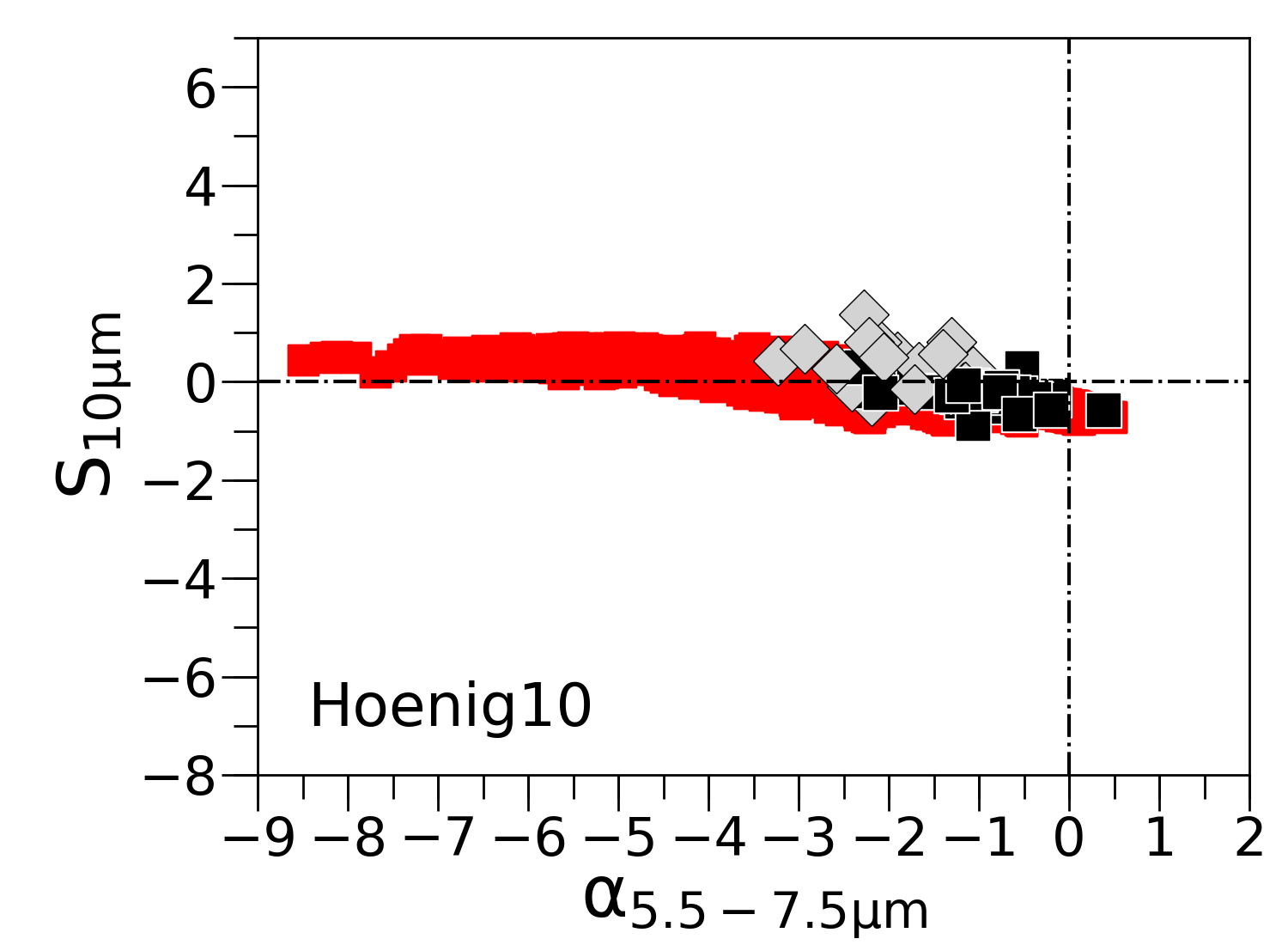}
\includegraphics[width=0.49\columnwidth, clip, trim=45 120 20 15]{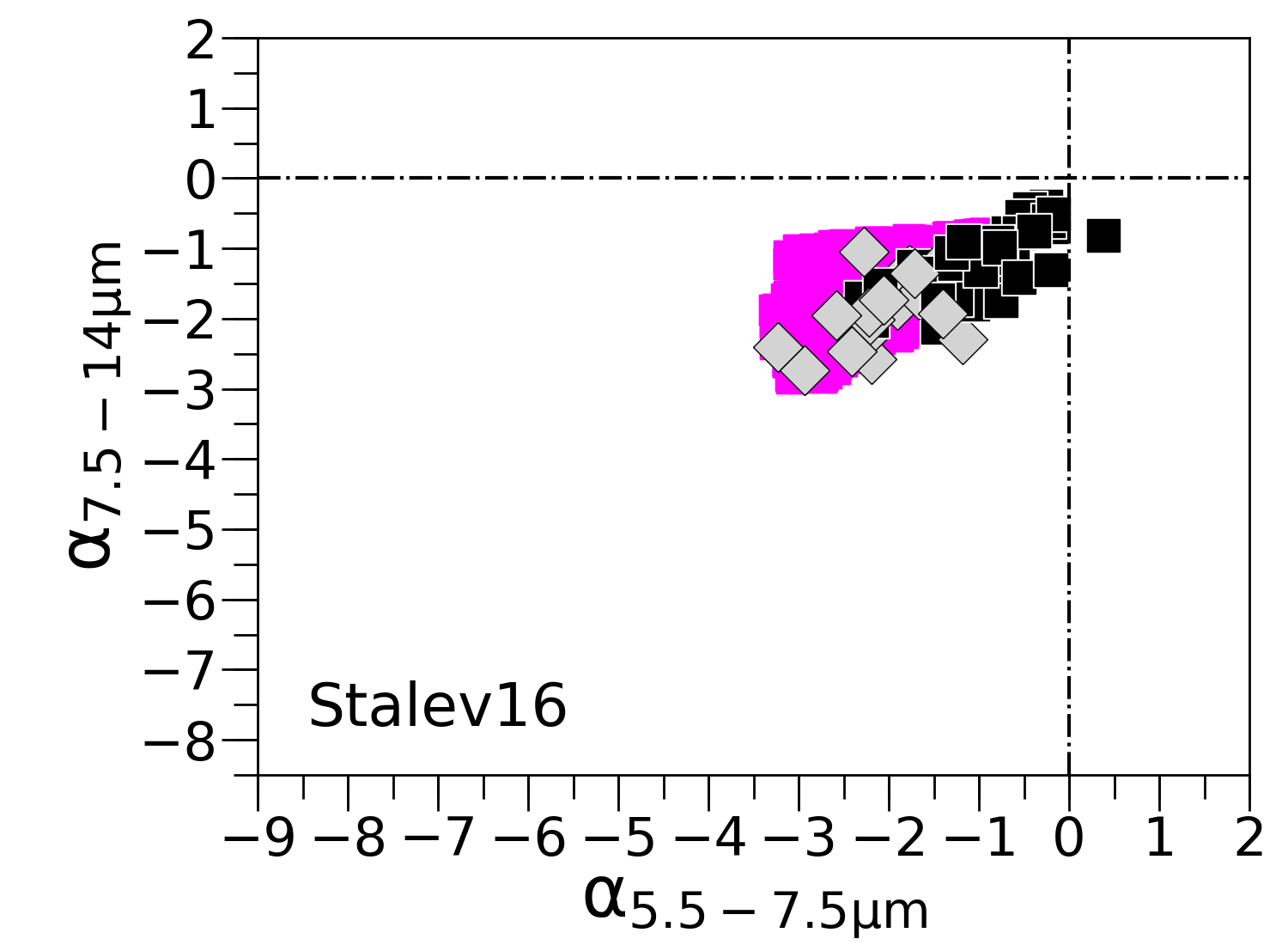}
\includegraphics[width=0.49\columnwidth, clip, trim=45 120 20 15]{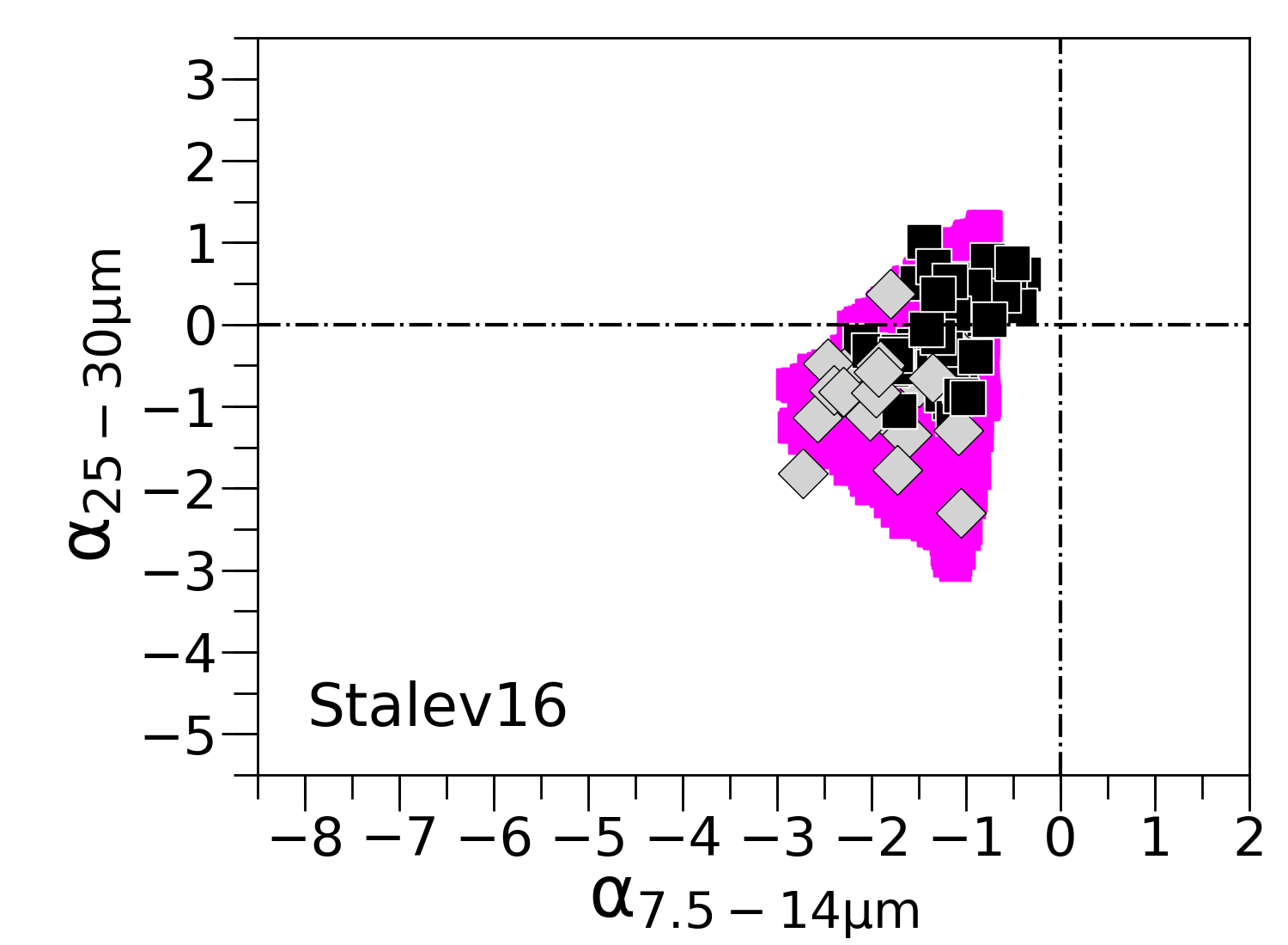}
\includegraphics[width=0.49\columnwidth, clip, trim=45 120 20 15]{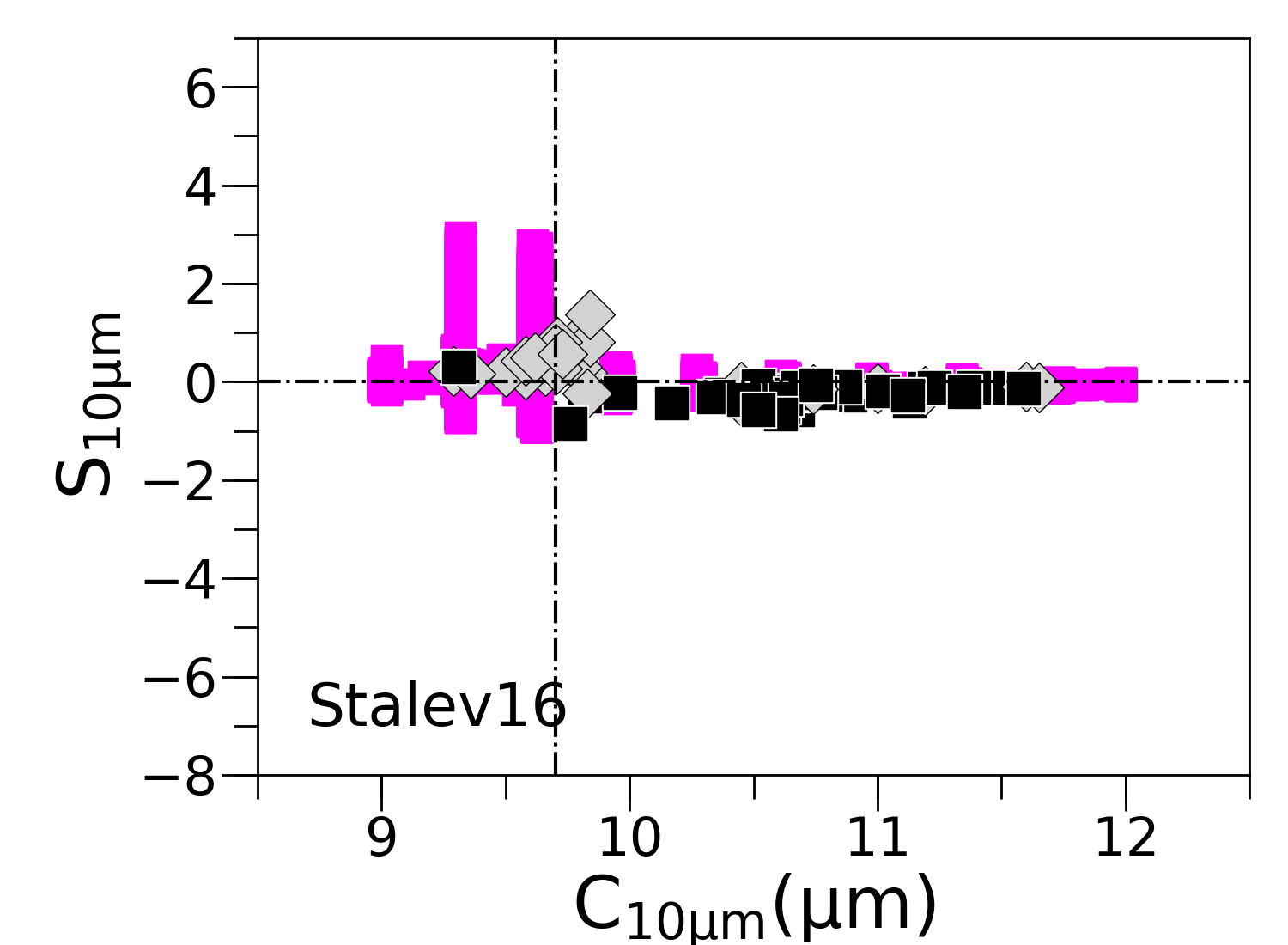}
\includegraphics[width=0.49\columnwidth, clip, trim=45 120 20 15]{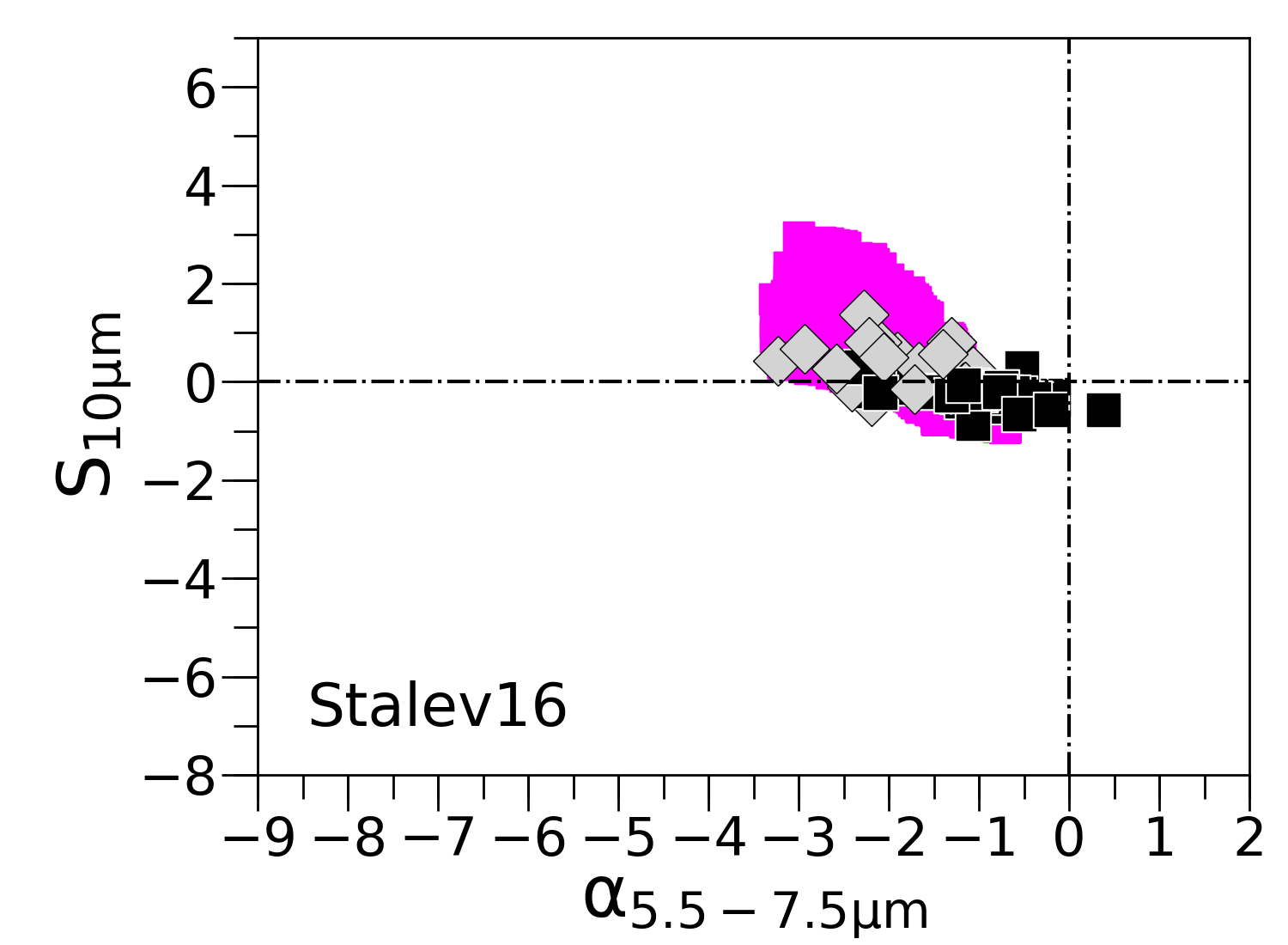}
\includegraphics[width=0.49\columnwidth, clip, trim=45 0 20 15]{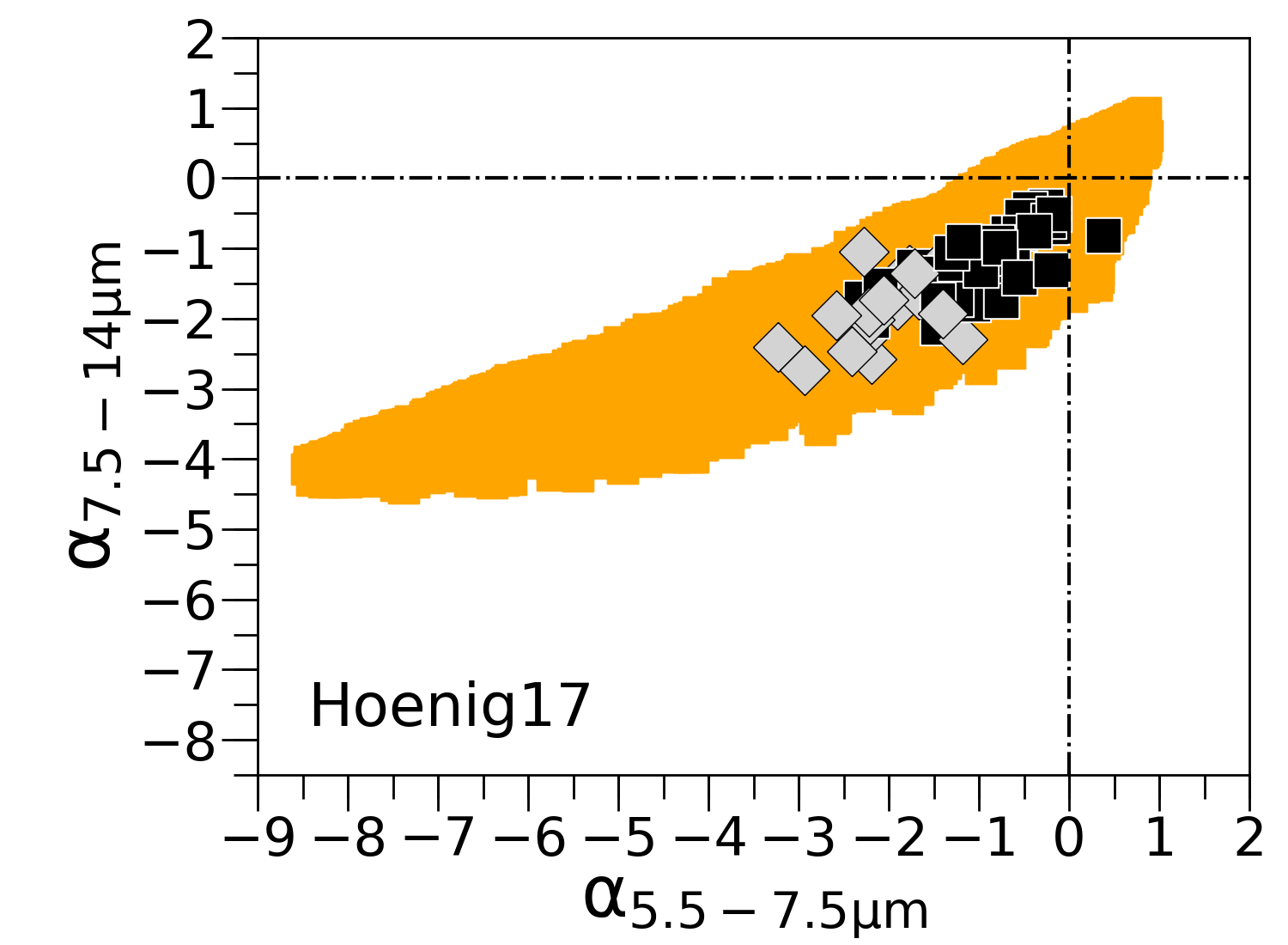}
\includegraphics[width=0.49\columnwidth, clip, trim=45 0 20 15]{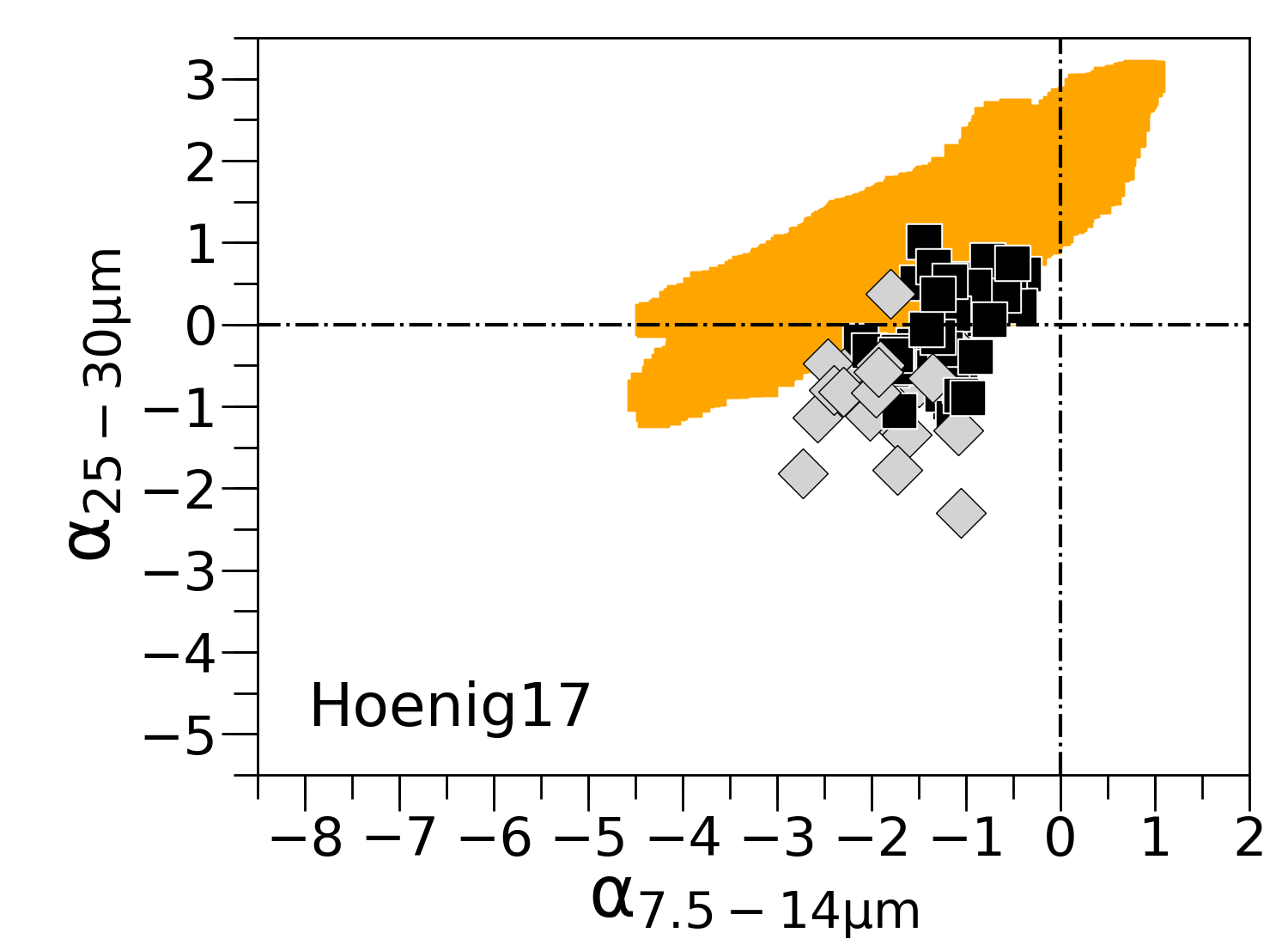}
\includegraphics[width=0.49\columnwidth, clip, trim=45 0 20 15]{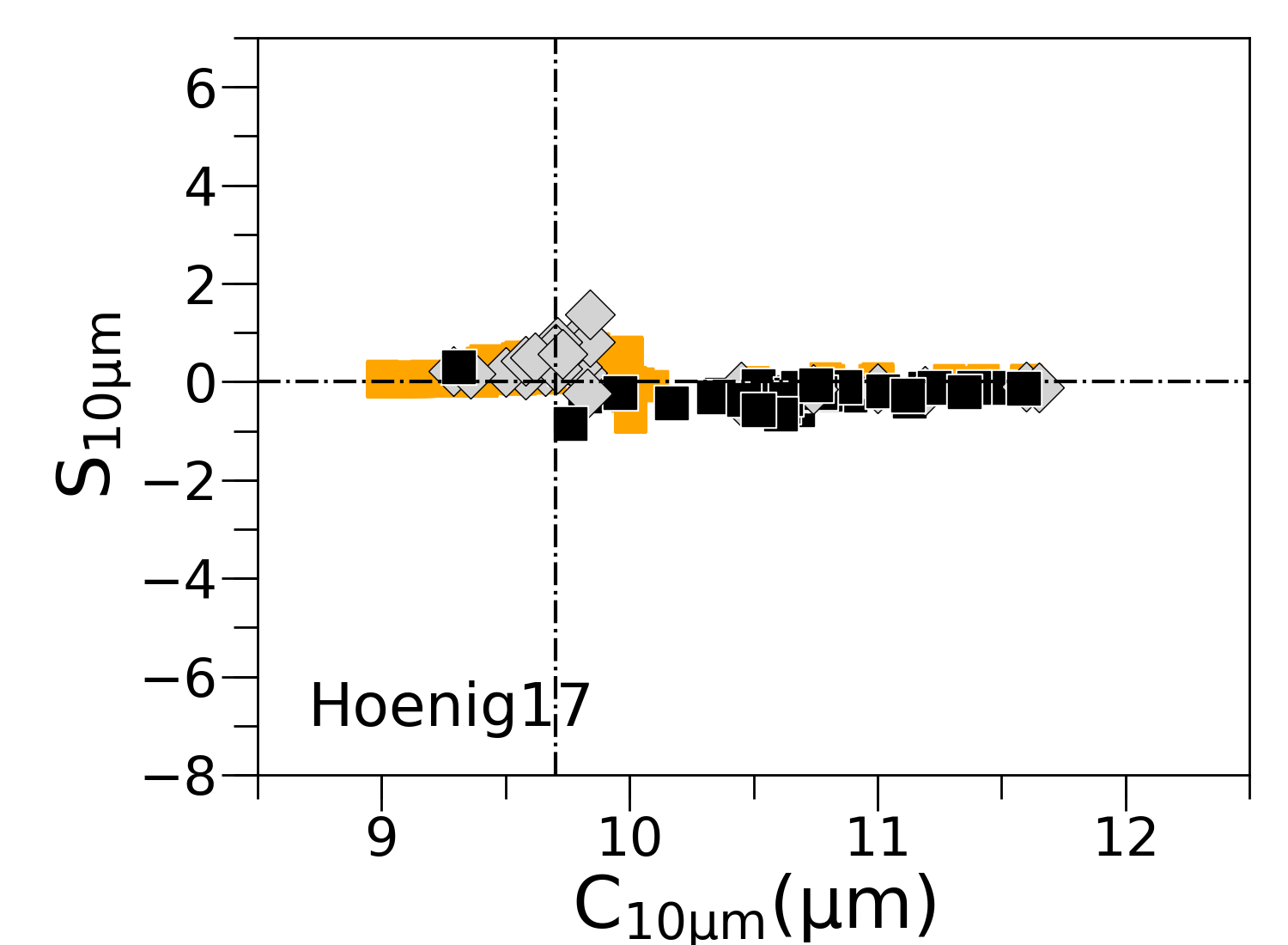}
\includegraphics[width=0.49\columnwidth, clip, trim=45 0 20 15]{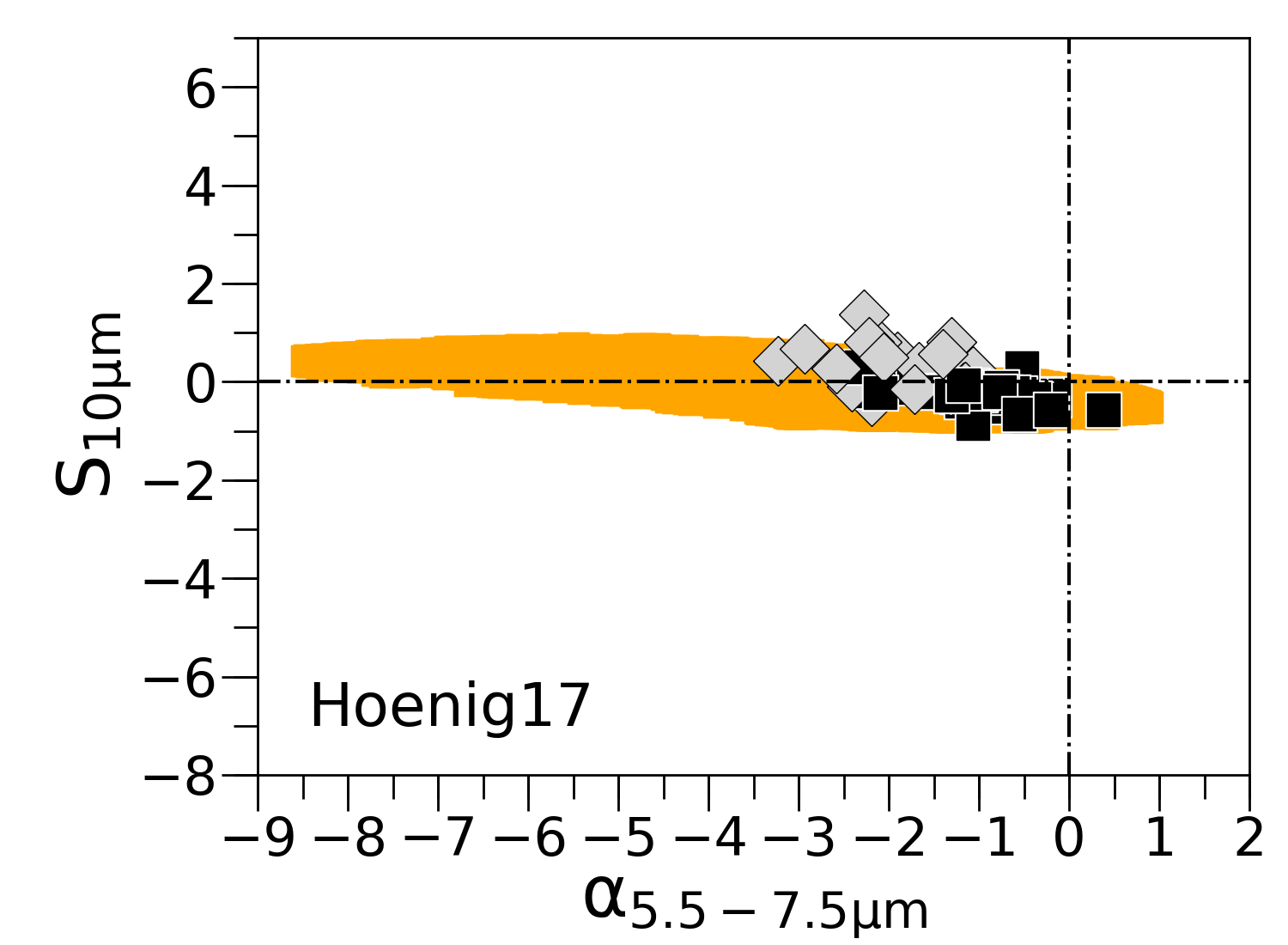}
\caption{Comparison between the spectral shapes of our AGN sample (black squares for type-1 AGN and gray diamonds for type-2 AGN) and the spectral shapes of the SED libraries. From left to right we show mid-infrared versus near-infrared slopes, far-infrared versus mid-infrared slopes, the 10\,$\rm{\mu m}$ feature silicate strength versus its center, and the 10\,$\rm{\mu m}$ feature silicate strength versus the near-infrared slope. From top to bottom we show the results for [GoMar23] (purple), [Fritz06] (blue), [Nenkova08] (green), [Hoenig10] (red), [Stalev16] (magenta), and [Hoenig17] (orange) models. Note that the tabulated values of the center of the silicate emission feature $\rm{C_{10\mu m}}$ in some of the models (e.g. [Fritz06] or [GoMar23] models) are an artifact of the spectral resolution used for to generate each SED library.}
\label{fig:modelshapes}
\end{center}
\end{figure*}

\subsection{Spectral fitting method}

We use the software XSPEC \citep{Arnaud96} to perform the spectral fitting of the \emph{Spitzer}/IRS spectra to the models. This is a command-driven, interactive, spectral-fitting program within the HEASOFT\footnote{https://heasarc.gsfc.nasa.gov} package. XSPEC provides a wide range of tools to perform spectral fittings to the data, being able to work in parallel processes in order to speed them up. Thus, it is ideal for our spectral fitting requirements. 

We converted the \emph{Spitzer}/IRS spectra into XSPEC format using the {\sc flx2xsp} task within HEASOFT. These files are easily read by XSPEC to perform statistical tests when fitting to models. In \citet{Gonzalez-Martin19A}, we already reported the conversion of several AGN dust models into XSPEC format. We also convert our new [GoMar23] model in the same way. Very briefly, we first created an additive one-parameter table (in fits format) associated with all the SEDs using the {\sc flx2tab} task within HEASOFT. We then used a Python routine used by \citet{Gonzalez-Martin19A} to change the headers associating each SED with a set of parameters. This model has seven free parameters, plus redshift (which is usually fixed to the known value) and normalization (linked to the luminosity of the system). Note that XSPEC internally interpolates among the given values of the model. However, this could lead to local rather than global solutions if the parameter space is not well sampled. This is the reason why we decided to improve the coverage of some parameters with respect to [Stalev16] model (e.g. Y which was only evaluated at three values by [Stalev16] model, see Section\,\ref{sec:NewModel}). 

We also added foreground extinction by dust grains to the dusty models using the {\sc zdust} component \citep{Pei92}, already included as a multiplicative component within XSPEC. Among the options of {\sc zdust}, we use the standard Galactic extinction curve. However, using the extinction curves of the Magellanic clouds does not change the main results of this paper. The main impact on the resulting spectrum of this extinction model is the attenuation of the near-infrared emission and the inclusion of silicate absorption features. The free parameter is the color excess E(B-V). Note that we test the importance of foreground extinction by comparing the final fit when including and excluding this component, and the results are presented in the following section. We use two definitions of $\rm{\chi^2}$ to get insights on the best model;  $\rm{\chi^2_{Model}}$ compares the residuals with the model and $\rm{\chi^2_{Error}}$ compares the residuals with the errors associated with the data. Furthermore, we use the f-test and AIC to compare spectral fits. Appendix\,\ref{sec:stats} gives further details on the statistical tools used for this analysis.

\begin{figure*}[!t]
\begin{center}
\includegraphics[width=0.55\columnwidth, clip, trim=50 0 20 15]{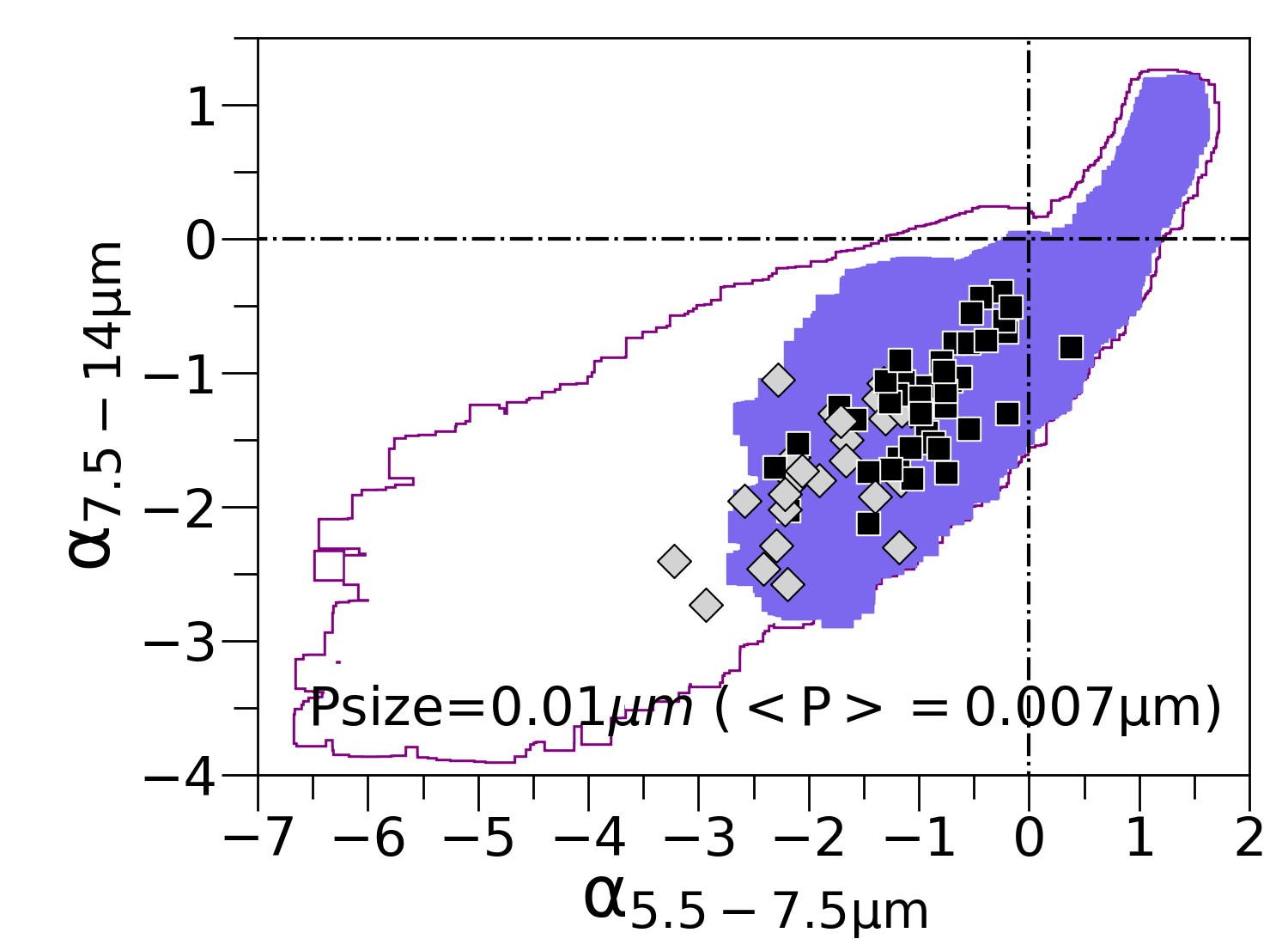}
\includegraphics[width=0.46\columnwidth, clip, trim=200 0 20 15]{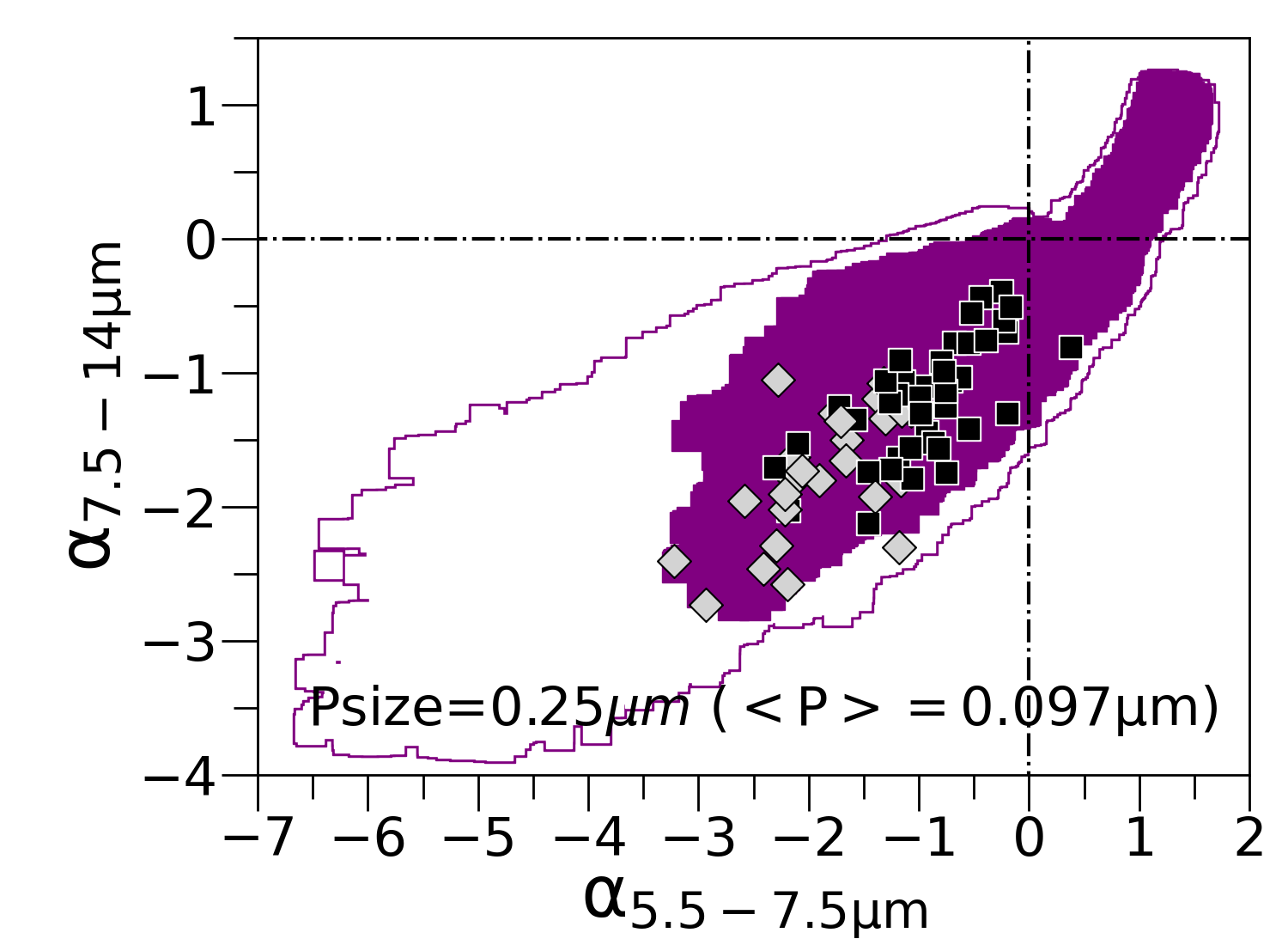}
\includegraphics[width=0.46\columnwidth, clip, trim=200 0 20 15]{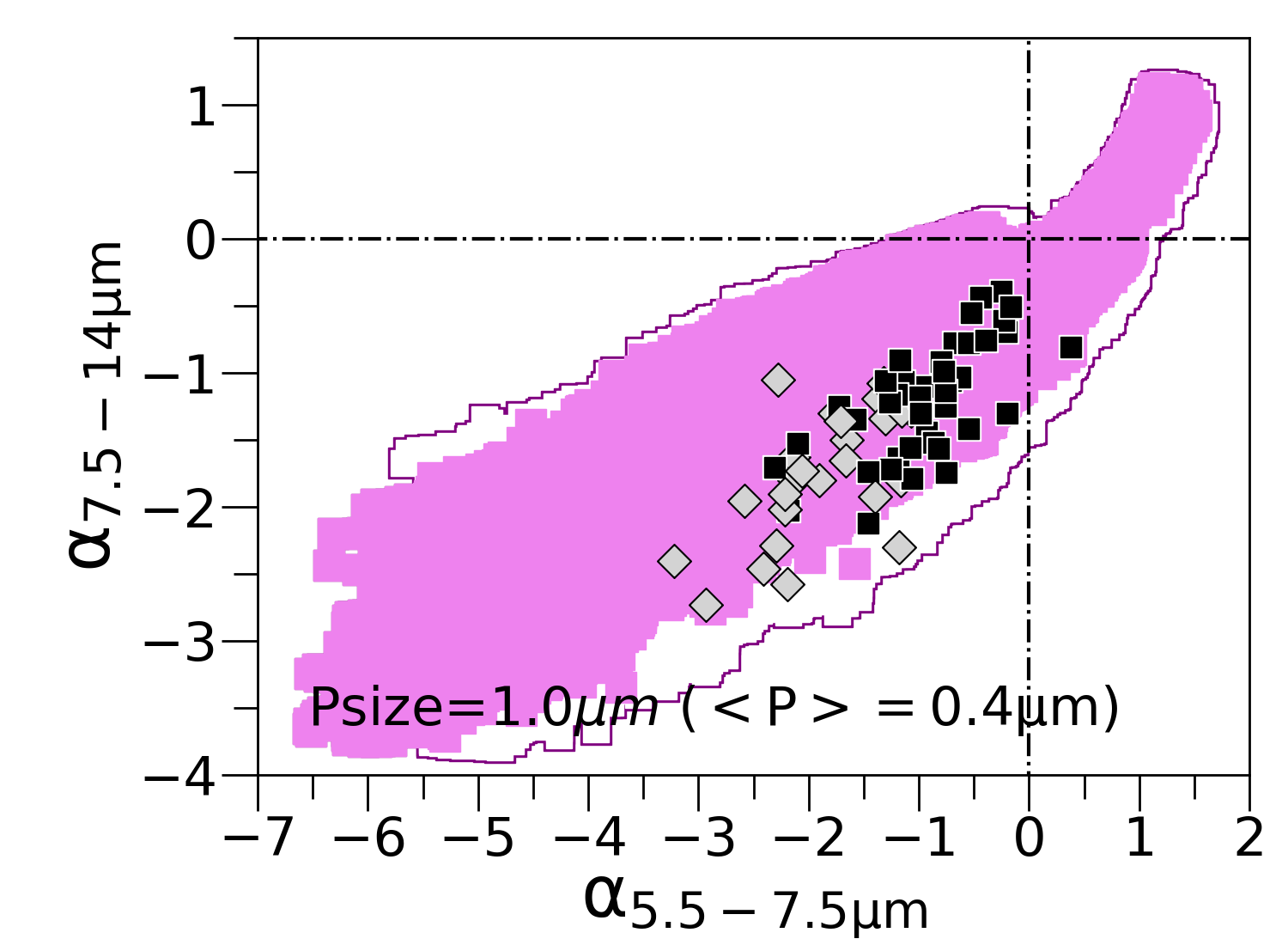}
\includegraphics[width=0.46\columnwidth, clip, trim=200 0 20 15]{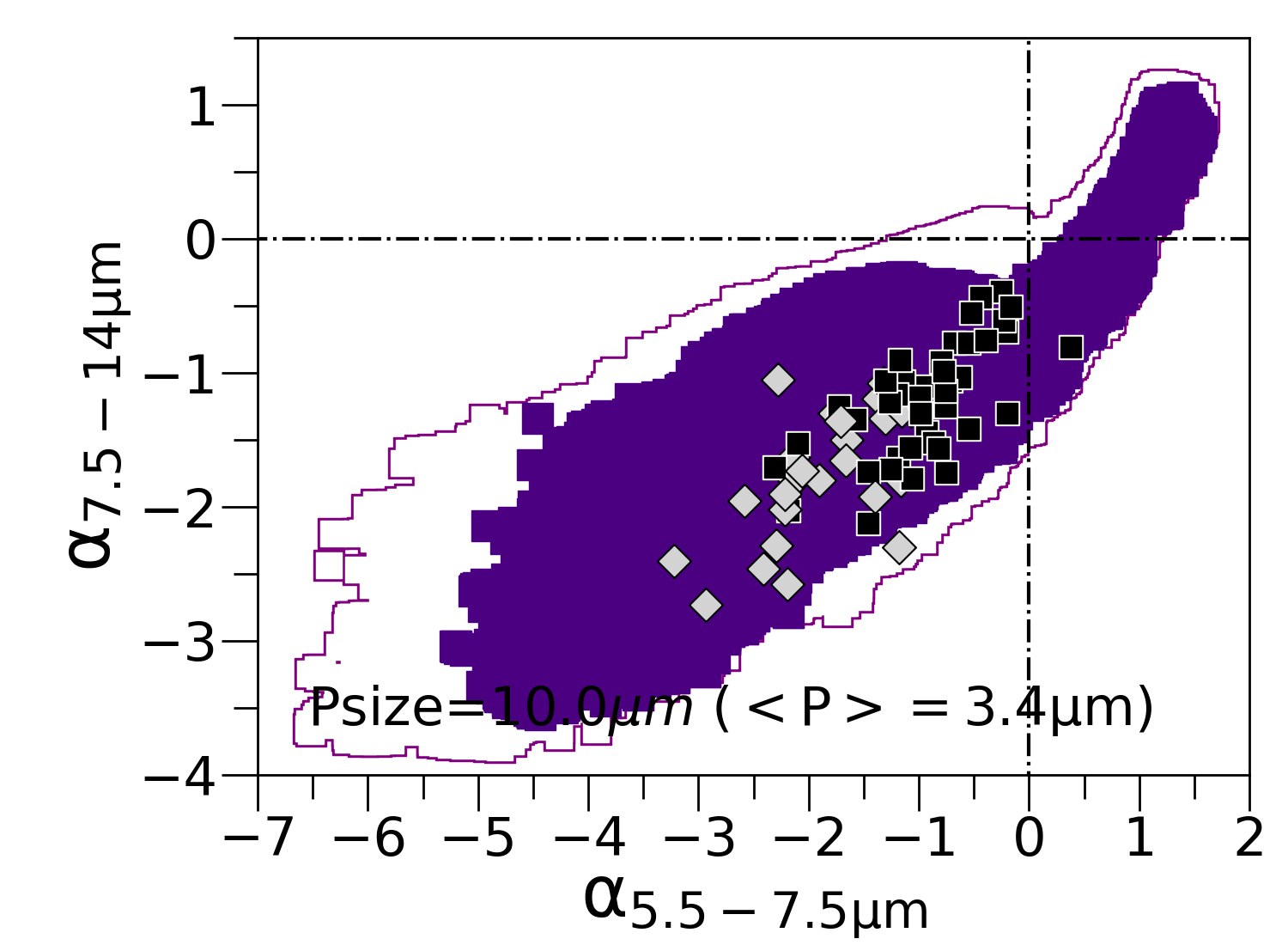}
\includegraphics[width=0.55\columnwidth, clip, trim=50 0 20 15]{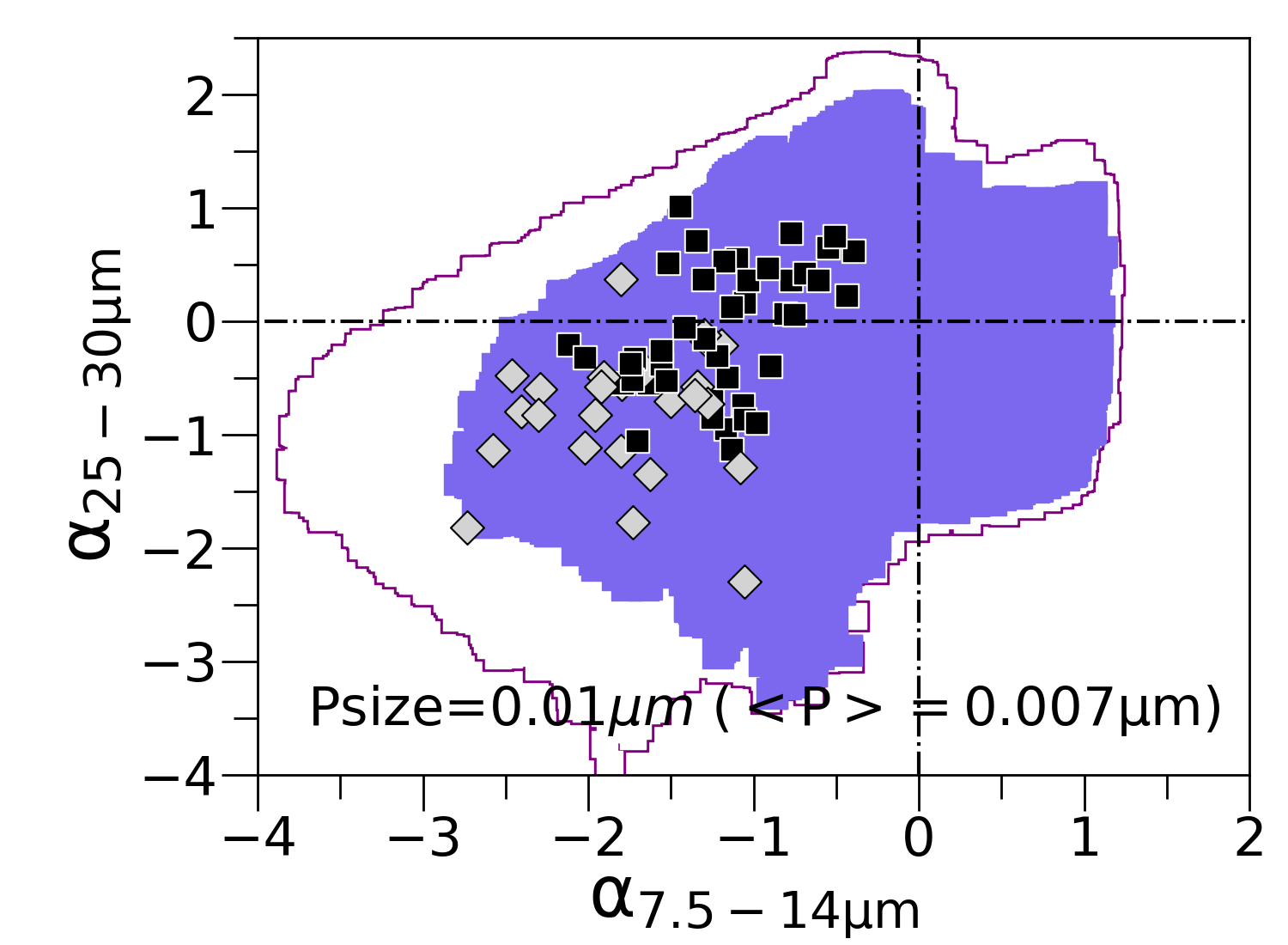}
\includegraphics[width=0.46\columnwidth, clip, trim=200 0 20 15]{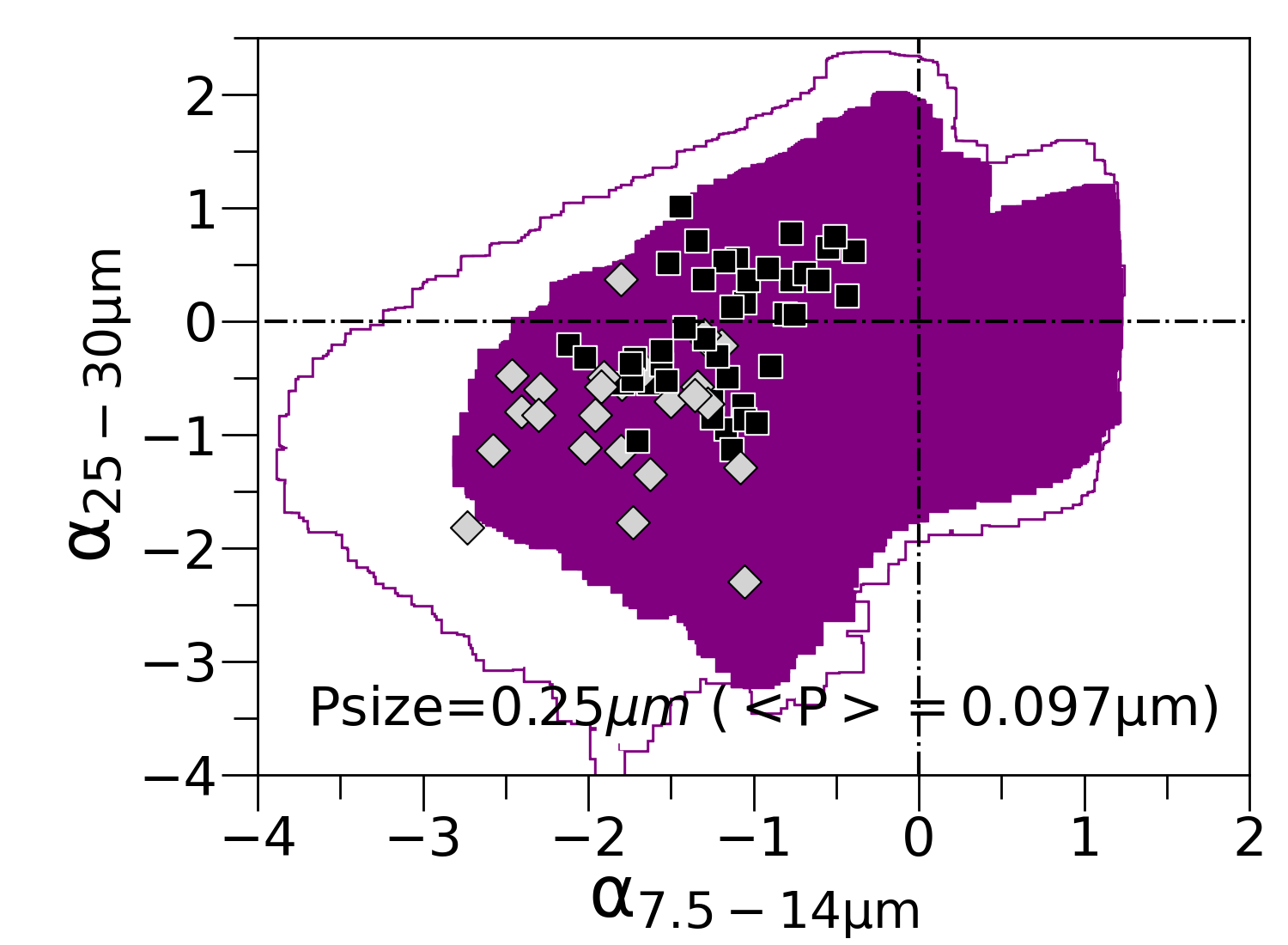}
\includegraphics[width=0.46\columnwidth, clip, trim=200 0 20 15]{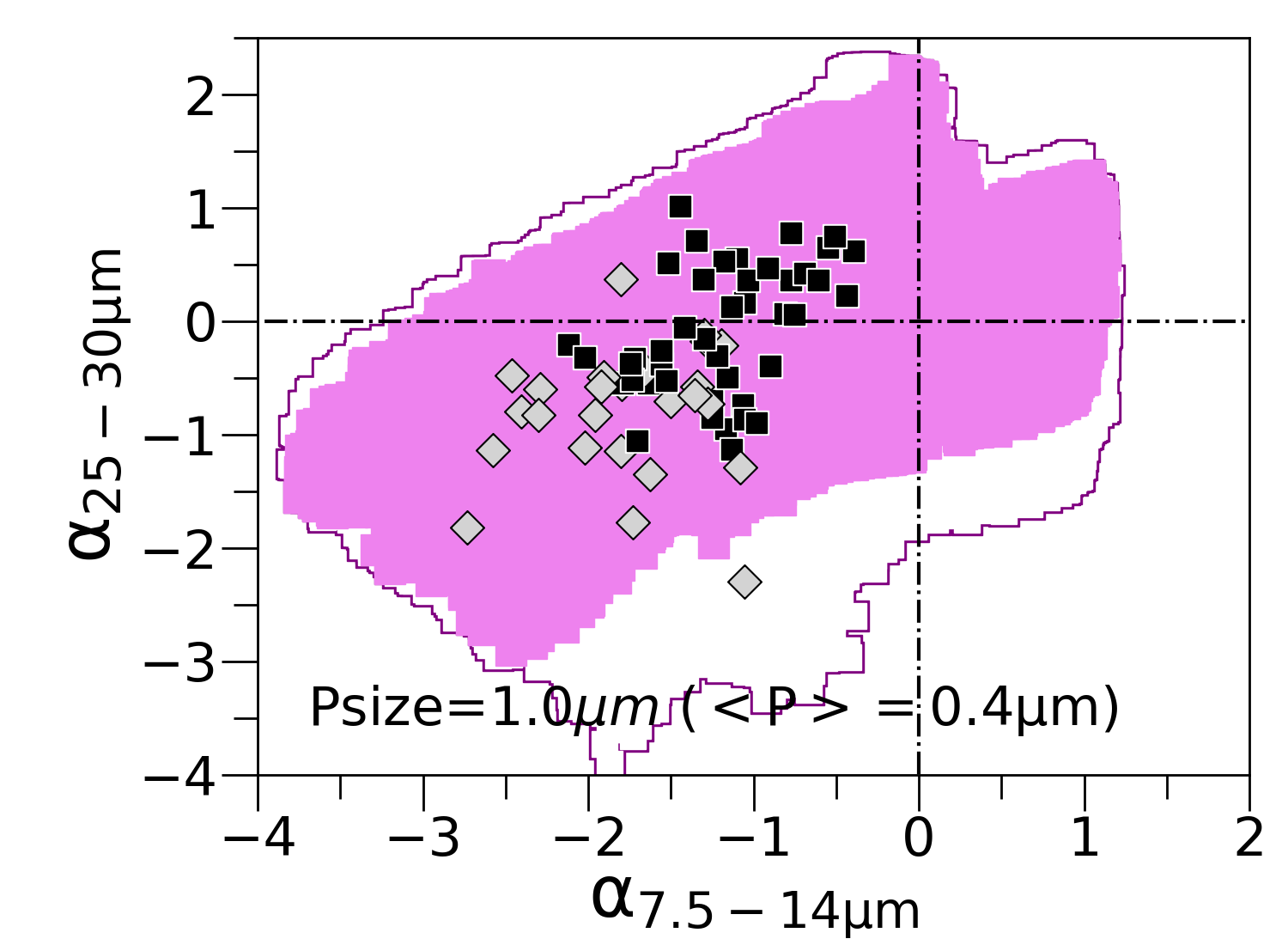}
\includegraphics[width=0.46\columnwidth, clip, trim=200 0 20 15]{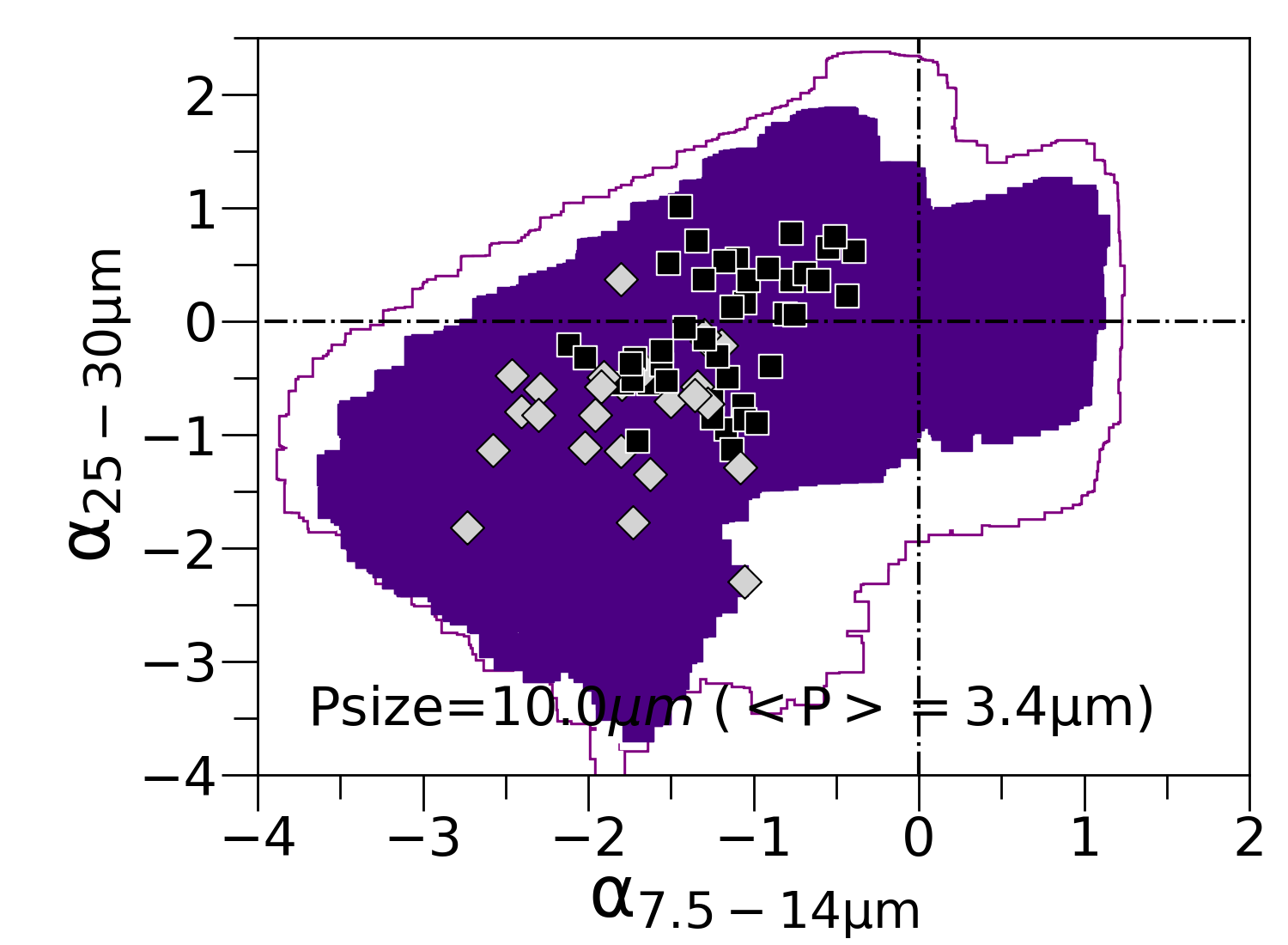}
\includegraphics[width=0.57\columnwidth, clip, trim=50 0 20 15]{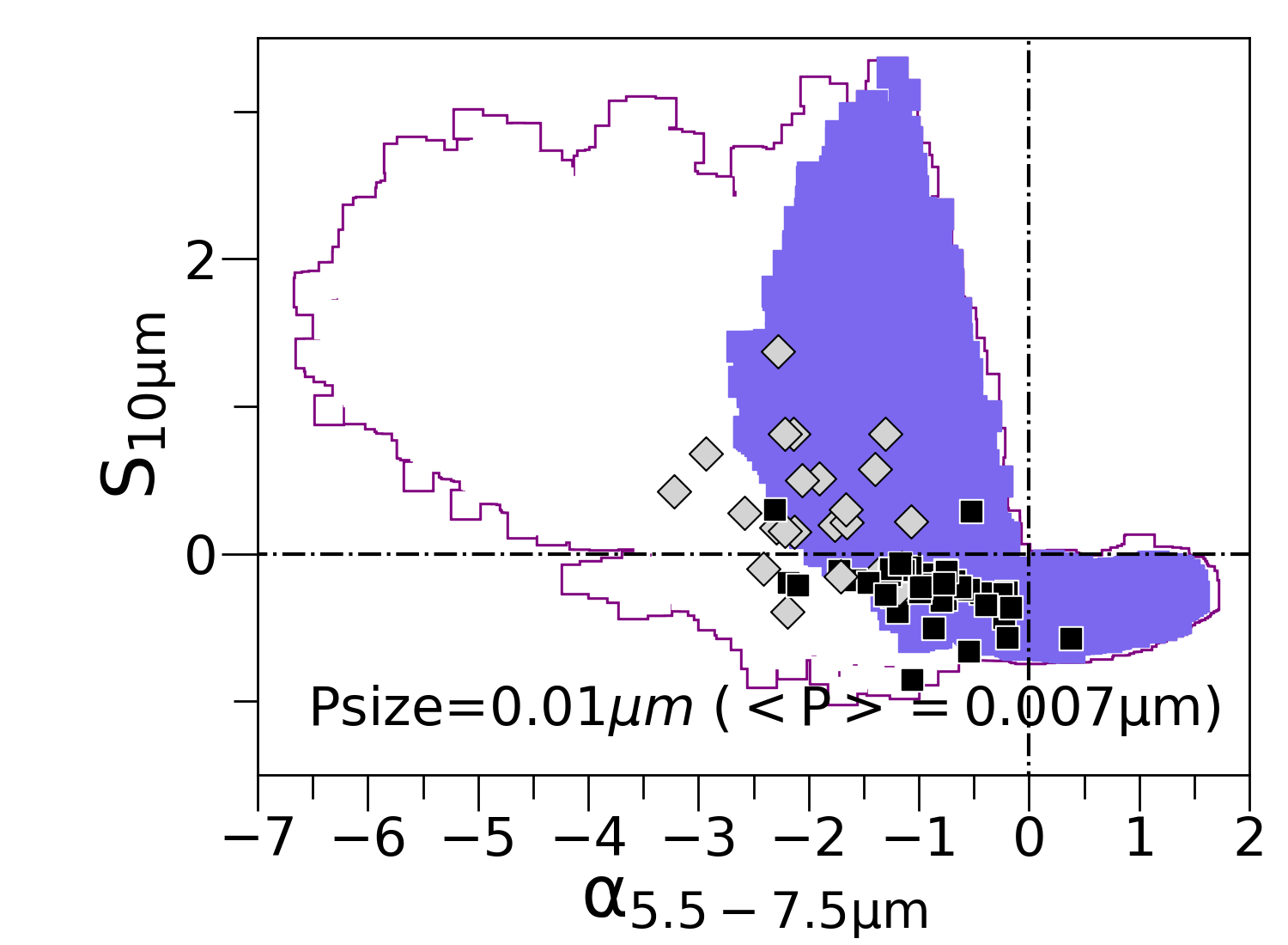}
\includegraphics[width=0.48\columnwidth, clip, trim=200 0 20 15]{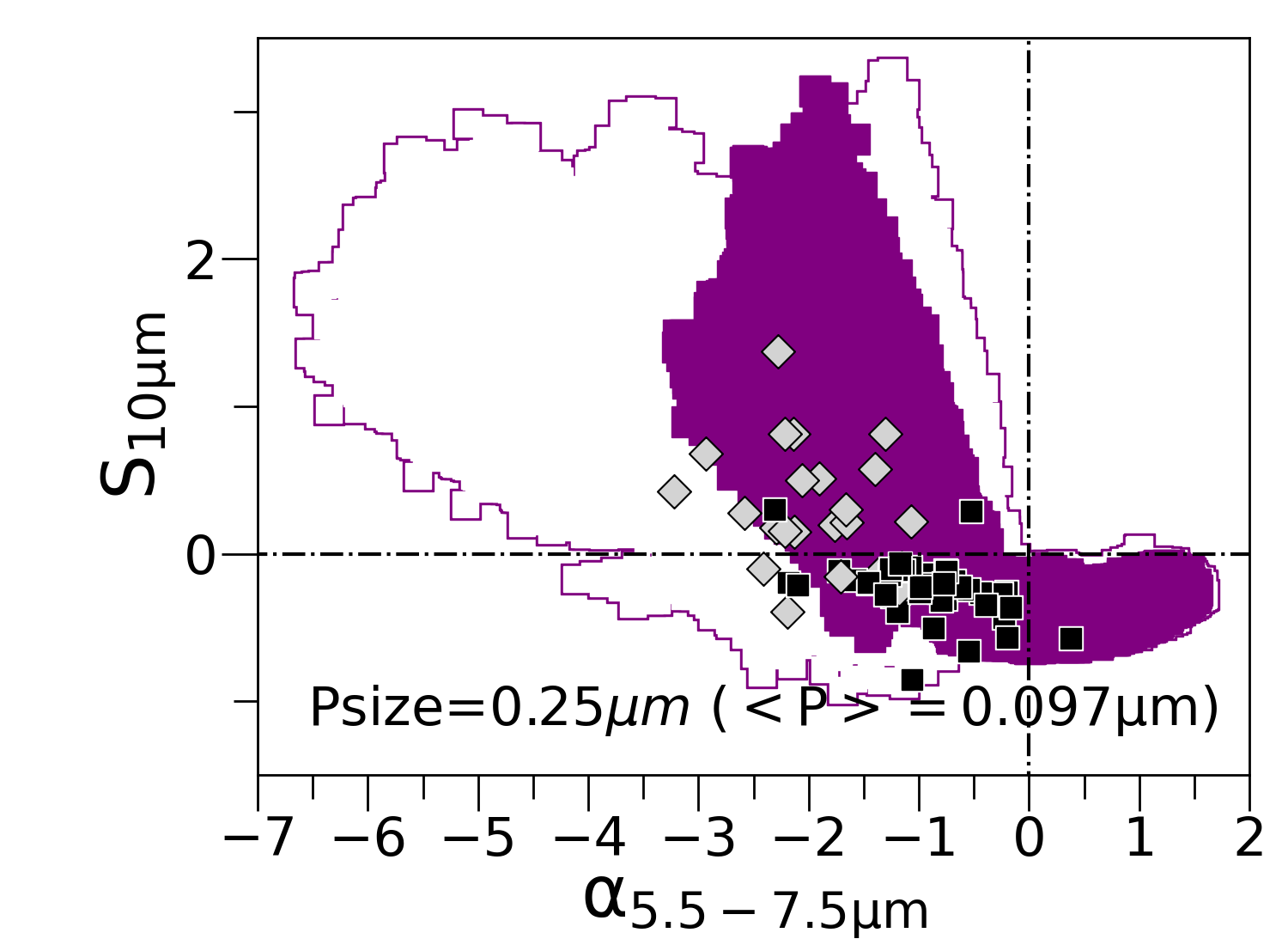}
\includegraphics[width=0.48\columnwidth, clip, trim=200 0 20 15]{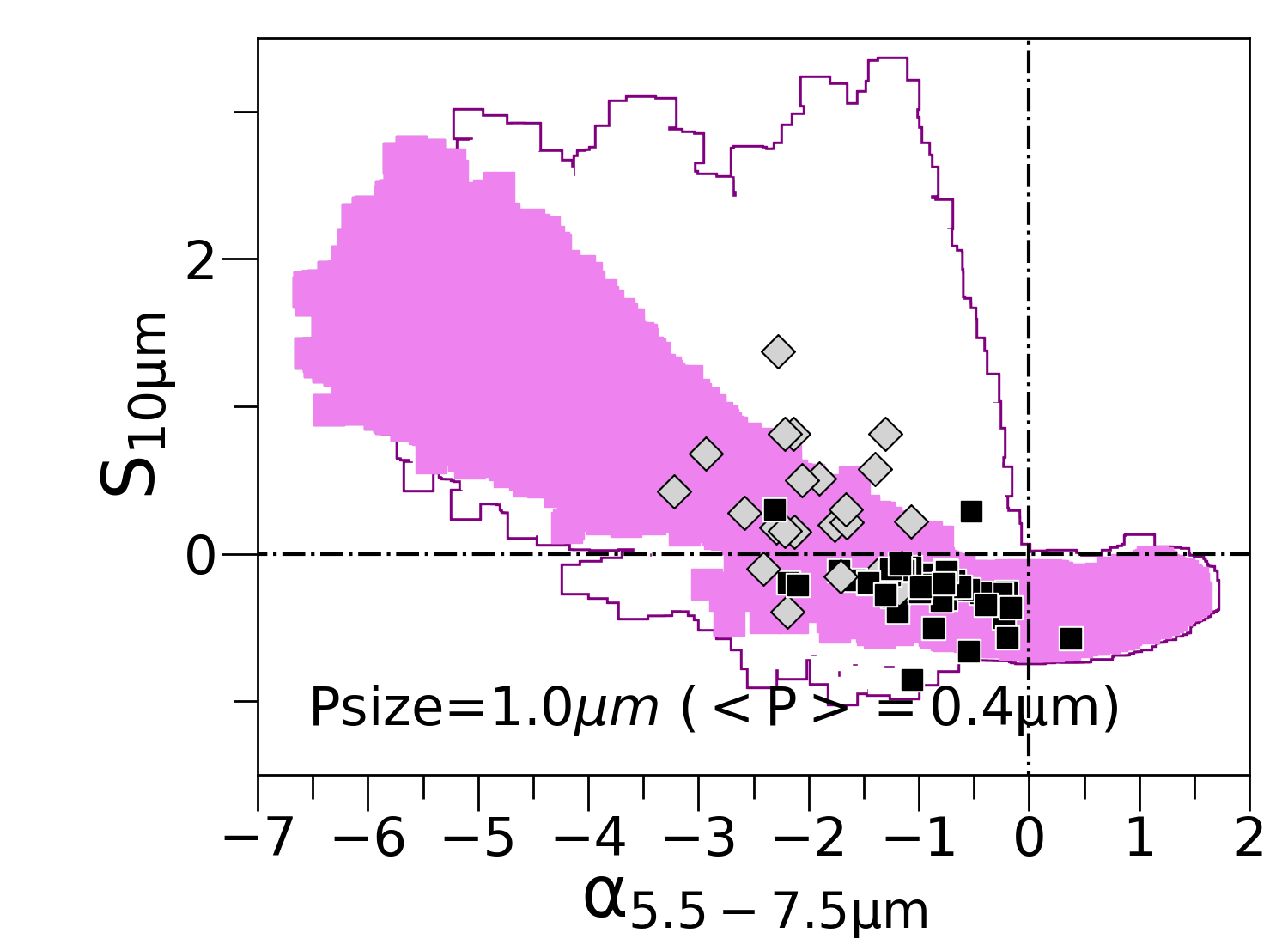}
\includegraphics[width=0.48\columnwidth, clip, trim=200 0 20 15]{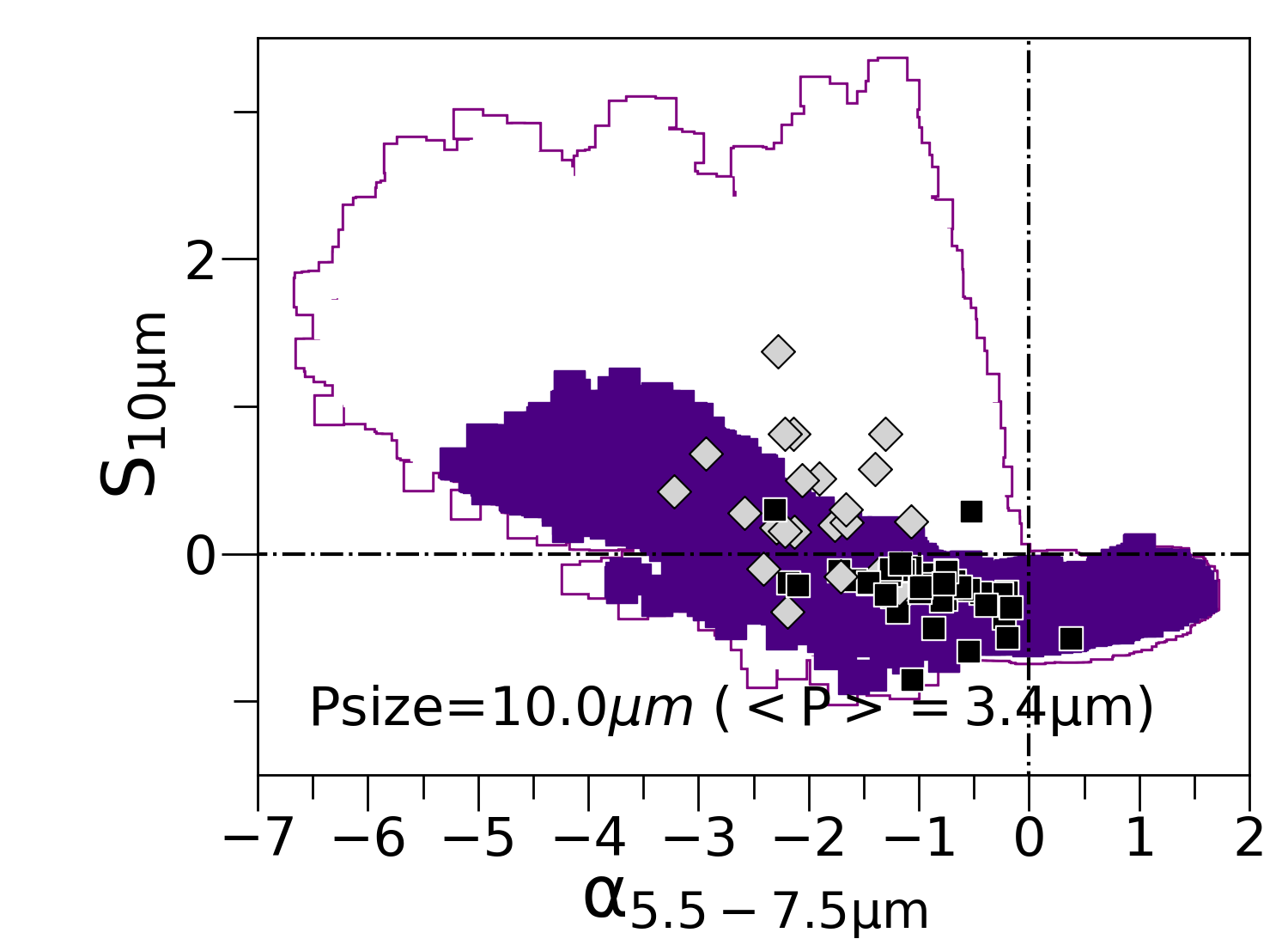}
\caption{The resulting diagrams for the [GoMar23] model when the maximum grain size of the dust particles is explored (colored area) compared with the overall space parameter (empty area). From left to right the areas show the locus of SEDs for $\rm{P_{size} = 0.01\mu m}$ ($\rm{ <P> = 0.007\mu m}$, blue), $\rm{P_{size} = 0.25\mu m}$ ($\rm{ <P> = 0.097\mu m}$, purple), $\rm{P_{size} = 1.0\mu m}$ ($\rm{ <P> = 0.36\mu m}$, pink), and $\rm{P_{size} = 10.0\mu m}$ ($\rm{ <P> = 3.41\mu m}$, violet). From top to bottom: $\rm{\alpha_{5.5-7.5\mu m}}$ versus $\rm{\alpha_{7.5-14\mu m}}$, $\rm{\alpha_{25-30\mu m}}$ versus $\rm{\alpha_{7.5-14 \mu m}}$, and $\rm{S_{10\mu m}}$ versus $\rm{\alpha_{5.5-7.5\mu m}}$. Black squares and gray diamonds show the locus for type-1 AGN and type-2 AGN, respectively.}
\label{fig:modelshapeGoMar22:Psize}
\end{center}
\end{figure*}

\begin{figure*}[!t]
\begin{center}
\includegraphics[width=0.445\columnwidth, clip, trim=50 0 20 15]{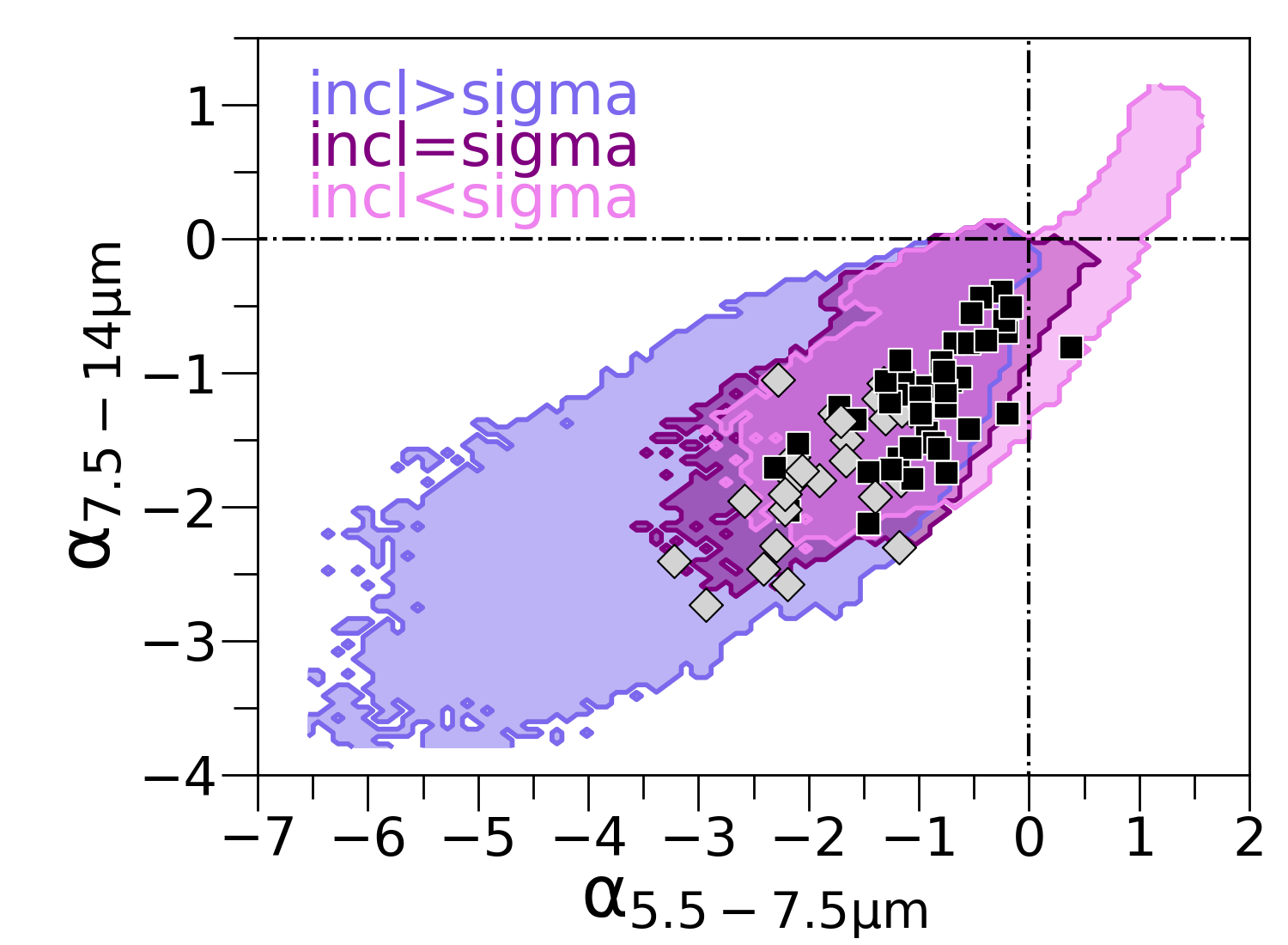}
\includegraphics[width=0.38\columnwidth, clip, trim=200 0 20 15]{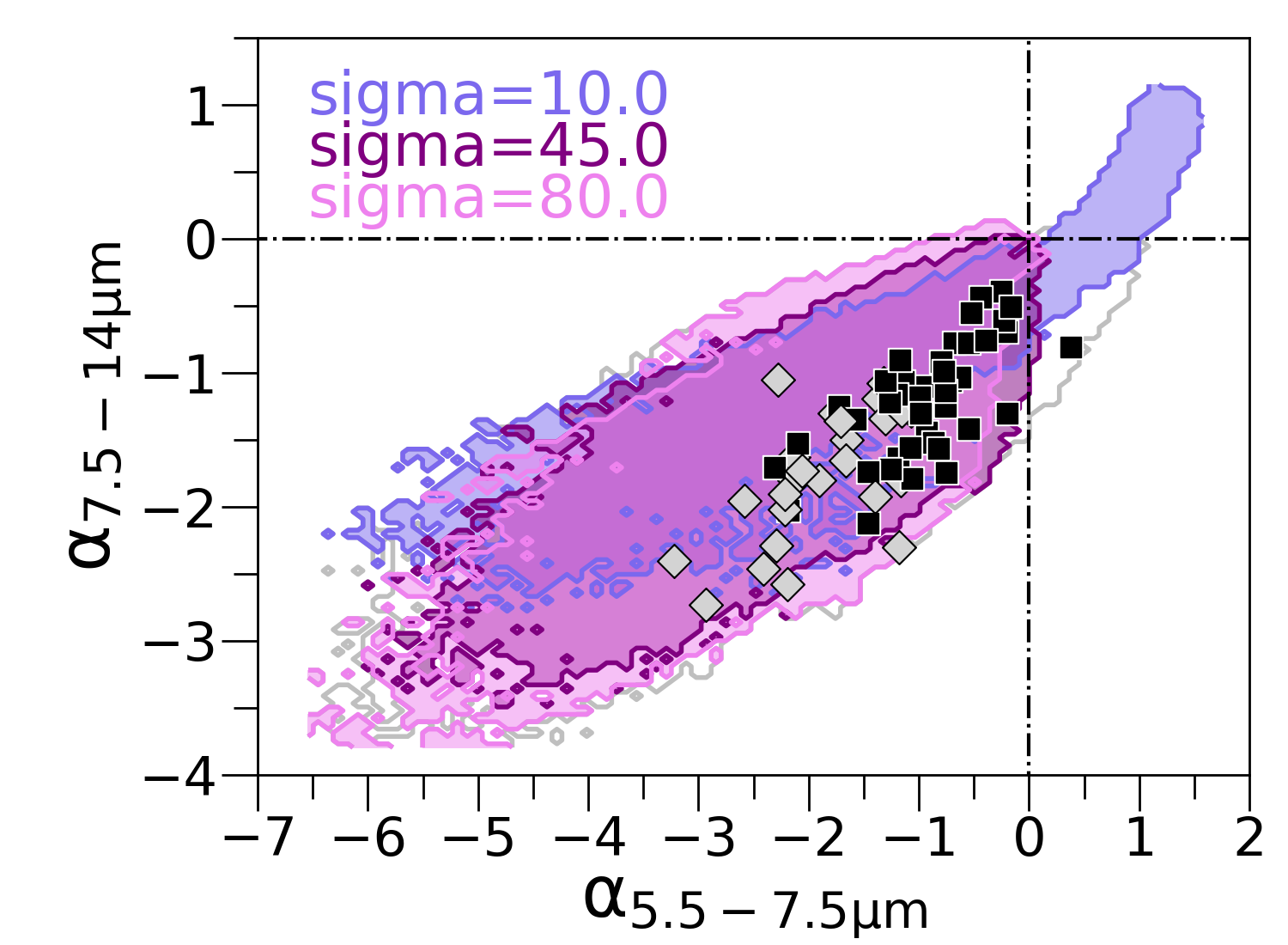}
\includegraphics[width=0.38\columnwidth, clip, trim=200 0 20 15]{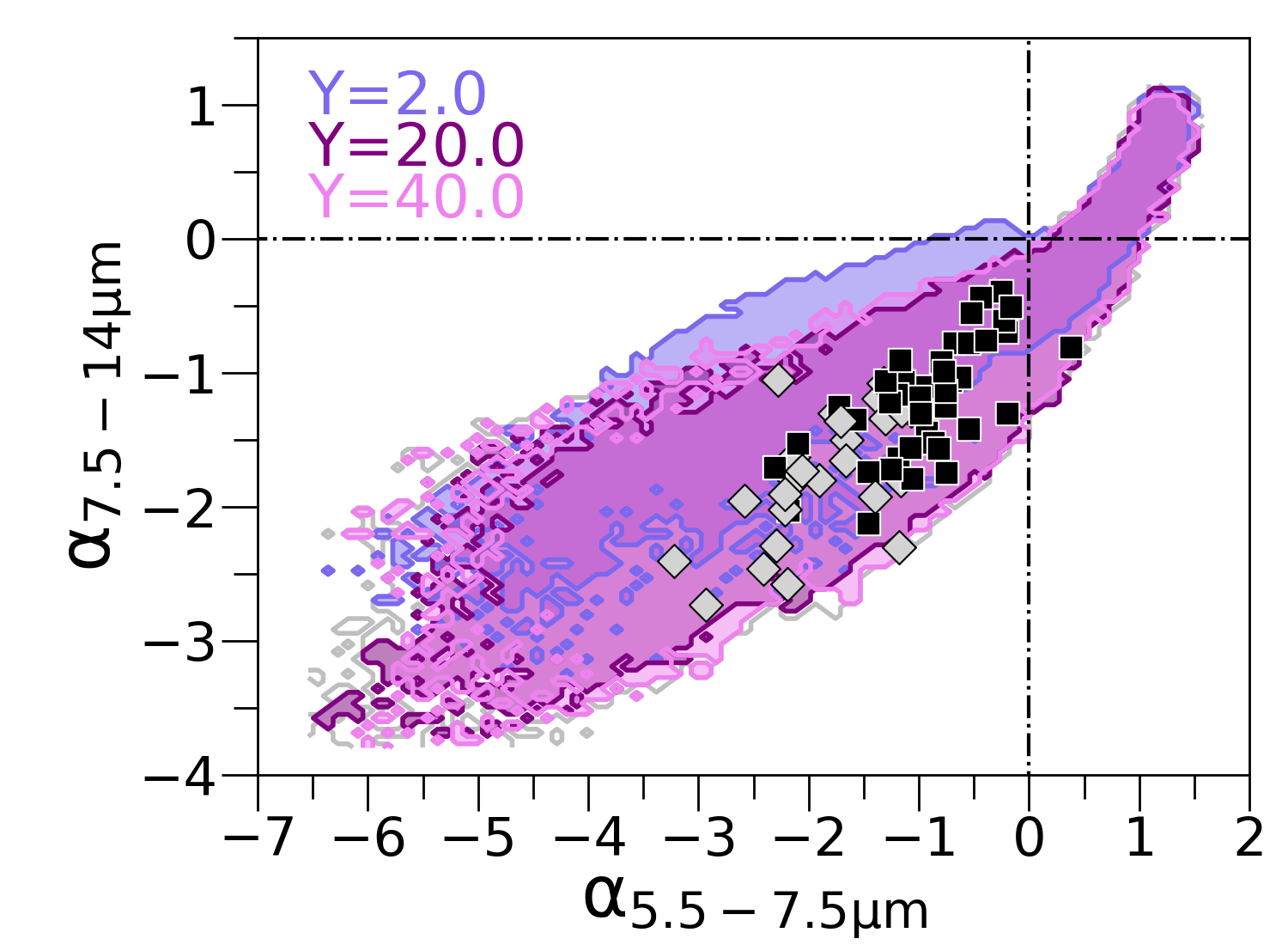}
\includegraphics[width=0.38\columnwidth, clip, trim=200 0 20 15]{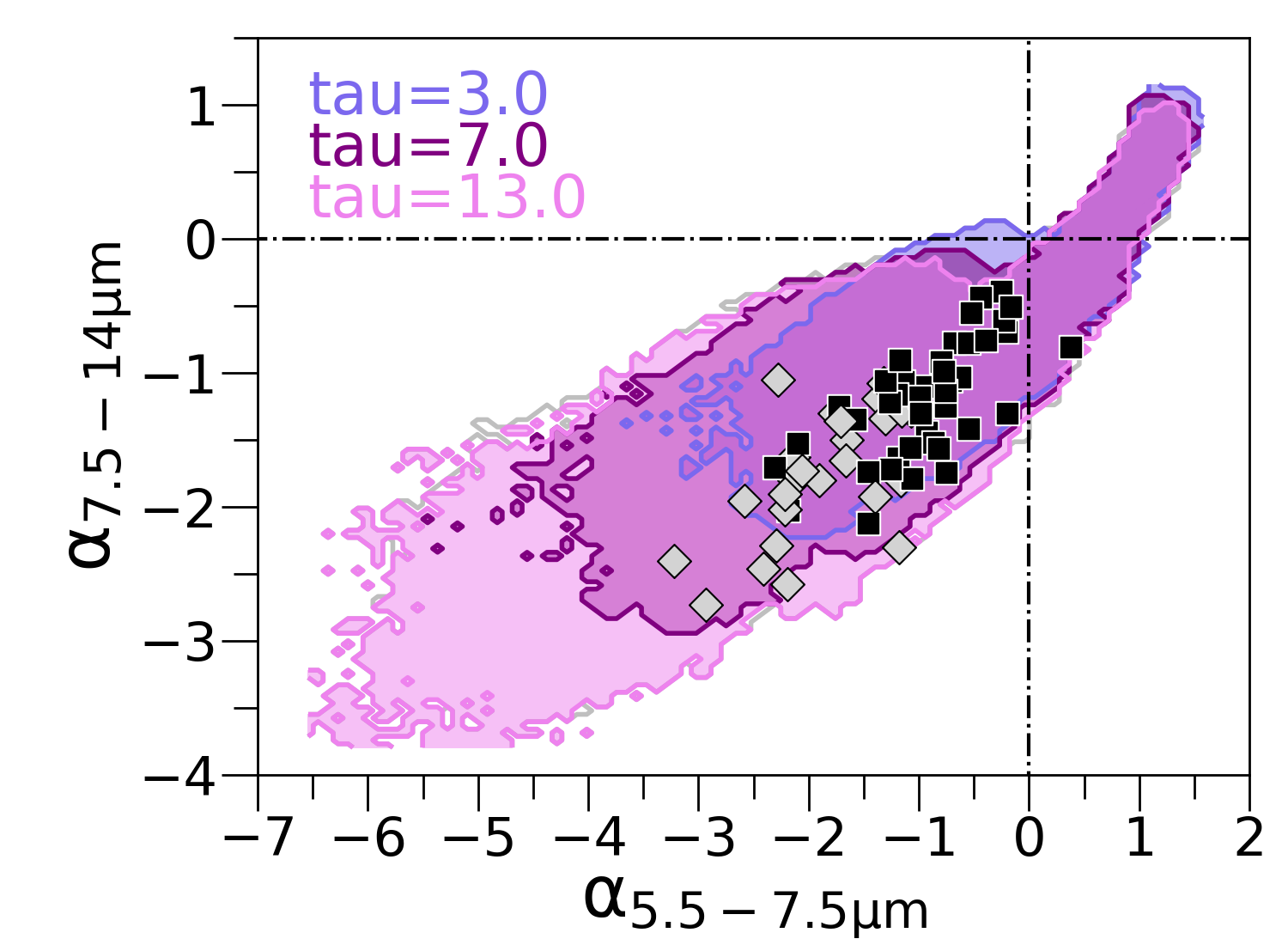}
\includegraphics[width=0.38\columnwidth, clip, trim=200 0 20 15]{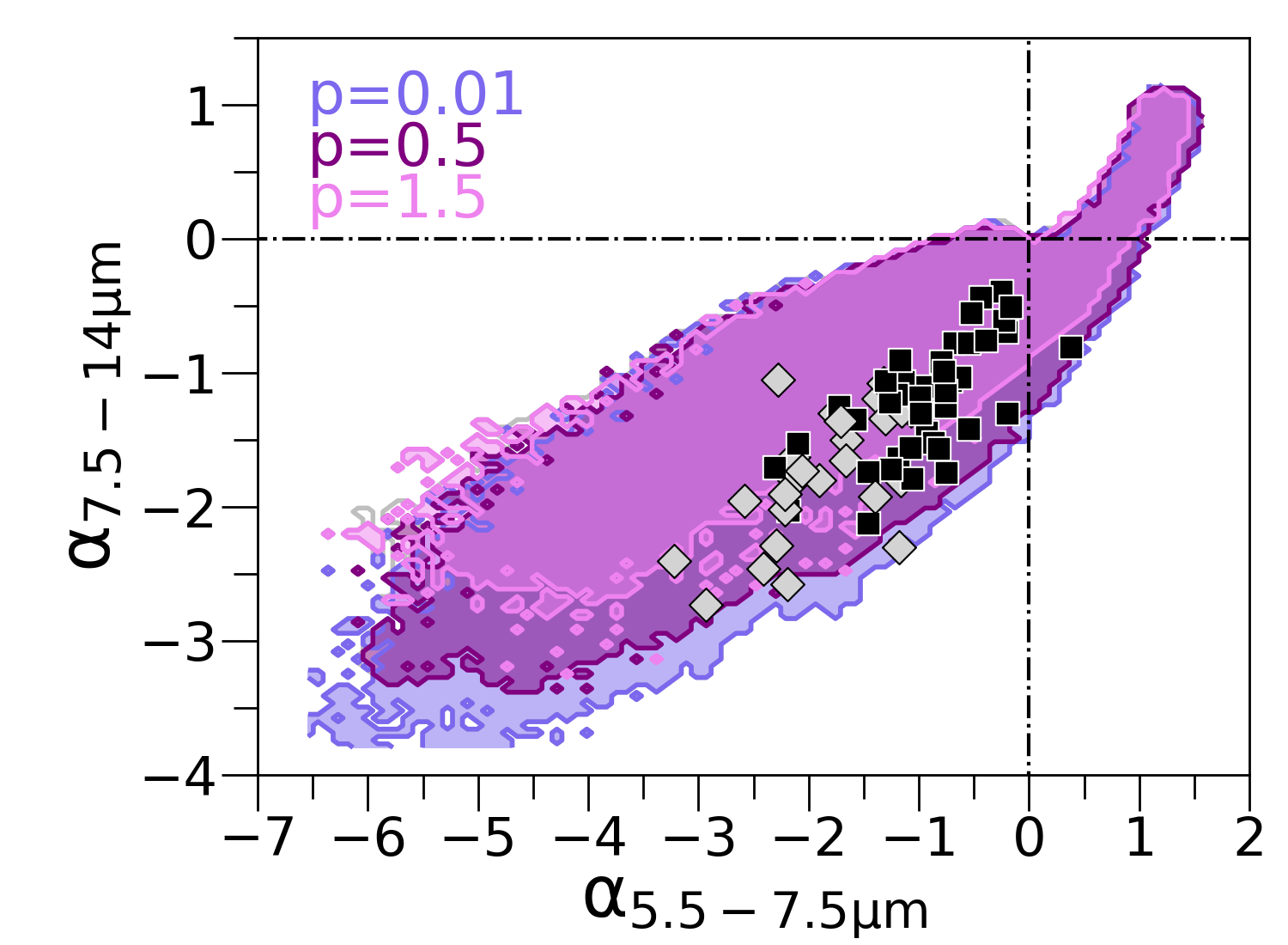}
\includegraphics[width=0.445\columnwidth, clip, trim=50 0 20 15]{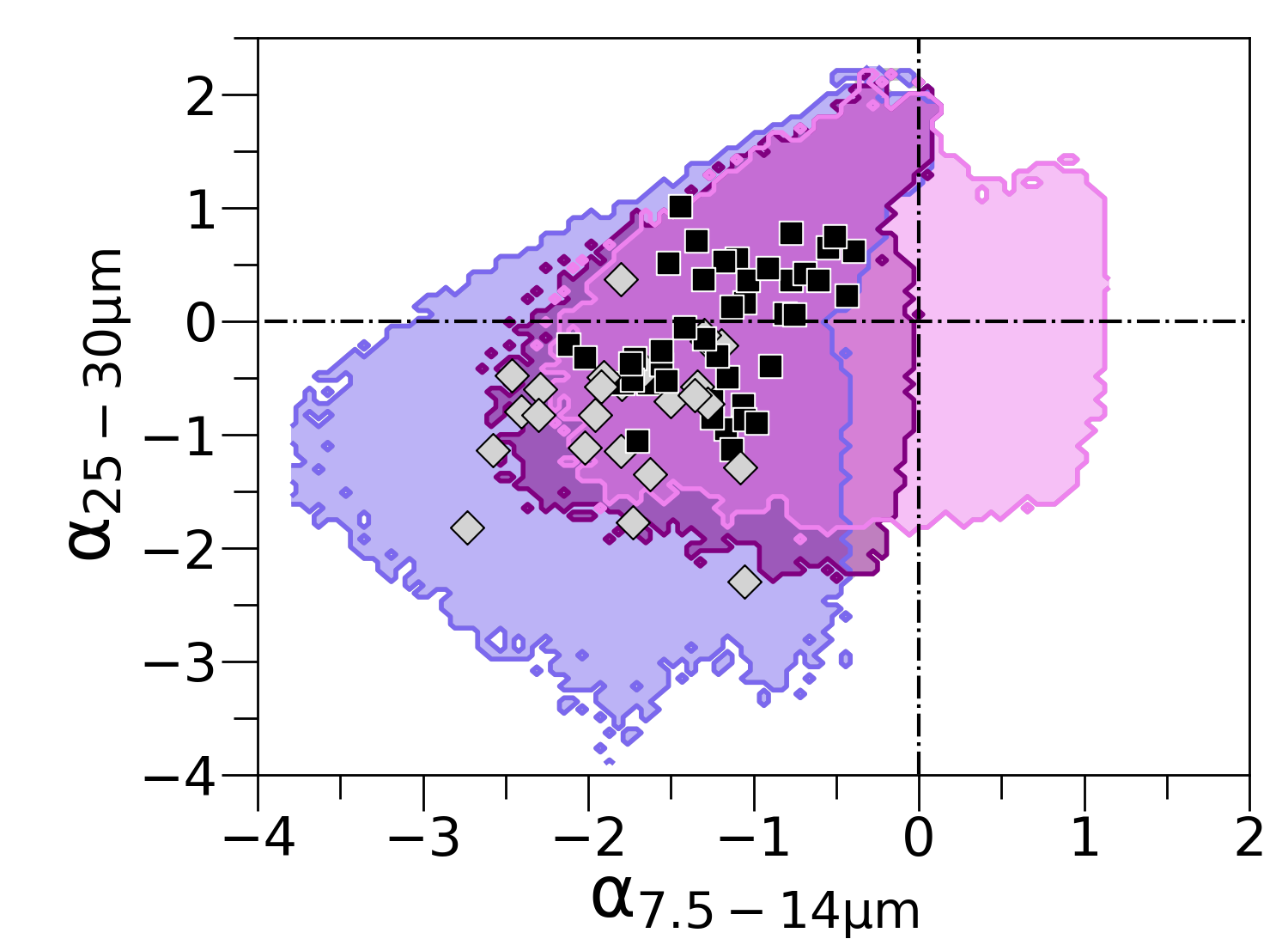}
\includegraphics[width=0.38\columnwidth, clip, trim=200 0 20 15]{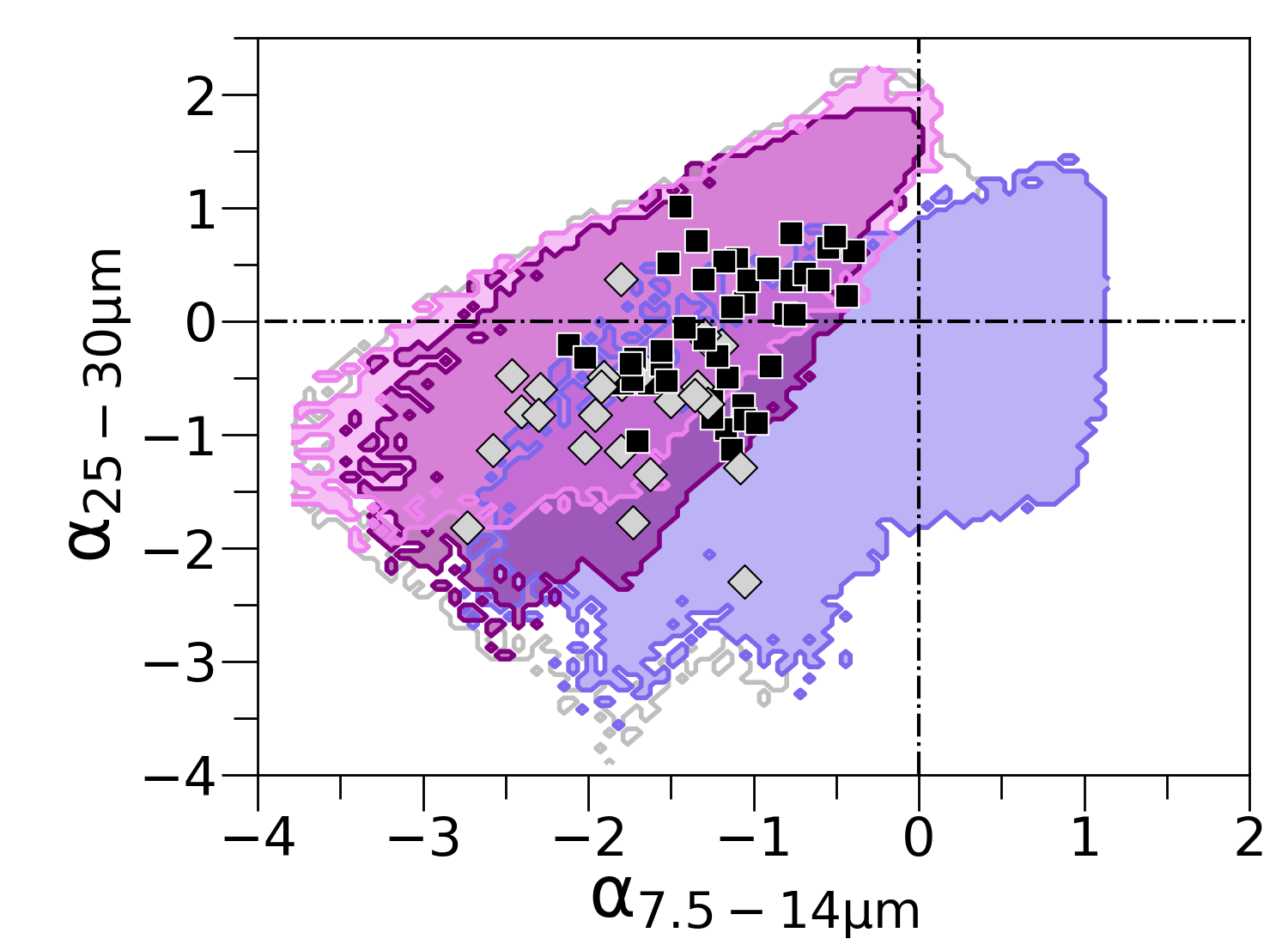}
\includegraphics[width=0.38\columnwidth, clip, trim=200 0 20 15]{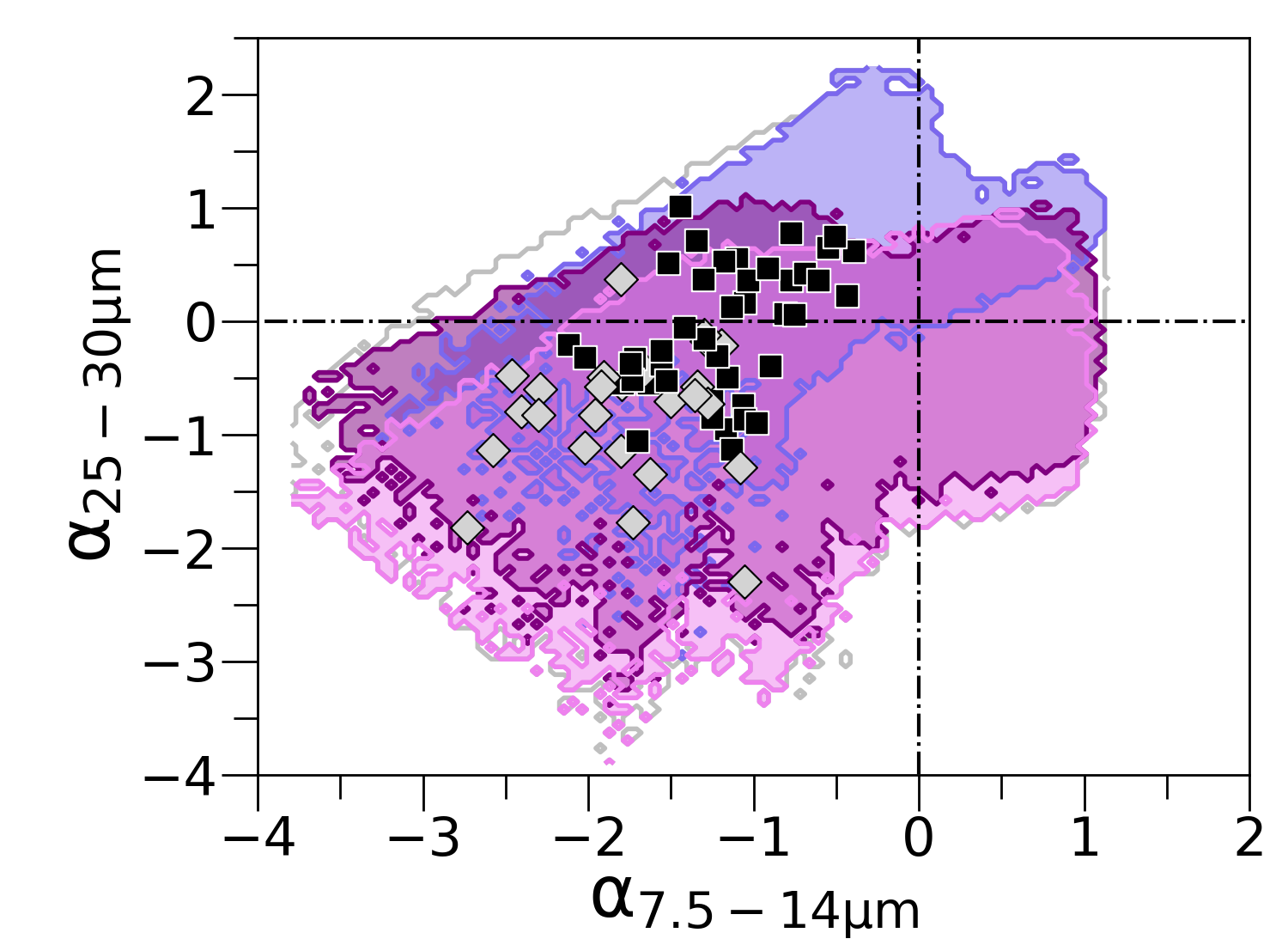}
\includegraphics[width=0.38\columnwidth, clip, trim=200 0 20 15]{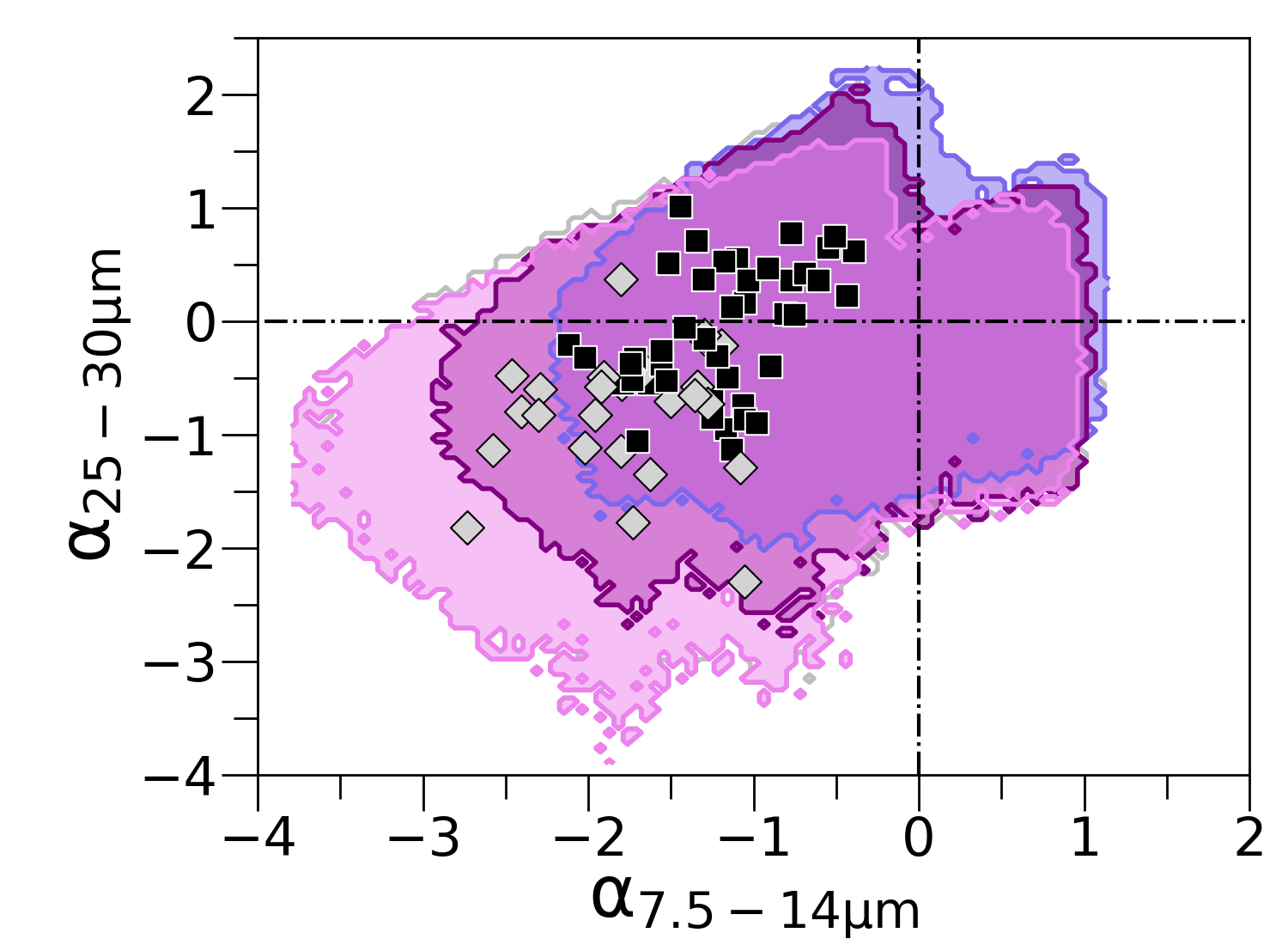}
\includegraphics[width=0.38\columnwidth, clip, trim=200 0 20 15]{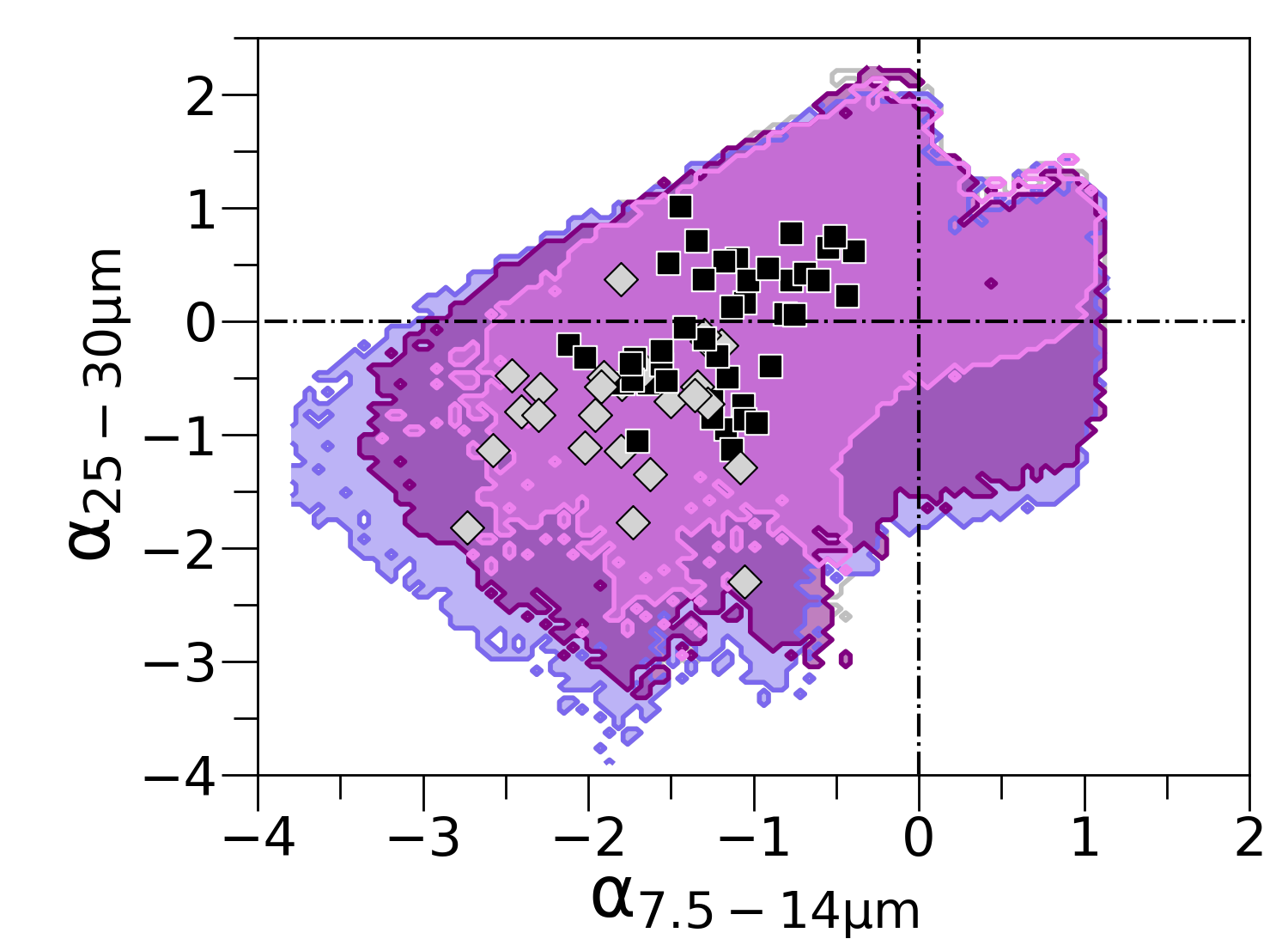}
\includegraphics[width=0.445\columnwidth, clip, trim=50 0 20 15]{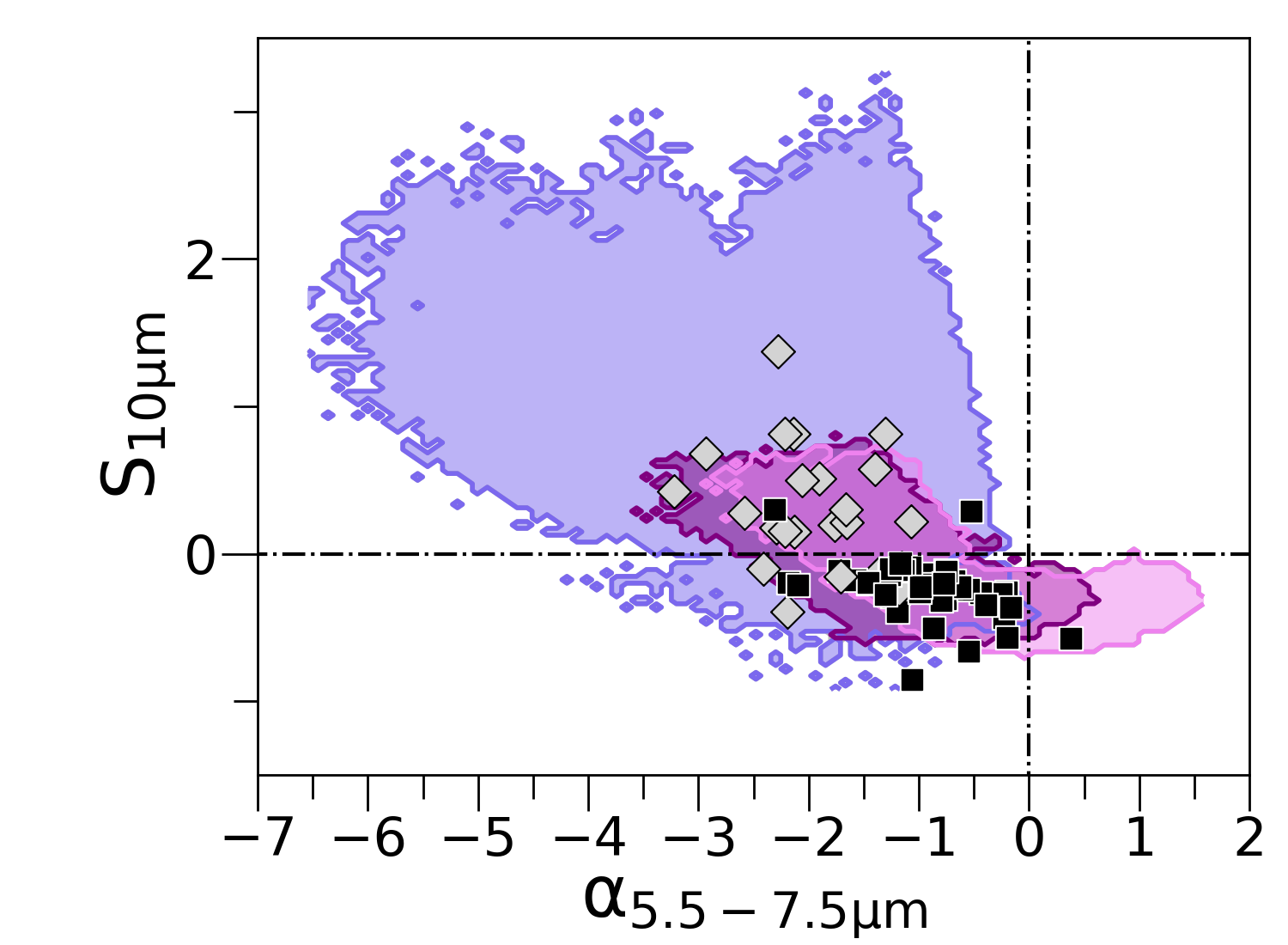}
\includegraphics[width=0.38\columnwidth, clip, trim=200 0 20 15]{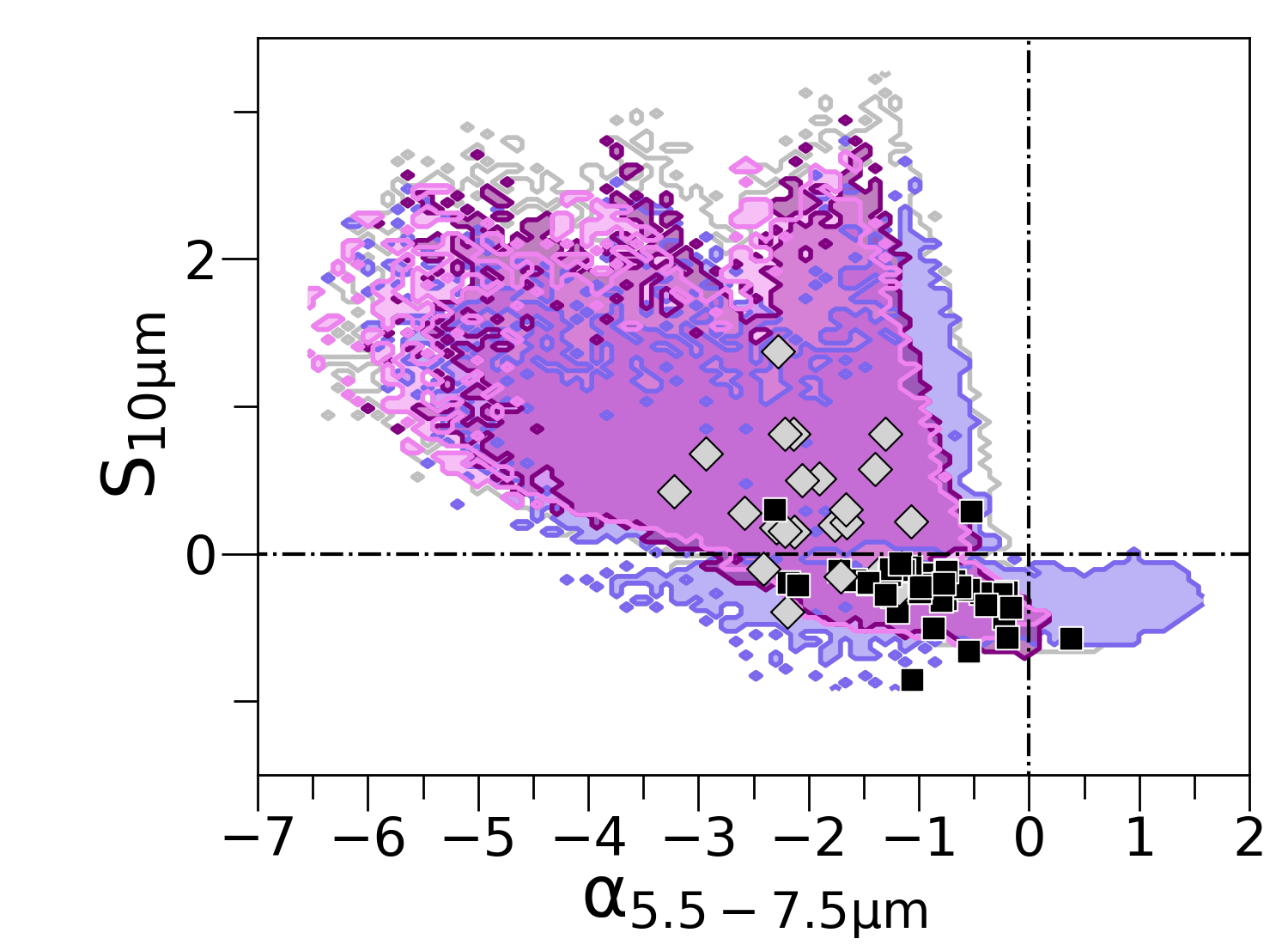}
\includegraphics[width=0.38\columnwidth, clip, trim=200 0 20 15]{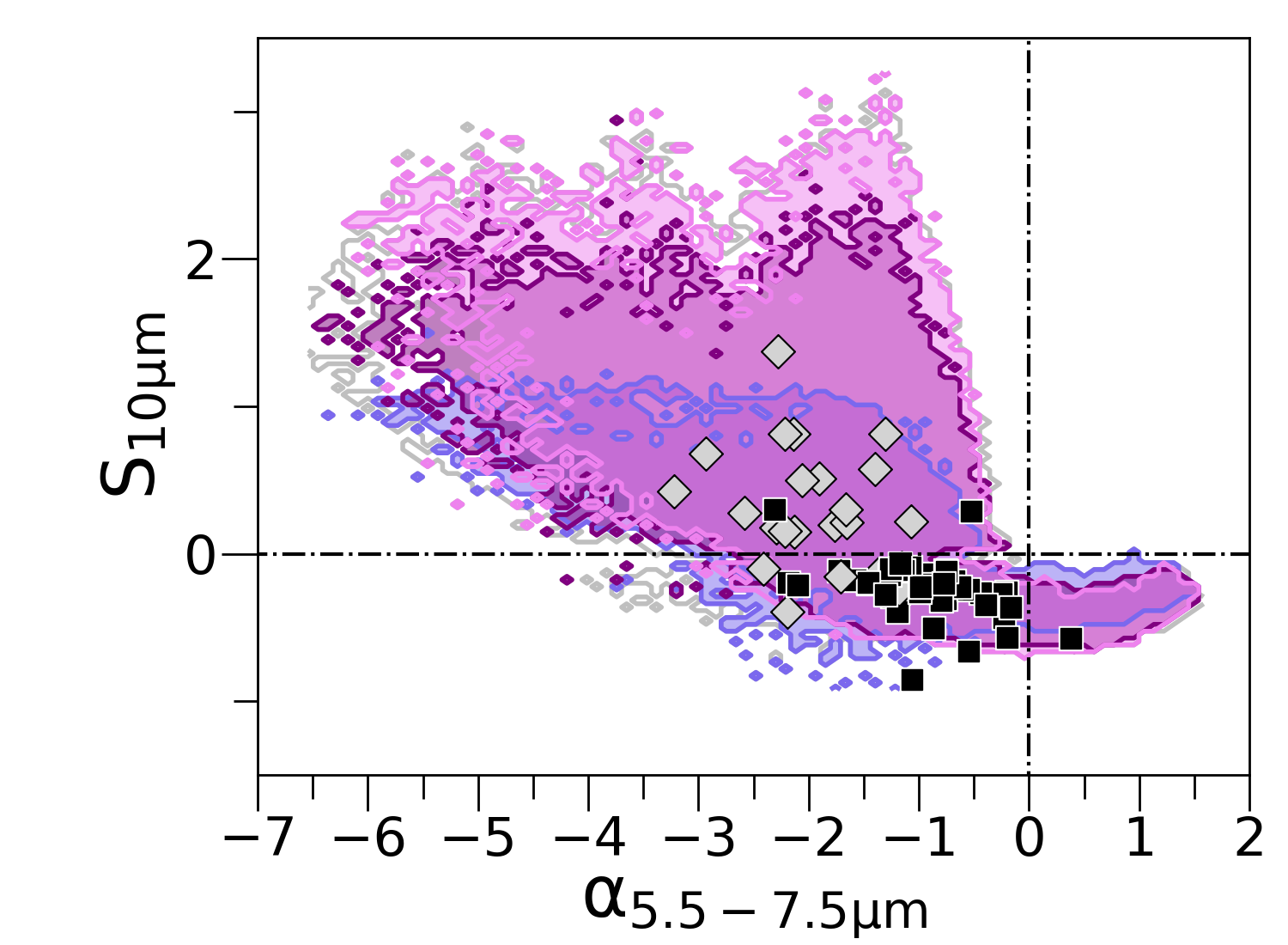}
\includegraphics[width=0.38\columnwidth, clip, trim=200 0 20 15]{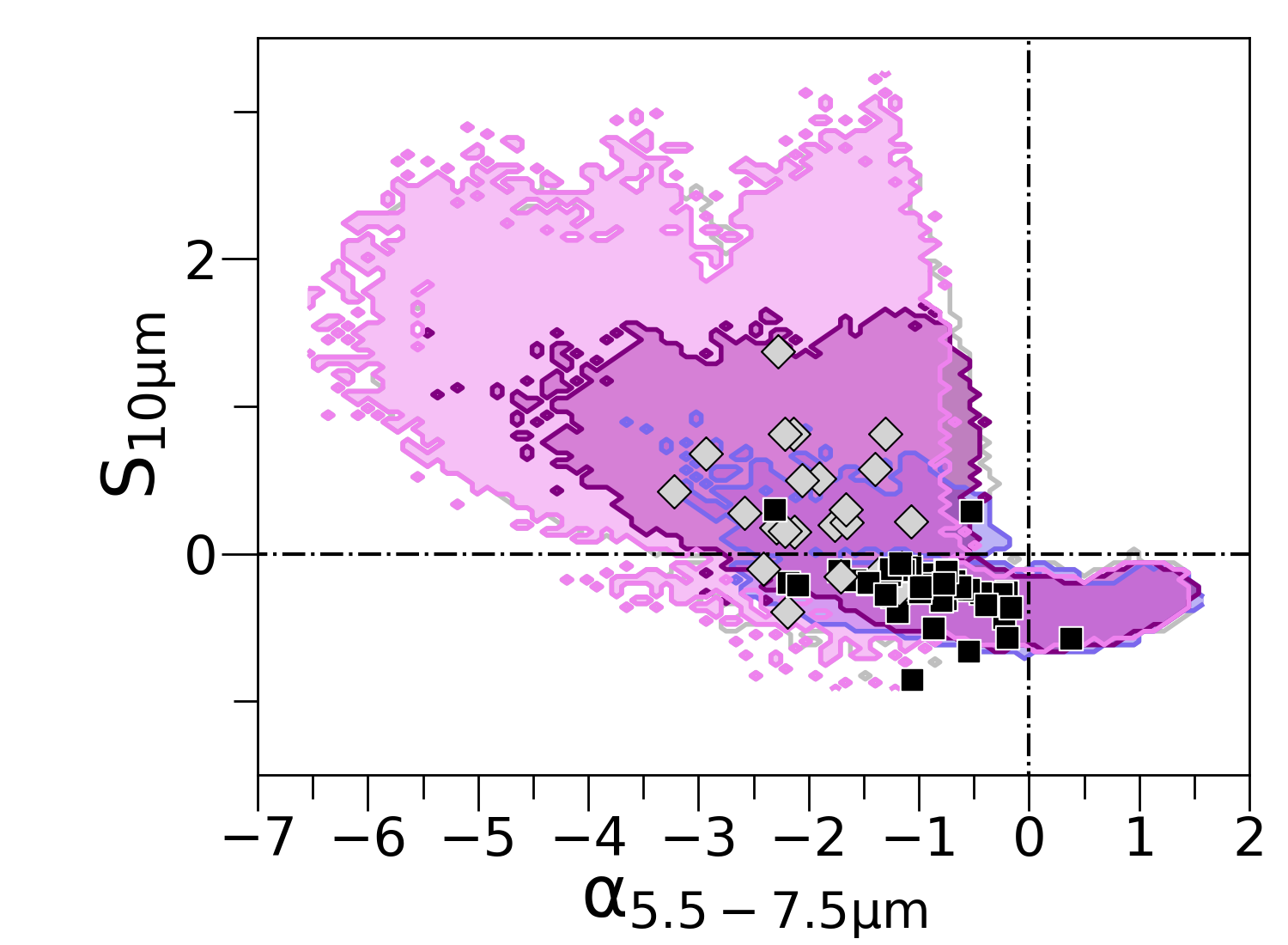}
\includegraphics[width=0.38\columnwidth, clip, trim=200 0 20 15]{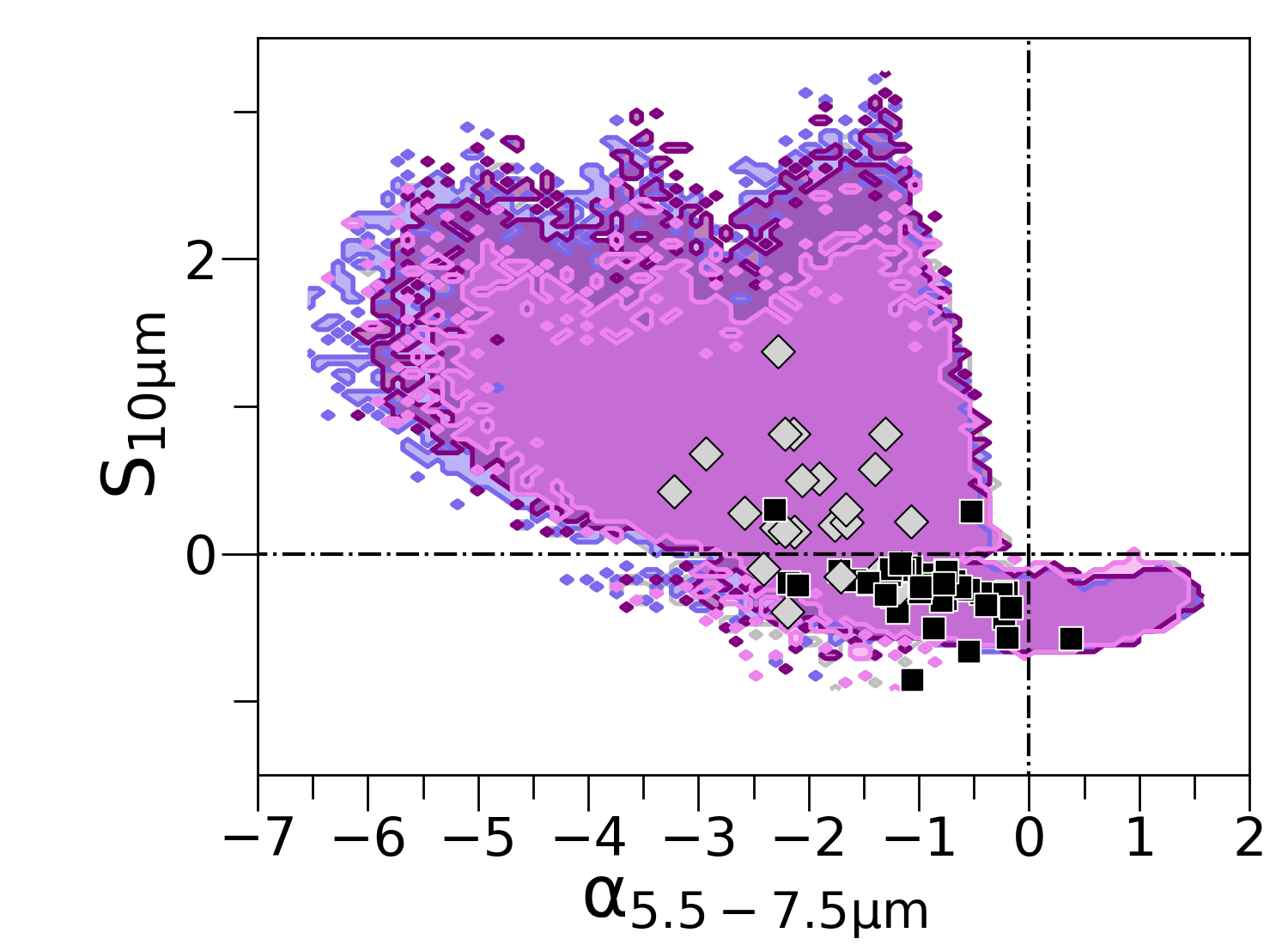}
\caption{The resulting diagrams for the [GoMar23] model when the following parameters vary (from left to right): the viewing angle, the half opening angle, the ratio between the outer and the inner radius, the optical extinction at the equator, and the radial distribution of the dust. From top to bottom: $\rm{\alpha_{5.5-7.5\mu m}}$ versus $\rm{\alpha_{7.5-14\mu m}}$, $\rm{\alpha_{25-30\mu m}}$ versus $\rm{\alpha_{7.5-14\mu m}}$, and $\rm{S_{10\mu m}}$ versus $\rm{\alpha_{5.5-7.5\mu m}}$. The gray contour shows the result for the entire SED library. The blue, purple, and pink areas show the locus of SEDs for the minimum, medium, and maximum values of the parameter, respectively (see legend in the top panels). The results for the maximum grain size are shown in Fig.\,\ref{fig:modelshapeGoMar22:Psize}. We skip the results for the polar density distribution of the dust since no differences are found.} 
\label{fig:modelshapeGoMar22:otherpar}
\end{center}
\end{figure*}

\section{Results}\label{sec:Results}

\subsection{Spectral shape}

Table\,\ref{tab:sample} reports the silicate central peak wavelength, silicate strength, and three continuum slopes measured from the \emph{Spitzer}/IRS spectra of the AGN sample used here. Fig.\,\ref{fig:observations} (top-left panel) shows the resulting optical depths for the type-1 and type-2 AGN. We have produced four diagrams comparing these five parameters (Fig.\,\ref{fig:observations}). In general type-1 AGN show silicate emission features (black lines) and type-2 AGN show silicate absorption features (gray lines). However, two type-1 AGN show weak absorption features, and eight type-2 AGN show emission features (top-left panel of Fig.\,\ref{fig:observations} and Table\,\ref{tab:sample}). This behavior is already reported in the literature \citep[][]{Hoenig10A, Gonzalez-Martin13}, and it is easily explained by the stochastic nature of the clumpiness of the system and its combination of viewing angle, density profile, and optical depths. The range of optical depths at $\rm{10\mu m}$ is $\rm{\tau_{10\mu m} \sim [-1, 1]}$ and absorption features are usually narrower than emission features, and the latter show a central plateau. The central wavelength of the feature is different when it is in absorption (around 9.7$\rm{\mu m}$) compared to when it is in emission. In the case of the emission features, they are usually shifted to 10.5$\rm{\mu m}$ except in a few objects. The continuum slopes are correlated, as expected, being mostly negative for $\rm{\alpha_{5.5-7.5\mu m}}$ and $\rm{\alpha_{7.5-14\mu m}}$ (both in the range [-3.5,0.5]) while $\rm{\alpha_{25-30\mu m}}$ show both negative and positive values in the range [-2.5,1.5]. Moreover, Type-1 AGN have flatter slopes than type-2 AGN. 

All these spectral shape properties should be reproduced by any SED library. To explore this, Fig.\,\ref{fig:modelshapes} shows these four diagrams for all the studied models, including our new [GoMar23] model (top panel in purple). Most models reproduce the spectral shape measurement of our AGN sample. The exception is the $\rm{\alpha_{5.5-7.5\mu m}}$ and/or $\rm{\alpha_{7.5-14\mu m}}$ for the [Hoenig10] model (fourth row from top to bottom). The [Hoenig17] model reproduces $\rm{\alpha_{7.5-14\mu m}}$ versus $\rm{\alpha_{5.5-7.5\mu m}}$ (sixth row, left panel) but fails to reproduce $\rm{\alpha_{25-30\mu m}}$ versus $\rm{\alpha_{7.5-14\mu m}}$, with a clear shift toward higher values of $\rm{\alpha_{25-30\mu m}}$ (fifth row, second panel from left to right). Interestingly, [Hoenig10] and [Hoenig17] models show the narrowest range for the silicate strengths and match well the observed range for emission features. Except for the [Stalev16] model, the shapes found in each SED library include a wider range of values compared to the observed AGN properties. In particular, the [Nenkova08] model shows quite prominent 10$\rm{\mu m}$ silicate emission features ($\rm{S_{10\mu m}<-2}$) which are not needed by our AGN sample. It also includes slopes $\rm{\alpha < -3}$ that are never observed (this is also the case for [Hoenig10] and [Hoenig17] models). On the other hand, [Fritz06] model shows large silicate absorption strengths that do not have an observational counterpart.

Our new [GoMar23] model is also able to reproduce all these spectral characteristics (top row in Fig.\,\ref{fig:modelshapes}) but it also shows a wider range of spectral shape measurements, compared with observations. In particular, it includes steeper slopes ($\rm{\alpha < -3}$) and deeper silicate absorption features than observed. This might indicate that the complexity of the model (i.e. number of parameters or parameter space) might not be needed by the data or that a particular range of configurations is required. In order to explore this, we study how these diagrams are populated depending on the parameter space. Fig.\,\ref{fig:modelshapeGoMar22:Psize} shows the impact of the maximum grain size $\rm{P_{size}}$ when fixed to 0.01, 0.25, 1.0 and 10$\rm{\mu m}$ (corresponding to a mass-weighted average of $\rm{<P>}$ 0.007, 0.097, 0.36, and 3.41$\rm{\mu m}$) and Fig.\,\ref{fig:modelshapeGoMar22:otherpar} shows the results for each of the other parameters of [GoMar23] model. For each of them, we show the results for the minimum, maximum, and intermediate values using a different color. Note that we exclude the $\rm{S_{10\mu m}}$ versus $\rm{C_{10\mu m}}$ diagram because all these plots are similar irrespective of the parameter space, except when $\rm{\tau_{9.7\mu m}}$ is large, which naturally produces large values of $\rm{S_{10\mu m}}$. We also exclude the polar slopes of the density distribution, $q$ parameter, because it shows a negligible impact on the diagrams. The main findings are:

\begin{itemize}
    \item We find that the area occupied in these diagrams is similar when the viewing angle is set to the outer wall of the torus (i.e. $\rm{i=\sigma}$) than when a theoretical type-1 view (i.e. $\rm{i<\sigma}$) is set. Type-2 views (i.e. $\rm{i>\sigma}$) can explain most of the AGN while type-1 views fail to reproduce some type-2 AGN because of the lack of the low values of the spectral slopes. Type-1 views also fail at reproducing the deepest silicate absorption features. Notice that type-2 (type-1) views can also produce silicate emission (absorption) features, as found in our AGN sample.
    \item Intermediate to high values of the half opening angle of the torus, $\rm{\sigma}$, seem to better match observations (in particular the trend of $\rm{\alpha_{25-30\mu m}}$ versus $\rm{\alpha_{7.5-14\mu m}}$), although low values are also required for some objects. 
    \item Relatively small values of the ratio between the outer and the inner radius, $Y$, could explain the lack of deep absorption features.  
    \item Low and intermediate values of the equatorial optical depth at 9.7$\rm{\mu m}$, $\rm{\tau_{9.7\mu m}}$, are able to explain most of the type-1 AGN spectra in our sample but the locus in $\rm{\alpha_{7.5-14\mu m}}$ versus $\rm{\alpha_{5.5-7.5\mu m}}$ and $\rm{\alpha_{25-30\mu m}}$ versus $\rm{\alpha_{7.5-14\mu m}}$ diagrams for some type-2 AGN are better explained with large values of $\rm{\tau_{9.7\mu m}}$. 
    \item The maximum size of the dust particles, $\rm{P_{size}}$, plays a crucial role on the resulting diagrams. Particles in the range $\rm{0.25<P_{size}<10 \mu m}$ (i.e. $\rm{<P> = [0.097-3.41] \mu m}$) better  matches the $\rm{S_{10\mu m}}$ versus $\rm{\alpha_{5.5-7.5\mu m}}$ diagram. In particular, $\rm{P_{size}=1\mu m}$ (i.e. $\rm{<P> = 0.36 \mu m}$) nicely matches the observed spectral shapes in this diagram (pink shaded area, the third column in Fig.\,\ref{fig:modelshapeGoMar22:Psize}). Although not shown here, we would like to emphasize that the grain size has not a preferential role on the center of the silicate feature at 9.7$\rm{\mu m}$ ($\rm{C_{10\mu m}}$). 
\end{itemize}

\begin{figure*}[!t]
\begin{center}
\includegraphics[width=1.0\columnwidth]{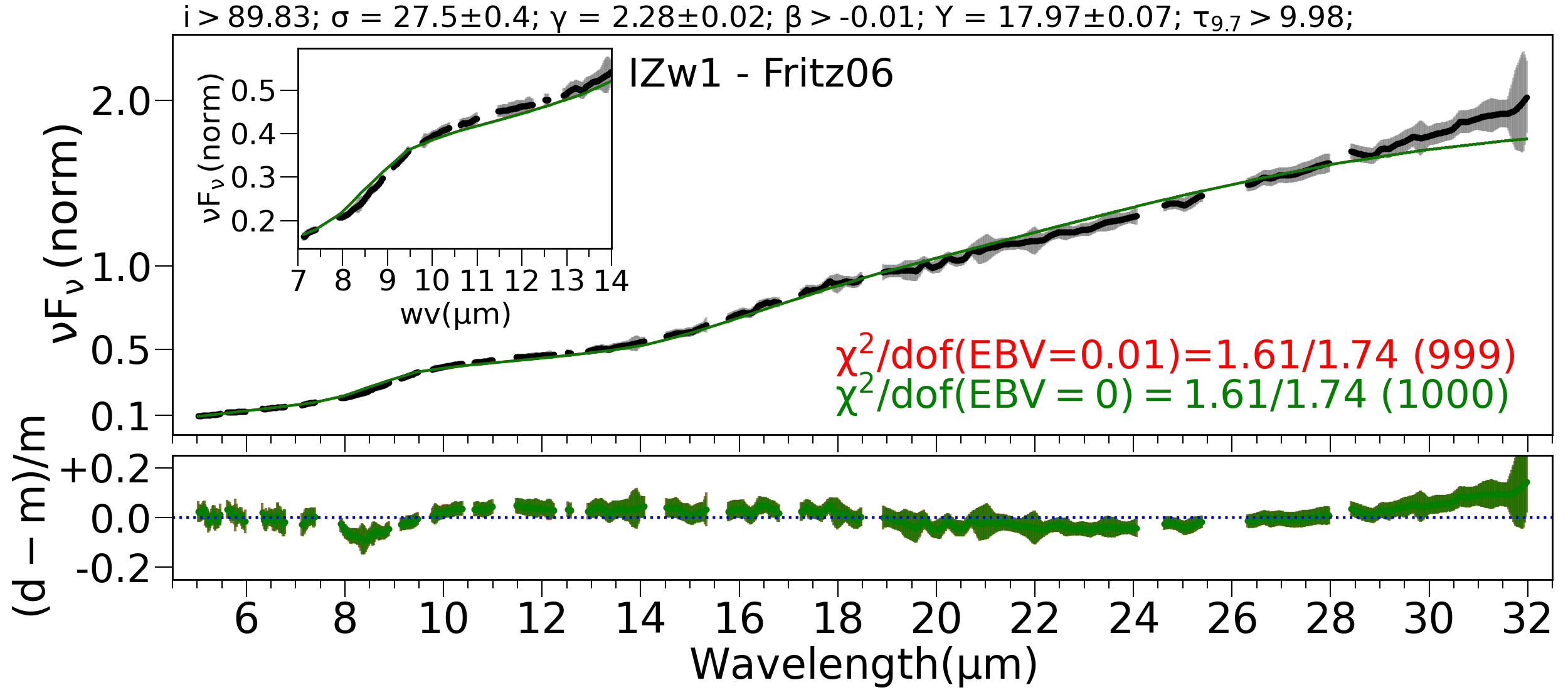} 
\includegraphics[width=1.0\columnwidth]{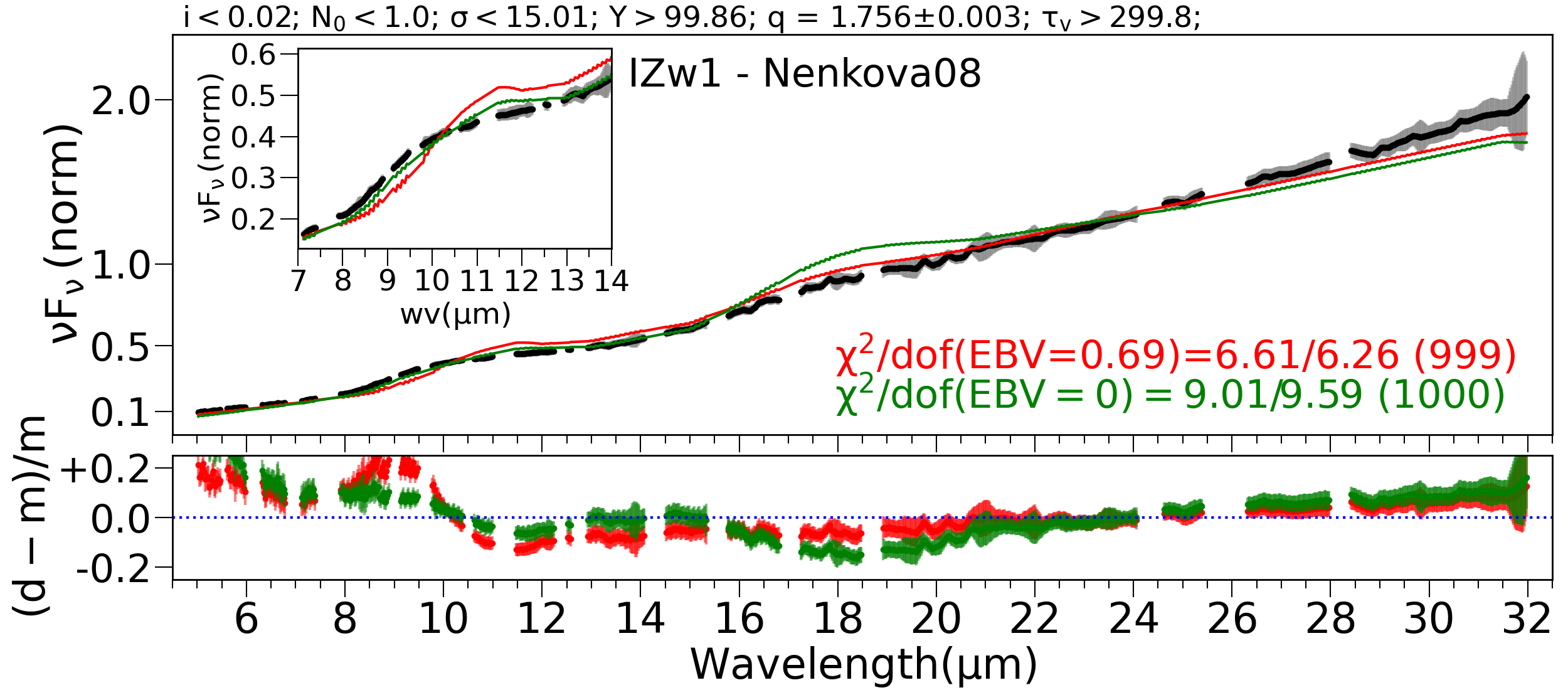}
\includegraphics[width=1.0\columnwidth]{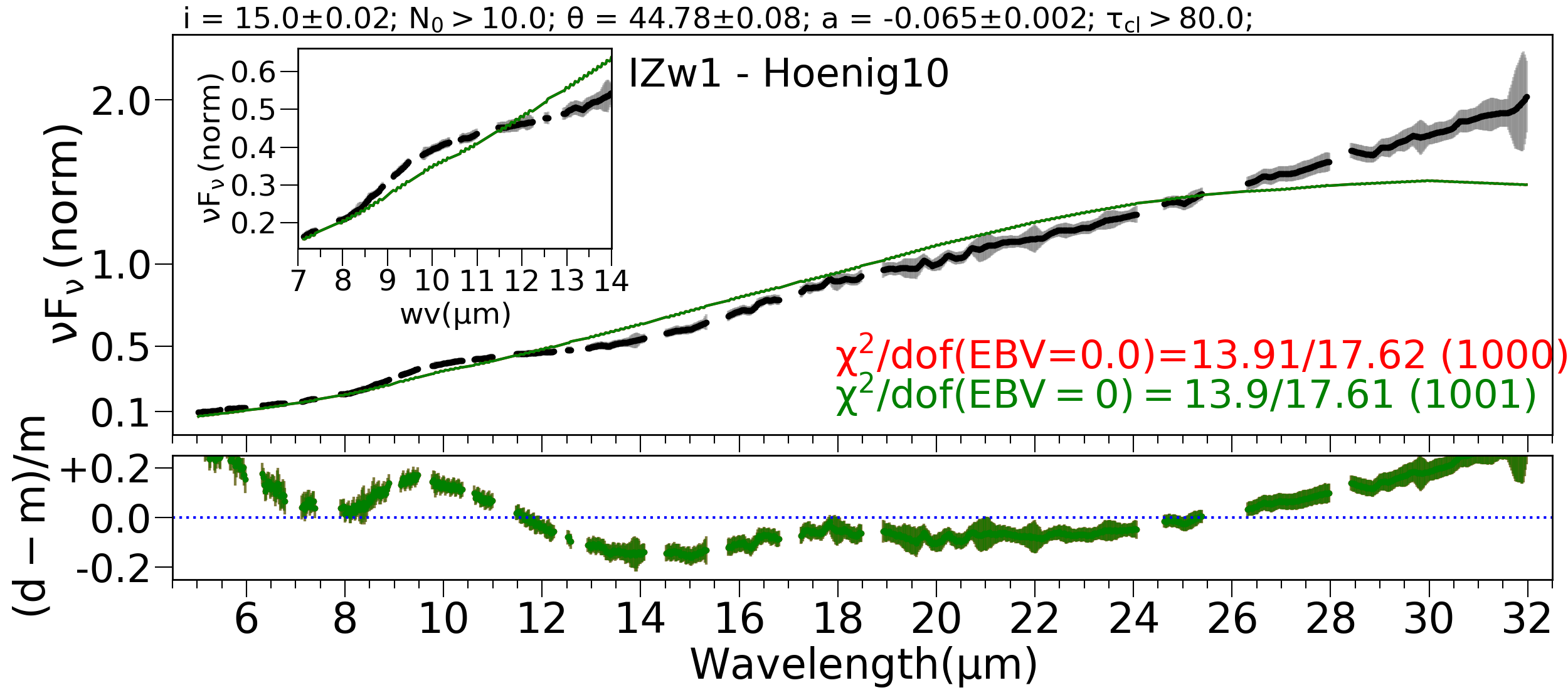} 
\includegraphics[width=1.0\columnwidth]{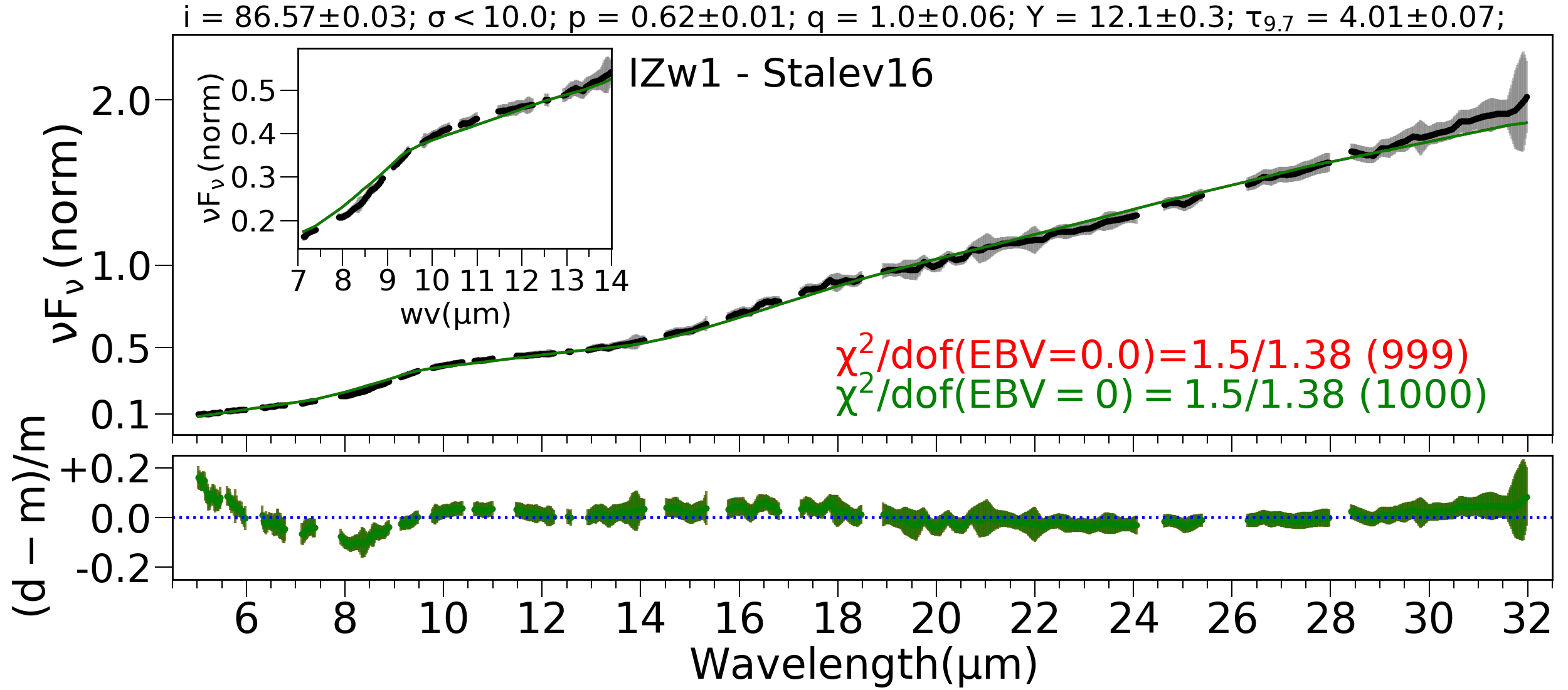} 
\includegraphics[width=1.0\columnwidth]{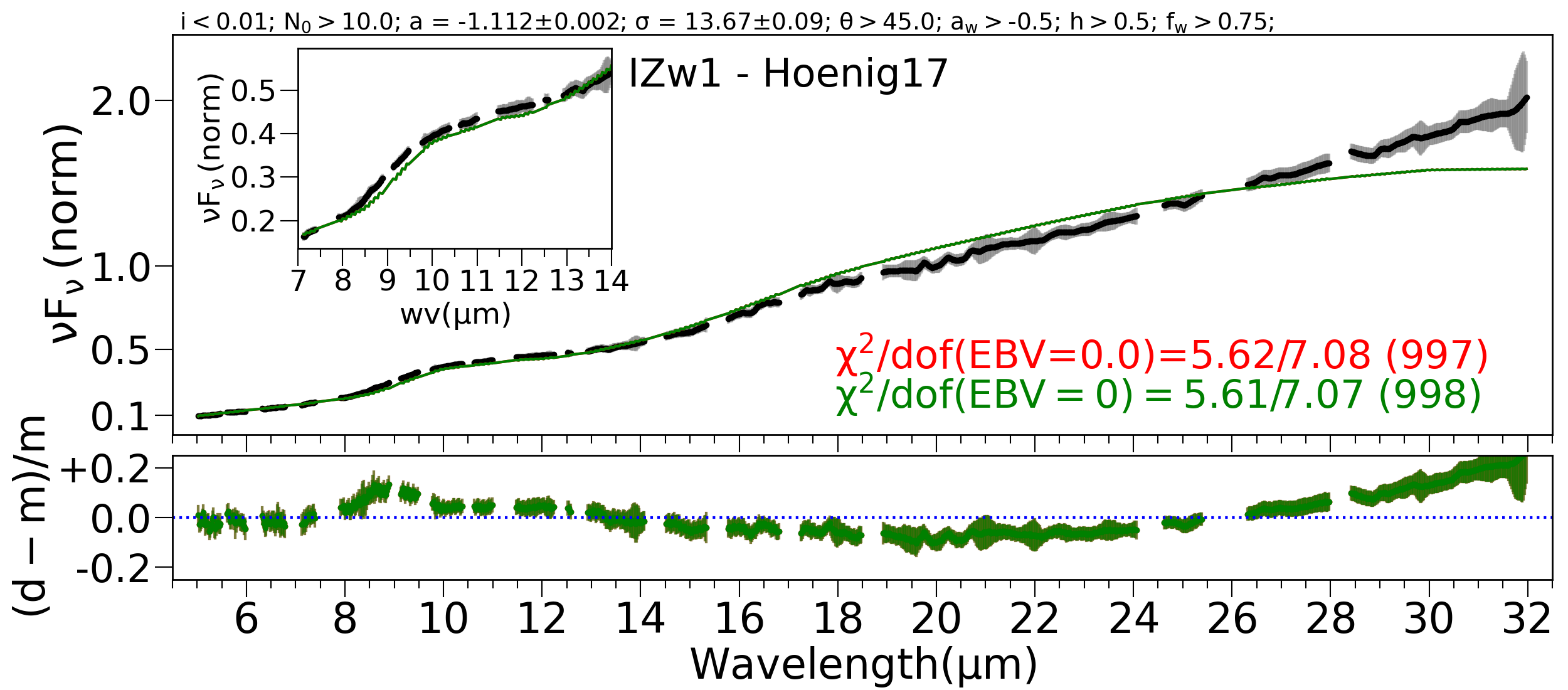} 
\includegraphics[width=1.0\columnwidth]{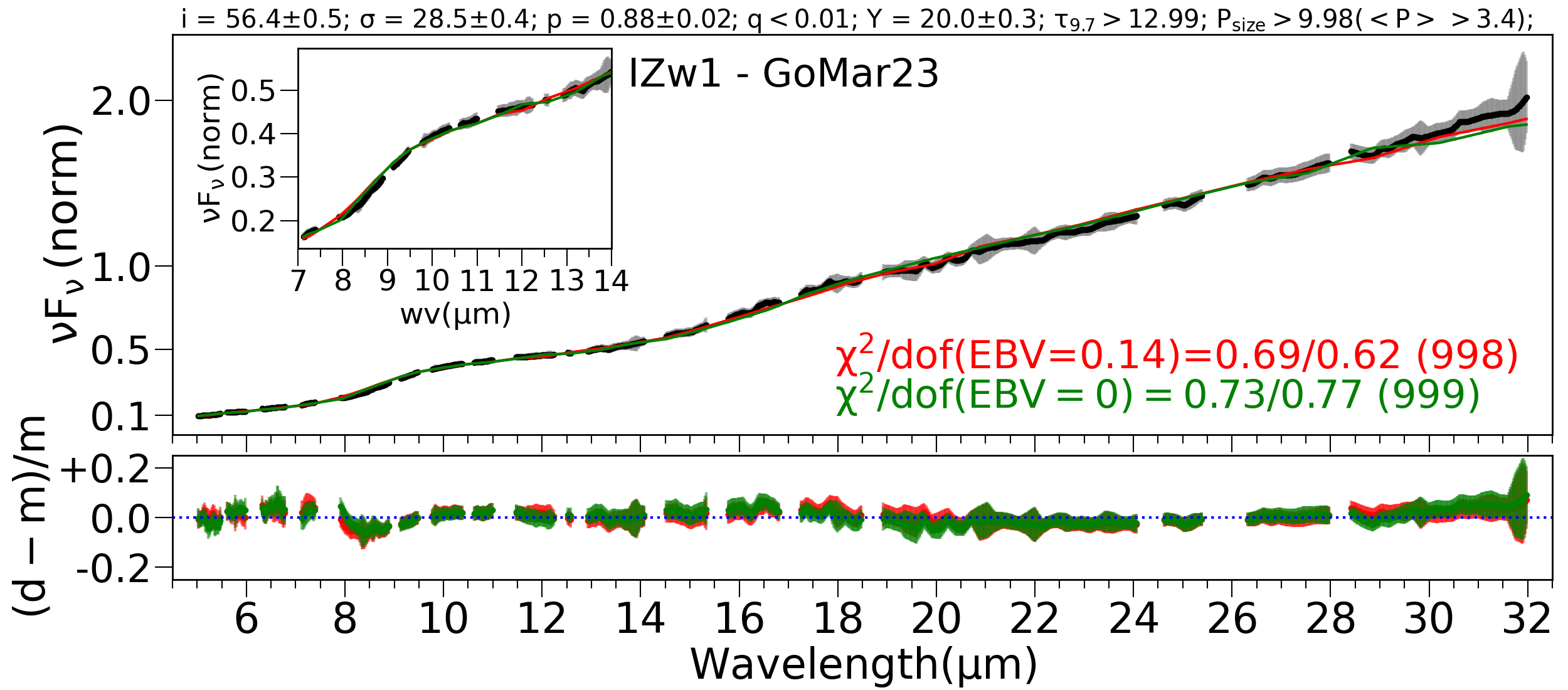} 
\caption{Spectral fits obtained for the type-1 AGN IZw1. The black line and gray shaded areas show the \emph{Spitzer} spectrum and its error bars, respectively. The red and green lines show the best SED for each model with and without foreground extinction, respectively (see text). The top panel shows the spectral fit and the bottom panel shows the residuals. The inset within the top panel shows zoom-in to the 7-14$\rm{\mu m}$ wavelength range associated with the 9.7$\rm{\mu m}$ silicate feature. Reduced $\rm{\chi^2}$ is given at the bottom-right corner of the top panel (error/model), including the degree of freedom within brackets (red and green for the SED with and without foreground extinction). Above each panel, we show the resulting parameters for each model when foreground extinction is included.}
\label{fig:IZw1fit}
\end{center}
\end{figure*}

Numerical works in the literature, yielding theoretical SEDs for AGN emission, usually provide thousands or more, and one quick way to check their reliability is through diagrams as those presented in this section \citep[e.g.][]{Garcia-Gonzalez17,Gonzalez-Martin19B}. Nevertheless, these diagrams are only partially informative of the ability of a model to represent the emission of real objects. What they can highlight, is the inability of a model to reproduce a given set of observational features. But a successful model should be able to properly reproduce all the spectral features with one single parameter set, something that such kinds of diagrams are unable to capture. An appropriate SED should hence at least show the same range of slope and silicate strength as observed, and to approach the problem we calculate the number of SEDs able to reproduce the observed spectral features for each object. Among the 68 objects in our sample, 25, 34, 6, 20, 9, and 59 objects can be reproduced (i.e. their spectral shape measurements are compatible with those of the SEDs, within errors) with at least one SED within [Fritz06], [Nenkova08], [Hoenig10], [Stalev16], [Hoenig17], and [GoMar23] models, respectively. Therefore, although the available models include SEDs with the appropriate range for each of these spectral characteristics, they lack the right combination to reproduce all the spectral shape features simultaneously. It is worth noticing that the model able to reproduce all the spectral characteristics for the largest number of objects is [GoMar23] model. However, this is just a crude analysis since the spectral fitting is required to interpolate among the SED libraries and add foreground extinction when appropriate. This is the main subject of the following section. 

\begin{figure*}[!t]
\begin{center}
\includegraphics[width=0.74\columnwidth, clip, trim=0 0 40 20]{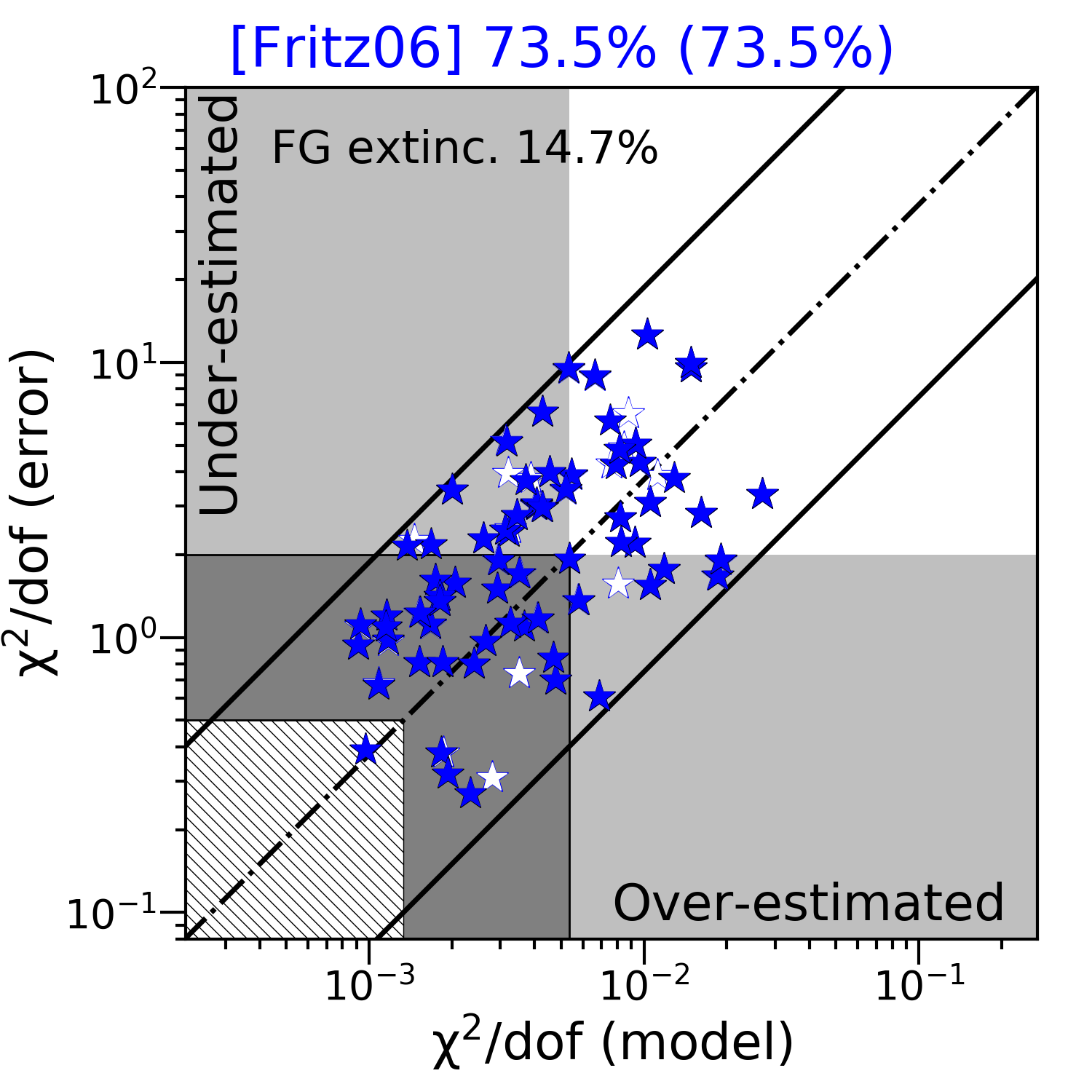} 
\includegraphics[width=0.625\columnwidth, clip, trim=160 0 40 20]{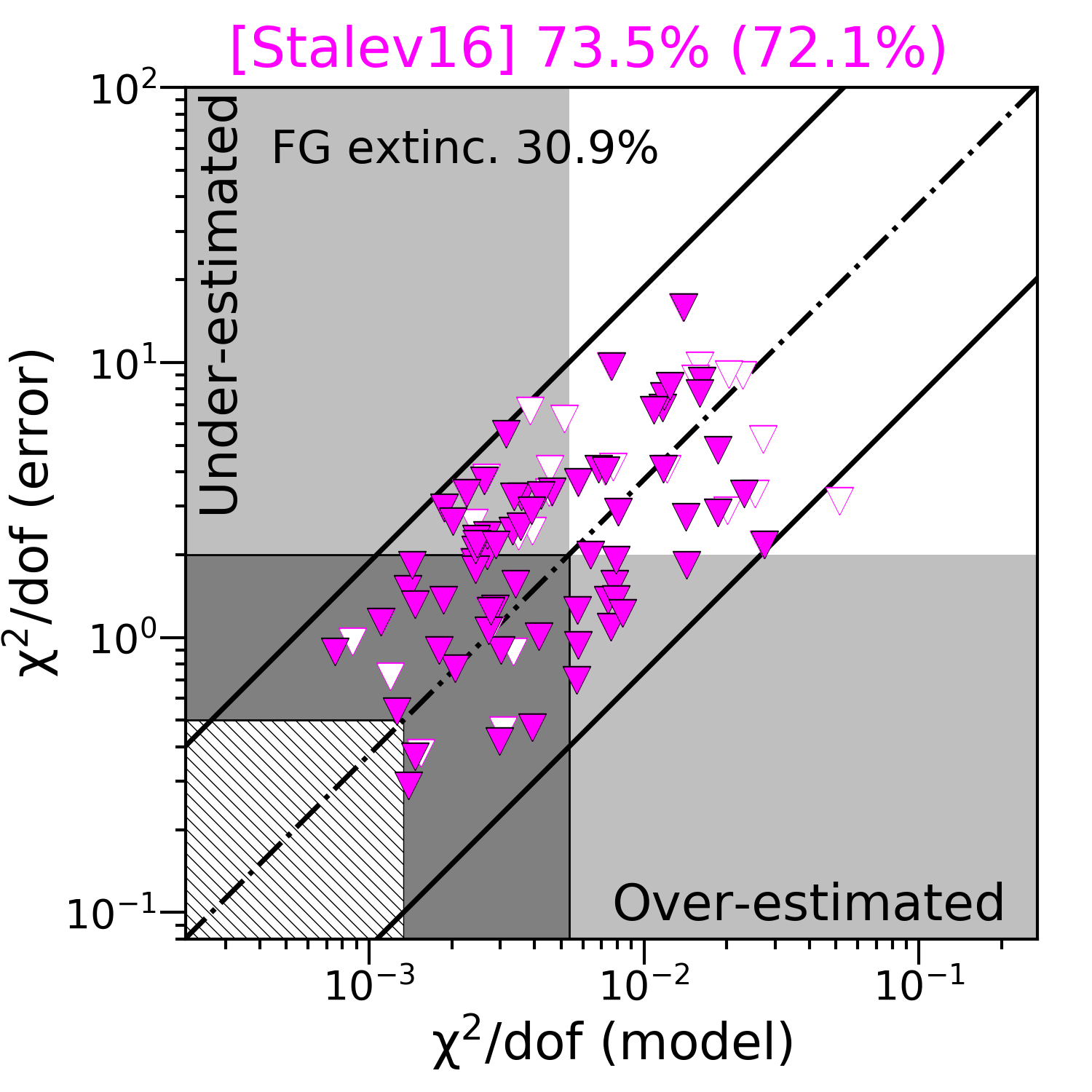} 
\includegraphics[width=0.625\columnwidth, clip, trim=160 0 40 20]{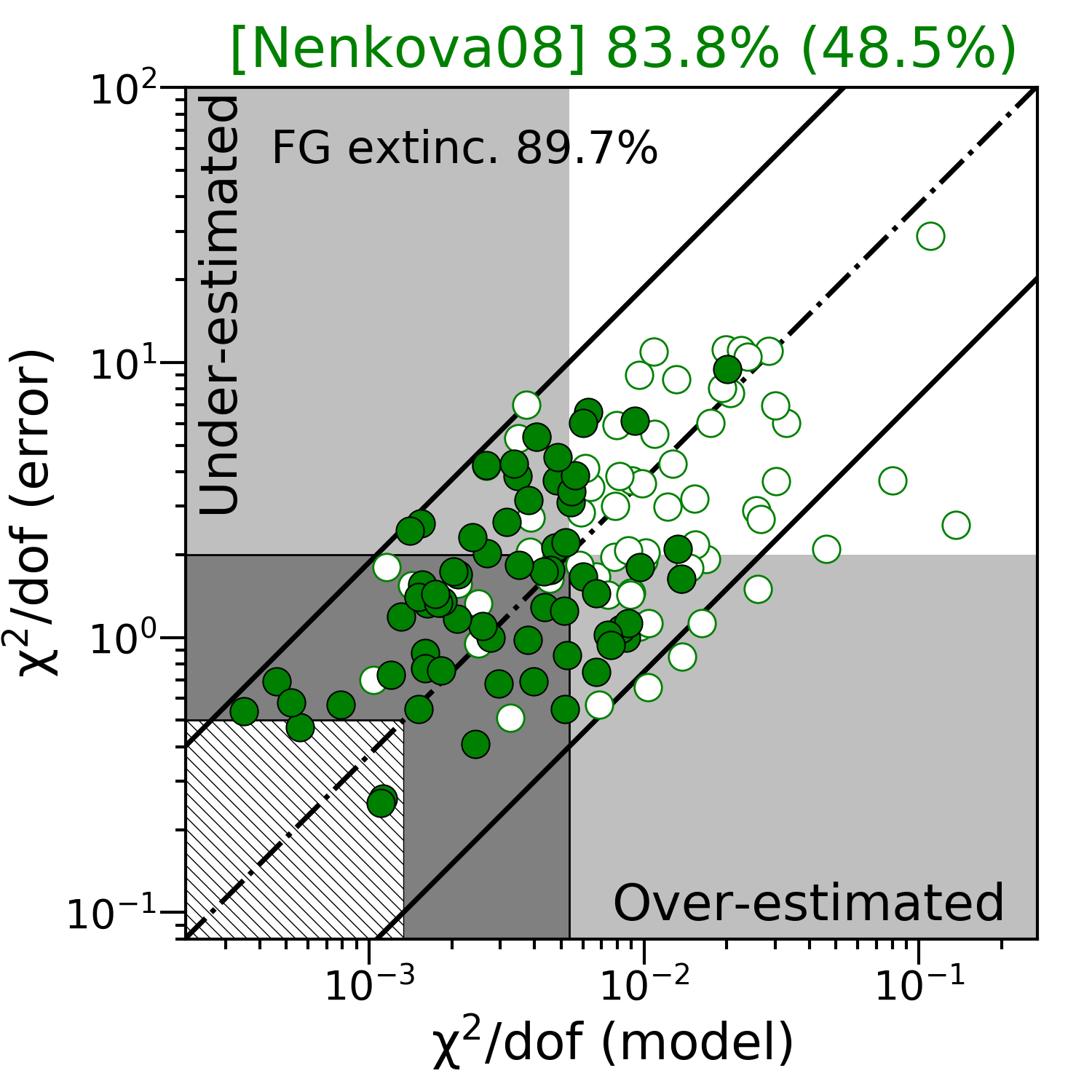}
\includegraphics[width=0.74\columnwidth, clip, trim=0 0 40 20]{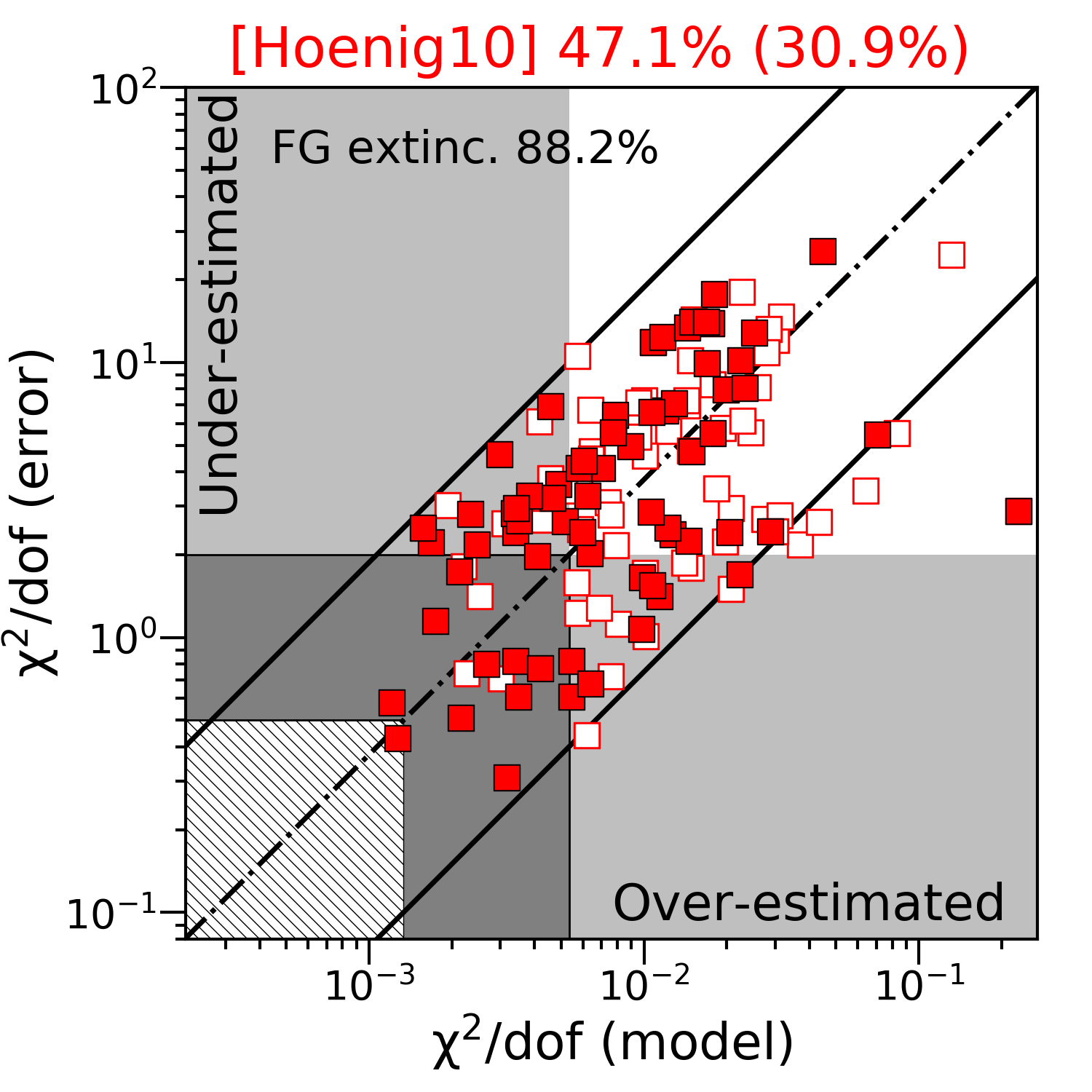}
\includegraphics[width=0.625\columnwidth, clip, trim=160 0 40 20]{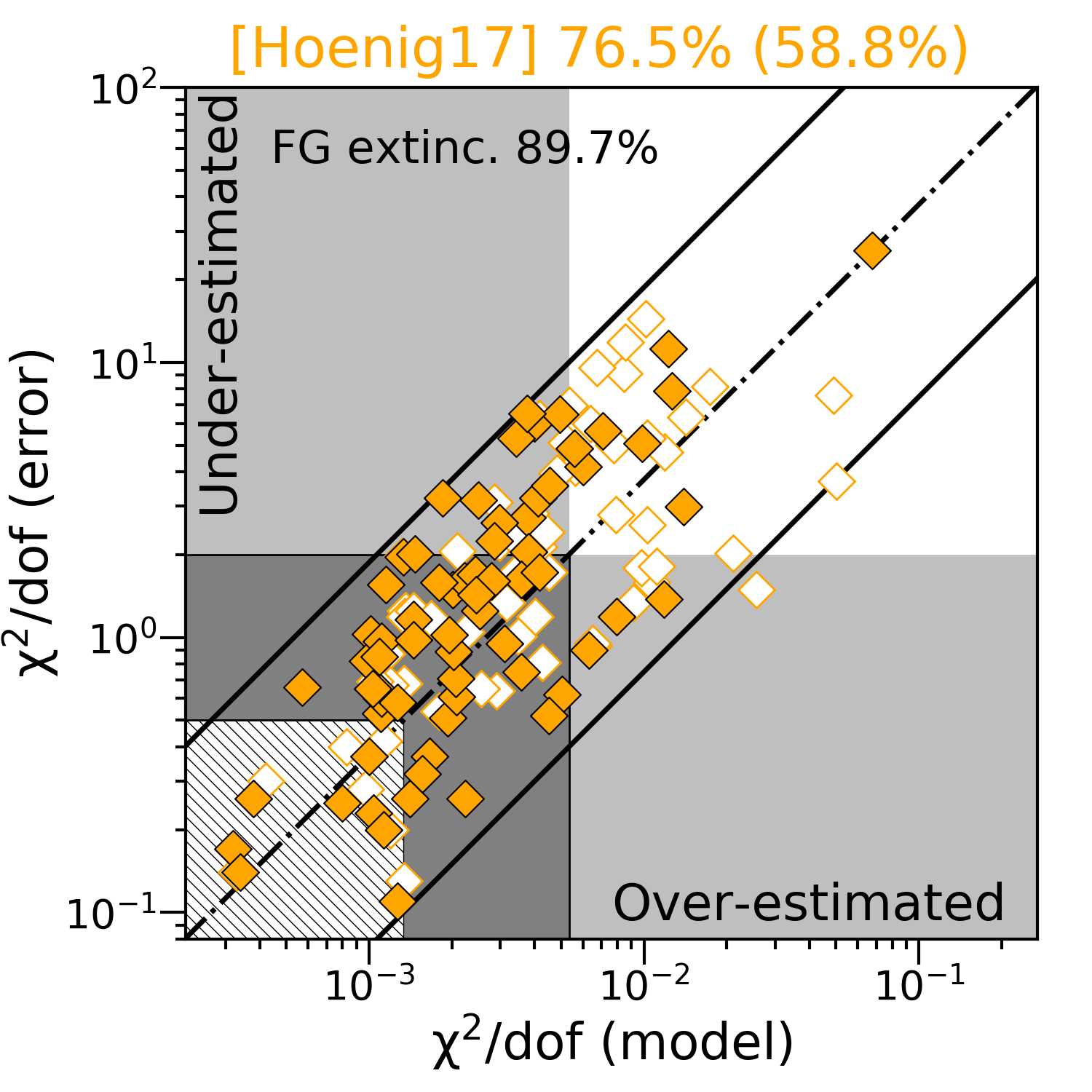}
\includegraphics[width=0.625\columnwidth, clip, trim=160 0 40 20]{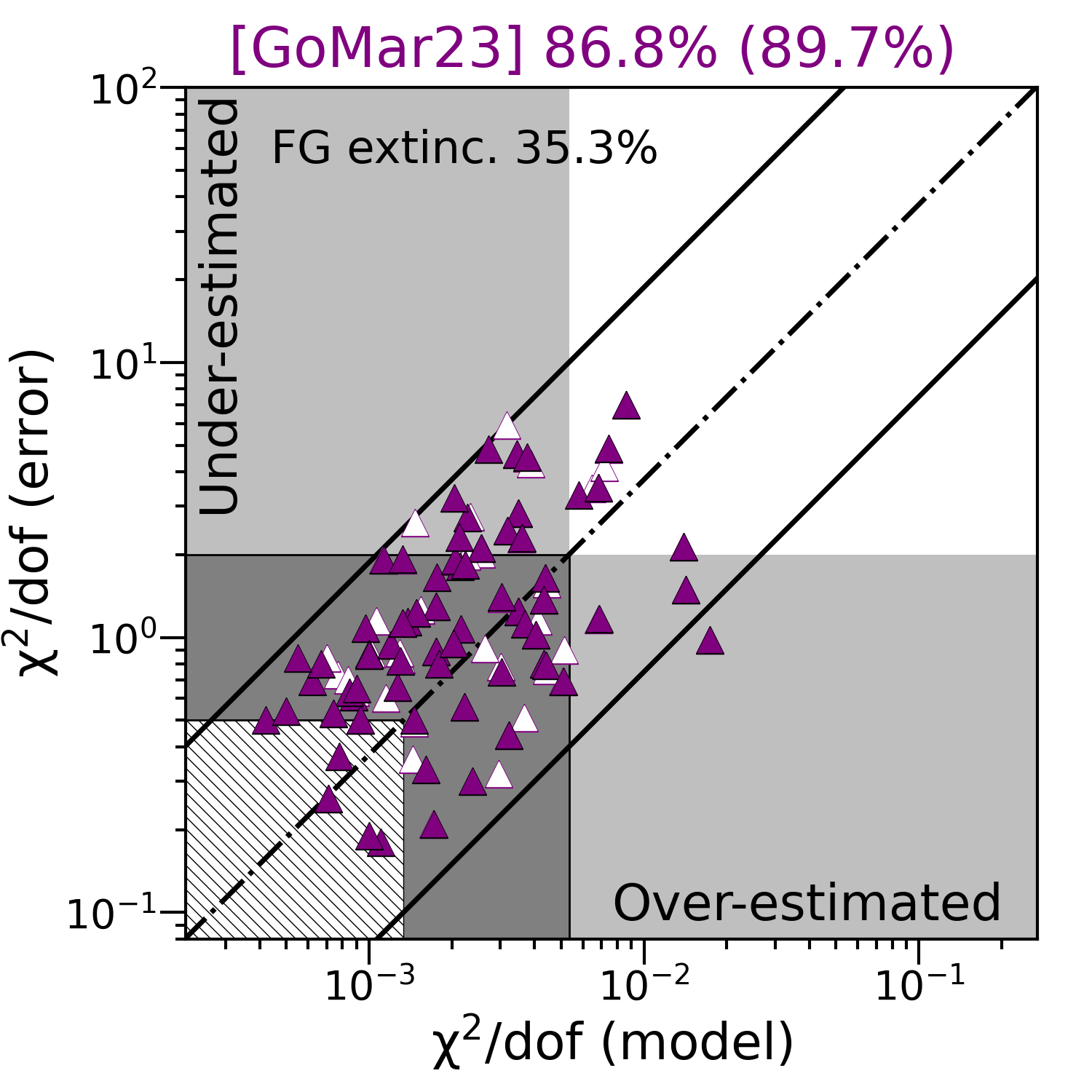}
\caption{The reduced $\rm{\chi^2}$ using the errors in the $\rm{\chi^2}$ definition ($\rm{\chi^2(error)/dof}$, see Eq.\,\ref{eq:chierror}), versus that using the model in the $\rm{\chi^2}$ definition ($\rm{\chi^2(model)/dof}$, see Eq.\,\ref{eq:chimodel}). Filled and empty symbols show the values when foreground extinction is included and neglected in the spectral fitting procedure, respectively. The dot-dashed line shows the locus expected if both measurements agree with each other. The continuous lines show a factor of five of agreement between both $\rm{\chi^2}$ definitions. An object located in the blank area shows complexity not well fitted by the model. The dashed area shows the locus when the model is too complex for the data ($\rm{\chi^2/dof<0.5}$). The dark-gray shaded area shows the locus where both reduced $\rm{\chi^2}$ indicate that the model provides a statistically acceptable fit (i.e. $\rm{\chi^2/dof (error) <2}$ and $\rm{\chi^2/dof (model) <1.5\times 10^{-3}}$). The light-gray shaded area shows the region where the object is expected to locate if the error bars of the spectra are overestimated and underestimated although the spectral fit is good. Each panel shows the results for one of the models. We show the percentage of good spectral fits per model, considering those objects in the gray areas (the result without foreground extinction is shown next to it). The percentage of objects statistically requesting foreground extinction (FG extinct.) is shown in the top-left corner of each figure.}
\label{fig:chis}
\end{center}
\end{figure*}

\subsection{Spectral fits}

We fitted the spectra of our AGN sample using the five models reported before and the new [GoMar23] model presented in this paper. We use $\rm{\chi^2 (error)}$ to minimize the errors and therefore obtain the best set of parameters among the SED libraries. As an example of a typical type-1 AGN, Fig.\,\ref{fig:IZw1fit} shows the resulting best fits for IZw1 using each of the tested models. Note that we performed the fits with (red) and without (green) foreground extinction to show its impact on the final fit. In the case of IZw1, it is clear that [Nenkova08], [Hoenig10], and [Hoenig17] models cannot reproduce the observations. Significantly better fits ($\rm{\chi^2/dof <2}$) are obtained by [Fritz06], [Stalev16], and [GoMar23] models, the last model being statistically preferred. The same conclusion is drawn from this object if no foreground extinction is included.  Interestingly, [Fritz06], [Stalev16], and [GoMar23] models converge to a thin torus with $\rm{\sigma \sim 10-30^{\circ}}$. In this case the maximum dust grain size is $\rm{P_{size}\sim 10 \mu m}$ ($\rm{ <P>\sim 3.41 \mu m}$), well above the canonical 0.25$\rm{\mu m}$ ($\rm{ <P>\sim 0.097 \mu m}$) assumed in previous models.  

We consider good fits those with $\rm{0.5<\chi^2_{r} (error)<2}$. Values of $\rm{\chi^2_{r} (error)<0.5}$ indicate that the model is too complex for the data and values of $\rm{\chi^2_{r} (error)>2}$ are found when the model is too simplistic for the data. However, as explained in Appendix\,\ref{sec:stats}, $\rm{\chi^2_{r} (error)}$ might also show low values due to the overestimation of the error bars while large $\rm{\chi^2_{r} (error)}$ might indicate an underestimation. For this reason, we explore the evaluation of the goodness of the fit using both $\rm{\chi^2 (error)}$ and $\rm{\chi^2 (model)}$ (see Eqs.\,\ref{eq:chierror}-\ref{eq:chimodel} in Appendix\,\ref{sec:stats}). Fig.\,\ref{fig:chis} shows the comparison between these two $\rm{\chi^2}$ for each of the models. Objects below $\rm{\chi^2 (error)=0.5}$ and $\rm{\chi^2 (model)=1.5 \times 10^{-3}}$ (dashed area) are over-fitted by the model. This is particularly relevant for [Hoenig17] model (with eight objects in the dashed area) which indeed is the model with the highest number of free parameters.
[GoMar23], [Fritz06], [Nenkova08] and [Hoenig10] have 4, 1, 1, and 1 objects in the dashed area of Fig.\,\ref{fig:chis}, respectively. $\rm{\chi^2_{r} (model)\sim 5.5 \times 10^{-3}}$ is equivalent to $\rm{\chi^2_{r} (error)\sim 2}$ and $\rm{\chi^2_{r} (model)\sim 1.5 \times 10^{-3}}$ is equivalent to $\rm{\chi^2_{r} (error)\sim 0.5}$, with both quantities correlated. Fig.\,\ref{fig:chis} guarantee that the use of $\rm{\chi^2_{r} (error)}$ is equivalent to $\rm{\chi^2_{r} (model)}$. We therefore use $\rm{\chi^2_{r} (error)}$ in the subsequent analysis.

\begin{figure*}[!t]
\begin{center}
\includegraphics[width=2.\columnwidth, clip, ,trim=0 30 15 20]{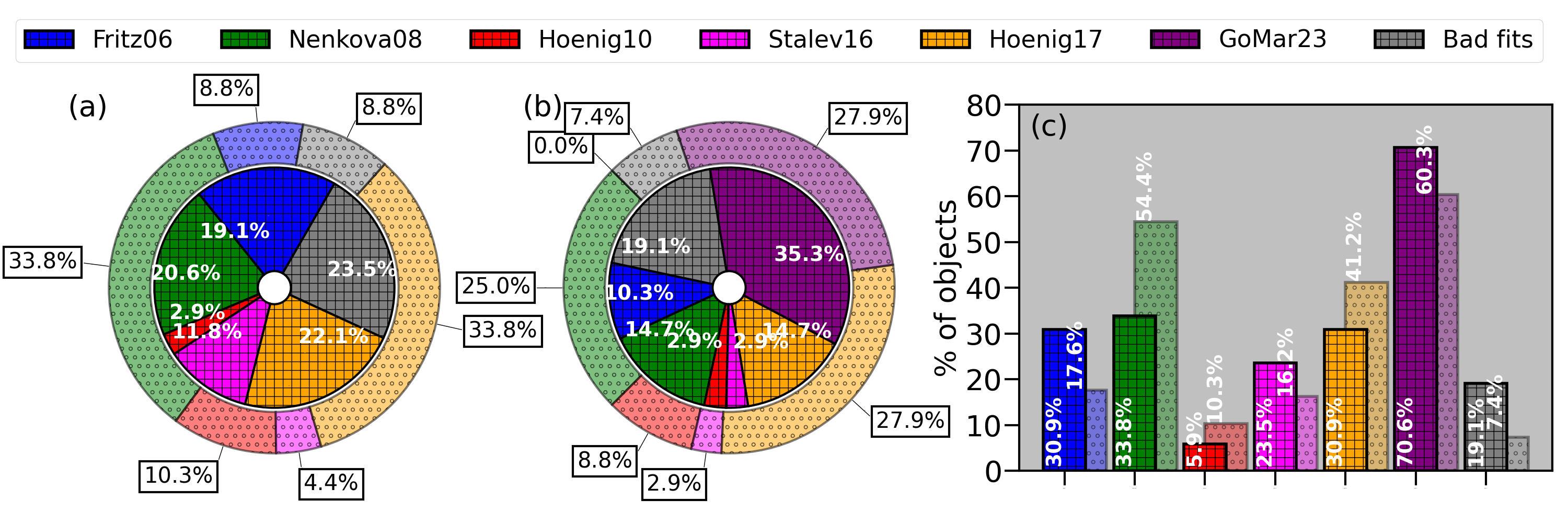} 
\caption{Panels (a) and (b) show the percentage of object best fitted with each of the models tested before (panel a) and after (panel b) the inclusion of [GoMar23] model. The inner (squared-filled) and outer (circle-filled) pie charts show the results without and with foreground extinction in the spectral fits, respectively. Panel (c) shows the final percentage of objects best fitted to each model when including statistically similar fits using AIC probability (see text). The square-filled and circle-filled areas show the results without and with foreground extinction in the spectral fits, respectively. 
}
\label{fig:GoodFitsHist}
\end{center}
\end{figure*}

Before attempting to obtain the final statistics of good fits, we explore the role of foreground extinction. Fig.\,\ref{fig:chis} shows the results when foreground extinction is excluded (empty symbols) and included (filled symbols) into the spectral fit. The percentage of objects well fitted with each model is included on top of each panel (the percentage within brackets shows the results without foreground extinction). Furthermore, we also quantify the number of objects requesting foreground extinction by comparing the $\rm{\chi^2}$ statistics with and without foreground extinction using the f-statistic test (f-value $\rm{<10^{-4}}$ implies statistically need for foreground extinction). [Nenkova08], [Hoenig10], and [Hoenig17] models require foreground extinction for $\rm{\sim 90\%}$ of the sample (i.e. the spectral fit is statistically better compared with the result when $\rm{E(B-V)=0}$) while [Fritz06], [Stalev16], and [GoMar23] keep this percentage at $\rm{\sim 15-35\%}$, with the lowest percentage obtained by the [Fritz06] model ($\rm{\sim}$15\%). Indeed, the use of foreground extinction strongly affects the percentage of objects well-fitted by the models. The most extreme example is [Nenkova08] model, which goes from $\rm{\sim}$50\% of successful fits up to above 80\% when foreground extinction is included. [Hoenig10] and [Hoenig17] also show an increment of $\rm{\sim}$15\% on the number of objects well fitted. In the opposite scenario, the percentage of good fits is not affected at all for [Fritz06] model, while a minor effect (below 5\%) is seen for [Stalev16] and [GoMar23] models. Therefore, we find that a smooth distribution of dust helps to mitigate the need for foreground extinction, as expected since smooth distribution produces deeper silicate absorption features. The role of foreground extinction is further explored in Appendix\,\ref{sec:FGextinction}.

We also studied which is the best model for each object. Table\,\ref{tab:fitresults} shows the best model among the already available SED libraries in Col.\,2 and the corresponding $\rm{\chi^2/dof}$ in Col.\,3. We mark with a dot when $\rm{\chi^2/dof>2}$. We then report in Col.\,4 the $\rm{\chi^2/dof}$ obtained for the [GoMar23] model and the final best-fit model in Col.\,5. Panels (a) and (b) of Fig.\,\ref{fig:GoodFitsHist} summarize these results. The results strongly depend on the inclusion of foreground extinction (inner square-filled and outer circle-filled pie-charts when foreground extinction is excluded and included, respectively). If no foreground extinction is included for the spectral fitting procedure and before the use of [GoMar23] model, [Fritz06], [Nenkova08], and [Hoenig17] models equally fit $\rm{\sim 20}$\% of the sample and [Stalev16] model describes $\rm{\sim 12}$\% of the sample, respectively. [Hoenig10] model fails to describe most of the objects, being the best fit only in 3\% of the sample. The inclusion of foreground extinction significantly helps to improve [Nenkova08], [Hoenig10], and [Hoenig17] models, with a percentage of best fits of $\rm{\sim 34}$\%, $\rm{\sim 10}$\%, and $\rm{\sim 34}$\%, respectively (outer circle-filled pie-chart of Fig.\,\ref{fig:GoodFitsHist} panel a). 

When we include into the analysis [GoMar23] model (panel b in Fig.\,\ref{fig:GoodFitsHist}), it is the best model for $\rm{\sim 35}$\% of the sample before the inclusion of foreground extinction (inner pie-chart). Among the others, the most relevant are [Nenkova08] and [Hoenig17] model, which is the best fit for $\rm{\sim 15}$\% of the sample. However, when foreground extinction is included in the fit, there are three models describing $\rm{\sim 25-28}$\% of the sample: [Nenkova08] ($\rm{\sim 25}$\%), [Hoenig17] ($\rm{\sim 28}$\%) and [GoMar23] ($\rm{\sim 28}$\%) (outer circle-filled pie-chart of panel b in Fig.\,\ref{fig:GoodFitsHist}). 

\begin{figure*}[!t]
\begin{center}
\includegraphics[height=0.415\columnwidth, clip, ,trim=5 160 8 0]{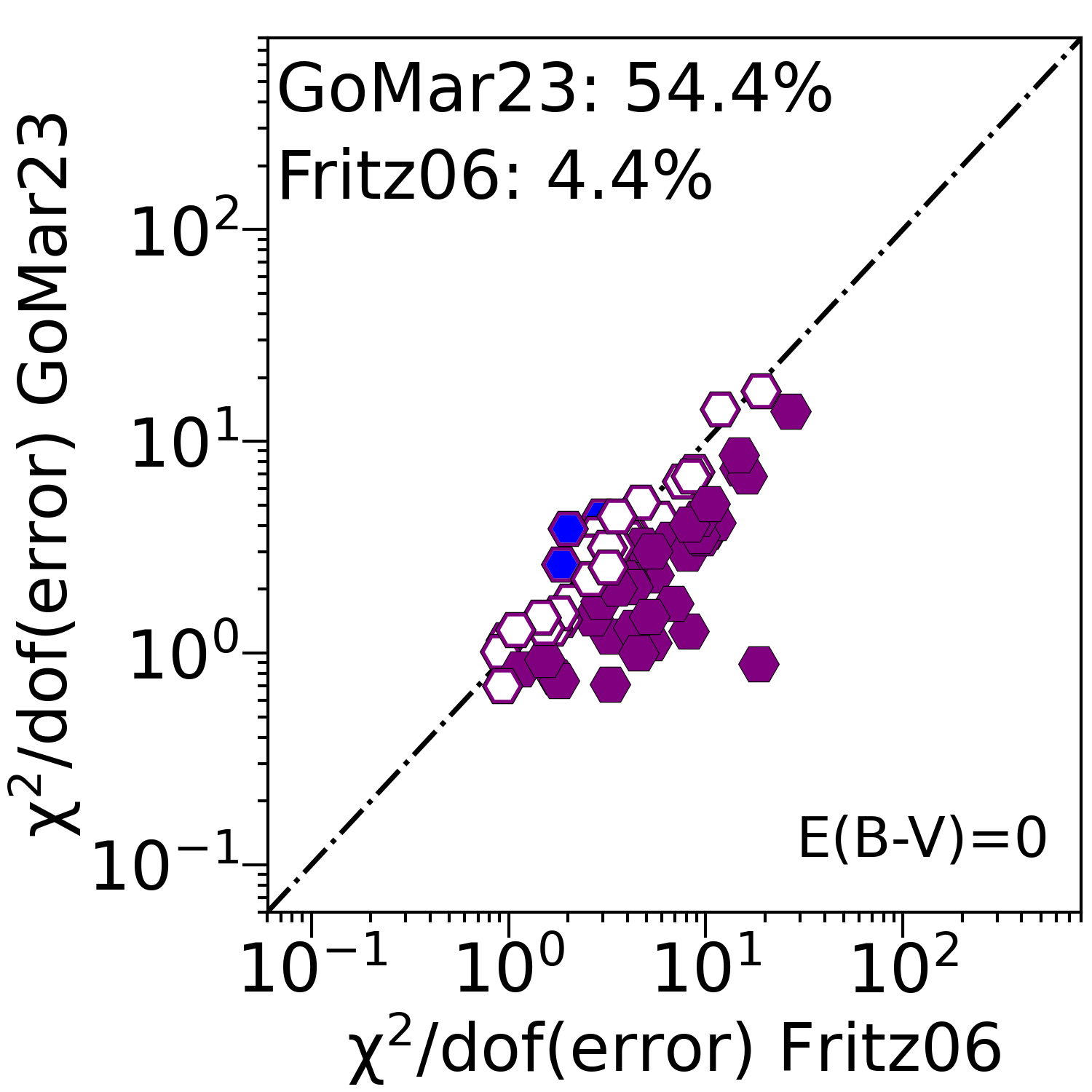} 
\includegraphics[height=0.415\columnwidth, clip, ,trim=245 160 8 0 ]{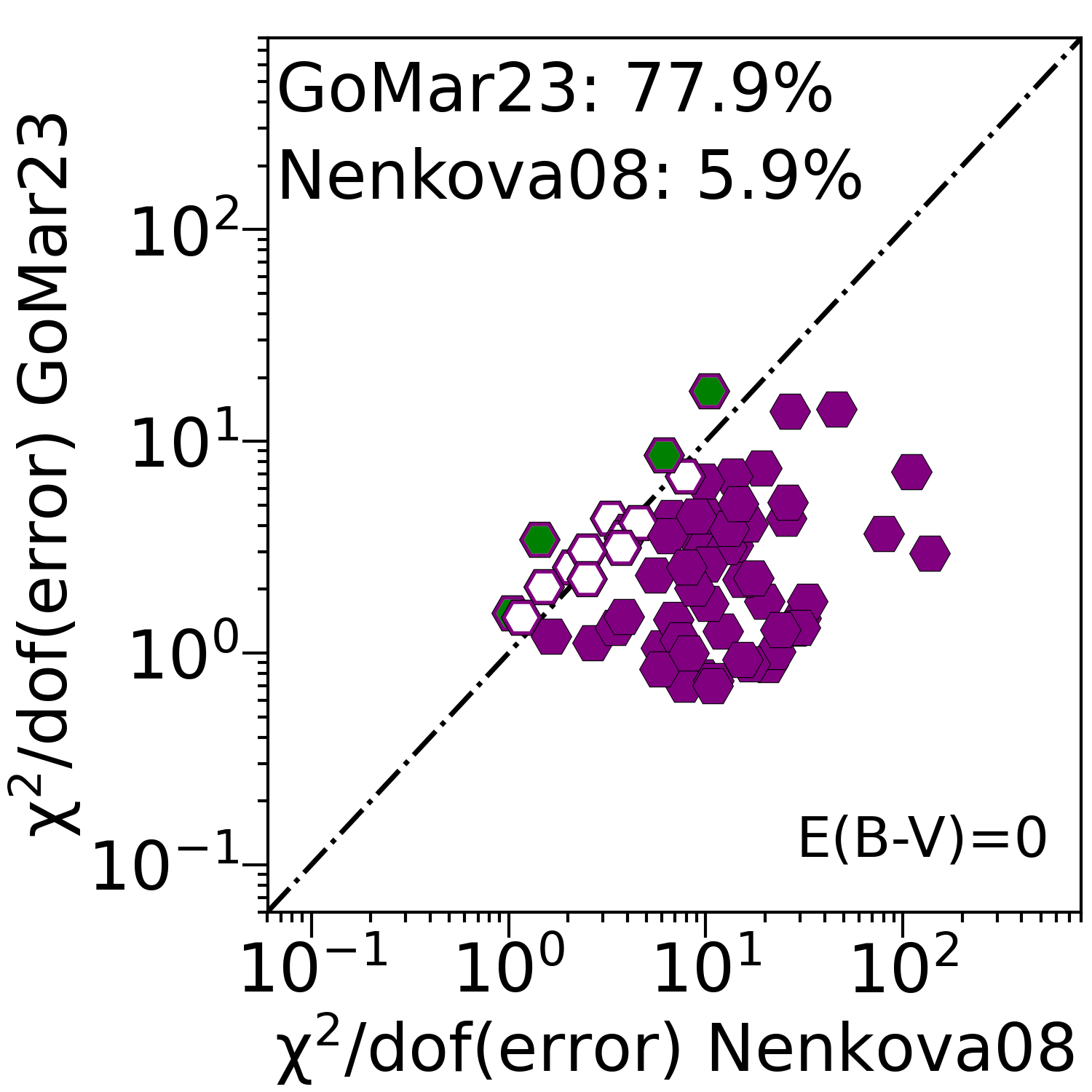}
\includegraphics[height=0.415\columnwidth, clip, ,trim=245 160 8 0]{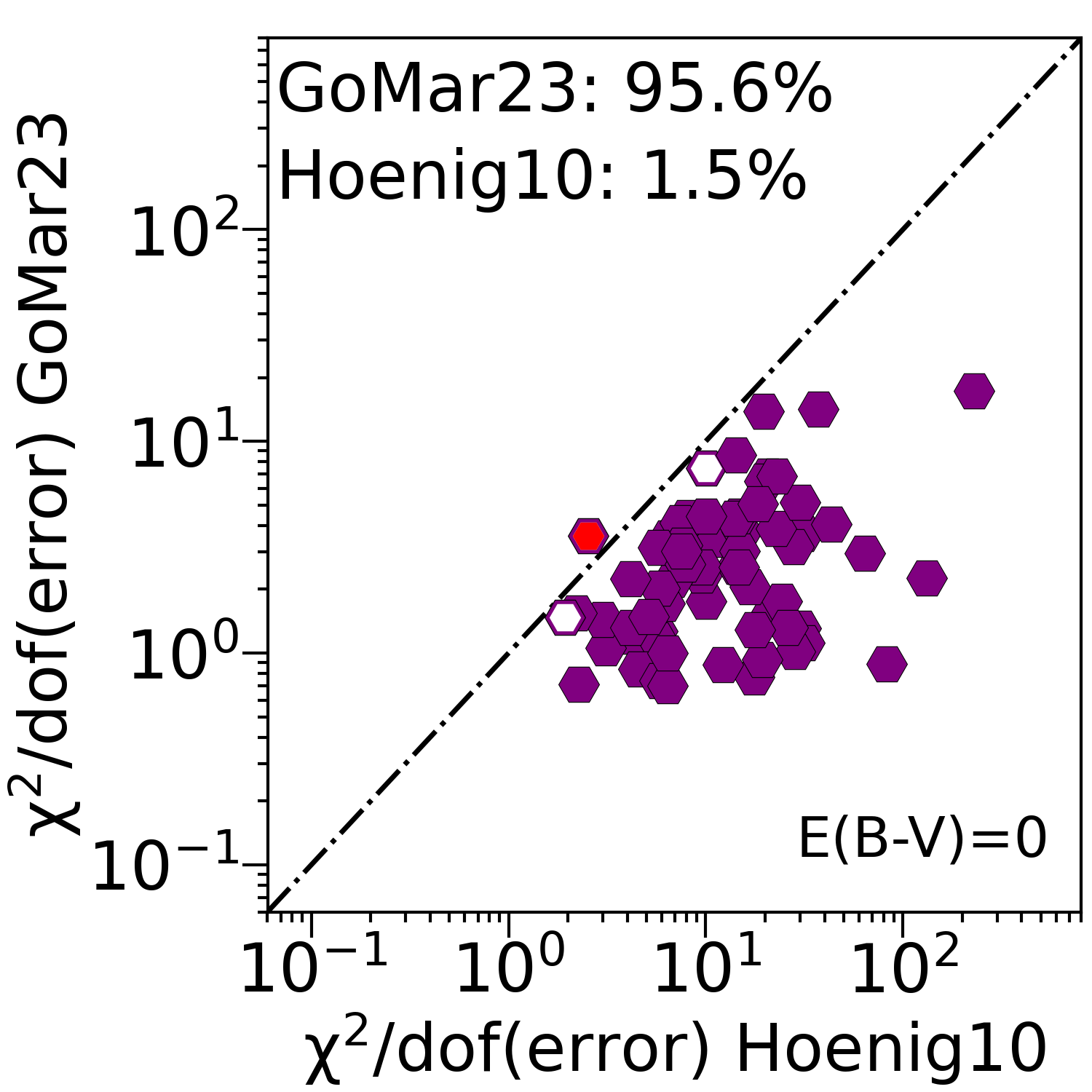} 
\includegraphics[height=0.415\columnwidth, clip, ,trim=245 160 8 0]{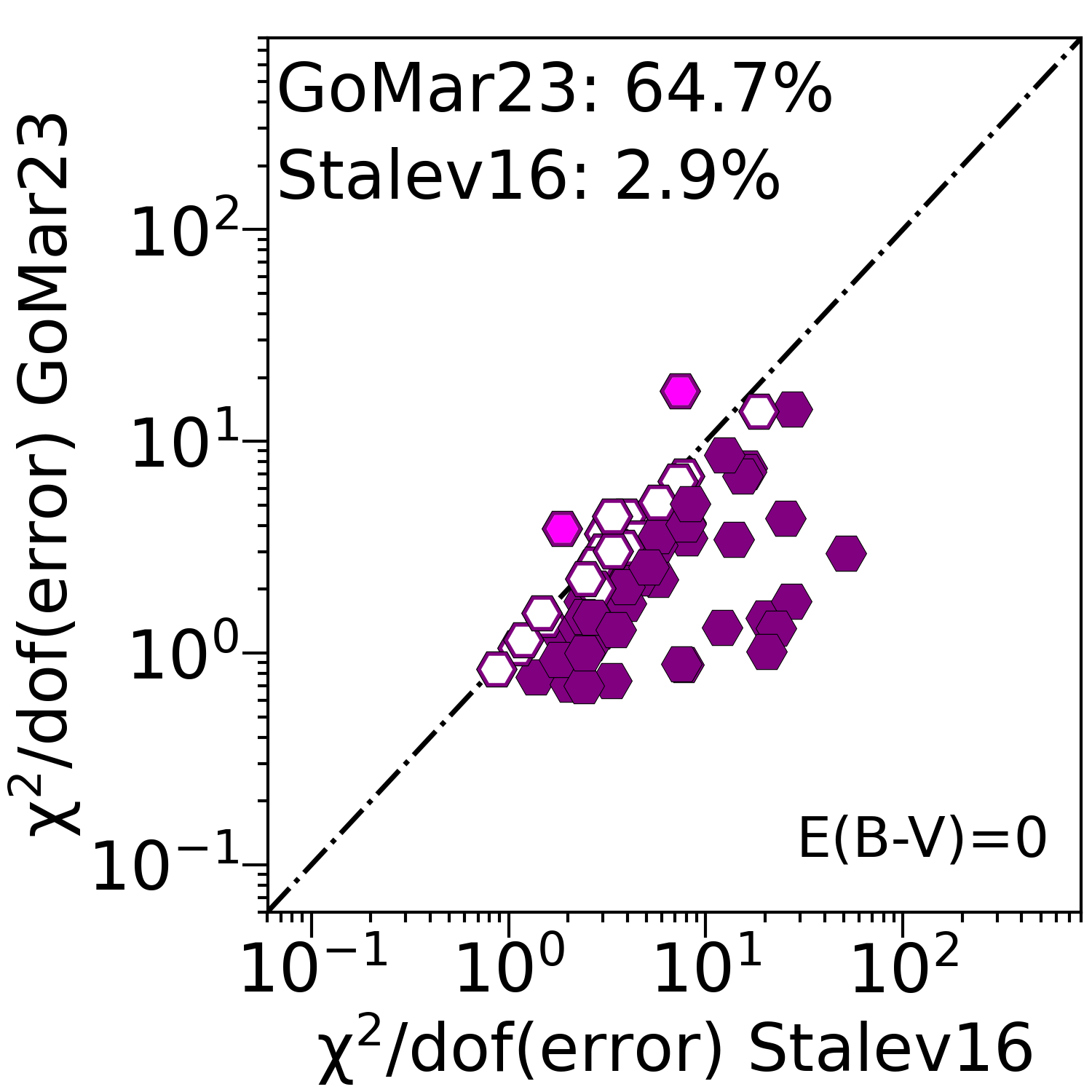} 
\includegraphics[height=0.415\columnwidth, clip, ,trim=245 160 8 0]{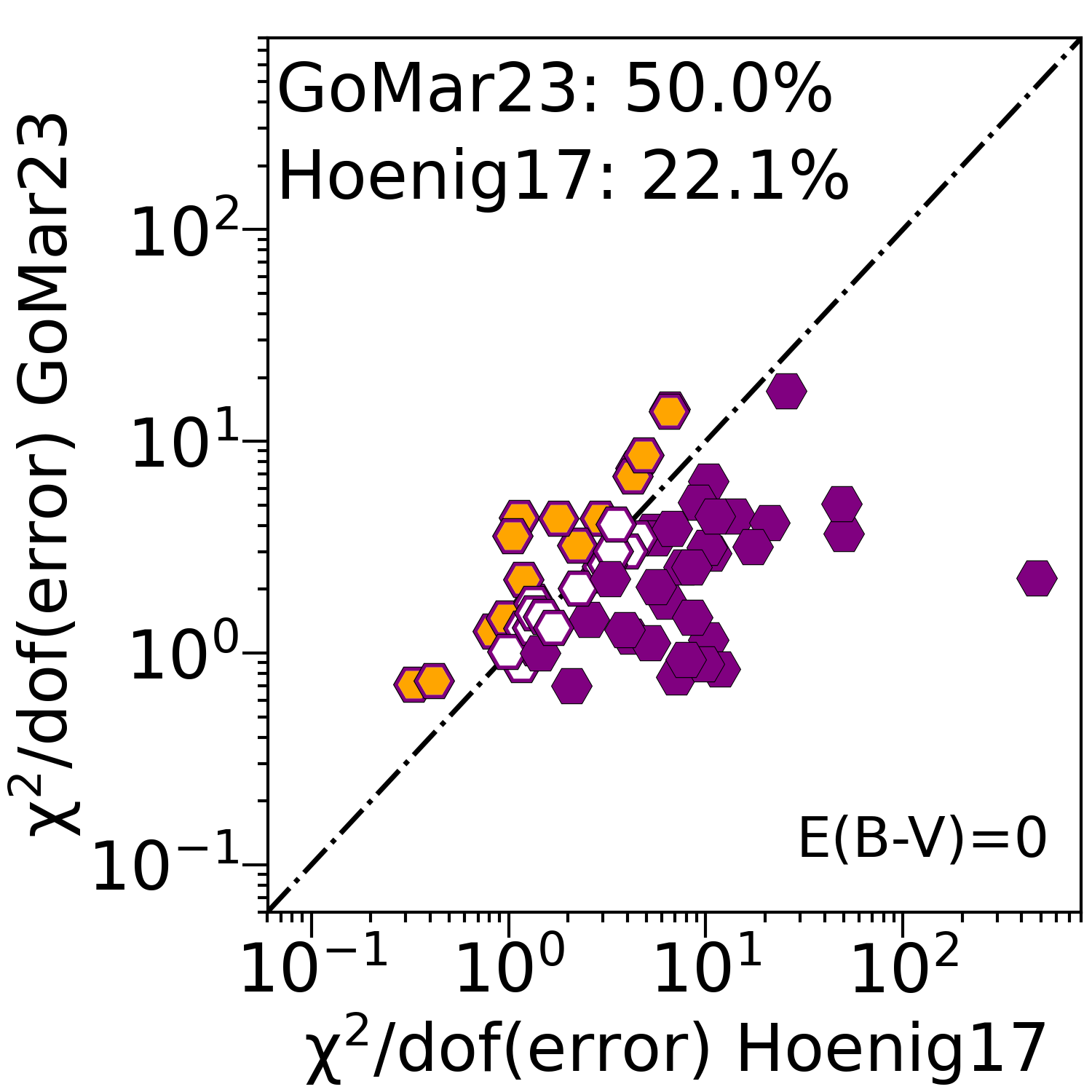} 
\includegraphics[height=0.485\columnwidth, clip, ,trim=5 0 8 0]{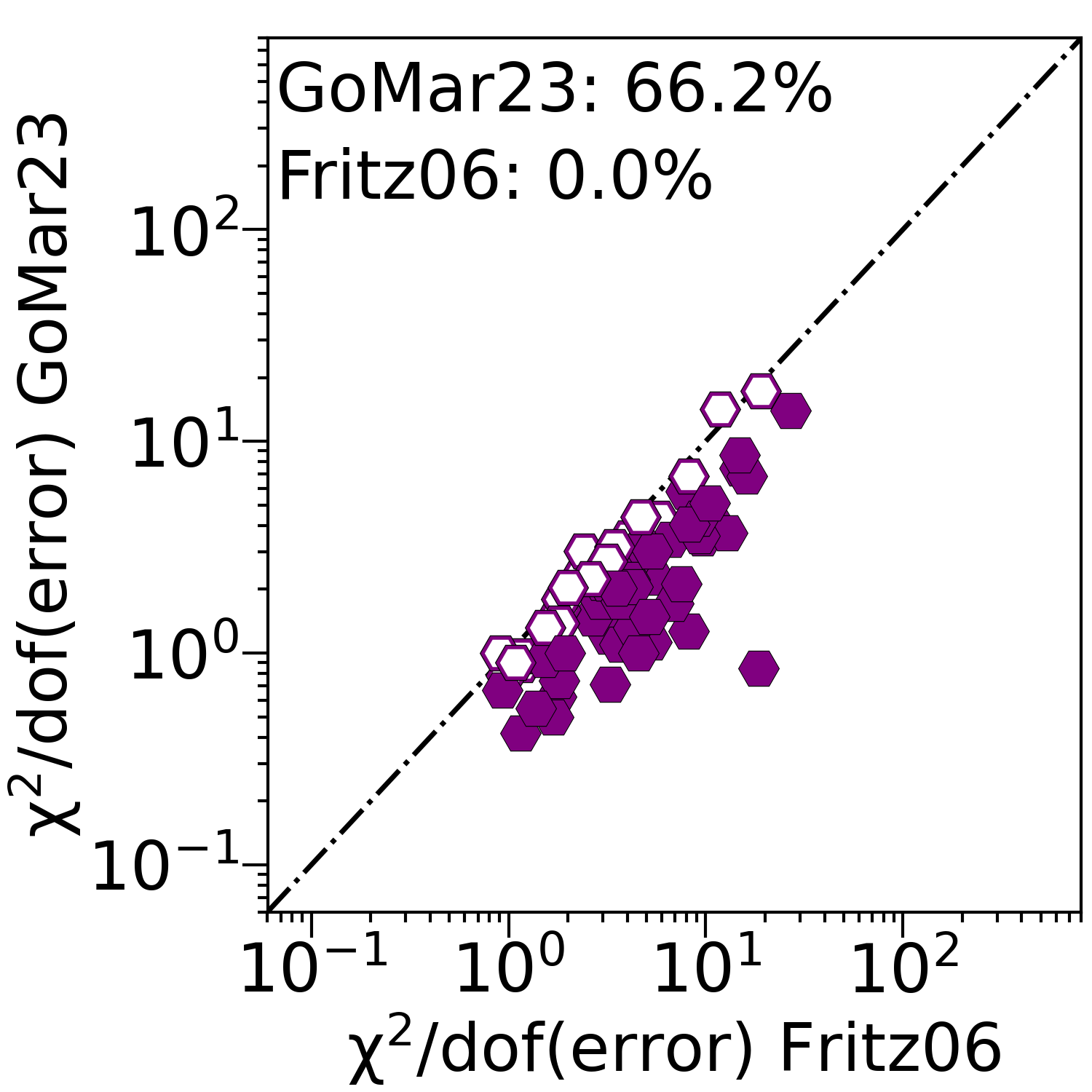} 
\includegraphics[height=0.485\columnwidth, clip, ,trim=245 0 8 0]{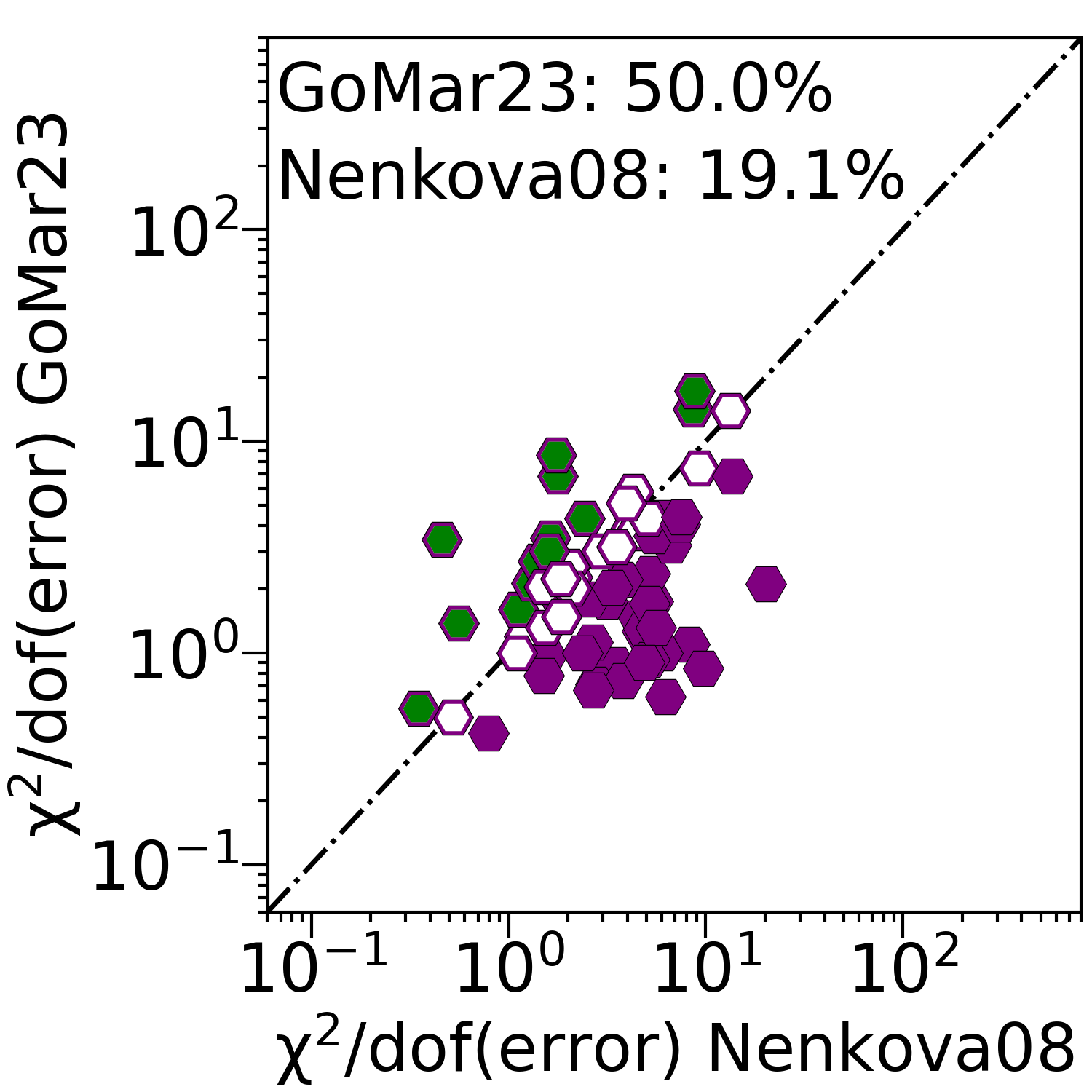}
\includegraphics[height=0.485\columnwidth, clip, ,trim=245 0 8 0]{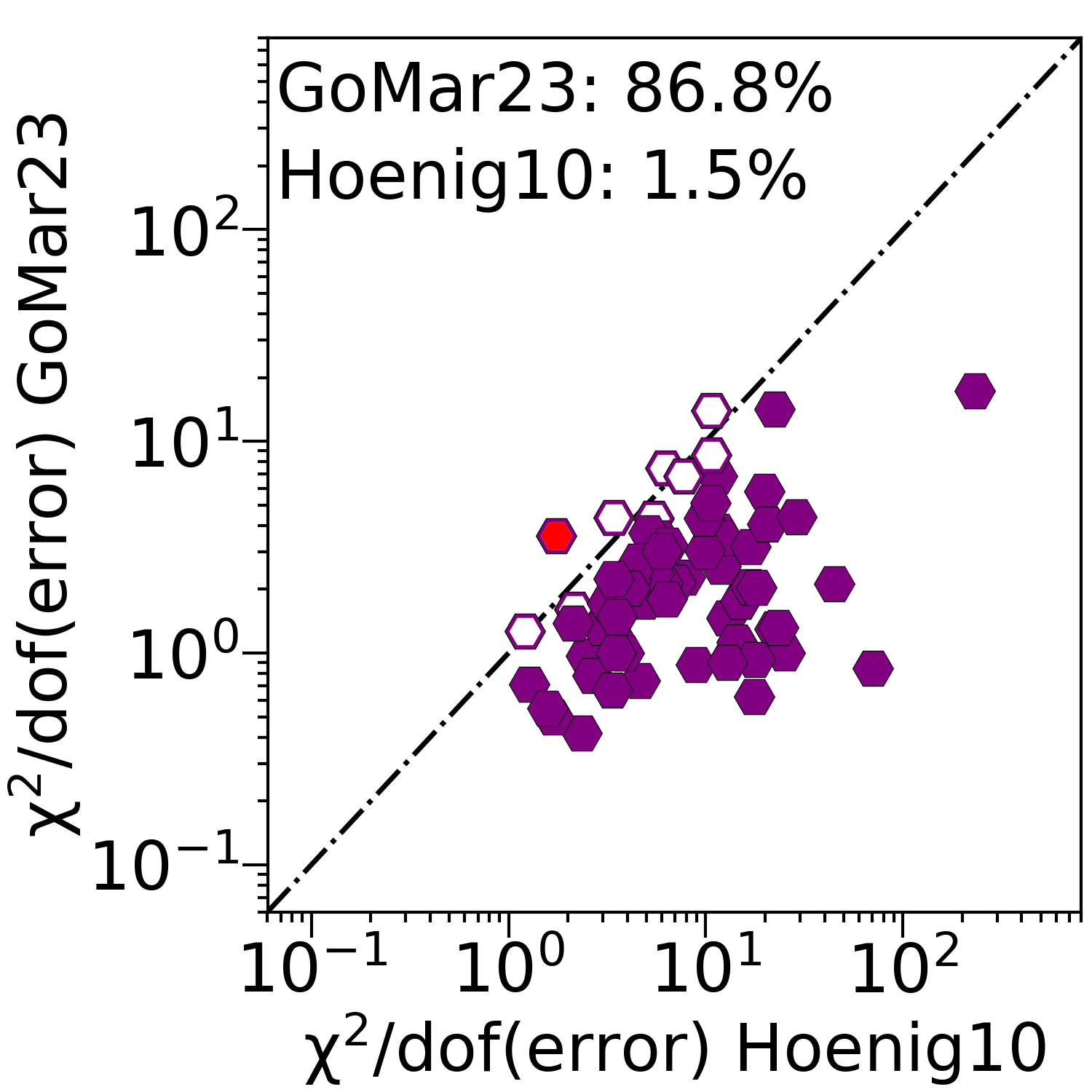} 
\includegraphics[height=0.485\columnwidth, clip, ,trim=245 0 8 0]{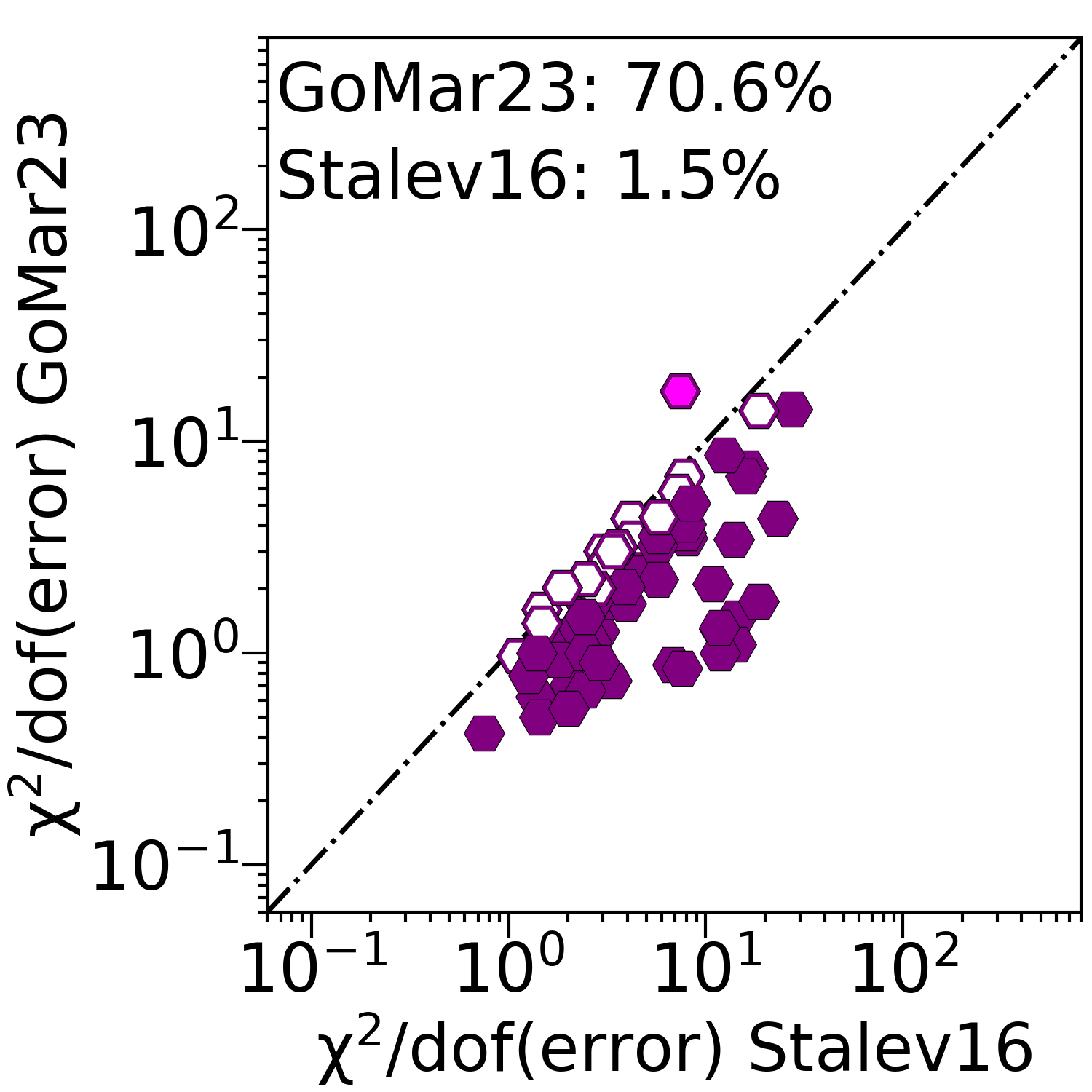} 
\includegraphics[height=0.485\columnwidth, clip, ,trim=245 0 8 0]{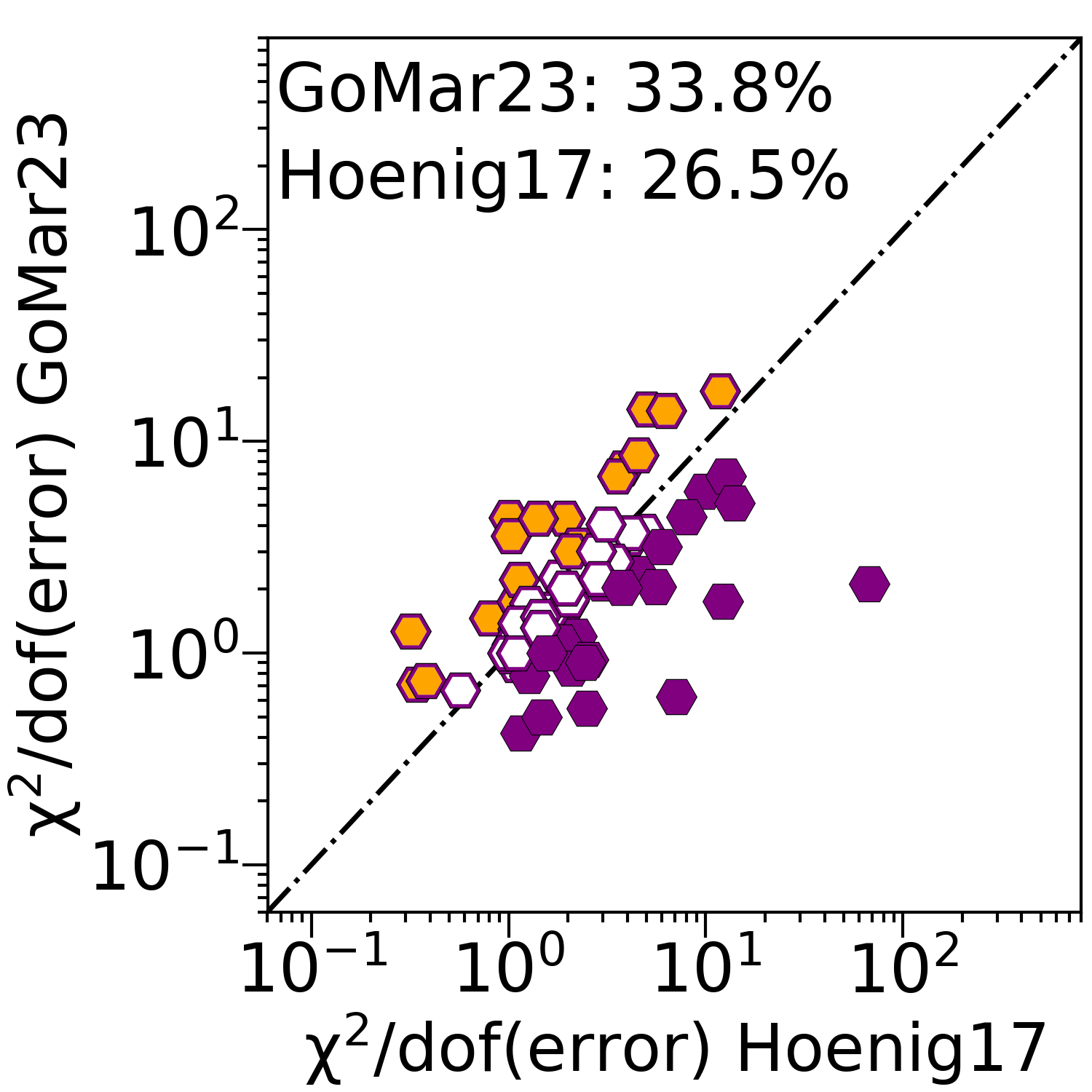} 
\caption{Comparison of $\rm{\chi^2_{r} (error)}$ obtained with our new [GoMar23] model and those previously reported in the literature. Open hexagons show results that are statistically similar according to the AIC probability. Purple-filled hexagons are marked when [GoMar23] model is preferred over the compared model and other colors-filled hexagons are those objects that statistically prefer the compared model instead of [GoMar23] model (color code as in previous plots). The dot-dashed line shows the 1:1 relation. In the left corner of each panel, we show the percentage of objects preferring [GoMar23] model over the other model (top row) and the percentage of objects preferring the other model rather than [GoMar23] model.}
\label{fig:GoMar22AIC}
\end{center}
\end{figure*}

The above results refer to the best fit (i.e. minimum $\rm{\chi^2_{r}}$). However, several models might give a statistically similar result. Fig.\,\ref{fig:GoMar22AIC} shows the comparison of $\rm{\chi^2_{r}}$ obtained with [GoMar23] model and those models previously reported in the literature. The top (bottom) row shows the results before (after) the inclusion of foreground extinction. Empty hexagons show statistically similar results according to the AIC probability (i.e. within less than 200 times as probable as [GoMar23] model, see Appendix\,\ref{sec:stats}). Purple-filled hexagons are marked when [GoMar23] model is preferred over the other model and hexagons with other colors (color code given in Fig.\,\ref{fig:GoodFitsHist}) are those objects that statistically prefer the other model instead of [GoMar23] model. The range given in the subsequent text refers to the inclusion or exclusion of the foreground extinction. As expected, [Stalev16] model is preferred over [GoMar23] model only in 1-2 objects because [GoMar23] model contains all the SEDs included by [Stalev16] model. This small percentage preferring [Stalev16] model might be the result of the stochastic nature of radiative transfer simulations or the inclusion of an anisotropic accretion disk. Furthermore, roughly $\rm{\sim}$65-71\% of the sample show statistically better results when using [GoMar23] instead of [Stalev16] model. As explained before, this is due to the better sampling of the parameter space (e.g. lower half opening angles) and/or the inclusion of the dust grain size into the model. The best-fit $\rm{\chi^2_{r}}$ for [Fritz06] model is correlated with that of [GoMar23] model, although in 54-66\% of the sample, the latter is statistically preferred. This is probably due to the fact that [Fritz06] model is geometrically similar to [Stalev16] and [GoMar23] models so somehow included with [GoMar23] model as well. [Nenkova08] and [Hoenig17] are preferred in 6-19\% and 22-26\% of the sample over the [GoMar23] model, respectively. Meanwhile, [GoMar23] model is preferred in 50-78\% and 34-50\% of the sample over [Nenkova08] and [Hoenig17] models, respectively. 

We repeat this analysis for all the model combinations, adding the AIC probability into the analysis to quantify how many models can describe each object. The last column of Table\,\ref{tab:fitresults} includes all the statistically similar best-fit models (other than the one providing the best fit, which is given in the previous column). The final percentage of objects (equally) best fitted with each model is shown in Fig.\,\ref{fig:GoodFitsHist} (panel c). This panel shows the results without (with) foreground extinction using squared-filled (circle-filled) bars. Note that the sum of the percentages is higher than 100\% because several models can equally fit a single object. Only in five objects (namely Mrk\,3, Mrk\,110, MCG-03-34-064, IC\,4518W, and Fairall\,51), none of the models produce a statistically good fit.

In 24 objects ($\rm{\sim}$35\%) a single model is preferred. Among them, nine prefer [Nenkova08] model, six [Hoenig17] model, six [GoMar23] model, and three of them prefers [Hoenig10] model. Among those objects only fitted with [GoMar23] model, we have the one with the highest silicate absorption feature, indicating that this model might be particularly useful for buried AGN. Five out of these six objects that are well-fitted only to [GoMar23] and [Hoenig17] models are type-1 AGN (the exceptions are 2MASSXJ\,10594361+6504063 and Mrk\,417, respectively). 
Without the inclusion of foreground extinction, except for [Hoenig10] model, most of the already available models are good at describing $\rm{\sim 20-30}$\% of the sample. Our new [GoMar23] model stands out as the best of them, giving the best fits for 71\% of the sample. Once foreground extinction is included, [Nenkova08] and [Hoenig17] models increase their percentages up to 54\% and 41\%, respectively, while [Fritz06] and [Stalev16] models decrease to 18 and 16\%, respectively. Although slightly decreasing after the inclusion of foreground extinction, [GoMar23] model still shows the largest percentage with 60\% of the sample. No relation between the model selection and the AGN luminosity is found for our sample. However, note that this AGN sample shows a relatively narrow luminosity range ($\rm{<log(L_{X})=43.5\pm 0.6>}$). A broader range of AGN luminosities is needed to explore the impact of AGN luminosity on the model selection.  

\begin{figure*}[!t]
\includegraphics[width=0.5\columnwidth, clip,trim=0 0 0 0]{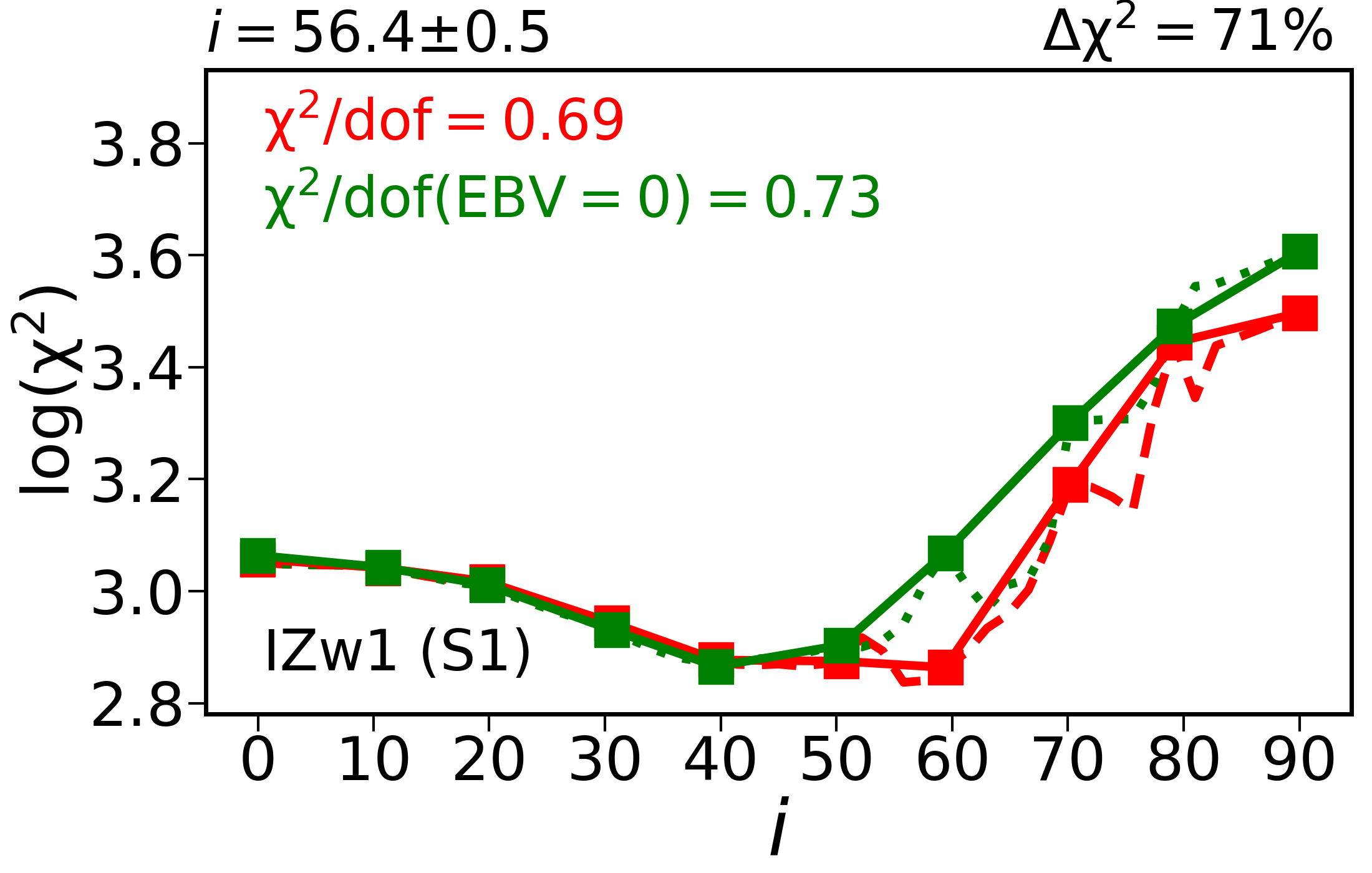} 
\includegraphics[width=0.5\columnwidth, clip,trim=0 0 0 0]{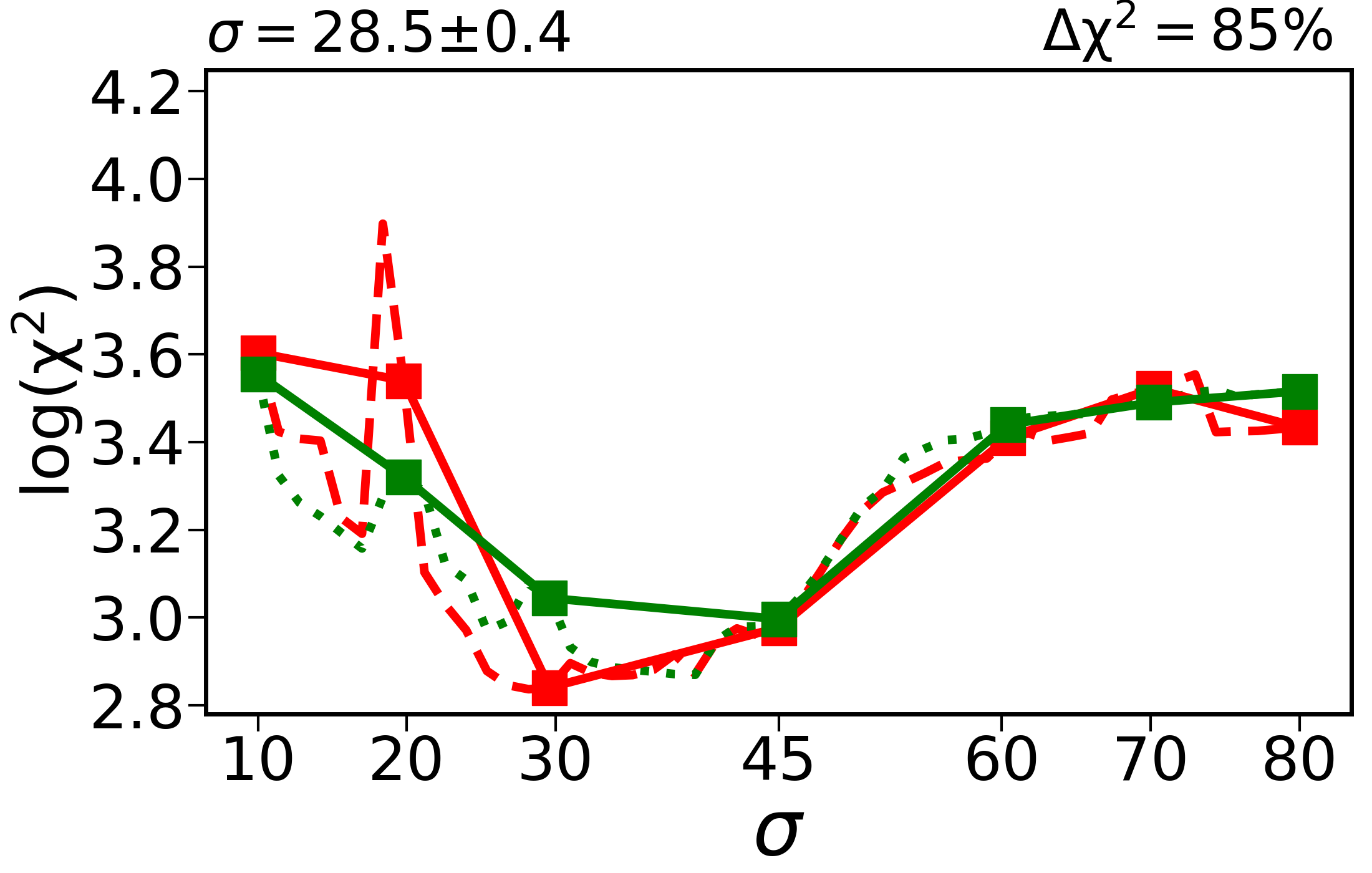} 
\includegraphics[width=0.5\columnwidth, clip,trim=0 0 0 0]{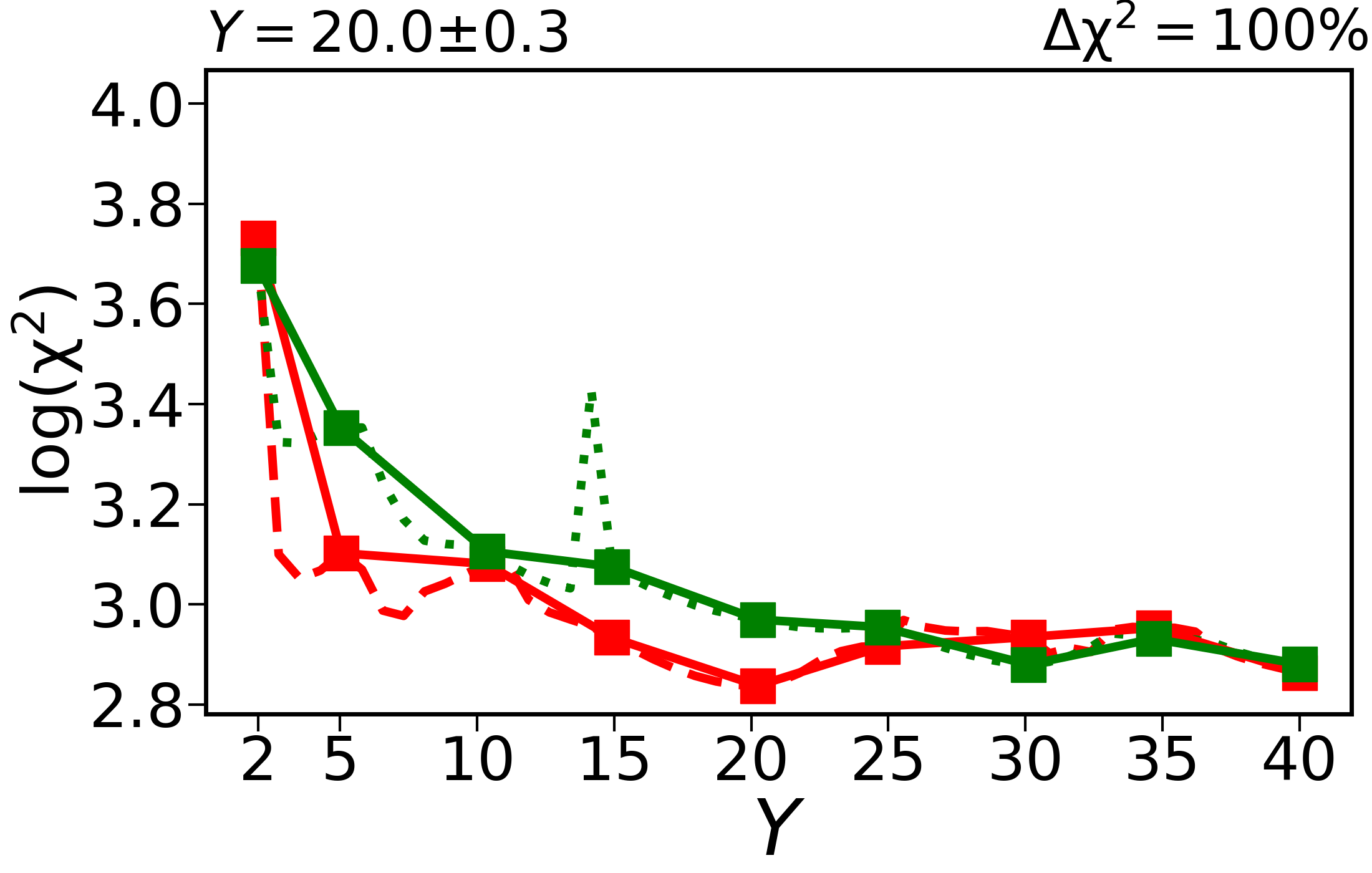} 
\includegraphics[width=0.5\columnwidth, clip,trim=0 0 0 0]{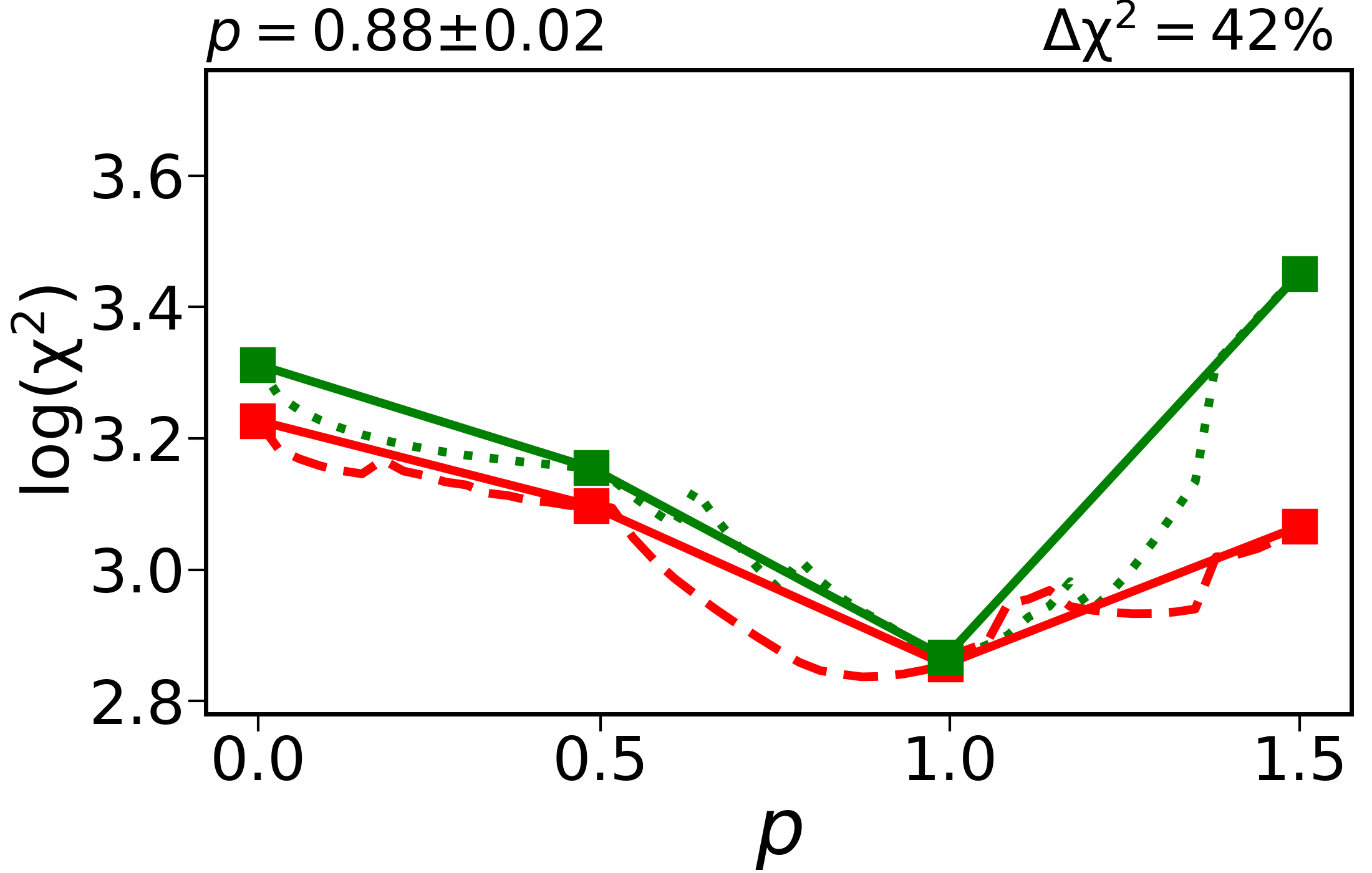} \\
\includegraphics[width=0.5\columnwidth, clip,trim=0 0 0 0]{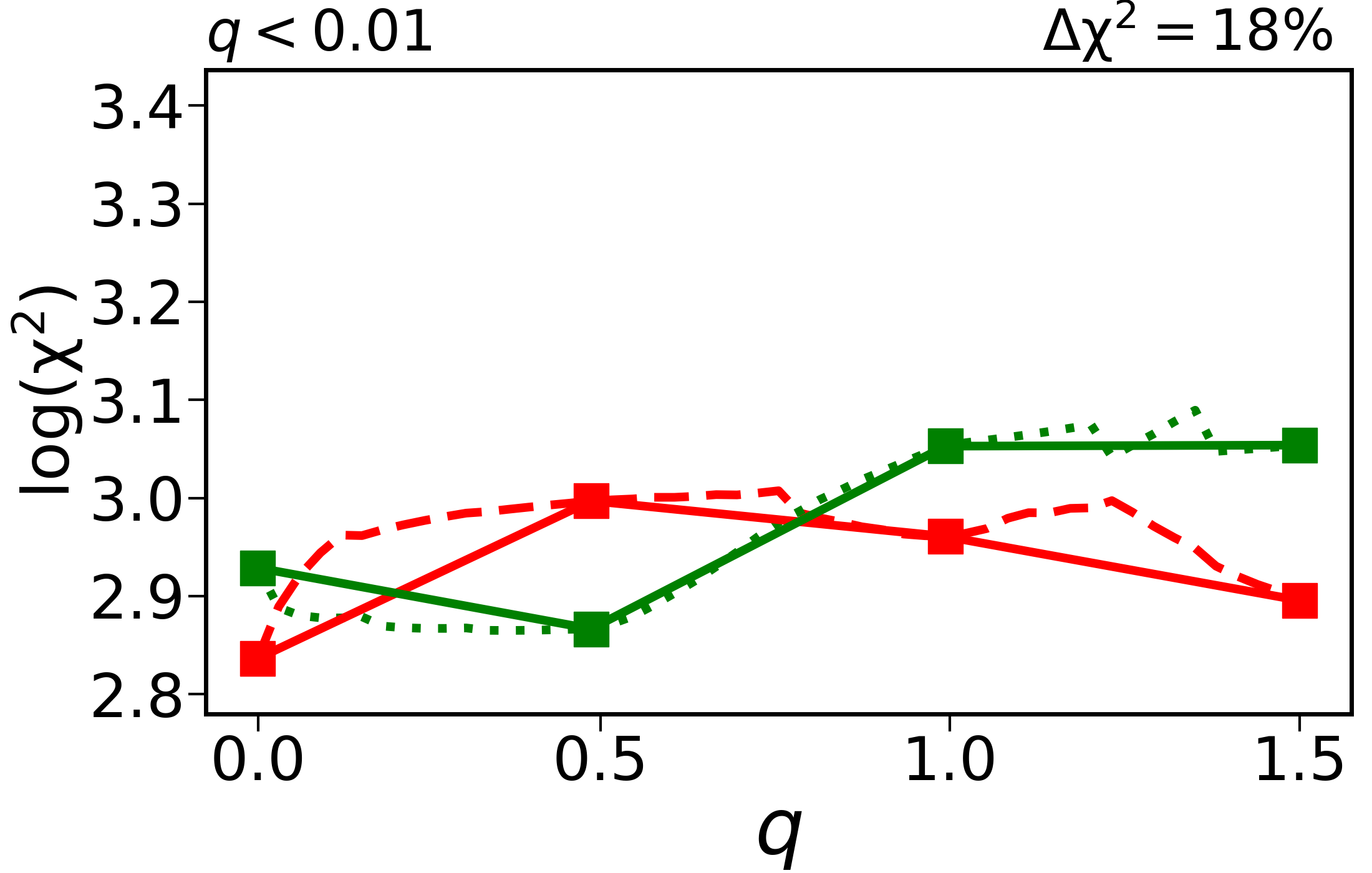} 
\includegraphics[width=0.5\columnwidth, clip,trim=0 0 0 0]{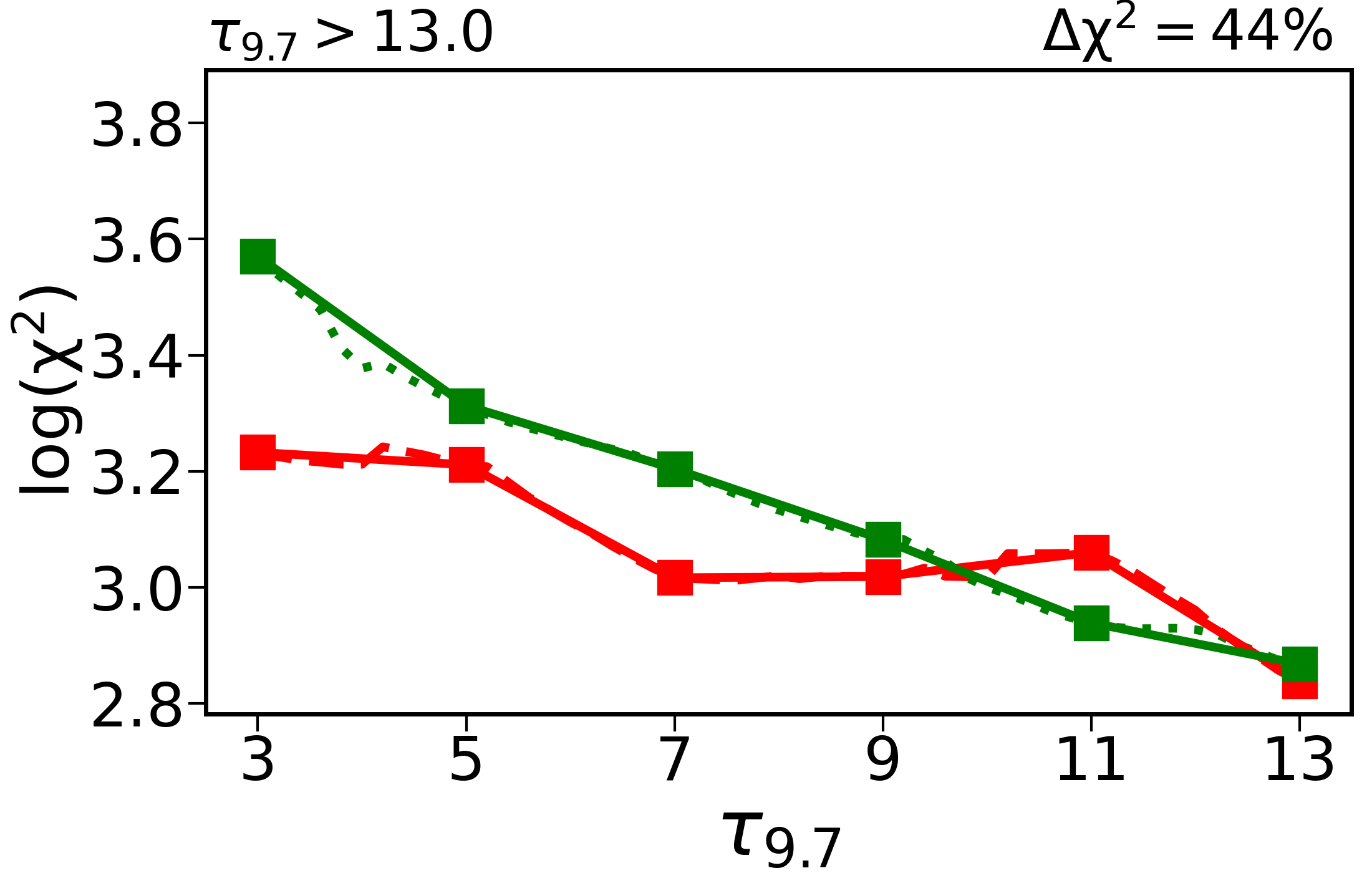} 
\includegraphics[width=0.5\columnwidth, clip,trim=0 0 0 0]{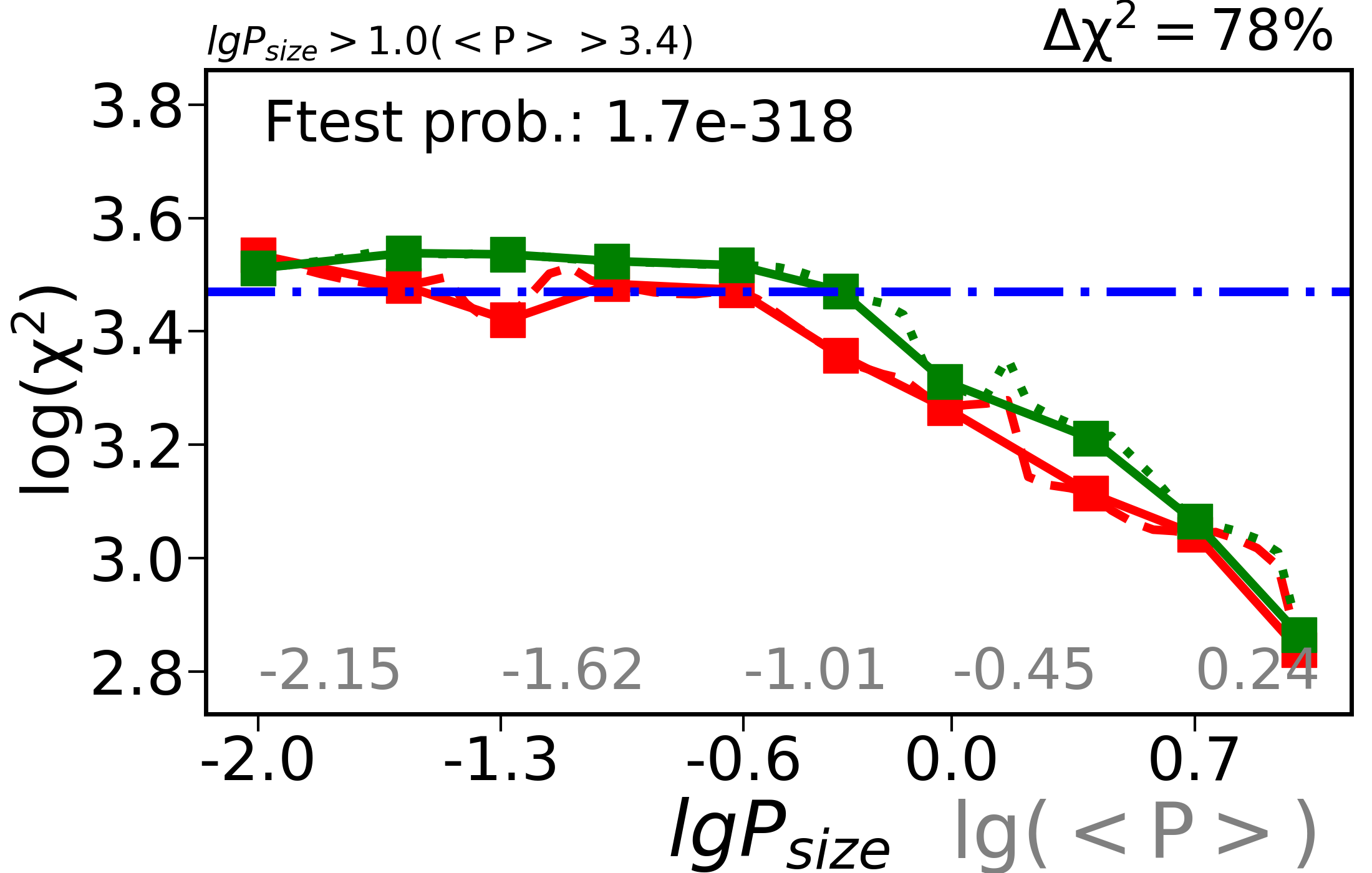} 
\includegraphics[width=0.5\columnwidth, clip,trim=0 0 0 0]{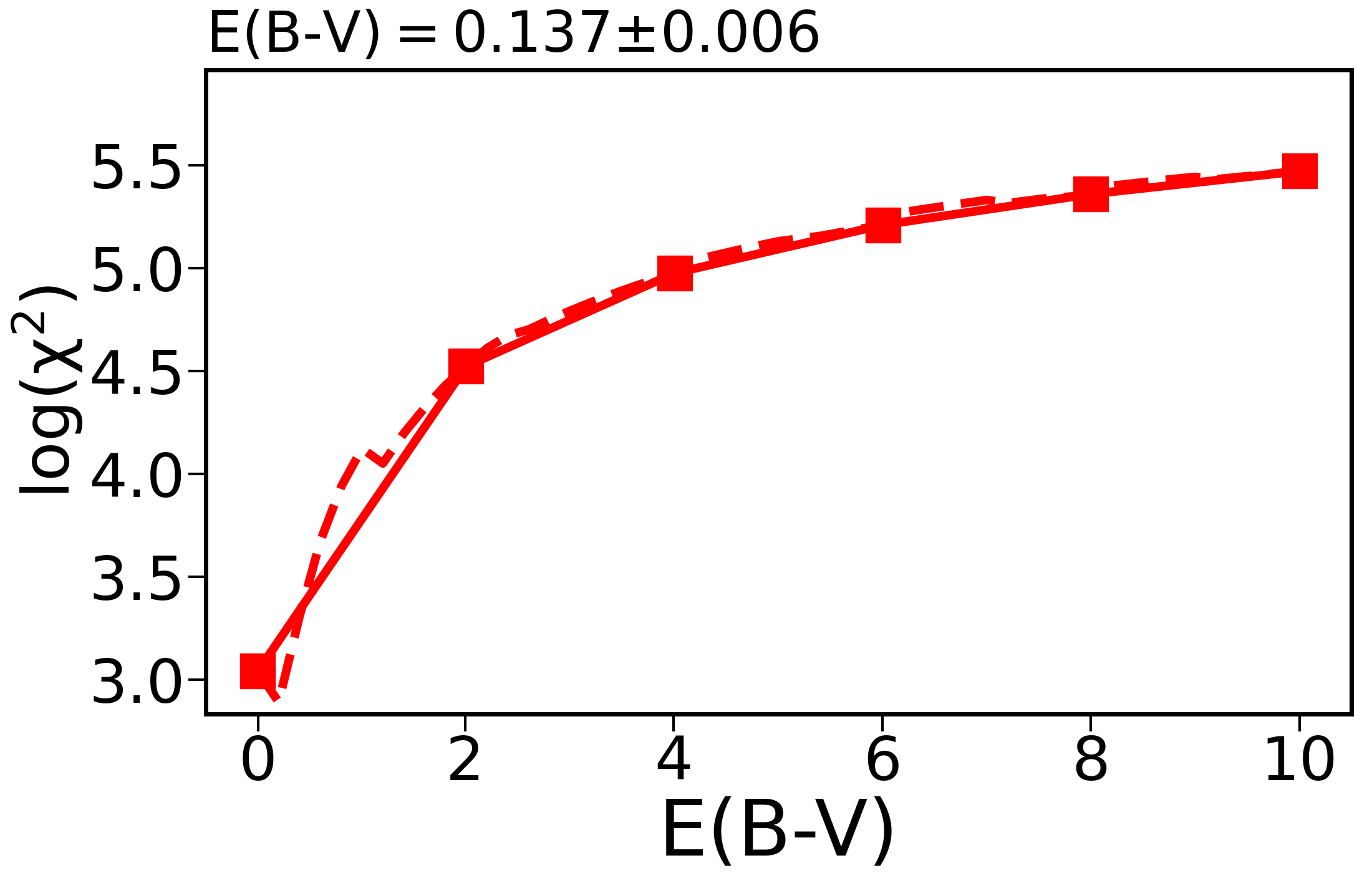} 
\caption{Convergence of $\rm{\chi^2}$ for each parameter of [GoMar23] model for IZw1. Squares show the value of $\rm{\chi^{2}}$ at the given values of the parameters sampled in [GoMar23] grid (continuous line linearly links these black squares). Dashed lines show the internal interpolation performed by {\sc XSPEC}. Green and red squares and lines report the results before and after the foreground extinction is included in the spectral fit procedure. The best value for each parameter is shown above the upper-left corner of each panel (when including foreground extinction). The top-left panel shows the resulting $\rm{\chi^2/dof(sigma)}$ in the upper-left corner. The percentage of variation of $\rm{\chi^2}$ compared to the minimum value is shown above the upper-right corner of each panel. The horizontal blue dot-dashed line in the third-column lower panel shows the $\rm{\chi^2}$ computed when the grain size is fixed to 0.25$\rm{\mu m}$. Note in this panel for $\rm{P_{size}}$ we also include the f-test probability that the grain size is consistent with 0.25$\rm{\mu m}$ (see text). This panel also shows the x-axis expressed in terms of the average grain size in light-gray colors. }
\label{fig:FAIRALL9Pars}
\end{figure*}

\subsection{[GoMar23] model parameters} \label{sec:GoMarparams}

Beyond the model selection, the infrared spectral fitting technique is commonly used to infer the parameters of these models with the aim of constraining the geometry and distribution of the dust \citep[e.g.][]{Ramos-Almeida09,Ramos-Almeida11,Alonso-Herrero11,Garcia-Bernete19,Gonzalez-Martin19B,Martinez-Paredes21,Esparza-Arredondo21,Garcia-Bernete22}. The resulting best-fit parameters and their associated errors through the $\rm{\chi^2}$ statistics for [GoMar23] model are reported in Table\,\ref{tab:parresults}. The XSPEC software also gives the opportunity to study the convergence of the parameter by computing how $\rm{\chi^2}$ changes across the parameter space. Fig.\,\ref{fig:FAIRALL9Pars} shows the posterior distribution of each parameter for [GoMar23] model for the object IZw1 as an example. The squares (linked with continuous lines) show the locus of the $\rm{\chi^2}$ for each point of the model grid. The dotted-green line shows the interpolation given by the XSPEC software, which in general agrees well with the continuous line. In general, all the parameters show monotonous behavior along the parameter space. Only viewing angle (i), half opening angle ($\rm{\sigma}$), the radial slope of the dust distribution ($p$), and the ratio between the outer and inner radius ($Y$) are restricted (at 3-$\rm{\sigma}$ level) with a clear minimum along the parameter space. 

Fig.\,\ref{fig:FAIRALL9Pars} also shows that not all the parameters equally contribute to obtaining a minimal $\rm{\chi^2}$. It is clear that variation on the slopes ($\rm{p}$ and $\rm{q}$) and the optical depth at 9.7$\rm{\mu m}$ ($\rm{\tau_{9.7}}$) of the density distribution contribute much less to the final fit compared to variations on the half opening angle, $\rm{\sigma}$, for instance. In order to quantify this, we compute the percentage of $\rm{\chi^2}$ variations compared to the parameter giving the maximal variation, denoted as $\rm{\Delta \chi^2}$:
\begin{equation}
\Delta \chi^2 (par) = \frac{\chi^2_{max}(par) - \chi^2_{min}(par)}{\chi^2_{max}(p_{ref)} - \chi^2_{min}(p_{ref})}
\end{equation}
\noindent where $par$ denote the parameter evaluated and $p_{ref}$ is the [GoMar23] model parameter giving the maximal $\rm{\chi^2}$ variation. This number is given in the upper-right corner of each panel of Fig.\,\ref{fig:FAIRALL9Pars}. In IZw1 the maximal $\rm{\chi^2}$ variation is obtained with the ratio between the outer and the inner radius of the torus ($Y$). 
The half opening angle ($\rm{\sigma}$), the maximum dust grain size ($\rm{P_{size}}$), the viewing angle ($i$), and significantly contribute to the minimization process with $\rm{\Delta \chi^2}$ of 85\,\%, 78\%, and 71\%, respectively. On the other hand, the optical depth at 9.7$\rm{\mu m}$ ($\rm{\tau_{9.7}}$) contributes 44\% while 18\% is found for the slopes of the density distribution ($p$). Thus, for this particular object, the maximum dust grain size, $\rm{P_{size}}$ (or equivalently the weighted-average grain size $\rm{<P>}$), is the third most relevant parameter (below $Y$ and $\sigma$). In order to further explore the need for this parameter in the final fit, we performed the same fit but kept the maximum dust grain size fixed to the canonical value of $\rm{P_{size}=0.25\mu m}$ ($\rm{<P>=0.097\mu m}$). The resulting minimum $\rm{\chi^2}$ for this fit is shown as the dot-dashed blue line in the panel associated with $\rm{P_{size}}$/$\rm{<P>}$ of Fig.\,\ref{fig:FAIRALL9Pars}. We then run the f-statistic test to compute the probability that allowing the maximum dust grain size to vary significantly improves the final fit (this number is quoted within the same panel). Indeed, the maximum dust grain size is requested by the data to significantly improve the fit for IZw1.

\begin{figure*}[!t]
\includegraphics[width=0.5\columnwidth, clip,trim=0 15 0 0]{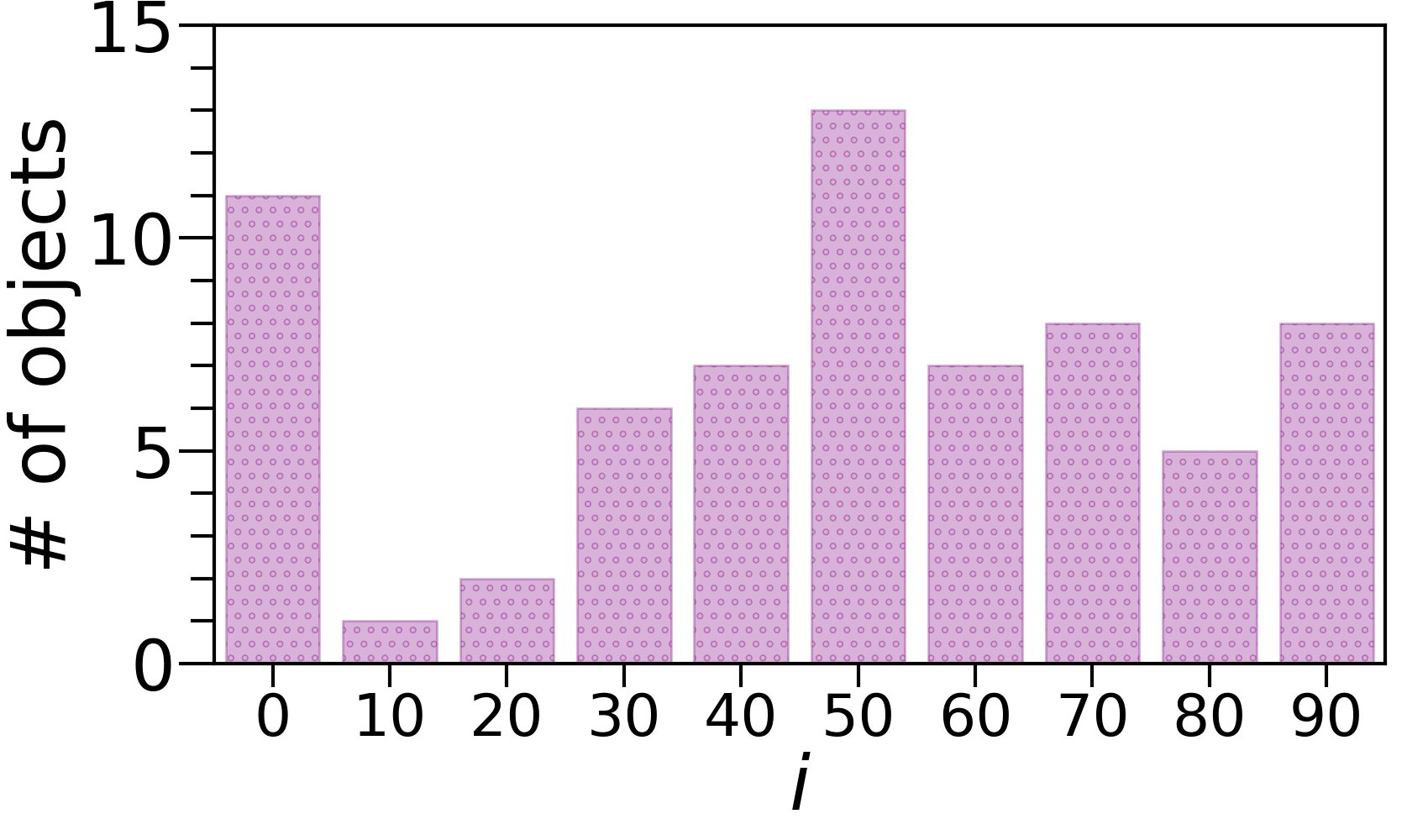}
\includegraphics[width=0.5\columnwidth, clip,trim=0 15 0 0]{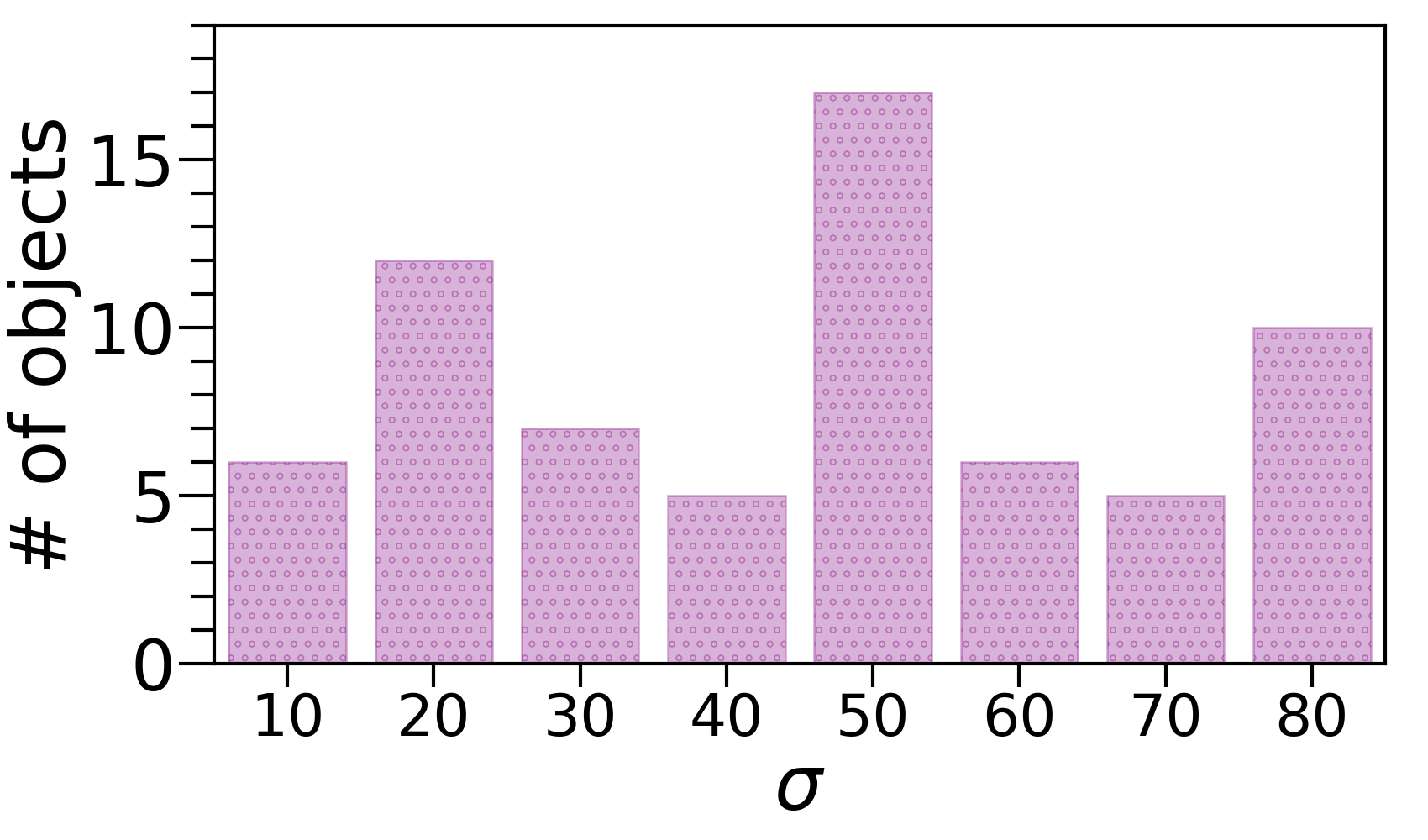} 
\includegraphics[width=0.5\columnwidth, clip,trim=0 15 0 0]{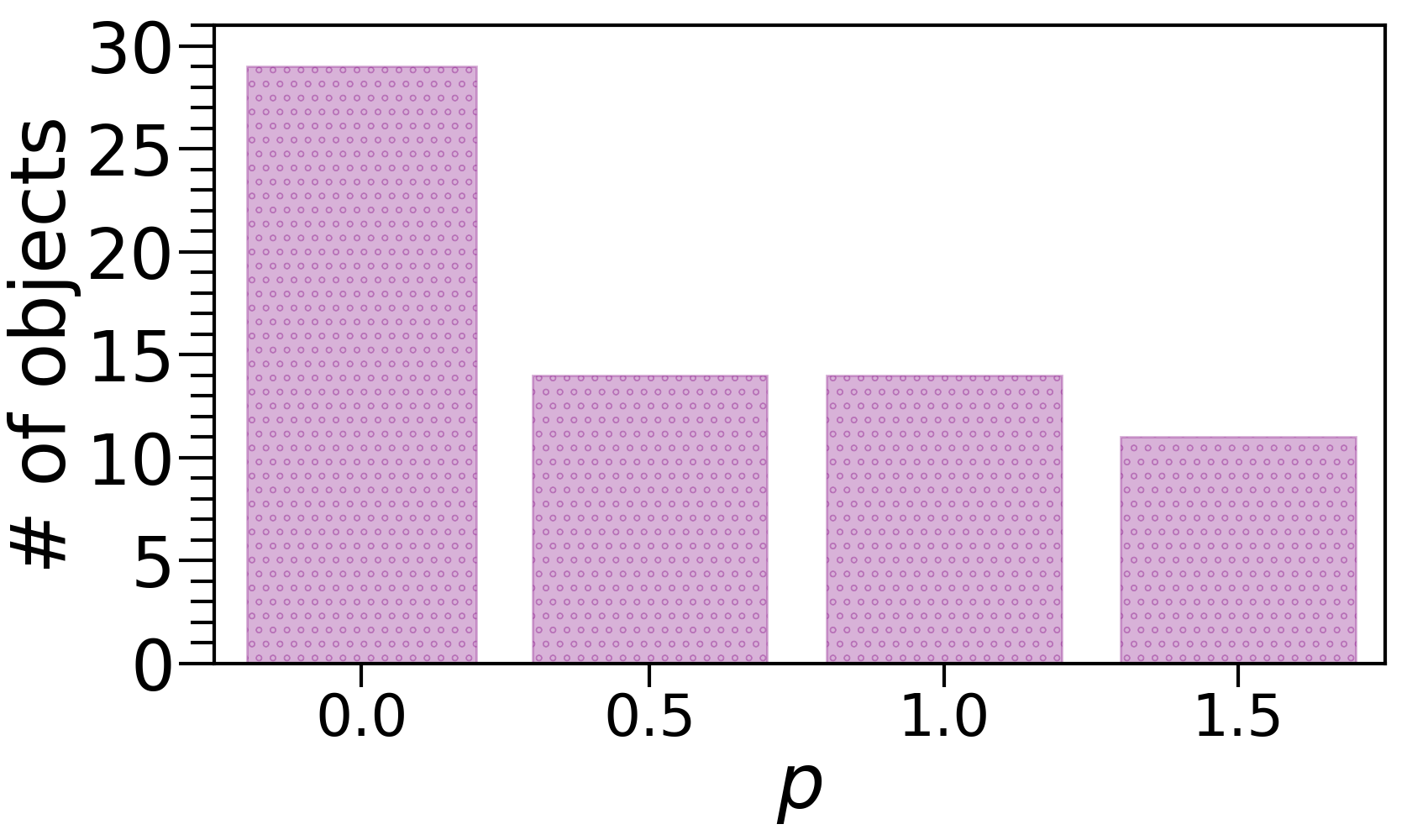} 
\includegraphics[width=0.5\columnwidth, clip,trim=0 15 0 0]{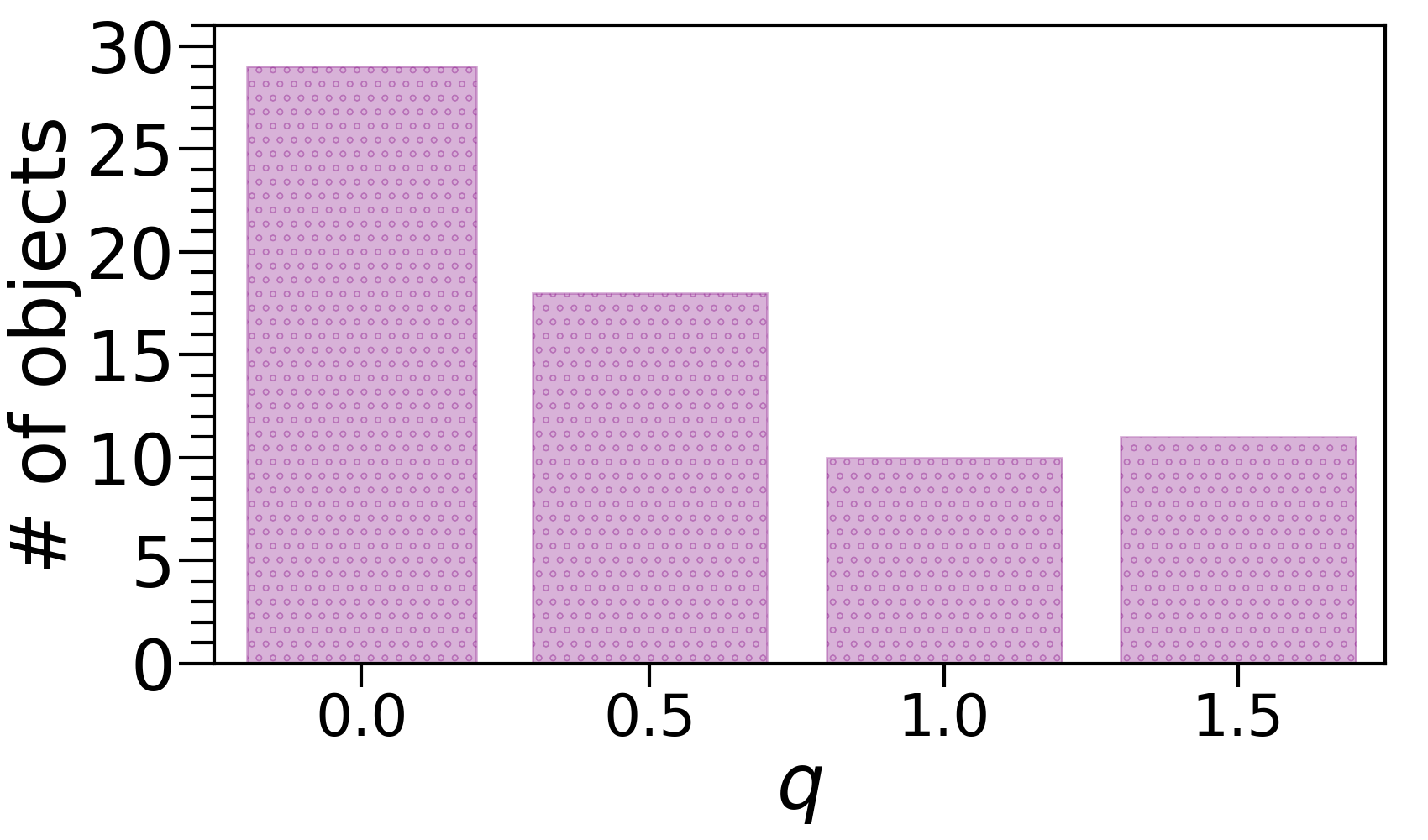} \\
\includegraphics[width=0.5\columnwidth, clip,trim=0 0 0 2]{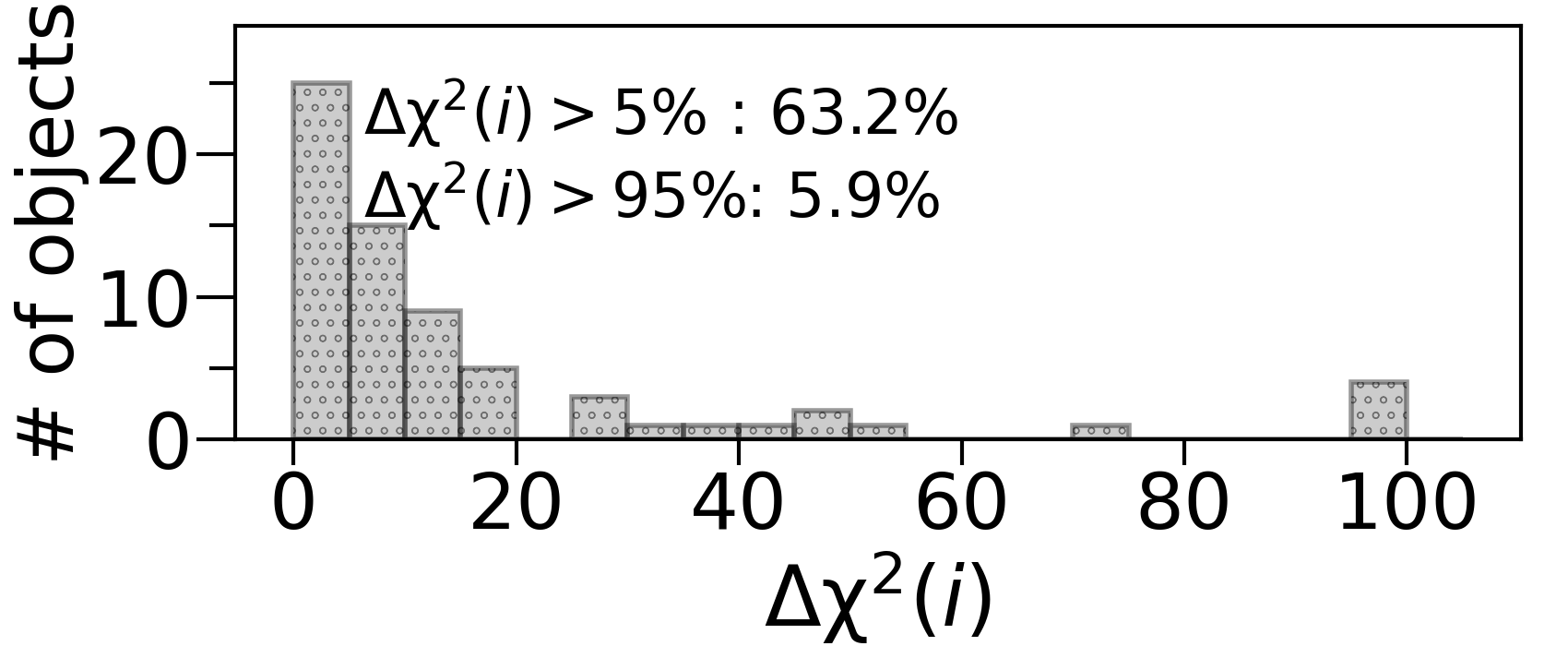} 
\includegraphics[width=0.5\columnwidth, clip,trim=0 0 0 2]{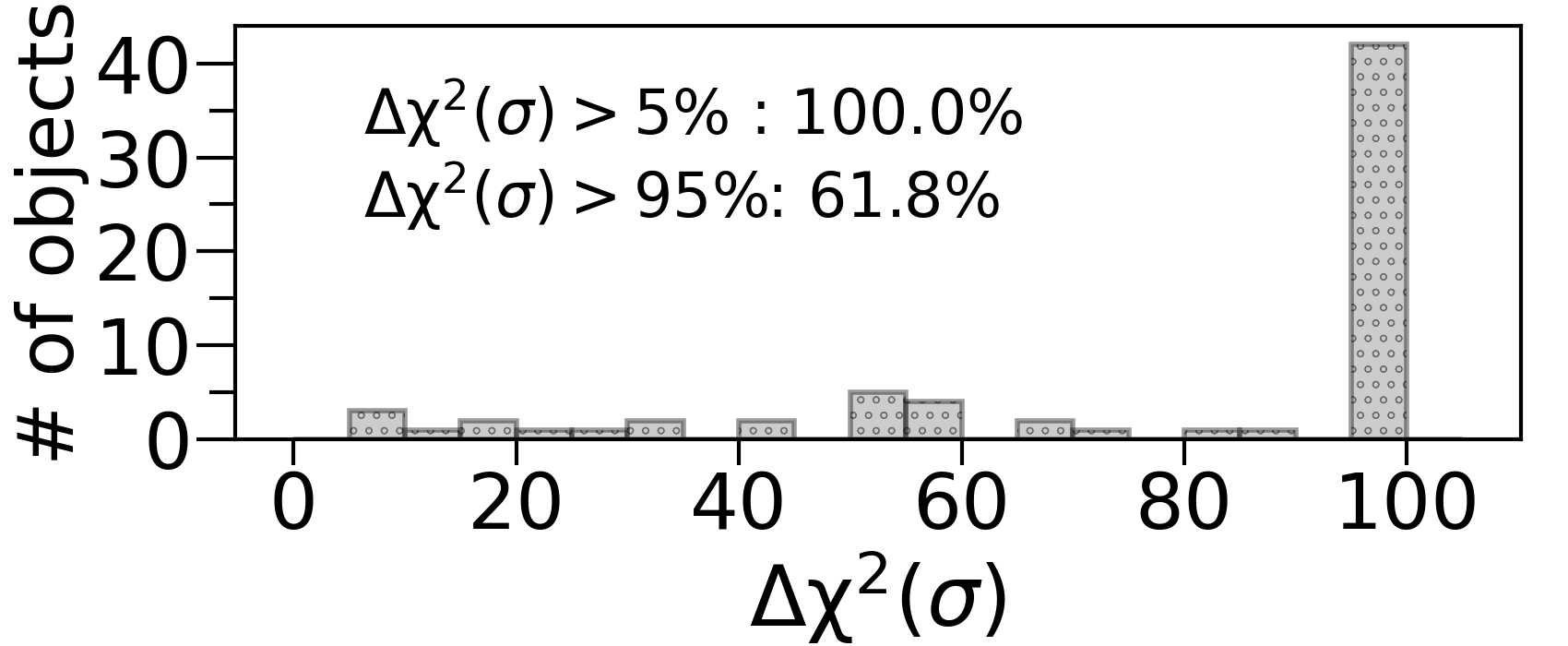} 
\includegraphics[width=0.5\columnwidth, clip,trim=0 0 0 2]{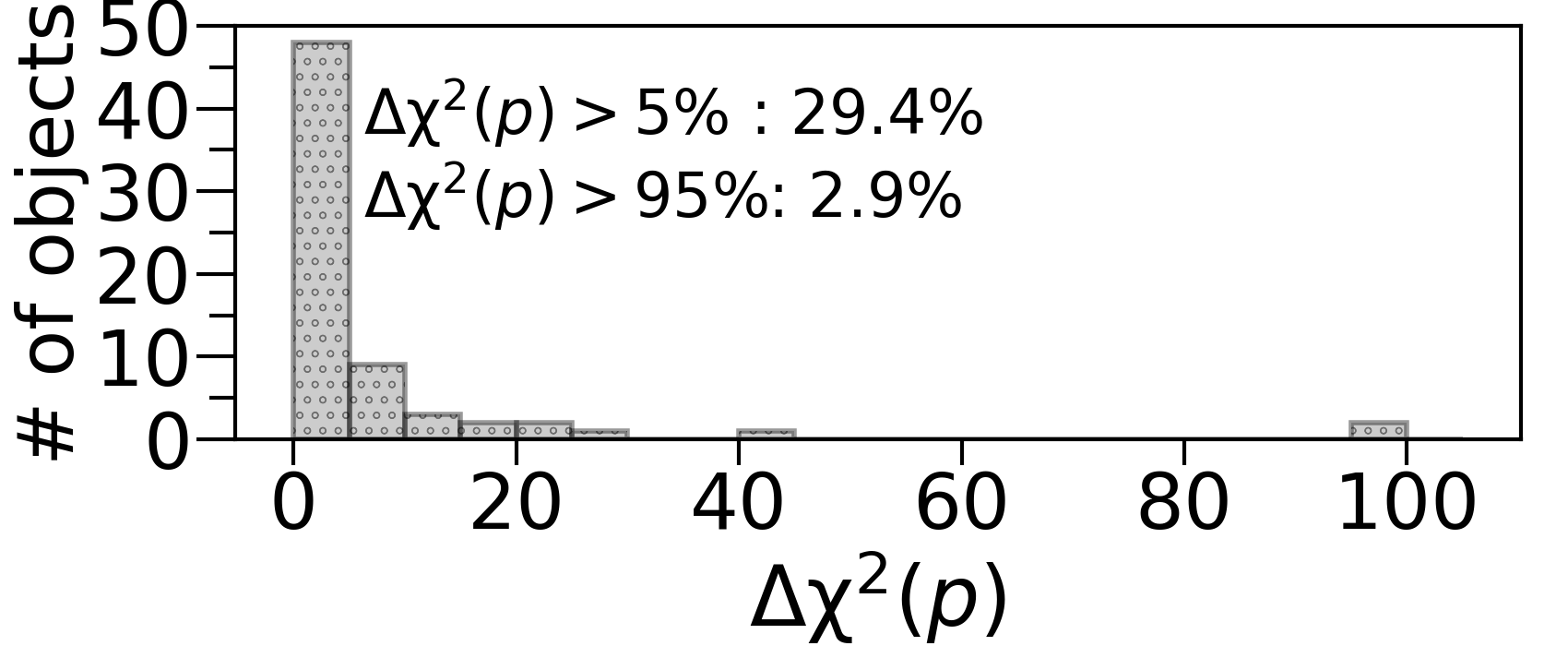} 
\includegraphics[width=0.5\columnwidth, clip,trim=0 0 0 2]{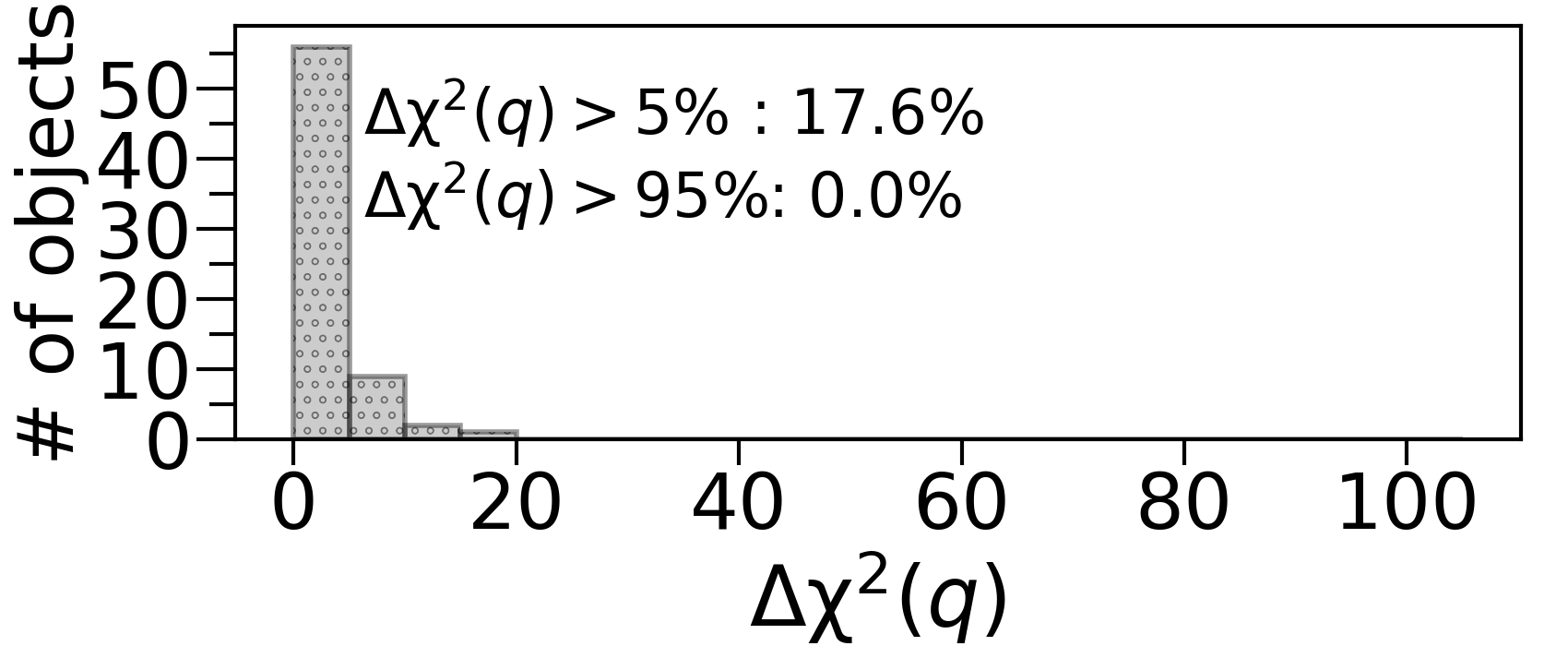} \\
\includegraphics[width=0.5\columnwidth, clip,trim=0 6 0 0]{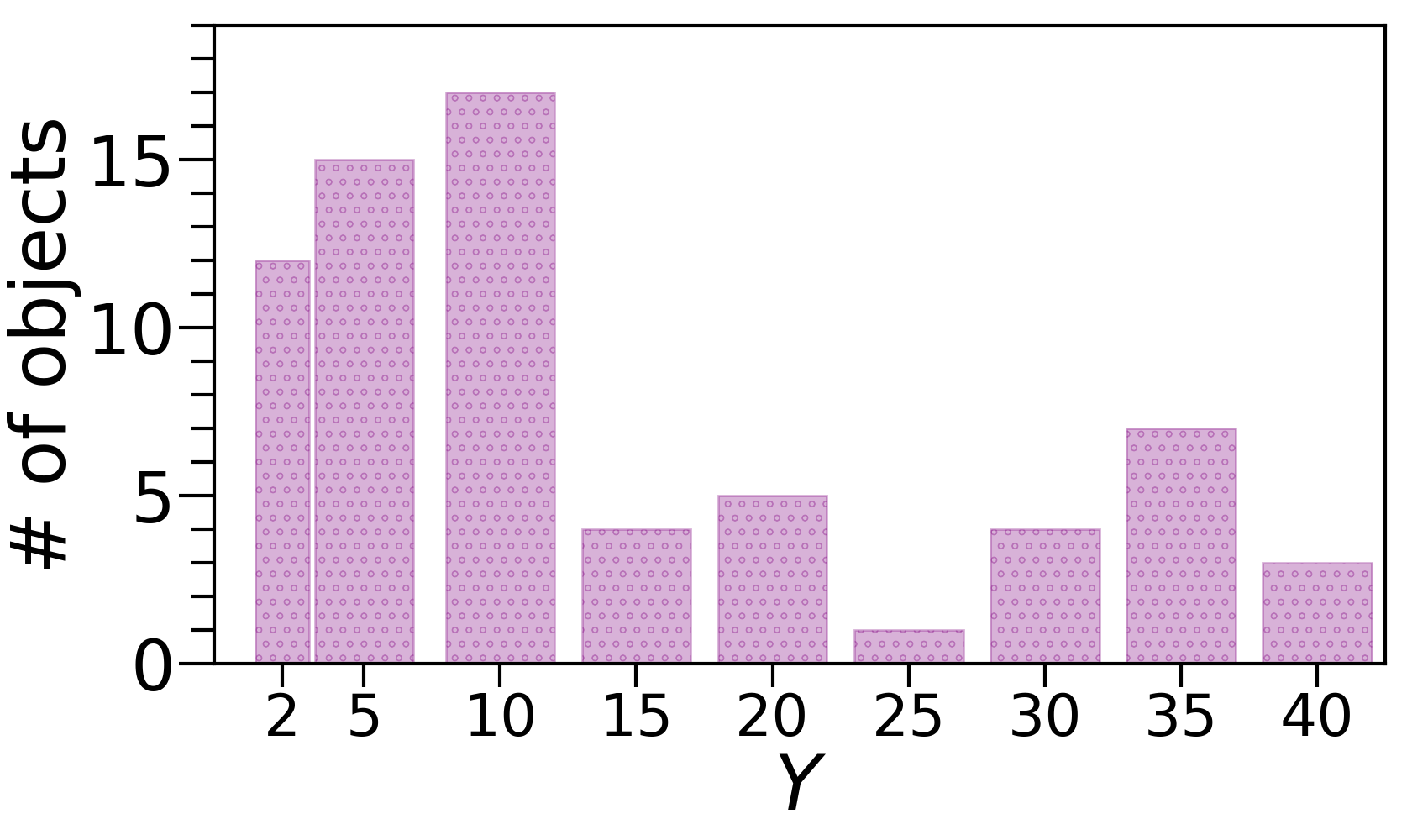} 
\includegraphics[width=0.5\columnwidth, clip,trim=0 6 0 0]{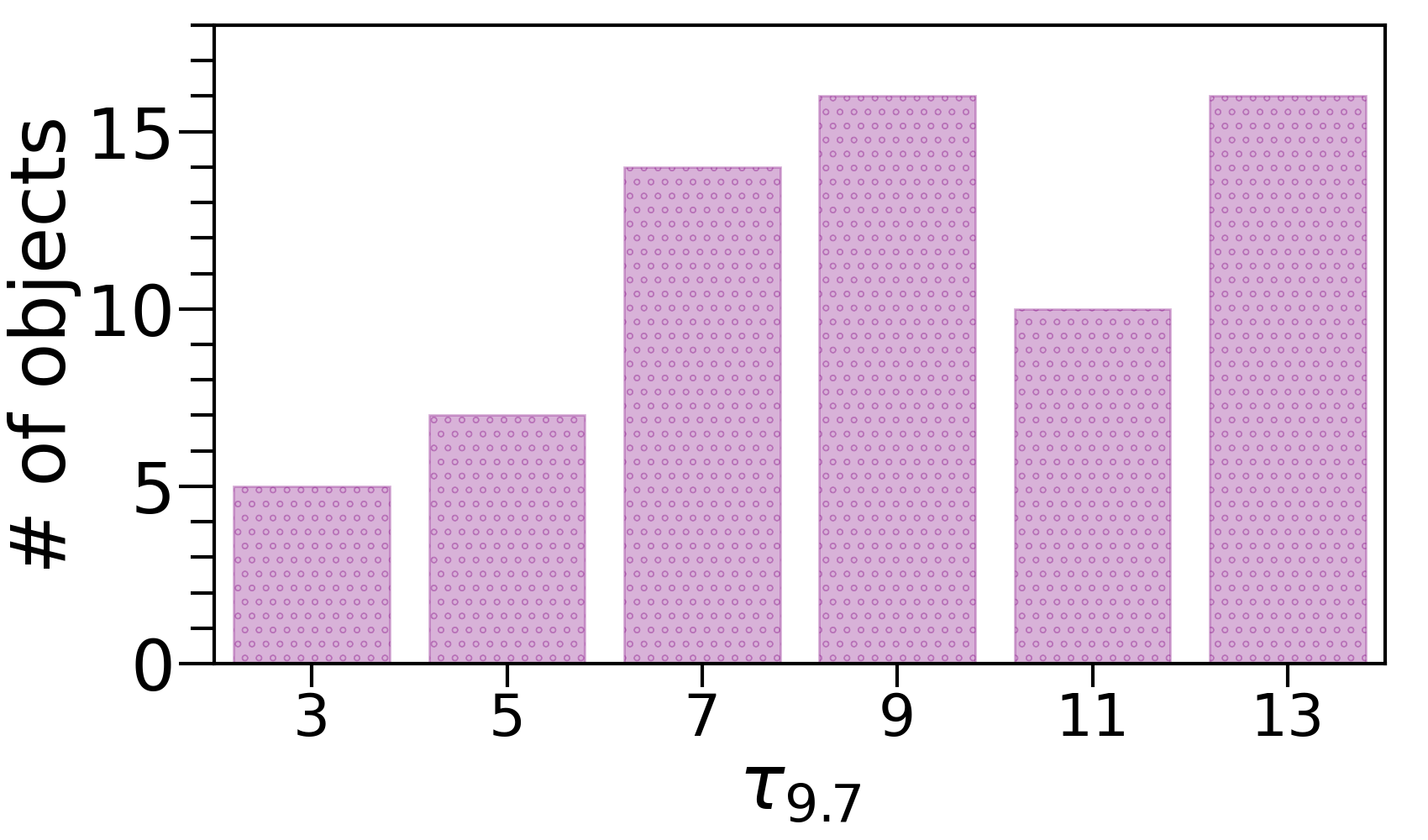} 
\includegraphics[width=0.5\columnwidth, clip,trim=0 6 0 0]{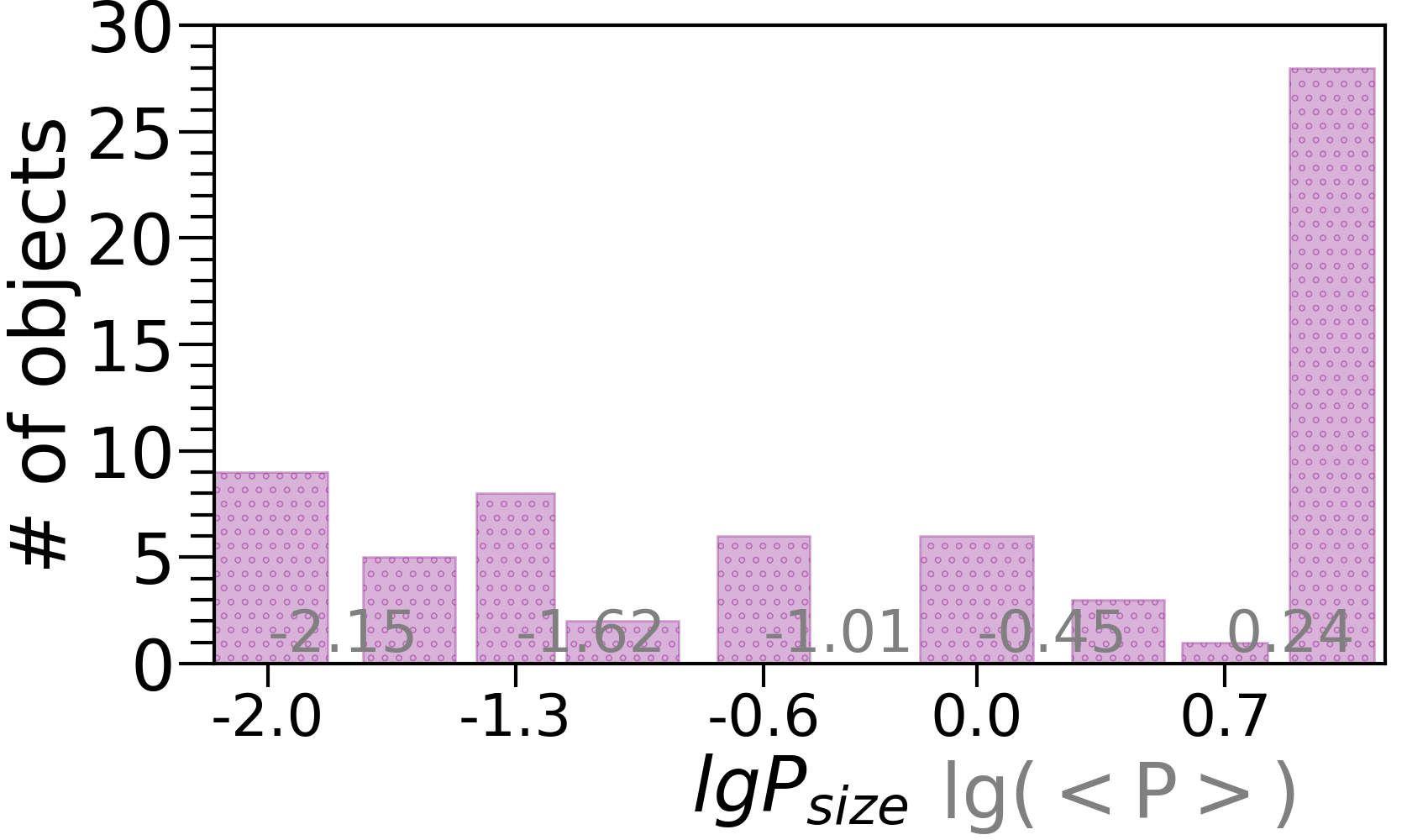}
\includegraphics[width=0.5\columnwidth, clip,trim=0 6 0 0]{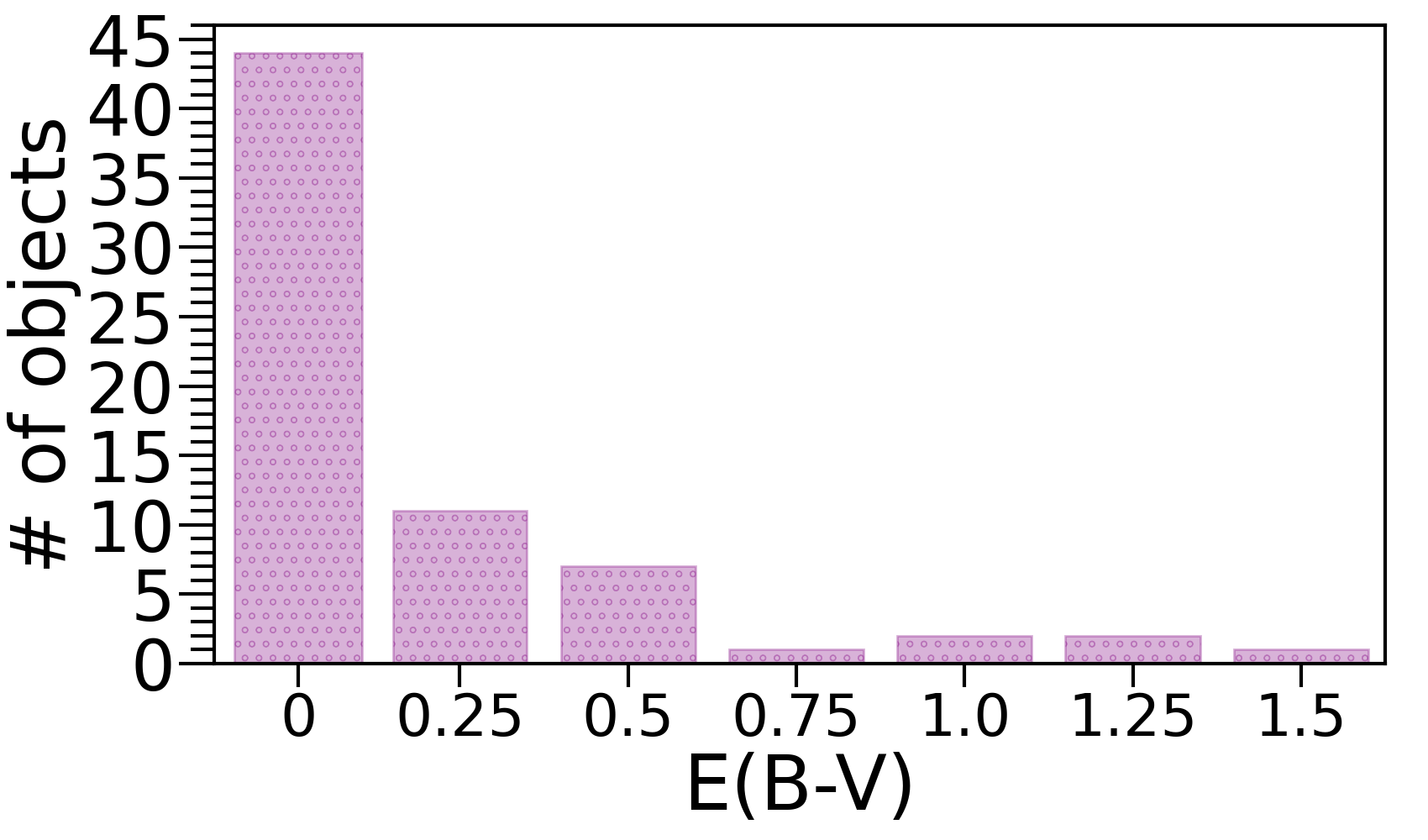} \\
\includegraphics[width=0.5\columnwidth, clip,trim=0 0 0 2]{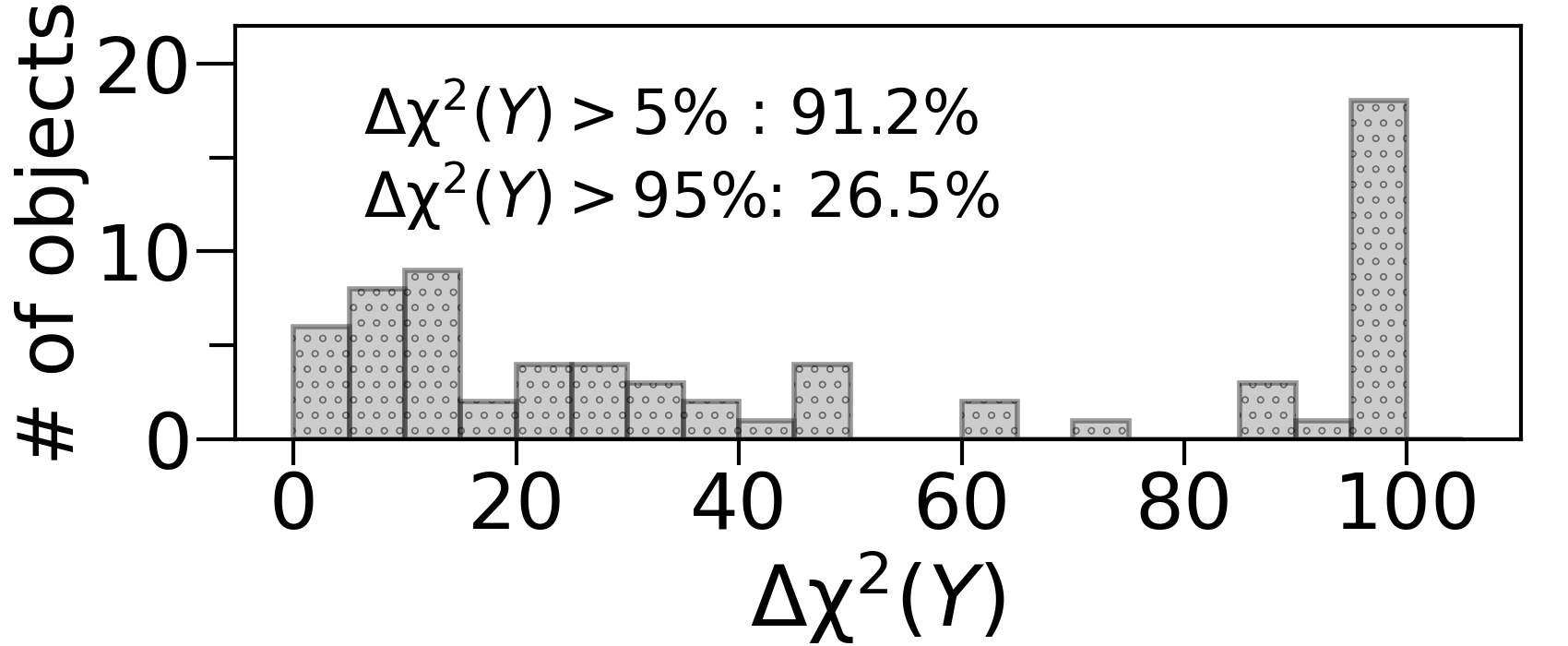} 
\includegraphics[width=0.5\columnwidth, clip,trim=0 0 0 2]{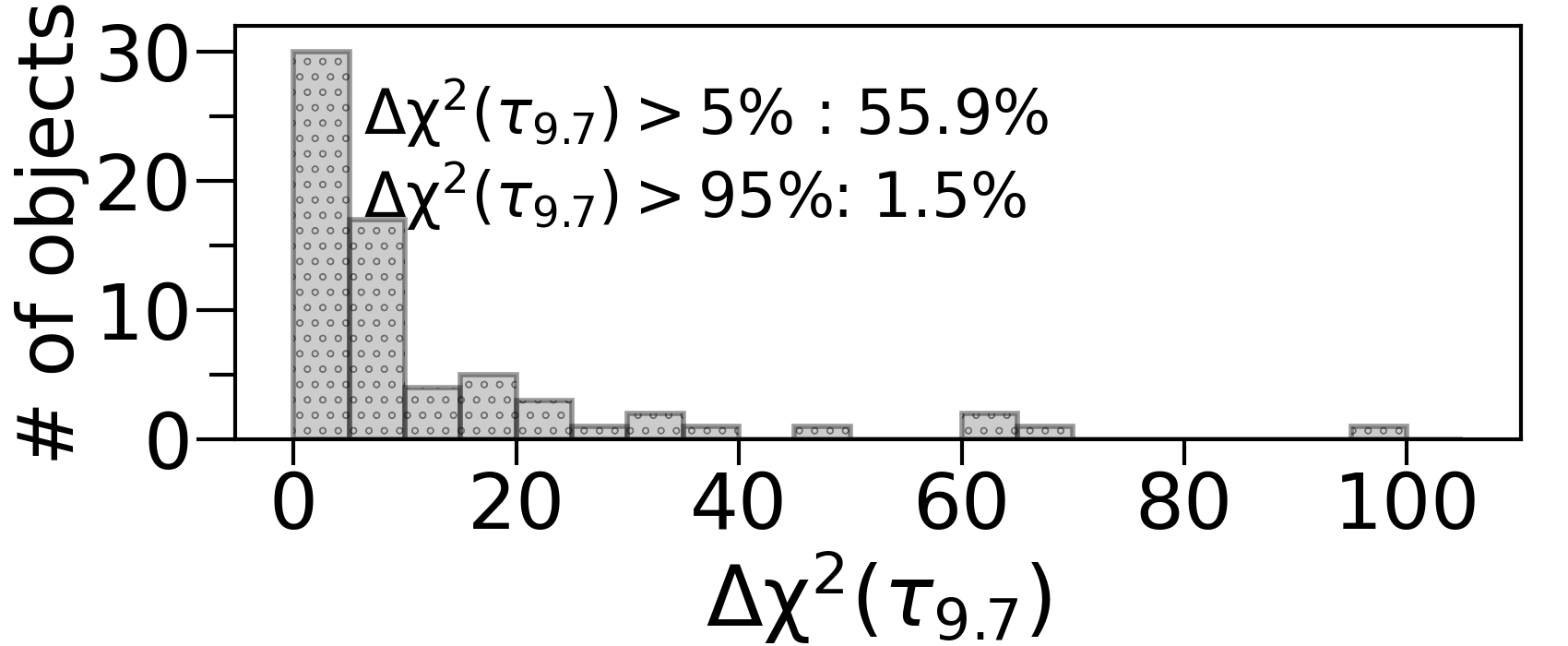}
\includegraphics[width=0.5\columnwidth, clip,trim=0 0 0 2]{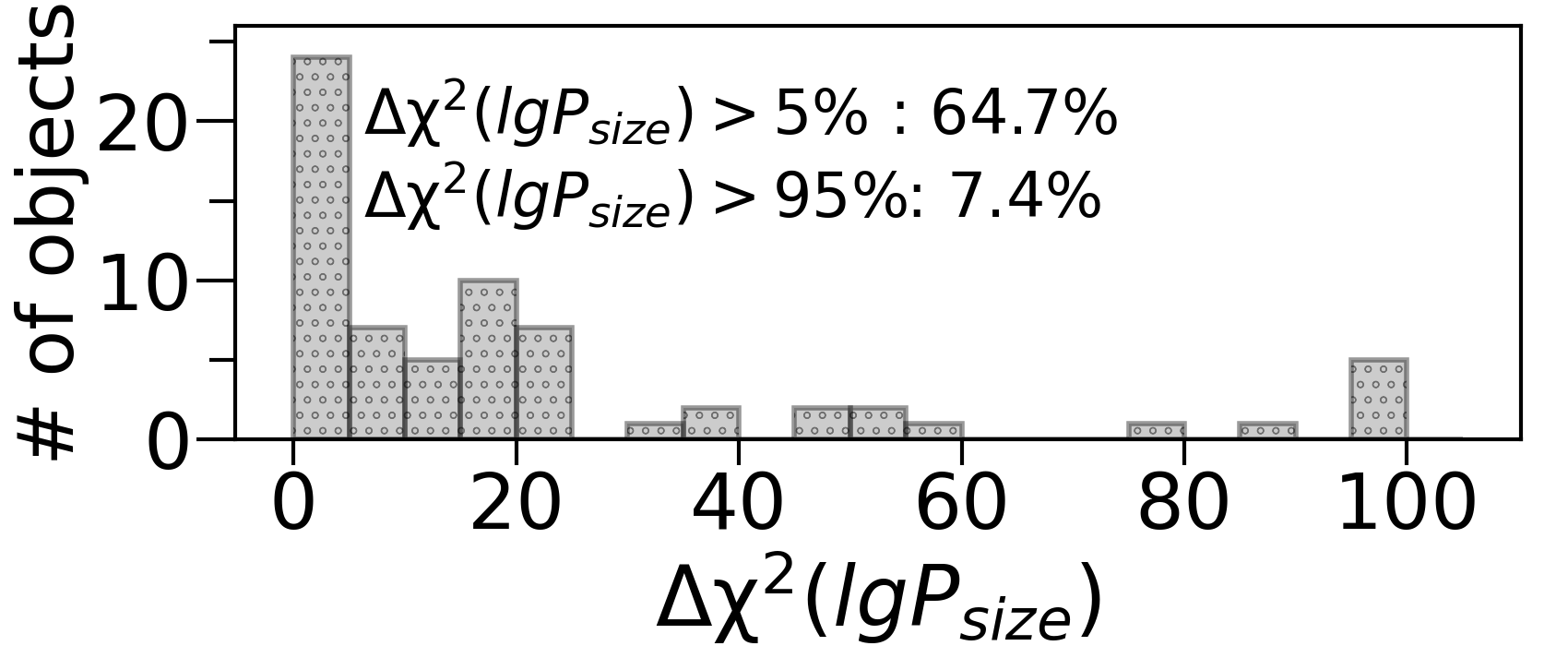} \\
\caption{Rows 1 and 3 show in purple the histograms of the resulting parameters for the AGN sample when using [GoMar23] model. We use the same bins as those used for the SED grid. Rows 2 and 4 show the histogram of the relative importance of each parameter in the convergence of the fit. It is obtained as the maximal variation of $\rm{\chi^2}$ for each parameter range compared to the parameter giving the maximum variation, denoted as $\rm{\Delta \chi^2}$ (see Fig.\,\ref{fig:FAIRALL9Pars}). Each panel also shows the percentage of objects with $\rm{\Delta \chi^2}$ above 5 and 95\%.}
\label{fig:GoMarParHist}
\end{figure*}

We then perform the above analysis for the full AGN sample. Rows 1 and 3 (from top to bottom) in Fig.\,\ref{fig:GoMarParHist} show the resulting distribution for each parameter for the AGN sample, and rows 2 and 4 show the distribution of $\rm{\Delta \chi^2}$ as computed above. Note that the bins of the parameter-space histograms are set to match the grid points. At first look, most of the parameters require the full range explored by the grid (i.e. no bins with zero objects are found). The parameters on the polar and radial density distribution of dust ($p$ and $q$) cluster to flat distributions of dust (i.e. $p=q=0$). In 28 out of the 68 objects (41\% of the sample) prefer very large particles with $\rm{P_{size}\sim 10\mu m}$ ($\rm{<P>\sim 3.41 \mu m}$). Only 6 out of the 68 objects (9\% of the sample) fall in the bin with the canonical value of $\rm{P_{size}\sim 0.25\mu m}$ ($\rm{<P>\sim 0.097 \mu m}$). After comparing the resulting best fit when fixing $\rm{P_{size}=0.25\mu m}$ ($\rm{<P>\sim 0.097 \mu m}$) to those in which the grain size is allowed to vary, we find that, according to f-test, 90\% of the objects in our sample statistically prefer.

Inside panels in rows 2 and 4, the percentage of objects giving $\rm{\Delta \chi^2}$ above 5 and 95\% of the variations is included. $\rm{\Delta \chi^2>95}$\% shows the percentage of objects where the considered parameter is the most fundamental one to obtain the fit, while above $\rm{\Delta \chi^2>5}$\% shows the percentage of objects where the considered parameter significantly improves the fit compared to the most important parameter. The parameter that systematically gave the maximal variation on the $\rm{\Delta \chi^2}$ (i.e. $\rm{\Delta \chi^2>95}$\%) for 62\% of the sample is the half opening angle, $\rm{\sigma}$. 
Attending to $\rm{\Delta \chi^2>5\%}$ (i.e. the significant improvement of the fit) the parameters can be sorted out from the most relevant to the less relevant as: $\rm{\sigma}$ $\rm{=>}$ $\rm{Y}$ $\rm{=>}$ $\rm{P_{size}}$ / $\rm{<P>}$ $\rm{=>}$ $\rm{i}$ $\rm{=>}$ $\rm{\tau_{9.7}}$ $\rm{=>}$ $p$ $\rm{=>}$ $q$ (ranging from 100 down to 18\% of the sample). Therefore, variations in the maximum dust grain size, $\rm{P_{size}}$ (or equivalently the mass-weighted average grain size $\rm{<P>}$), produces $\rm{\Delta \chi^2>5\%}$ for 65\% of the sample, which makes it the third most relevant parameter for improving the fits.

Finally, although the study of the parameters is out of the scope of this paper, we warn the reader that the spectral coverage and/or sensitivity of the \emph{Spitzer}/IRS spectra might not be enough to recover all the parameters at the same time. In particular, we find a clear degeneracy between the viewing angle and the half-opening angle of the torus ($\rm{r=0.9}$), which is also found for [Stalev16] model. We plan a future investigation on model constrain using \emph{JWST} near and mid-infrared observations.

\section{Discussion} \label{sec:Discussion}

\subsection{Adequacy of [GoMar23] model to the AGN dust continuum}

The thermal infrared SED emission from the central region of an AGN depends on the geometry, distribution, and composition of the dust. Although quite challenging from the observational point of view, it is key to study AGN nuclear dust to explain the AGN classes and possible interplay between the AGN and its host, acting as the outer layer of the nurturing/feedback mechanism \citep[see the most recent reviews][]{Ramos-Almeida17,Hoenig19}. 
The infrared observations of nearby AGN show a broad continuum that could be interpreted as the sum of two different components: one of them producing a $\rm{\sim 3\mu m}$ bump, which corresponds to dust close to the sublimation temperature, and another component that peaks at wavelengths longer than $\rm{10-20\mu m}$. These components were interpreted by \citet{Mor12} respectively as the hot pure graphite dust outside the BLR \citep{Garcia-Bernete22}, and the clumpy torus, which resides on larger scales and produces the mid-infrared emission. Furthermore, the mid-infrared spectra show moderate silicate absorption features for most of the type-2 AGN, and weak silicate emission features for most of the type-1 AGN \citep[][see also Fig.\,\ref{fig:observations}]{Hoenig10A,Gonzalez-Martin13,Alonso-Herrero16,Garcia-Bernete19}, although exceptions to this are also found \citep[][]{Mason12,Martinez-Paredes20}.

From the theoretical point of view, reproducing the infrared SED of AGN has turned to be challenging, particularly the silicate features and the overall infrared continuum slopes \citep[][and references therein]{Gonzalez-Martin19B,Martinez-Paredes21,Garcia-Bernete22}. We show that the currently available models are good at providing the range of infrared slopes and silicate strengths (see Fig.\,\ref{fig:modelshapes}) but they fail at reproducing the whole mid-infrared spectrum of a given AGN with a single SED. Between 25-50\% of our sample cannot be well matched to the available SED libraries, unless foreground extinction is included (Fig.\,\ref{fig:chis}). 

\subsubsection{Smooth versus clumpy models}

If we focus on the differences found between smooth and clumpy models in our analysis, the main one is associated with the need for foreground extinction applied to the data (Fig.\,\ref{fig:chis}). Foreground extinction has little impact on the models including a smooth dust distribution provided by \citet{Fritz06} ([Fritz06]) and \citet{Stalevski16} ([Stalev16]), with less than $\rm{\sim}$30\% of the sample requiring this component. On the other hand, the purely clumpy models by \citet{Nenkova08B} ([Nenkova08]), \citet{Hoenig10B} ([Hoenig10]), and \citet{Hoenig17} ([Hoenig17]) require foreground extinction for $\rm{\sim}$90\% of the sample analysed in this work. The most extreme result is obtained for the clumpy torus model by \citet{Nenkova08B} which is able to provide good fits for 50\% of the sample without the inclusion of foreground extinction while this percentage increases up to 84\% otherwise. Therefore, we conclude that purely clumpy models may have insufficient self-absorption to reproduce the observed shallow silicate absorption features and steeper near-infrared slopes. Although the nature of the foreground extinction is out of the scope of this analysis, it is worth mentioning that the E(B-V) values listed in this analysis are well above the expected values for the ISM of the Milky Way, when using the $\rm{N_{H}}$ hydrogen column densities derived from the full sky HI survey by the HI4PI collaboration \citep[][]{BenBekhti16} and the standard gas-to-dust ratio of $\rm{N_{H}/A_{V}= 1.8\times 10^{21}\, atoms\, cm^{-2}\, mag^{-1}}$. Instead, they are roughly consistent with the E(B-V) of the host galaxies as derived from the $\rm{H\alpha/H\beta}$ Balmer decrement \citep[we compiled Balmer decrements for 62 out of the 68 AGN, mostly from the measurement of the BASS survey,][]{Koss22,Oh22}. A detailed analysis of the foreground extinction and its role in the SED fitting technique is planned for a future investigation.

Perhaps, the two-phase dust distribution models \citep[e.g. smooth+clumpy torus model developed by][]{Stalevski16} are preferred since they might provide a more realistic treatment of the distribution of dust from the stability point of view because a smooth distribution of dust alone could suffer from the self-destruction of the dust due to its temperature. However, some fine-tuning is still needed because [Stalev16] models have moderate success explaining the overall observed mid-infrared spectra, accounting only for 25-40\% of the sample \citep[Fig.\,\ref{fig:GoodFitsHist}, see also][although it might be due to the somewhat limited number of SEDs computed]{Gonzalez-Martin19B}. A smooth distribution for the disk/torus and a clumpy distribution for the wind might also be a good representation according to MHD simulations \citep{Wada15, Williamson20}.

The new SED library presented by this work is based on the two-phase clumpy torus model described by \citet{Stalevski16} ([Stalev16]) with the inclusion of the maximum grain size as a free parameter. This new SED library provides good fits for 85-88\% of the sample, irrespective of the inclusion of foreground extinction, and provides the best model for 60-70\% of the sample (being the best among the models, see Figs.\,\ref{fig:chis} and \ref{fig:GoodFitsHist}). Indeed, foreground extinction is required only for 34\% of the sample, consistent with the other models including a smooth distribution of dust. [GoMar23] model gives a statistically better fit than [Stalev16] model in 65-70\% of the sample (see Fig.\,\ref{fig:GoMar22AIC}) with only 1-3\% preferring the [Stalev16] model. This is mostly due to the fact that [Stalev16] model is nested within [GoMar23] model, while this small percentage prefers [Stalev16] model might be the result of the stochastic nature of radiative transfer simulations or the inclusion of anisotropy for the accretion disk emission in [Stalev16] model.

\subsubsection{Torus versus wind models}

The most recent torus SED library was produced in an attempt to study the impact of a different geometry of the dust, which is assumed to be a combination of a (relatively) thin disk plus a hollow cone-like structure \citep{Hoenig17}. In fact, it is important to bear in mind that, besides the dust torus explored here, molecular lines show the presence of large and massive disks of cold gas \citep{Garcia-Burillo16,Alonso-Herrero18,Combes19,Garcia-Burillo21} and infrared observations reveal polar component at least for some objects \citep[interpreted as a dusty wind, see][]{Hoenig19,Alonso-Herrero21}. When we compare the spectral fits of our new SED library ([GoMar23]) with the disk+wind clumpy model provided by \citet{Hoenig17} ([Hoenig17]), we find that [GoMar23] model is statistically preferred in 34-50\% of the sample while [Hoenig17] model is statistically better in 22-26\% (see Fig.\,\ref{fig:GoMar22AIC}). Thus, this disk+wind geometry might be key to explain at a non-negligible fraction of nearby AGN. According to the recent work by \citet{Alonso-Herrero21}, the presence of polar dust in the mid-infrared might depend on the column density and Eddington ratio \citep[see also][]{Venanzi20,Garcia-Bernete22}. Evidence for a diffuse mid-infrared emitting polar component has been found in six AGN, including both type-1 and type-2 sources using interferometric observations \citep[Circinus Galaxy, NGC\,424, NGC\,1068, NGC\,3783, ESO\,323-G77, and NGC\,5506, see][and references therein]{ Lopez-Gonzaga16,Leftley18,Hoenig19}. Only NGC\,3783 is included in our sample. Indeed this object favors the disk+wind [Hoenig17] model, which is preferred given the spatially resolved data, although our torus-like [GoMar23] model also provides an equally acceptable fit according to the AIC statistics. 

We want to stress that these two models (torus- versus wind-like configurations) are not competing scenarios but probably two simplifications of a more dynamical scenario. \citet{Wada15} proposed that the nuclear dust in AGN is not static, but rather a dynamic moving failed dusty outflow or ``fountain" that produces the geometrically thick polar-elongated structure. Indeed, \citet{Williamson20} showed that a radiation-driven polar wind is almost inevitable for luminous AGN. The fountain is supported by the radiation pressure from the geometrically thin gas/dust disk. \citet{Baskin18} also explored the idea that the BLR is also part of this failed dust wind conformed by large-size ($\rm{>0.3\mu m}$) graphite grains \citep[see also][]{Czerny11}. The radiation pressure from the AGN naturally produces an inflated torus-like structure, with a predicted peak height at the outer radius of the BLR. \citet{Sarangi19} present detailed calculations of the chemistry of silicate dust formation in (magnetohydrodynamic) winds launched by AGN accretion disk. They show that the resultant distribution of the dense dusty gas resembles a failed wind (which could mimic a toroidal shape), with high column density and optical depths along the equatorial viewing angles, in agreement with the AGN unification picture. This explains why both torus-like and wind-like geometries are able to fit some of the AGN (even for the same object), as found for several authors already \citep[][]{Gonzalez-Martin19B,Martinez-Paredes21,Garcia-Bernete22}. The goal we must pursue is to establish under which conditions each geometry dominates because it will shed light on the evolutionary stages of this failed wind and the role of dust in this cycle.

\begin{figure}[!t]
\begin{center}
\includegraphics[width=1.\columnwidth,clip,trim=0 10 0 55]{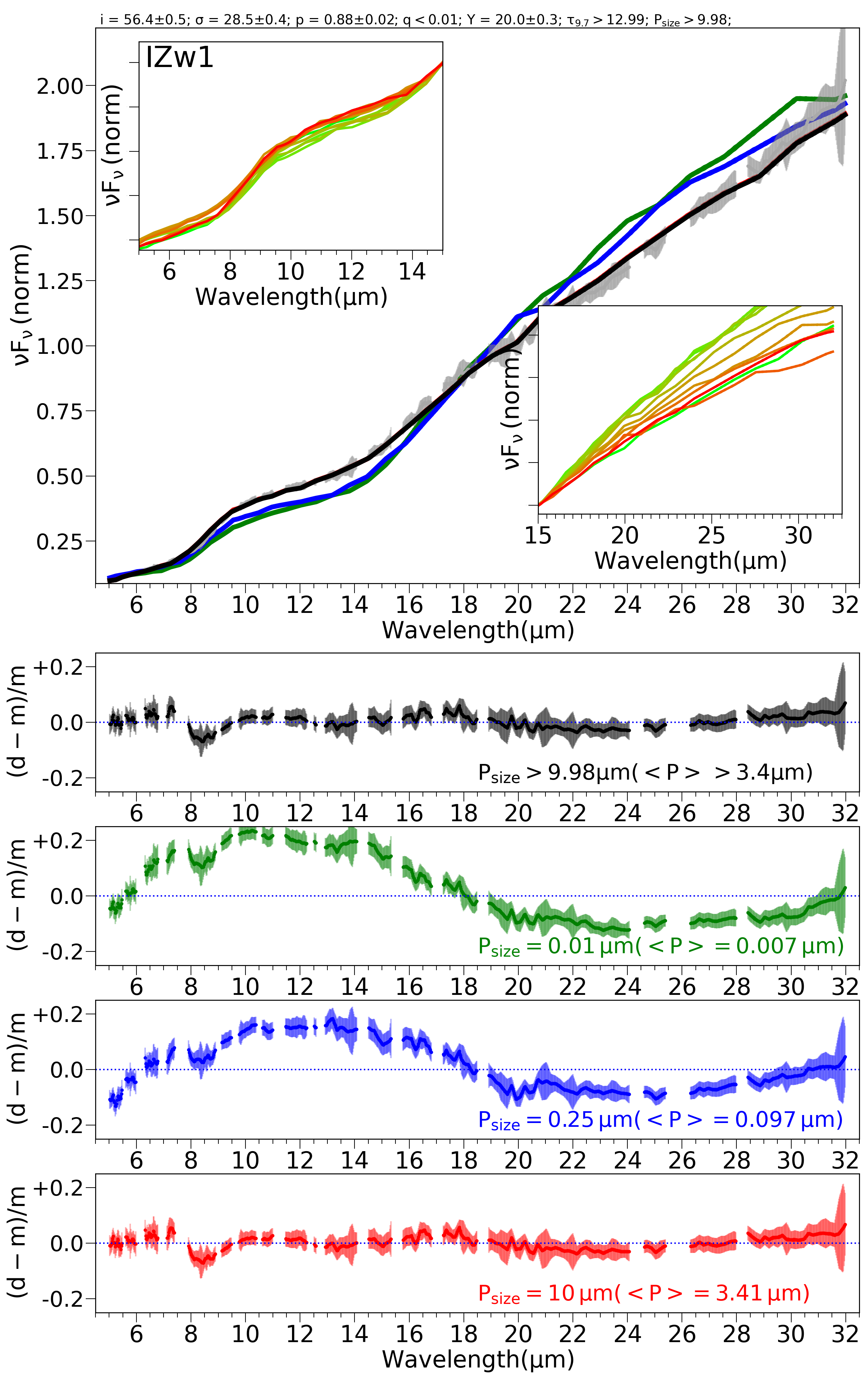} 
\caption{Effect of the maximum grain size on the spectral fit. We use as examples the type-1 AGN IZw1.
The first row shows the best spectral fit using [GoMar23] model (black continuous line). We also show the fit to three fixed maximum grain size $\rm{P_{size}}$ of 0.01$\rm{\mu m}$ ($\rm{<P>=0.007\mu m}$, green), 0.25$\rm{\mu m}$ ($\rm{<P>=0.097\mu m}$, blue), and 10$\rm{\mu m}$ ($\rm{<P>=3.41\mu m}$, red), while other parameters are kept as those obtained from the best fit. The two insets in these top panels show the set of SEDs with the same parameters and varying the maximum grain size $\rm{P_{size}}$ from 0.01$\rm{\mu m}$ ($\rm{<P>=0.007\mu m}$,  green) to 10$\rm{\mu m}$ ($\rm{<P>=3.41\mu m}$, red). Note that in these insets the SEDs are normalized at 15$\rm{\mu m}$. Bottom panels show the residuals of the best fit (black) and those using dust maximum grain size $\rm{P_{size}}$ of 0.01$\rm{\mu m}$ ($\rm{<P>=0.007\mu m}$, green), 0.25$\rm{\mu m}$ ($\rm{<P>=0.097\mu m}$, blue), and 10$\rm{\mu m}$ ($\rm{<P>=3.41\mu m}$, red). 
}
\label{fig:specfitPsize}
\end{center}
\end{figure}

\begin{figure*}[!t]
\begin{center}
\includegraphics[width=0.67\columnwidth,clip,trim=50 5 20 15]{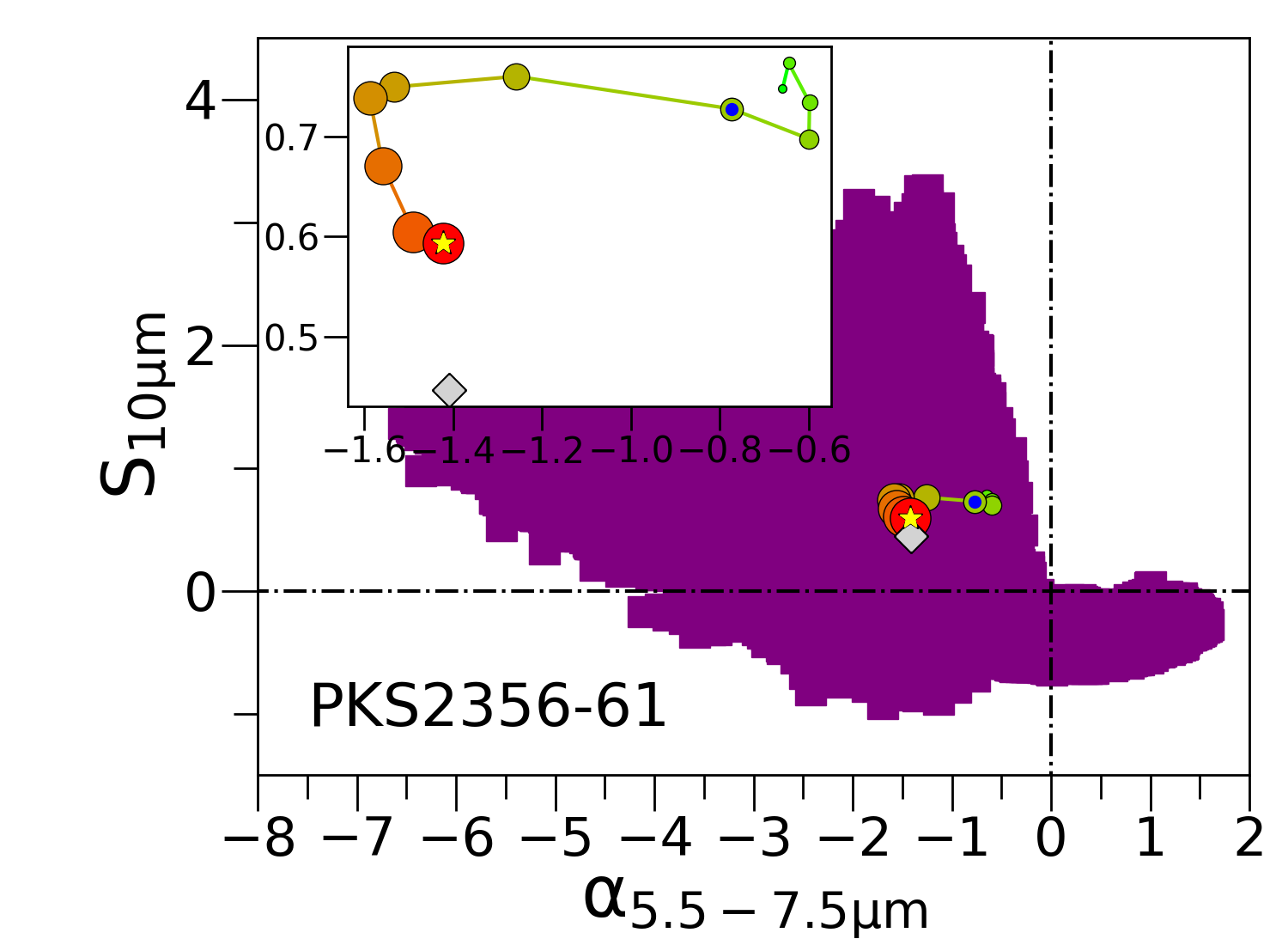} 
\includegraphics[width=0.67\columnwidth,clip,trim=50 5 20 15]{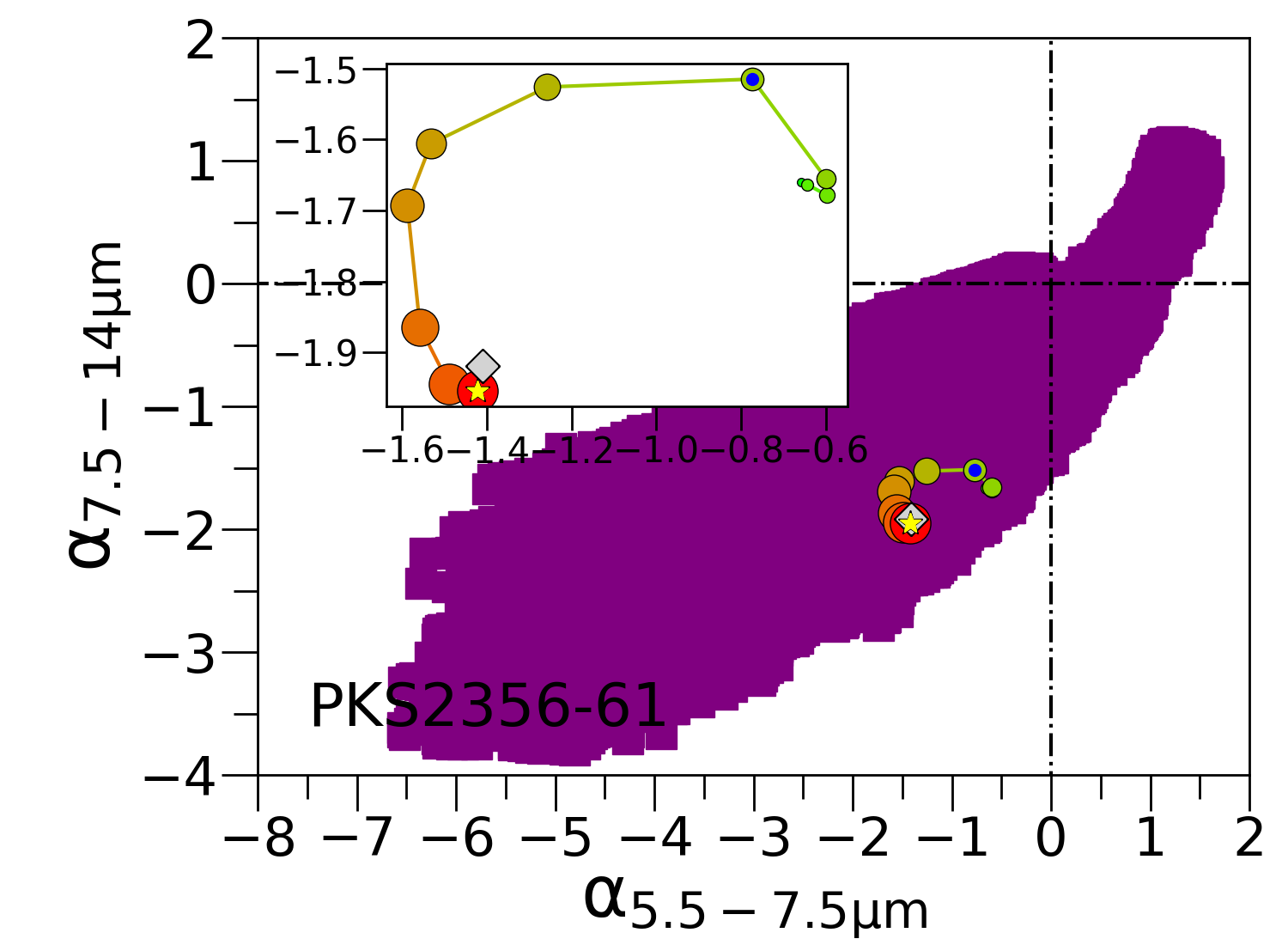} 
\includegraphics[width=0.67\columnwidth,clip,trim=50 5 20 15]{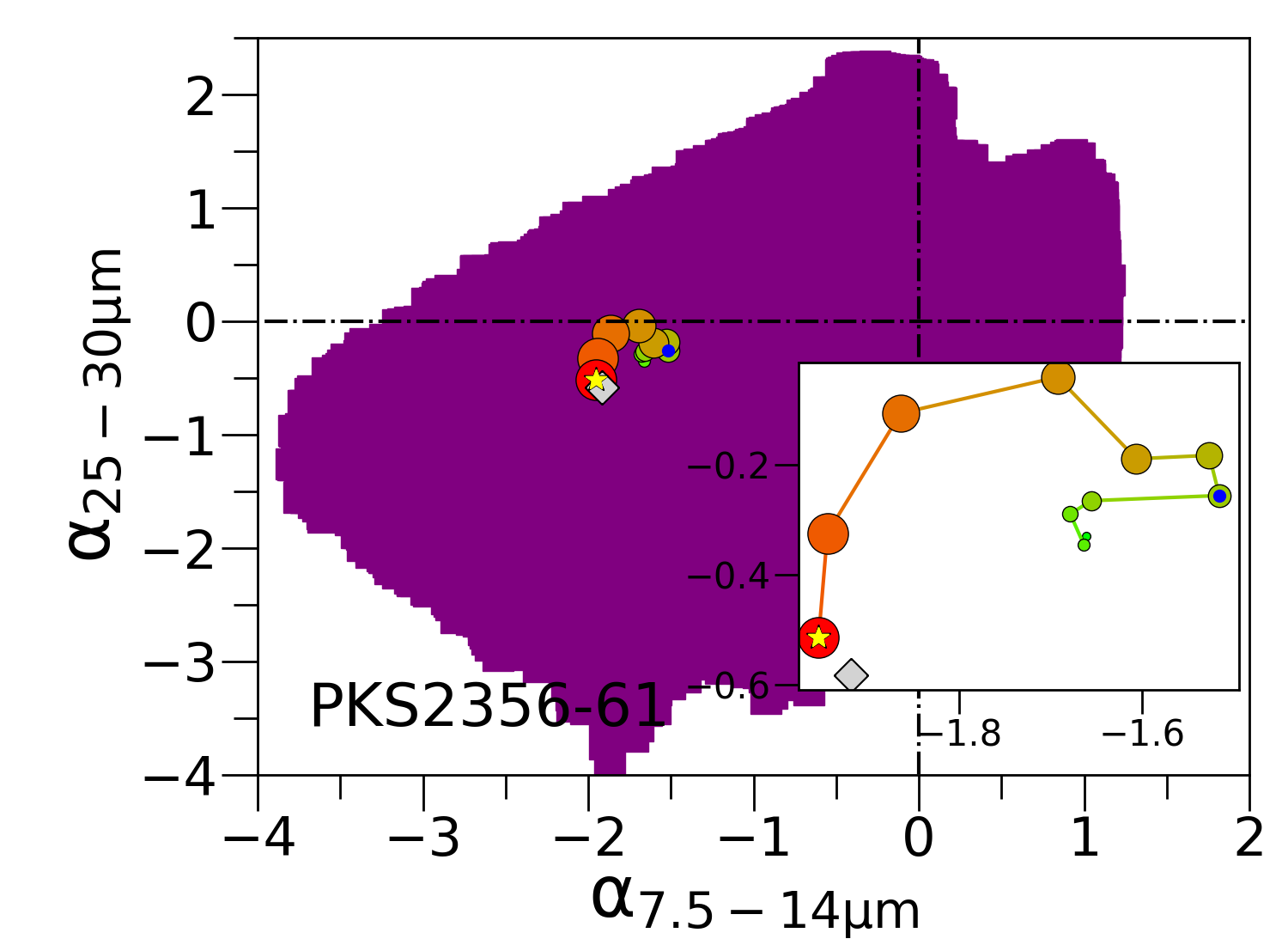} 
\caption{Effect of the maximum grain size on the diagrams of spectral shapes for the type-2 AGN PKS\,2356-61. It shows the locus of the SEDs computed using the best-fit parameters and varying the maximum grain size $\rm{P_{size}}$ from 0.01 to 10 $\rm{\mu m}$ (i.e. $\rm{<P>}$ from 0.007 to 3.41 $\rm{\mu m}$). The spectral shape diagrams are Silicate depth versus $\rm{\alpha_{5.5-7.5\mu m}}$ (left panels), $\rm{\alpha_{7.5-14\mu m}}$ versus $\rm{\alpha_{5.5-7.5\mu m}}$ slopes (middle panels), and $\rm{\alpha_{25-30\mu m}}$ versus $\rm{\alpha_{7.5-14\mu m}}$ (right panels). The empty-filled distribution is that from the entire SED library for [GoMar23] model. Green small dots correspond to smaller grain sizes and redder and larger dots correspond to larger grain sizes (linked with lines). We mark as a blue dot the locus of the resulting values for the SEDs with $\rm{P_{size=0.25\mu m}}$ ($\rm{<P>=0.097\mu m}$) and as a yellow star the locus of the resulting values for the best fitted SED. The type-1 AGN are marked as black squares and type-2 AGN as gray diamonds.} 
\label{fig:specshapePsize}
\end{center}
\end{figure*}

\subsection{The role of the grain size}

As discussed in the introduction, many of the previous AGN dust models used the dust grain size distribution of the Galactic diffuse interstellar medium. However, the assumption that the properties of dust in the extreme conditions of the circumnuclear region of AGN are similar to those of the Galactic diffuse ISM is somewhat naive, given that even in dense clouds of our Galaxy the extinction curve already differs from that of the ISM, suggesting that a universal grain size distribution is not feasibleÊ\citep{Mathis90}. The improvement of the [GoMar23] model presented here, compared to similar models, resides in the addition of the grain size (tabulated as the maximum or the mass-weighted average) as a parameter of the model. 

In this work, we showed that the grain size, as a parameter properly included in the SED library, is required to significantly improve the final spectral fit in $\rm{\sim}$90\% of the sample. Furthermore, among the seven parameters considered in our model, the grain size is the third most important parameter to achieve the best fit (see Fig.\,\ref{fig:GoMarParHist}), only after the half opening angle and the ratio between the outer and the inner radius of the torus. However, note that the lower relevance of the viewing angle might be related to its degeneracy with the half-opening angle of the torus which is most probably associated with the narrow wavelength range or low sensitivity of \emph{Spitzer} spectra. The grain size helps to reproduce the central wavelength of the silicate feature at 9.7$\rm{\mu m}$, which is shifted to longer wavelengths for emission features in AGN. It also helps to reproduce the excess toward near-infrared wavelengths, called the near-infrared bump (see below). 

Previous studies already showed that the removal of small grains and/or the prevalence of large grains makes the silicate feature less prominent \citep{Laor93}. We show that variations in the grain size help to better reproduce silicate strength and infrared slopes at the same time in our sample (see Fig.\,\ref{fig:modelshapeGoMar22:Psize}). In order to further explore the effect of the grain size on the spectral shape, we compare in Fig.\,\ref{fig:specfitPsize} the best spectral fit with those obtained using the maximum grain size of $\rm{P_{size}=0.01\mu m}$ ($\rm{<P>=0.007\mu m}$, green), $\rm{P_{size}=0.25\mu m}$ ($\rm{<P>=0.097\mu m}$, blue), and $\rm{P_{size}=10\mu m}$ ($\rm{<P>=3.41\mu m}$, red) for IZw1. IZw1 requires a maximum grain size of $\rm{P_{size}\sim 10 \mu m}$ ($\rm{<P>=3.41\mu m}$). Note that all the parameters except for the grain size are fixed to those found for the best-fitted model (black line) so that the effect of the grain size can be isolated and better highlighted in the fits. It is clear that the role of grain size is significant. Smaller grains produce deeper silicate absorption features and a steeper slope at longer wavelengths. 

The net effect on the spectral shape can be further explored in Fig.\,\ref{fig:specshapePsize}, where we show the locus of the SEDs with different dust grain sizes in the spectral shape diagrams for PKS\,2356-61. In these plots, green/small dots are the results for small particles (i.e. from $\rm{P_{size}=0.01\mu m}$ or $\rm{<P>=0.007\mu m}$) while red/large dots show the effect of increasing the grain size explored in this analysis (i.e. up to $\rm{P_{size}=10 \mu m}$ or $\rm{<P>=3.41\mu m}$, ). Note that the locus of the observational values for PKS\,2356-61 is shown as a gray diamond.  The spectral shapes of this object are well reproduced after including the grain size as a free parameter, where the best fit (yellow star) is close to the observed value. Indeed, the canonical maximum grain size of $\rm{P_{size}=0.25\mu m}$ ($\rm{<P>=0.097\mu m}$, blue dot) is far from the best value. Note that in most of the objects, the effect on the spectral shape is similar. As long as the grain size increases up to $\rm{P_{size}=1\mu m}$ ($\rm{<P>=0.36\mu m}$) the $\rm{\alpha_{5.5-7.5\mu m}}$ and $\rm{\alpha_{7.5-14\mu m}}$ become steeper, while the $\rm{\alpha_{25-30\mu m}}$ and the silicate strength remains the same. This is easily explained by the size of graphite grains since they contribute to the enhancement of the near-infrared continuum compared to small grains as the central source efficiency to heat the grains increases when the grain size decreases. However, when we use particle distributions with maximum sizes above $\rm{1\mu m}$ ($\rm{<P>=0.36\mu m}$) the silicate strength decreases. This combined effect is crucial in the model to explain the weaker silicate absorption features together with the spectral slopes. 

It is important to remark that our [GoMar23] SED libraries are created by varying the maximum size of both graphite and silicate grains at the same time. However, the dust sublimation radius depends on the dust composition and the grain size \citep[][and references therein]{Baskin18}. Therefore, silicate grain size distribution might differ from that of graphite grains. This was already shown by \citet[e.g.][]{Garcia-Gonzalez17} for the clumpy torus model by \citet{Hoenig10A}, which indeed is one of the key improvements of the disk+wind model by \citet{Hoenig17}. Although out of the scope of this work, this improvement in the model might further help to better reproduce the infrared slopes (mostly affected by graphite grains) and the silicate strengths (associated with the silicate grains). 

\subsection{The role of large grains}

Depletion of small grains or excess of large grains in a sizable fraction of the AGN is well established in our sample (41\% is clustered at $\rm{P_{size}=10\mu m}$ ($\rm{<P>=3.41\mu m}$) according to the results shown in Fig.\,\ref{fig:GoMarParHist}). \citet{Gaskell04} provided evidence for flat extinction curves in AGN, indicative of a larger fraction of larger dust grains in the dusty population. \citet{Lyu14} also suggested that the low $\rm{A_{V}/\tau_{9.7\mu m}}$ ratio found in their analysis could be due to the predominance of large grains in the AGN torus. \citet{Xie17} already suggested that $\rm{\mu m}$-sized particles can help to reproduce up to 90\% of the type-1 AGN explored in their analysis \citep[see also][]{Smith10}, although other aspects as geometry were not considered. Moreover, \citet{Maiolino01} already suggested that the dust composition in the circumnuclear region of AGN could be dominated by large grains ($\rm{P_{size}\sim 1 \mu m}$ or equivalently $\rm{<P>=0.36\mu m}$), making the extinction curve flatter and featureless to explain the reduction of the $\rm{E(B-V)/N_{H}}$ and $\rm{A_{V}/N_{H}}$ ratios compared to the ISM value \citep[see also][]{Shao17}. The decrement of $\rm{E(B-V)/N_{H}}$ was recently confirmed by \citet{Esparza-Arredondo21} from a detailed analysis of the SEDs considering the opacity of the torus.

Large grains ($\rm{P_{size}=0.1-1 \mu m}$ or equivalently $\rm{<P>=0.042-0.36\mu m}$) were preferred for the polar component found for Circinus galaxy \citep{Stalevski17,Stalevski19}. Within the disk+wind geometry proposed by \citet{Hoenig17}, the hot dust is assumed to be primarily comprised of larger grains in the range 0.075--1$\rm{\mu m}$ (or equivalently $\rm{<P>=0.033-0.36\mu m}$), because they are located in inner radii where large grains better survive. The near-infrared excess is explained under the assumption that it might trace dust at temperatures that can only be survived by the largest graphite grains \citep{Garcia-Gonzalez17}. Indeed, this model is better at reproducing spectra shortwards $\rm{\sim 7\mu m}$ when compared with observations \citep[see also][]{Gonzalez-Martin19B,Martinez-Paredes21,Garcia-Bernete22}. Recently, \citet{Garcia-Bernete22} studied the available SED dust libraries in a volume-limited sample of well-isolated and nearby AGN, finding that type-1/bright AGN are best fitted to the disk+wind [Hoenig17] model due to a better performance in reproducing the near-infrared bump. They argue that it is likely due to the combination of removing most silicate and small graphite grains due to sublimation and self-absorption effects due to the wind at intermediate and high inclinations \citep[see also][]{Martinez-Paredes21}. Thus, large grains are again a required element in these models. 

Even for the ISM, micron-sized graphite grains account for 2.5\% of the total IR emission \citep{Wang15}, which helped to unify distance estimates derived from spectroscopy and parallax by considering extinction by large grains \citep{Siebenmorgen23}. However, what is the origin of such an enhancement of large particles? Several effects could explain a preponderance of large grains in the dusty structure near the AGN. We gather them into two classes: grain destruction and grain growth. An obvious form of dust destruction in AGN is due to its radiation field. This trend has also been found for PAH emission of AGN \citep{Garcia-Bernete22a}. At the outer envelope of the sublimation radius of silicates (which sublimate at lower temperatures than graphite grains), we expect that the grain size distribution might be biased to large grains since they better resist the radiation field of the AGN. Thus, grains size distributions skewed to large grains might be due to the preferential destruction of small grains in the AGN radiation field \citep{Draine79}. 

If not destroyed, AGN radiation pressure combined with radiation pressure on dust can sweep out the particles away from the nucleus with an increasing speed that is inverse proportional to grain size, as long as grains are larger than $\rm{\sim 0.1 \mu m}$ \citep[so the radiation pressure efficiency can be approximated to one, see][]{Tazaki20}. Therefore, small grains require less time to reach the terminal velocity. After reaching a terminal velocity, the destruction of dust grains occurs due to sputtering by shock-heated gas \citep{Draine79}. This sputtering makes the grain lifetime proportional to the grain size \citep[i.e. sputtering destroys small grains faster than the large ones][]{Tazaki20}, producing a preponderance of large grains in the wind where this destruction process of small grains enhances the shift of the distribution toward larger sizes. Larger grains may also form either by accretion from the gas phase or by grain coagulation \citep{Li01}. \citet{Sarangi19}, from their calculations of the chemistry of silicate dust in AGN winds, reported that these winds have physical conditions suitable to form significant amounts of dust, especially for objects with an accretion rate close to their Eddington limit, making luminous AGN a source of dust in the universe. Since the grain coagulation rate increases with density \citep[proportional to the square root of the density][]{Draine84}, the effect is expected to be dramatic in the circumnuclear clouds of AGN \citep{Maiolino01}. \citet{Maiolino01} showed that dust grains have time to coagulate before being replaced by gas coming from the outer regions. 

Therefore, a dust distribution biased in favor of large grains, probably as a consequence of a trade-off between coagulation and accretion processes, is also naturally expected in the extreme physical conditions of the gas in the circumnuclear region of AGN. Furthermore, a significant variation of the maximum grain size for each object is expected depending on the evolutionary stage of the AGN (including wind phase) and its local conditions \citep{Sarangi19}, which is also consistent with our results.

\section{Summary} \label{sec:Summary}

The role of grain size in geometrical models of AGN dust continuum is poorly explored in the literature. In order to fill this gap we created a new SED library (including over 700,000 SEDs) based on the two-phase (smooth+clumpy) torus model described by \citet{Stalevski16}, including the grain size (denoted by the maximum grain size $\rm{P_{size}}$ or equivalently by the mass-weighted average grain size $\rm{<P>}$) as a model parameter. We then compared our new and previous models with a sample of nearby AGN (mainly Seyfert galaxies with intermediate luminosities) observed with \emph{Spitzer}/IRS (carefully selected to provide spectra dominated by the AGN component). The explored models are: the smooth torus model by \citet{Fritz06}, the clumpy torus models by \citet{Nenkova08B} and by \citet{Hoenig10B}, the two-phase torus model by \citet{Stalevski16}, and the clumpy disk+wind model by \citet{Hoenig17}. We named our new model [GoMar23], following the last name of the first author of the paper and the year of its submission (as we did for the other models). The main results are summarized as:

\begin{itemize}
    \item Although the range of spectral shapes of the objects in our sample is recovered with most of the models, a unique SED capable of reproducing the whole mid-infrared spectra of AGN is not found for a non-negligible fraction of objects (at least 15\%) with the already available models. The [GoMar23] model represents a significant improvement of this, providing good fits for 85\% of the sample.
    \item We demonstrate (with statistical tests) that allowing the maximum dust grain size up to sizes of 10$\rm{\mu m}$ (or the mass-weighted average grain size of 3.41$\rm{\mu m}$) is required to successfully reproduce the SEDs of 90\% of the sample. The canonical value of $\rm{P_{size}\sim 0.25\mu m}$ ($\rm{<P>\sim 0.097 \mu m}$) is preferred for $\rm{\sim}$9\% of the sample. 41\% of the sample prefers very large particles with $\rm{P_{size}\sim 10\mu m}$ ($\rm{<P>\sim 3.41 \mu m}$).
    \item The dust grain size is the third parameter in the order of its importance to obtain a good spectral fit. Only the half-opening angle and the ratio between the outer and inner radius of the dusty structure are above the dust grain size in this prioritized list of parameters. 
    \item The inclusion of foreground extinction for the smooth or two-phase models (i.e. [Fritz06], [GoMar23], and [Hoenig17] models) is irrelevant in 70-90\% of the objects while it is required in $\rm{\sim}$90\% of the objects when a purely clumpy distribution is used (i.e. [Nenkova08], [Hoenig10], and [Hoenig17] models). Therefore, the incorporation of a smooth dust distribution (smooth or smooth+clumpy) helps to reproduce the spectra without the need for additional foreground extinction for most of the objects.
  
\end{itemize}

As a concluding remark, these geometrical models gain in importance with the upcoming new data from the \emph{JWST}. Its superior sensitivity will allow us to obtain near- and mid-infrared spectra of fainter and/or more distant AGN that could be compared with these models to infer the properties of AGN nuclear dust. The [GoMar23] SED library is the first step to providing an improved geometrical model including the maximum grain size into the parameter space. However, there are several aspects that need to be further explored, including the geometry of the components (e.g. the polar/wind dusty component), dust composition (e.g. the fraction of graphite versus silicate grains), grain shape, or the slope of the grain size distribution. These aspects should be the focus of future investigations.

\begin{acknowledgements}
We thank the anonymous referee for his/her useful comments. We thank Dr. Peter Camps for his useful suggestions on the SKIRT's convergency tests. This research is mainly funded by the UNAM PAPIIT project IN105720 (PI OG-M). This research has made use of dedicated servers (IRyAGN2, Galaxias and Arambolas servers, and Calzonzin and Mouruka clusters) maintained by Daniel D\'iaz- Gonz\'alez, Miguel Espejel, Alfonso Ginori Gonz\'alez, and Gilberto Zavala at IRyA-UNAM. All of them are gratefully acknowledged. CRA acknowledges financial support from the project ``Feeding and feedback in active galaxies'', with reference PID2019-106027GB-C42, funded by MICINN-AEI/10.13039/501100011033. AAH acknowledges support from grant PID2021-124665NB-I00 Êfunded by the Spanish Ministry of Science and Innovation and the State Agency of Research MCIN/AEI/10.13039/501100011033. D.E.-A. acknowledges support from the Spanish Ministry of Science, Innovation, and Universities (MCIU), Agencia Estatal de Investigación (AEI), and the Fondo Europeo de Desarrollo Regional (EU-FEDER) under projects with references AYA2015-68217-P and PID2019- 107010GB-100. IGB acknowledges support from STFC through grant ST/S000488/1. NOC acknowledges support from CONACYT scholarship 897887. M.S. acknowledges support by the Science Fund of the Republic of Serbia, PROMIS 6060916, BOWIE, and by the Ministry of Education, Science and Technological Development of the Republic of Serbia through the contract No. 451-03-9/2022-14/200002. CV-C acknowledges support from a CONACyT scholarship. 
\end{acknowledgements}

%
%

\begin{table*}[ht!]
\scriptsize 
\renewcommand{\arraystretch}{0.8806}
\begin{center}
\begin{tabular}{lccccccccccccc}\hline \hline
Object name        &  z     &Class& $\rm{log(L_{X})}$ & $\rm{log(M_{BH})}$ &$\rm{C_{10\mu m}}$ & $\rm{S_{10\mu m}}$ &  $\rm{\alpha_{5.5-7.5\mu m}}$ &  $\rm{\alpha_{7.5-14\mu m}}$ &  $\rm{\alpha_{25-30\mu m}}$ \\
      (1)                &  (2)   & (3) &   (4)  &    (5)    &      (6)             &     (7)                &          (8)         &       (9)            &       (10)           \\ \hline
Mrk348                   &  0.005 & S2  &  43.86 &    7.39       &   9.66$\rm{\pm}$0.14 &   0.220$\rm{\pm}$0.008 &  -1.07$\rm{\pm}$0.04 &  -1.28$\rm{\pm}$0.02 &  -0.18$\rm{\pm}$0.06 \\ 
IZw1                     &  0.059 & S1  &  43.85 &    7.26       &   9.82$\rm{\pm}$0.12 &  -0.291$\rm{\pm}$0.008 &  -1.24$\rm{\pm}$0.05 &  -1.17$\rm{\pm}$0.04 &  -0.93$\rm{\pm}$0.08 \\ 
FAIRALL9                 &  0.047 & S1  &  44.09 &    8.21       &  10.60$\rm{\pm}$0.14 &  -0.184$\rm{\pm}$0.004 &  -0.68$\rm{\pm}$0.04 &  -0.78$\rm{\pm}$0.03 &   0.38$\rm{\pm}$0.06 \\ 
NGC526A                  &  0.019 & S1  &  43.78 &    8.05       &  11.30$\rm{\pm}$0.80 &  -0.136$\rm{\pm}$0.004 &  -0.93$\rm{\pm}$0.04 &  -1.44$\rm{\pm}$0.02 &   1.02$\rm{\pm}$0.05 \\ 
NGC788                   &  0.014 & S2  &  43.51 &    7.74       &   9.57$\rm{\pm}$0.17 &   0.151$\rm{\pm}$0.004 &  -2.13$\rm{\pm}$0.05 &  -1.80$\rm{\pm}$0.02 &  -0.55$\rm{\pm}$0.03 \\ 
Mrk1018                  &  0.042 & S1  &  42.82 &    8.21       &  10.37$\rm{\pm}$0.03 &  -0.260$\rm{\pm}$0.002 &  -0.20$\rm{\pm}$0.01 &  -0.69$\rm{\pm}$0.01 &   0.42$\rm{\pm}$0.04 \\ 
Mrk590                   &  0.021 & S1  &  43.23 &    7.85       &  10.56$\rm{\pm}$0.06 &  -0.200$\rm{\pm}$0.004 &  -1.46$\rm{\pm}$0.03 &  -2.12$\rm{\pm}$0.01 &  -0.20$\rm{\pm}$0.04 \\ 
NGC1052                  &  0.005 & S1  &  42.24 &    8.39       &  10.89$\rm{\pm}$0.04 &  -0.265$\rm{\pm}$0.002 &  -1.18$\rm{\pm}$0.04 &  -1.64$\rm{\pm}$0.01 &  -0.54$\rm{\pm}$0.05 \\ 
NGC1275                  &  0.016 & S2  &  43.76 &    9.00       &  10.46$\rm{\pm}$0.21 &  -0.393$\rm{\pm}$0.008 &  -2.18$\rm{\pm}$0.15 &  -2.58$\rm{\pm}$0.06 &  -1.14$\rm{\pm}$0.13 \\ 
ESO548-G081              &  0.014 & S1  &  43.29 &    8.34       &  10.70$\rm{\pm}$0.05 &  -0.271$\rm{\pm}$0.003 &  -0.25$\rm{\pm}$0.02 &  -0.39$\rm{\pm}$0.01 &   0.62$\rm{\pm}$0.03 \\ 
3C120                    &  0.033 & S1  &  44.38 &    7.85       &  10.65$\rm{\pm}$0.06 &  -0.305$\rm{\pm}$0.004 &  -0.76$\rm{\pm}$0.04 &  -1.25$\rm{\pm}$0.02 &  -0.70$\rm{\pm}$0.05 \\ 
MCG-01-13-025            &  0.016 & S1  &  43.29 &               &  10.68$\rm{\pm}$0.02 &  -0.575$\rm{\pm}$0.001 &   0.38$\rm{\pm}$0.01 &  -0.81$\rm{\pm}$0.01 &   0.07$\rm{\pm}$0.02 \\ 
CGCG420-015              &  0.029 & S2  &  43.74 &    7.88       &  10.98$\rm{\pm}$0.40 &  -0.138$\rm{\pm}$0.005 &  -1.37$\rm{\pm}$0.04 &  -1.19$\rm{\pm}$0.02 &  -0.21$\rm{\pm}$0.05 \\ 
2MASXJ05054575-2351139   &  0.035 & S2  &  44.22 &               &   9.50$\rm{\pm}$0.05 &   0.197$\rm{\pm}$0.004 &  -1.76$\rm{\pm}$0.02 &  -1.30$\rm{\pm}$0.01 &  -0.12$\rm{\pm}$0.03 \\ 
Ark120                   &  0.032 & S1  &  43.78 &    8.24       &  10.60$\rm{\pm}$0.13 &  -0.266$\rm{\pm}$0.006 &  -0.44$\rm{\pm}$0.02 &  -0.43$\rm{\pm}$0.01 &   0.22$\rm{\pm}$0.05 \\ 
PICTORA                  &  0.035 & S1  &  44.03 &    7.60       &  10.62$\rm{\pm}$0.02 &  -0.502$\rm{\pm}$0.002 &  -0.87$\rm{\pm}$0.02 &  -1.52$\rm{\pm}$0.01 &   0.51$\rm{\pm}$0.02 \\ 
2MASXJ05580206-3820043   &  0.034 & S1  &  43.86 &    8.06       &   9.74$\rm{\pm}$0.33 &   0.289$\rm{\pm}$0.025 &  -0.51$\rm{\pm}$0.14 &  -0.55$\rm{\pm}$0.04 &   0.65$\rm{\pm}$0.11 \\ 
Mrk3                     &  0.014 & S2  &  43.79 &    8.42       &   9.82$\rm{\pm}$0.28 &   0.182$\rm{\pm}$0.012 &  -2.28$\rm{\pm}$0.18 &  -2.30$\rm{\pm}$0.09 &  -0.59$\rm{\pm}$0.15 \\ 
ESO426-G002              &  0.022 & S2  &  43.44 &               &   9.29$\rm{\pm}$0.02 &   0.212$\rm{\pm}$0.003 &  -1.65$\rm{\pm}$0.02 &  -1.50$\rm{\pm}$0.01 &  -0.70$\rm{\pm}$0.02 \\ 
Mrk78                    &  0.037 & S2  &  43.47 &    6.66       &   9.71$\rm{\pm}$0.03 &   0.508$\rm{\pm}$0.004 &  -1.91$\rm{\pm}$0.04 &  -1.80$\rm{\pm}$0.02 &  -1.14$\rm{\pm}$0.04 \\ 
Mrk1210                  &  0.013 & S1  &  43.37 &    6.09       &  11.04$\rm{\pm}$0.77 &  -0.191$\rm{\pm}$0.010 &  -2.19$\rm{\pm}$0.09 &  -2.02$\rm{\pm}$0.03 &  -0.32$\rm{\pm}$0.08 \\ 
PG0804+761               &  0.100 & S1  &  44.46 &    8.50       &  10.16$\rm{\pm}$0.04 &  -0.433$\rm{\pm}$0.002 &  -0.23$\rm{\pm}$0.02 &  -0.61$\rm{\pm}$0.01 &   0.36$\rm{\pm}$0.05 \\ 
MCG+04-22-042            &  0.032 & S1  &  43.73 &    7.46       &  10.45$\rm{\pm}$0.11 &  -0.207$\rm{\pm}$0.005 &  -0.97$\rm{\pm}$0.02 &  -1.10$\rm{\pm}$0.01 &   0.55$\rm{\pm}$0.04 \\ 
Mrk110                   &  0.033 & S1  &  44.25 &    7.31       &  10.49$\rm{\pm}$0.04 &  -0.243$\rm{\pm}$0.002 &  -0.55$\rm{\pm}$0.01 &  -0.78$\rm{\pm}$0.01 &   0.78$\rm{\pm}$0.02 \\ 
Mrk705                   &  0.029 & S1  &  43.41 &    7.62       &  11.54$\rm{\pm}$0.19 &  -0.112$\rm{\pm}$0.004 &  -1.73$\rm{\pm}$0.02 &  -1.25$\rm{\pm}$0.01 &  -0.85$\rm{\pm}$0.03 \\ 
MCG-05-23-016            &  0.008 & S2  &  43.53 &    7.43       &   9.77$\rm{\pm}$0.32 &   0.298$\rm{\pm}$0.024 &  -1.67$\rm{\pm}$0.12 &  -1.66$\rm{\pm}$0.05 &  -0.44$\rm{\pm}$0.11 \\ 
NGC3081                  &  0.006 & S2  &  43.07 &    7.72       &   9.36$\rm{\pm}$0.12 &   0.157$\rm{\pm}$0.007 &  -2.22$\rm{\pm}$0.08 &  -2.02$\rm{\pm}$0.03 &  -1.12$\rm{\pm}$0.05 \\ 
ESO374-G044              &  0.028 & S2  &  43.64 &               &  11.61$\rm{\pm}$0.79 &  -0.100$\rm{\pm}$0.002 &  -2.41$\rm{\pm}$0.04 &  -2.46$\rm{\pm}$0.01 &  -0.47$\rm{\pm}$0.03 \\ 
Mrk417                   &  0.033 & S2  &  43.91 &    8.08       &  11.65$\rm{\pm}$0.13 &  -0.107$\rm{\pm}$0.004 &  -1.17$\rm{\pm}$0.04 &  -1.80$\rm{\pm}$0.02 &   0.37$\rm{\pm}$0.05 \\ 
2MASSXJ10594361+6504063  &  0.084 & S2  &  43.48 &    8.45       &   9.78$\rm{\pm}$0.01 &   0.811$\rm{\pm}$0.003 &  -1.31$\rm{\pm}$0.01 &  -1.34$\rm{\pm}$0.01 &  -0.57$\rm{\pm}$0.02 \\ 
ESO439-G009              &  0.025 & S2  &  43.27 &               &   9.58$\rm{\pm}$0.06 &   0.426$\rm{\pm}$0.003 &  -3.22$\rm{\pm}$0.06 &  -2.40$\rm{\pm}$0.02 &  -0.80$\rm{\pm}$0.03 \\ 
NGC3783                  &  0.011 & S1  &  43.56 &    7.30       &  10.81$\rm{\pm}$0.31 &  -0.258$\rm{\pm}$0.010 &  -0.81$\rm{\pm}$0.29 &  -1.56$\rm{\pm}$0.11 &  -0.39$\rm{\pm}$0.12 \\ 
UGC6728                  &  0.007 & S1  &  42.40 &    6.44       &   9.97$\rm{\pm}$0.10 &  -0.223$\rm{\pm}$0.002 &  -0.71$\rm{\pm}$0.02 &  -1.05$\rm{\pm}$0.01 &   0.16$\rm{\pm}$0.02 \\ 
2MASXJ11454045-1827149   &  0.033 & S1  &  44.08 &    7.01       &  10.48$\rm{\pm}$0.24 &  -0.223$\rm{\pm}$0.003 &  -0.63$\rm{\pm}$0.01 &  -1.03$\rm{\pm}$0.01 &   0.36$\rm{\pm}$0.04 \\ 
Ark347                   &  0.022 & S2  &  43.52 &    7.98       &  10.45$\rm{\pm}$0.17 &  -0.095$\rm{\pm}$0.027 &  -1.15$\rm{\pm}$0.03 &  -1.28$\rm{\pm}$0.01 &  -0.73$\rm{\pm}$0.03 \\ 
NGC4151                  &  0.002 & S1  &  43.17 &    7.42       &  10.54$\rm{\pm}$0.41 &  -0.092$\rm{\pm}$0.156 &  -1.06$\rm{\pm}$0.18 &  -1.57$\rm{\pm}$0.09 &  -0.28$\rm{\pm}$0.32 \\ 
PG1211+143               &  0.090 & S1  &  43.70 &    7.65       &  10.34$\rm{\pm}$0.05 &  -0.315$\rm{\pm}$0.003 &  -0.80$\rm{\pm}$0.02 &  -0.91$\rm{\pm}$0.01 &   0.47$\rm{\pm}$0.04 \\ 
M106                     &  0.002 & S1  &  41.06 &    7.54       &  11.13$\rm{\pm}$0.04 &  -0.394$\rm{\pm}$0.003 &  -1.19$\rm{\pm}$0.04 &  -1.13$\rm{\pm}$0.02 &  -1.13$\rm{\pm}$0.04 \\ 
NGC4388                  &  0.005 & S2  &  43.64 &    6.98       &   9.85$\rm{\pm}$0.07 &   0.816$\rm{\pm}$0.017 &  -2.14$\rm{\pm}$0.12 &  -1.62$\rm{\pm}$0.06 &  -1.34$\rm{\pm}$0.10 \\ 
NGC4507                  &  0.012 & S2  &  43.76 &    7.87       &  11.23$\rm{\pm}$0.65 &  -0.193$\rm{\pm}$0.015 &  -1.32$\rm{\pm}$0.08 &  -1.08$\rm{\pm}$0.03 &  -1.29$\rm{\pm}$0.06 \\ 
NGC4939                  &  0.009 & S2  &  42.81 &    7.91       &   9.83$\rm{\pm}$0.10 &  -0.243$\rm{\pm}$0.001 &  -1.19$\rm{\pm}$0.05 &  -2.30$\rm{\pm}$0.02 &  -0.82$\rm{\pm}$0.02 \\ 
IISZ010                  &  0.034 & S1  &  43.52 &    7.34       &  10.68$\rm{\pm}$0.03 &  -0.177$\rm{\pm}$0.002 &  -1.58$\rm{\pm}$0.01 &  -1.35$\rm{\pm}$0.01 &   0.71$\rm{\pm}$0.02 \\ 
MCG-03-34-064            &  0.020 & S1  &  43.28 &    8.12       &   9.30$\rm{\pm}$0.16 &   0.301$\rm{\pm}$0.015 &  -2.31$\rm{\pm}$0.14 &  -1.71$\rm{\pm}$0.07 &  -1.07$\rm{\pm}$0.17 \\ 
MCG-06-30-015            &  0.008 & S1  &  42.82 &    6.94       &  10.77$\rm{\pm}$0.39 &  -0.112$\rm{\pm}$0.009 &  -1.15$\rm{\pm}$0.04 &  -1.06$\rm{\pm}$0.02 &  -0.74$\rm{\pm}$0.06 \\ 
IC4329A                  &  0.016 & S1  &  43.77 &    7.51       &  11.17$\rm{\pm}$1.21 &  -0.116$\rm{\pm}$0.048 &  -0.76$\rm{\pm}$0.12 &  -1.13$\rm{\pm}$0.06 &   0.13$\rm{\pm}$0.17 \\ 
UM614                    &  0.033 & S1  &  41.74 &    7.49 (HL)  &  10.64$\rm{\pm}$0.02 &  -0.261$\rm{\pm}$0.001 &  -0.99$\rm{\pm}$0.01 &  -1.18$\rm{\pm}$0.01 &   0.53$\rm{\pm}$0.03 \\ 
Mrk279                   &  0.030 & S1  &  43.87 &    7.86       &  11.42$\rm{\pm}$0.74 &  -0.108$\rm{\pm}$0.006 &  -1.21$\rm{\pm}$0.05 &  -1.16$\rm{\pm}$0.02 &  -0.49$\rm{\pm}$0.06 \\ 
PG1351+640               &  0.088 & S1  &  43.10 &    8.07       &   9.76$\rm{\pm}$0.01 &  -0.854$\rm{\pm}$0.002 &  -1.06$\rm{\pm}$0.03 &  -1.79$\rm{\pm}$0.01 &  -0.54$\rm{\pm}$0.04 \\ 
NGC5548                  &  0.025 & S1  &  43.76 &    7.79       &  10.71$\rm{\pm}$0.10 &  -0.212$\rm{\pm}$0.005 &  -2.10$\rm{\pm}$0.07 &  -1.52$\rm{\pm}$0.03 &  -0.52$\rm{\pm}$0.05 \\ 
ESO511-G030              &  0.015 & S1  &  43.65 &    7.84       &  10.46$\rm{\pm}$0.02 &  -0.365$\rm{\pm}$0.002 &  -0.17$\rm{\pm}$0.01 &  -0.51$\rm{\pm}$0.01 &   0.75$\rm{\pm}$0.06 \\ 
PG1448+273               &  0.065 & S1  &  43.30 &    6.92       &  11.55$\rm{\pm}$0.37 &  -0.140$\rm{\pm}$0.002 &  -1.26$\rm{\pm}$0.01 &  -1.22$\rm{\pm}$0.01 &  -0.30$\rm{\pm}$0.02 \\ 
IC4518W                  &  0.016 & S2  &  43.19 &    7.71       &   9.84$\rm{\pm}$0.05 &   1.370$\rm{\pm}$0.013 &  -2.27$\rm{\pm}$0.07 &  -1.05$\rm{\pm}$0.03 &  -2.31$\rm{\pm}$0.05 \\ 
Mrk841                   &  0.036 & S1  &  44.01 &    8.08       &  10.85$\rm{\pm}$0.53 &  -0.101$\rm{\pm}$0.005 &  -1.26$\rm{\pm}$0.04 &  -1.72$\rm{\pm}$0.02 &  -0.33$\rm{\pm}$0.06 \\ 
Mrk1392                  &  0.036 & S1  &  43.74 &    8.30 (HL)  &  10.75$\rm{\pm}$0.05 &  -0.198$\rm{\pm}$0.003 &  -1.46$\rm{\pm}$0.02 &  -1.74$\rm{\pm}$0.01 &  -0.51$\rm{\pm}$0.02 \\ 
Mrk1393                  &  0.054 & S1  &  43.80 &    8.61       &  10.69$\rm{\pm}$0.19 &  -0.118$\rm{\pm}$0.001 &  -0.75$\rm{\pm}$0.01 &  -1.74$\rm{\pm}$0.01 &  -0.37$\rm{\pm}$0.01 \\ 
Mrk290                   &  0.030 & S1  &  43.68 &    7.46       &  10.77$\rm{\pm}$0.04 &  -0.225$\rm{\pm}$0.002 &  -0.98$\rm{\pm}$0.01 &  -1.30$\rm{\pm}$0.01 &  -0.15$\rm{\pm}$0.02 \\ 
ESO138-G001              &  0.009 & S2  &  42.55 &    6.66       &  10.74$\rm{\pm}$0.50 &  -0.147$\rm{\pm}$0.161 &  -1.70$\rm{\pm}$0.13 &  -1.36$\rm{\pm}$0.07 &  -0.67$\rm{\pm}$0.17 \\ 
ESO103-G35               &  0.013 & S2  &  43.63 &    7.48       &   9.71$\rm{\pm}$0.05 &   0.816$\rm{\pm}$0.016 &  -2.22$\rm{\pm}$0.13 &  -1.91$\rm{\pm}$0.06 &  -0.49$\rm{\pm}$0.11 \\ 
Fairall51                &  0.011 & S1  &  43.22 &    6.84       &  11.12$\rm{\pm}$0.13 &  -0.277$\rm{\pm}$0.007 &  -1.30$\rm{\pm}$0.05 &  -1.06$\rm{\pm}$0.02 &  -0.86$\rm{\pm}$0.04 \\ 
ESO141-G055              &  0.037 & S1  &  44.25 &    7.50       &  10.63$\rm{\pm}$0.05 &  -0.344$\rm{\pm}$0.003 &  -0.39$\rm{\pm}$0.03 &  -0.75$\rm{\pm}$0.01 &   0.06$\rm{\pm}$0.03 \\ 
NGC6814                  &  0.003 & S1  &  42.59 &    7.08       &  11.31$\rm{\pm}$0.53 &  -0.203$\rm{\pm}$0.008 &  -0.77$\rm{\pm}$0.05 &  -0.98$\rm{\pm}$0.02 &  -0.89$\rm{\pm}$0.04 \\ 
MCG+07-41-03             &  0.056 & S2  &  44.57 &    9.11       &   9.70$\rm{\pm}$0.02 &   0.680$\rm{\pm}$0.005 &  -2.93$\rm{\pm}$0.09 &  -2.73$\rm{\pm}$0.04 &  -1.82$\rm{\pm}$0.05 \\ 
IC5063                   &  0.009 & S2  &  43.29 &    7.63       &   9.70$\rm{\pm}$0.47 &   0.279$\rm{\pm}$0.026 &  -2.58$\rm{\pm}$0.13 &  -1.95$\rm{\pm}$0.06 &  -0.82$\rm{\pm}$0.18 \\ 
IIZw136                  &  0.078 & S1  &  43.50 &    7.61       &  10.71$\rm{\pm}$1.00 &  -0.068$\rm{\pm}$0.004 &  -1.17$\rm{\pm}$0.03 &  -0.90$\rm{\pm}$0.01 &  -0.40$\rm{\pm}$0.04 \\ 
NGC7213                  &  0.005 & S1  &  42.46 &    7.71       &  10.61$\rm{\pm}$0.03 &  -0.659$\rm{\pm}$0.003 &  -0.55$\rm{\pm}$0.05 &  -1.42$\rm{\pm}$0.02 &  -0.05$\rm{\pm}$0.04 \\ 
NGC7314                  &  0.004 & S2  &  42.47 &    6.59       &   9.62$\rm{\pm}$0.06 &   0.497$\rm{\pm}$0.009 &  -2.05$\rm{\pm}$0.05 &  -1.73$\rm{\pm}$0.02 &  -1.77$\rm{\pm}$0.02 \\ 
PG2304+042               &  0.042 & S1  &  43.40 &               &  10.52$\rm{\pm}$0.01 &  -0.569$\rm{\pm}$0.001 &  -0.20$\rm{\pm}$0.01 &  -1.30$\rm{\pm}$0.01 &   0.37$\rm{\pm}$0.02 \\ 
PKS2356-61               &  0.096 & S2  &  43.77 &    8.96       &   9.73$\rm{\pm}$0.02 &   0.573$\rm{\pm}$0.001 &  -1.40$\rm{\pm}$0.01 &  -1.92$\rm{\pm}$0.01 &  -0.58$\rm{\pm}$0.01 \\ 
\hline\hline
\end{tabular}
\end{center}
\caption{Observational details of the sample of AGN observed by \emph{Spitzer}, including spectral shape measurements (see text).}
\label{tab:sample}
\end{table*}

\begin{table*}[ht!]
\scriptsize 
\renewcommand{\arraystretch}{0.9}
\begin{center}
\begin{tabular}{llccccl}\hline \hline
\# & Object name        &  Best model     &$\rm{\chi^2/dof (dof)}$&  $\rm{\chi^2/dof (dof)}$ & Final model  & Equally good models \\
                  &    &  & Best model & [GoMar23]  &     &  \\ \hline
1 & Mrk348  &  [Hoenig17]  &  1.49  (948) & 1.3  (949) & [GoMar23] & [Fritz06] [Hoenig17]  \\
2 & IZw1  &  [Stalev16]  &  1.5  (999) & 0.69  (998) & [GoMar23] &  \\
3 & FAIRALL9  &  [Hoenig17]  &  0.64  (986) & 0.61  (987) & [GoMar23] & [Hoenig17]  \\
4 & NGC526A  &  [Hoenig10]  &  0.58  (955) & 0.66  (953) & [Hoenig10] & [GoMar23]  \\
5 & NGC788  &  [Nenkova08]  &  0.73  (946) & 0.94  (945) & [Nenkova08] & [GoMar23]  \\
6 & Mrk1018  &  [Fritz06]  &  0.97  (983) & 0.5  (982) & [GoMar23] &  \\
7 & Mrk590  &  [Hoenig17]  &  0.51  (954) & 0.8  (955) & [Hoenig17] &  \\
8 & NGC1052  &  [Hoenig17]  &  1.59  (935) & 2.72  (936) & [Hoenig17] & [Nenkova08]  \\
9 & NGC1275  &  [Nenkova08]  &  0.69  (952) & 4.64  (951) & [Nenkova08] &  \\
10 & ESO548-G081  &  [Hoenig17]  &  0.53  (946) & 0.89  (947) & [Hoenig17] &  \\
11 & 3C120  &  [Hoenig17]  &  1.7  (971) & 2.11  (972) & [Hoenig17] & [Nenkova08] [GoMar23]$\rm{^\bullet}$  \\
12 & MCG-01-13-025  &  [Hoenig17]  &  0.62  (951) & 1.49  (952) & [Hoenig17] &  \\
13 & CGCG420-015  &  [Hoenig17]  &  1.03  (965) & 1.08  (966) & [Hoenig17] & [Fritz06] [Stalev16] [GoMar23]  \\
14 & 2MASXJ05054575-2351139  &  [Hoenig10]  &  0.51  (974) & 1610.0  (972) & [Hoenig10] &  \\
15 & Ark120  &  [Hoenig17]  &  0.6  (969) & 0.82  (970) & [Hoenig17] & [GoMar23]  \\
16 & PICTORA  &  [Hoenig10]  &  0.82  (973) & 1.64  (971) & [Hoenig10] &  \\
17 & 2MASXJ05580206-3820043  &  [Nenkova08]  &  1.07  (971) & 1100.0  (970) & [Nenkova08] & [Fritz06]  \\
18 & Mrk3  &  [Nenkova08]$\rm{^\bullet}$  &  5.37  (948) & 4.51  (947) & [GoMar23]$\rm{^\bullet}$ & [Nenkova08]$\rm{^\bullet}$  \\
19 & ESO426-G002  &  [Stalev16]  &  0.54  (958) & 780.0  (957) & [Stalev16] & [Nenkova08] [Hoenig17]  \\
20 & Mrk78  &  [Nenkova08]  &  0.88  (979) & 1.07  (978) & [Nenkova08] & [GoMar23]  \\
21 & Mrk1210  &  [Hoenig17]$\rm{^\bullet}$  &  3.21  (945) & 1.91  (946) & [GoMar23] &  \\
22 & PG0804+761  &  [Hoenig17]  &  0.82  (1040) & 0.87  (1041) & [Hoenig17] & [Fritz06] [GoMar23]  \\
23 & MCG+04-22-042  &  [Stalev16]  &  0.77  (972) & 710.0  (971) & [Stalev16] & [Nenkova08]  \\
24 & Mrk110  &  [Hoenig17]$\rm{^\bullet}$  &  2.05  (970) & 4.86  (971) & [Hoenig17]$\rm{^\bullet}$ &  \\
25 & Mrk705  &  [Nenkova08]  &  1.33  (966) & 2.82  (965) & [Nenkova08] &  \\
26 & MCG-05-23-016  &  [Nenkova08]  &  0.57  (941) & 0.5  (940) & [GoMar23] & [Nenkova08]  \\
27 & NGC3081  &  [Nenkova08]  &  1.19  (938) & 1.81  (937) & [Nenkova08] &  \\
28 & ESO374-G044  &  [Nenkova08]  &  1.45  (963) & 1.24  (962) & [GoMar23] & [Nenkova08] [Hoenig17]  \\
29 & Mrk417  &  [Hoenig17]  &  0.75  (966) & 1.17  (967) & [Hoenig17] &  \\
30 & 2MASSXJ10594361+6504063  &  [Hoenig17]  &  0.52  (1023) & 2370.0  (1024) & [Hoenig17] & [Hoenig10]  \\
31 & ESO439-G009  &  [Nenkova08]  &  1.29  (960) & 1.12  (959) & [GoMar23] & [Nenkova08]  \\
32 & NGC3783  &  [Hoenig17]  &  0.61  (943) & 0.63  (944) & [Hoenig17] & [GoMar23]  \\
33 & UGC6728  &  [Hoenig10]  &  0.68  (943) & 3220.0  (941) & [Hoenig10] & [Fritz06] [Nenkova08] [Stalev16]  \\
34 & 2MASXJ11454045-1827149  &  [Hoenig10]  &  0.77  (973) & 0.56  (971) & [GoMar23] &  \\
35 & Ark347  &  [Nenkova08]  &  0.68  (958) & 0.75  (957) & [Nenkova08] & [Fritz06] [Stalev16] [Hoenig17] [GoMar23]  \\
36 & NGC4151  &  [Hoenig17]  &  1.96  (930) & 1.93  (931) & [GoMar23] & [Nenkova08]$\rm{^\bullet}$ [Hoenig17]  \\
37 & PG1211+143  &  [Fritz06]  &  1.41  (1031) & 0.53  (1030) & [GoMar23] &  \\
38 & M106  &  [Nenkova08]  &  1.74  (934) & 3.29  (933) & [Nenkova08] &  \\
39 & NGC4388  &  [Nenkova08]  &  1.34  (942) & 1.65  (941) & [Nenkova08] & [GoMar23]  \\
40 & NGC4507  &  [Nenkova08]  &  1.83  (945) & 2.44  (944) & [Nenkova08] & [Stalev16]$\rm{^\bullet}$ [GoMar23]$\rm{^\bullet}$  \\
41 & NGC4939  &  [Nenkova08]  &  1.13  (941) & 0.98  (940) & [GoMar23] & [Nenkova08]  \\
42 & IISZ010  &  [Hoenig17]  &  0.65  (969) & 2.3  (970) & [Hoenig17] &  \\
43 & MCG-03-34-064  &  [Nenkova08]$\rm{^\bullet}$  &  2.44  (954) & 4.83  (953) & [Nenkova08]$\rm{^\bullet}$ &  \\
44 & MCG-06-30-015  &  [Fritz06]  &  0.81  (940) & 0.5  (939) & [GoMar23] &  \\
45 & IC4329A  &  [Hoenig17]  &  0.66  (950) & 0.8  (951) & [Hoenig17] & [GoMar23]  \\
46 & UM614  &  [Nenkova08]  &  0.55  (967) & 1720.0  (966) & [Nenkova08] & [Fritz06]  \\
47 & Mrk279  &  [Fritz06]  &  0.81  (967) & 0.8  (966) & [GoMar23] & [Fritz06] [Nenkova08] [Stalev16] [Hoenig17]  \\
48 & PG1351+640  &  [Nenkova08]  &  1.4  (1024) & 1.9  (1023) & [Nenkova08] & [GoMar23]  \\
49 & NGC5548  &  [Hoenig17]  &  0.97  (961) & 1.14  (962) & [Hoenig17] & [Stalev16] [GoMar23]  \\
50 & ESO511-G030  &  [Hoenig10]  &  1.07  (951) & 1.37  (949) & [Hoenig10] & [Nenkova08] [GoMar23]  \\
51 & PG1448+273  &  [Hoenig17]  &  1.02  (1002) & 0.95  (1003) & [GoMar23] & [Nenkova08] [Stalev16] [Hoenig17]  \\
52 & IC4518W  &  [Fritz06]$\rm{^\bullet}$  &  6.12  (950) & 2.32  (949) & [GoMar23]$\rm{^\bullet}$ &  \\
53 & Mrk841  &  [Hoenig17]  &  1.16  (975) & 1.23  (976) & [Hoenig17] & [Nenkova08] [GoMar23]  \\
54 & Mrk1392  &  [Nenkova08]  &  0.77  (975) & 1.4  (974) & [Nenkova08] &  \\
55 & Mrk1393  &  [Hoenig17]  &  0.95  (989) & 1.02  (990) & [Hoenig17] & [Nenkova08] [GoMar23]  \\
56 & Mrk290  &  [Hoenig17]  &  0.85  (968) & 0.86  (969) & [Hoenig17] & [GoMar23]  \\
57 & ESO138-G001  &  [Nenkova08]  &  0.76  (944) & 1.84  (943) & [Nenkova08] &  \\
58 & ESO103-G35  &  [Nenkova08]  &  0.58  (950) & 0.54  (949) & [GoMar23] & [Nenkova08]  \\
59 & Fairall51  &  [Stalev16]$\rm{^\bullet}$  &  2.97  (944) & 3.22  (943) & [Stalev16]$\rm{^\bullet}$ & [Fritz06]$\rm{^\bullet}$ [GoMar23]$\rm{^\bullet}$  \\
60 & ESO141-G055  &  [Hoenig17]  &  0.98  (977) & 1.13  (978) & [Hoenig17] & [Fritz06] [GoMar23]  \\
61 & NGC6814  &  [Fritz06]  &  0.84  (933) & 0.79  (932) & [GoMar23] & [Fritz06] [Nenkova08] [Stalev16]  \\
62 & MCG+07-41-03  &  [Nenkova08]  &  1.34  (997) & 3.51  (996) & [Nenkova08] &  \\
63 & IC5063  &  [Nenkova08]  &  0.54  (942) & 0.84  (941) & [Nenkova08] &  \\
64 & IIZw136  &  [Fritz06]  &  0.67  (1003) & 0.65  (1002) & [GoMar23] & [Fritz06]  \\
65 & NGC7213  &  [Nenkova08]  &  1.44  (940) & 7.02  (939) & [Nenkova08] &  \\
66 & NGC7314  &  [Nenkova08]  &  0.69  (933) & 0.69  (932) & [GoMar23] & [Nenkova08]  \\
67 & PG2304+042  &  [Hoenig17]  &  0.9  (978) & 2.14  (979) & [Hoenig17] &  \\
68 & PKS2356-61  &  [Hoenig10]  &  0.61  (1036) & 1000.0  (1034) & [Hoenig10] &  \\
\hline\hline
\end{tabular}
\end{center}
\caption{Best fit model and statistic before (Col.\,2 \& 3) and after (Col.\,4 \& 5) our new [GoMar23] model is included into the analysis. Col.\,5 gives all the statistically similar models according to AIC probability (see text). Poor spectral fits (i.e. $\rm{\chi^2/dof >2}$) are marks with dots next to the model name in Cols.\,4\,\&\,5. }
\label{tab:fitresults}
\end{table*}

\begin{table*}[ht!]
\scriptsize
\renewcommand{\arraystretch}{0.9}
\renewcommand{\tabcolsep}{0.05cm}
\begin{center}
\begin{tabular}{lccccccccccccc}\hline \hline
Object name        &  $\rm{\chi^2_{r,err}}$  & $\rm{E(B-V)}$  & i ($\rm{^\circ}$)& $\rm{\sigma\,(^\circ)}$ & $\rm{p}$ &$\rm{q}$ & $\rm{Y}$ &  $\rm{\tau_{9.7}}$ &  $\rm{log(P_{size})}$ & $\rm{log(<P>)}$ & $\rm{M_{BH}}$ & $\rm{L_{bol}}$ & $\rm{\frac{L_{bol}}{L_{Edd}}}$  \\ \hline 
Mrk348 & 1.3 & $ \rm{<}$0.01 & $ \rm{<}$0.02 & 73.3$ \rm{\pm}$0.2 & $ \rm{>}$1.5 & $ \rm{<}$0.01 & 24.7$ \rm{\pm}$0.4 & $ \rm{>}$12.9 & -0.62$ \rm{\pm}$0.04 & -1.02$ \rm{\pm}$0.04 & 7.39 & 44.9 & -0.63 \\ 
IZw1 & 0.69 & 0.14$ \rm{\pm}$0.01 & 56.4$ \rm{\pm}$0.9 & 28.5$ \rm{\pm}$0.9 & 0.88$ \rm{\pm}$0.03 & $ \rm{<}$0.01 & 20.0$ \rm{\pm}$0.4 & $ \rm{>}$12.98 & $ \rm{>}$1.0 & $ \rm{>}$0.53 & 7.26 & 44.8 & -0.51 \\ 
FAIRALL9 & 0.61 & $ \rm{<}$0.01 & 69.8$ \rm{\pm}$0.1 & 21.9$ \rm{\pm}$0.2 & 0.5$ \rm{\pm}$0.1 & $ \rm{>}$1.45 & 2.52$ \rm{\pm}$0.02 & 7.0$ \rm{\pm}$0.03 & -1.6$ \rm{\pm}$0.02 & -1.85$ \rm{\pm}$0.02 & 8.21 & 45.1 & -1.18 \\ 
NGC526A & 0.66 & $ \rm{<}$0.01 & 53.3$ \rm{\pm}$0.3 & 47.1$ \rm{\pm}$0.5 & $ \rm{<}$0.01 & 0.18$ \rm{\pm}$0.04 & 5.0$ \rm{\pm}$0.01 & 7.0$ \rm{\pm}$0.04 & $ \rm{<}$-1.99 & $ \rm{<}$-2.14 & 8.05 & 44.8 & -1.38 \\ 
NGC788 & 0.94 & $ \rm{<}$0.01 & $ \rm{<}$0.01 & 78.3$ \rm{\pm}$0.3 & $ \rm{<}$0.02 & $ \rm{<}$0.01 & 10.0$ \rm{\pm}$0.2 & 11.8$ \rm{\pm}$0.2 & $ \rm{>}$1.0 & $\rm{>}$0.53 & 7.74 & 44.5 & -1.38 \\ 
Mrk1018 & 0.5 & $ \rm{<}$0.01 & 72.0$ \rm{\pm}$0.2 & 20.0$ \rm{\pm}$0.4 & 0.5$ \rm{\pm}$0.03 & $ \rm{<}$0.01 & $ \rm{<}$2.01 & 3.7$ \rm{\pm}$0.1 & $ \rm{<}$-1.99 & $ \rm{<}$-2.14 & 8.21 & 43.7 & -2.61 \\ 
Mrk590 & 0.8 & $ \rm{<}$0.01 & 28.8$ \rm{\pm}$2.5 & 65.3$ \rm{\pm}$1.6 & $ \rm{<}$0.01 & $ \rm{<}$0.01 & 12.4$ \rm{\pm}$0.2 & 6.9$ \rm{\pm}$0.5 & $ \rm{>}$1.0 & $\rm{>}$0.53 & 7.85 & 44.1 & -1.81 \\ 
NGC1052 & 2.72 & 0.2$ \rm{\pm}$0.01 & 43.4$ \rm{\pm}$0.1 & 45.0$ \rm{\pm}$0.1 & $ \rm{<}$0.01 & 0.2$ \rm{\pm}$0.1 & 10.44$ \rm{\pm}$0.05 & $ \rm{>}$12.98 & -1.6$ \rm{\pm}$0.01 & -1.85$ \rm{\pm}$0.01 & 8.39 & 43.1 & -3.41 \\ 
NGC1275 & 4.64 & $ \rm{<}$0.01 & 50.0$ \rm{\pm}$0.02 & 45.0$ \rm{\pm}$0.1 & $ \rm{<}$0.01 & $ \rm{<}$0.01 & 13.92$ \rm{\pm}$0.05 & $ \rm{>}$12.99 & $ \rm{>}$1.0 & $\rm{>}$0.53 & 9.0 & 44.7 & -2.36 \\ 
ESO548-G081 & 0.89 & $ \rm{<}$0.04 & 71.0$ \rm{\pm}$0.3 & 18.3$ \rm{\pm}$0.6 & 0.51$ \rm{\pm}$0.04 & $ \rm{<}$0.01 & $ \rm{<}$2.0 & $ \rm{<}$3.01 & $ \rm{<}$-1.99 & $ \rm{<}$-2.14 & 8.34 & 44.2 & -2.23 \\ 
3C120 & 2.11 & 0.01$ \rm{\pm}$0.01 & 42.0$ \rm{\pm}$0.1 & 50.0$ \rm{\pm}$0.1 & $ \rm{>}$1.5 & $ \rm{<}$0.01 & 30.0$ \rm{\pm}$0.04 & 7.0$ \rm{\pm}$0.01 & -1.301$ \rm{\pm}$0.003 & -1.620$ \rm{\pm}$0.003 & 7.85 & 45.5 & -0.47 \\ 
MCG-01-13-025 & 1.49 & $ \rm{<}$0.02 & 20.1$ \rm{\pm}$1.5 & $ \rm{>}$79.87 & 1.05$ \rm{\pm}$0.03 & $ \rm{<}$0.01 & 10.0$ \rm{\pm}$0.1 & $ \rm{<}$3.02 & -1.0$ \rm{\pm}$0.01 &  -1.38$ \rm{\pm}$0.01 & $\dots$ & 44.2 & $\dots$ \\ 
CGCG420-015 & 1.08 & 0.1$ \rm{\pm}$0.01 & 58.8$ \rm{\pm}$0.3 & 33.3$ \rm{\pm}$0.2 & 0.73$ \rm{\pm}$0.05 & 0.5$ \rm{\pm}$0.02 & 5.3$ \rm{\pm}$0.1 & $ \rm{>}$12.98 & -0.60$ \rm{\pm}$0.01 & -1.01$ \rm{\pm}$0.01 & 7.88 & 44.7 & -1.26 \\ 
2MASXJ05054575-2351139 & 0.33 & 0.21$ \rm{\pm}$0.04 & $ \rm{<}$0.02 & $ \rm{>}$78.85 & 1.0$ \rm{\pm}$0.1 & 0.9$ \rm{\pm}$0.1 & 11.2$ \rm{\pm}$0.3 & $ \rm{>}$12.88 & $ \rm{>}$1.0 & $\rm{>}$0.53 & $\dots$ & 45.3 & $\dots$ \\ 
Ark120 & 0.82 & 0.04$ \rm{\pm}$0.01 & 70.6$ \rm{\pm}$0.1 & 14.8$ \rm{\pm}$0.1 & 1.0$ \rm{\pm}$0.01 & $ \rm{<}$0.02 & $ \rm{<}$2.02 & 7.0$ \rm{\pm}$0.05 & -0.6$ \rm{\pm}$0.01 & -1.01$ \rm{\pm}$0.01 & 8.24 & 44.8 & -1.57 \\ 
PICTORA & 1.64 & $ \rm{<}$0.01 & 50.0$ \rm{\pm}$0.1 & 45.0$ \rm{\pm}$0.1 & $ \rm{<}$0.01 & $ \rm{<}$0.01 & 5.02$ \rm{\pm}$0.02 & 6.8$ \rm{\pm}$0.1 & $ \rm{<}$-1.97 & $ \rm{<}$-2.1 & 7.6 & 45.1 & -0.64 \\ 
2MASXJ05580206-3820043 & 0.18 & 0.6$ \rm{\pm}$0.1 & $ \rm{<}$1.57 & 59.0$ \rm{\pm}$2.4 & $ \rm{<}$0.02 & $ \rm{<}$0.03 & 2.6$ \rm{\pm}$0.1 & 8.3$ \rm{\pm}$0.5 & -0.9$ \rm{\pm}$0.2 & -1.28$ \rm{\pm}$0.2 & 8.06 & 44.9 & -1.3 \\ 
Mrk3 & 4.51 & $ \rm{<}$0.01 & $ \rm{>}$89.84 & 49.0$ \rm{\pm}$0.2 & 0.15$ \rm{\pm}$0.01 & 1.48$ \rm{\pm}$0.02 & 10.0$ \rm{\pm}$0.02 & 11.0$ \rm{\pm}$0.01 & 0.66$ \rm{\pm}$0.01 & 0.20$ \rm{\pm}$0.01 & 8.42 & 44.8 & -1.74 \\ 
ESO426-G002 & 0.37 & 0.4$ \rm{\pm}$0.02 & 46.9$ \rm{\pm}$0.2 & 45.0$ \rm{\pm}$0.2 & $ \rm{>}$1.48 & $ \rm{<}$0.01 & 35.0$ \rm{\pm}$0.1 & 9.4$ \rm{\pm}$0.2 & $ \rm{>}$1.0 & $\rm{>}$0.53 & $\dots$ & 44.4 & $\dots$ \\ 
Mrk78 & 1.07 & 0.43$ \rm{\pm}$0.01 & 62.2$ \rm{\pm}$0.6 & 45.0$ \rm{\pm}$0.1 & $ \rm{>}$1.5 & $ \rm{<}$0.01 & 35.0$ \rm{\pm}$0.1 & 8.96$ \rm{\pm}$0.04 & $ \rm{>}$1.0 & $\rm{>}$0.53 & 6.66 & 44.4 & -0.35 \\ 
Mrk1210 & 1.91 & $ \rm{<}$0.01 & 39.7$ \rm{\pm}$0.2 & 60.0$ \rm{\pm}$0.1 & $ \rm{<}$0.01 & 0.5$ \rm{\pm}$0.01 & 20.0$ \rm{\pm}$0.03 & 6.67$ \rm{\pm}$0.04 & 0.14$ \rm{\pm}$0.02 & -0.31$ \rm{\pm}$0.02 & 6.09 & 44.3 & 0.11 \\ 
PG0804+761 & 0.87 & 0.03$ \rm{\pm}$0.01 & $ \rm{<}$0.64 & 24.0$ \rm{\pm}$0.3 & $ \rm{<}$0.01 & 0.5$ \rm{\pm}$0.1 & 6.1$ \rm{\pm}$0.1 & $ \rm{>}$12.84 & 0.43$ \rm{\pm}$0.04 & -0.03$ \rm{\pm}$0.04 & 8.5 & 45.6 & -1.02 \\ 
MCG+04-22-042 & 0.26 & $ \rm{<}$0.01 & 54.6$ \rm{\pm}$0.7 & 41.3$ \rm{\pm}$0.5 & 0.7$ \rm{\pm}$0.2 & 0.5$ \rm{\pm}$0.1 & 4.8$ \rm{\pm}$0.2 & 6.9$ \rm{\pm}$0.3 & $ \rm{<}$-1.93 & $ \rm{<}$-2.10 & 7.46 & 44.7 & -0.85 \\ 
Mrk110 & 4.86 & $ \rm{<}$0.01 & 40.1$ \rm{\pm}$1.4 & $ \rm{>}$79.92 & 0.999$ \rm{\pm}$0.005 & 0.4$ \rm{\pm}$0.1 & 5.0$ \rm{\pm}$0.01 & 5.0$ \rm{\pm}$0.01 & $ \rm{<}$-2.0 & $ \rm{<}$-2.15 & 7.31 & 45.3 & -0.09 \\ 
Mrk705 & 2.82 & $ \rm{<}$0.03 & $ \rm{<}$0.01 & $ \rm{>}$79.21 & $ \rm{>}$1.5 & 0.7$ \rm{\pm}$0.02 & 30.0$ \rm{\pm}$0.1 & $ \rm{>}$12.96 & $ \rm{>}$1.0 & $\rm{>}$0.53 & 7.62 & 44.3 & -1.38 \\ 
MCG-05-23-016 & 0.5 & 0.17$ \rm{\pm}$0.02 & 0.02$ \rm{\pm}$0.01 & $ \rm{>}$79.89 & 1.12$ \rm{\pm}$0.02 & $ \rm{<}$0.01 & 20.2$ \rm{\pm}$0.2 & 11.0$ \rm{\pm}$0.1 & $ \rm{>}$1.0 & $\rm{>}$0.53 & 7.43 & 44.5 & -1.05 \\ 
NGC3081 & 1.81 & 0.39$ \rm{\pm}$0.01 & 45.7$ \rm{\pm}$0.1 & 50.0$ \rm{\pm}$0.04 & $ \rm{<}$0.01 & $ \rm{<}$0.01 & 12.2$ \rm{\pm}$0.1 & 9.0$ \rm{\pm}$0.01 & $ \rm{>}$1.0 & $\rm{>}$0.53 & 7.72 & 44.0 & -1.85 \\ 
ESO374-G044 & 1.24 & $ \rm{<}$0.01 & $ \rm{>}$89.34 & 73.0$ \rm{\pm}$2.0 & $ \rm{<}$0.01 & $ \rm{>}$1.46 & 11.4$ \rm{\pm}$0.1 & $ \rm{>}$12.77 & $ \rm{>}$1.0 & $\rm{>}$0.53 & $\dots$ & 44.6 & $\dots$ \\ 
Mrk417 & 1.17 & $ \rm{<}$0.01 & 0.0$ \rm{\pm}$0.1 & $ \rm{>}$79.91 & 0.2$ \rm{\pm}$0.05 & $ \rm{<}$0.01 & 10.01$ \rm{\pm}$0.05 & 8.0$ \rm{\pm}$0.1 & 0.398$ \rm{\pm}$0.004 &  -0.06$ \rm{\pm}$0.004 & 8.08 & 44.9 & -1.26 \\ 
2MASSXJ10594361+6504063 & 0.3 & 1.24$ \rm{\pm}$0.05 & 78.5$ \rm{\pm}$0.1 & 15.3$ \rm{\pm}$0.6 & 1.0$ \rm{\pm}$0.04 & $ \rm{>}$1.08 & 4.1$ \rm{\pm}$0.6 & 11.0$ \rm{\pm}$0.3 & 0.01$ \rm{\pm}$0.04 & -0.44$ \rm{\pm}$0.04 & 8.45 & 44.4 & -2.13 \\ 
ESO439-G009 & 1.12 & 0.14$ \rm{\pm}$0.04 & 88.5$ \rm{\pm}$0.4 & 23.2$ \rm{\pm}$0.7 & $ \rm{<}$0.01 & 0.5$ \rm{\pm}$0.02 & 2.09$ \rm{\pm}$0.02 & $ \rm{>}$12.98 & $ \rm{>}$1.0 & $\rm{>}$0.53 & $\dots$ & 44.2 & $\dots$ \\ 
NGC3783 & 0.63 & 0.02$ \rm{\pm}$0.01 & 46.4$ \rm{\pm}$0.2 & 45.0$ \rm{\pm}$0.2 & 1.03$ \rm{\pm}$0.01 & $ \rm{<}$0.01 & 20.0$ \rm{\pm}$0.1 & 11.01$ \rm{\pm}$0.04 & -1.29$ \rm{\pm}$0.02 &  -1.62$ \rm{\pm}$0.02 & 7.3 & 44.5 & -0.89 \\ 
UGC6728 & 0.44 & $ \rm{<}$0.01 & 58.4$ \rm{\pm}$0.7 & 38.0$ \rm{\pm}$1.8 & 0.5$ \rm{\pm}$0.1 & $ \rm{>}$1.34 & 5.0$ \rm{\pm}$0.2 & 7.0$ \rm{\pm}$0.4 & -1.5$ \rm{\pm}$0.1 & -1.82$ \rm{\pm}$0.1 & 6.44 & 43.2 & -1.29 \\ 
2MASXJ11454045-1827149 & 0.56 & $ \rm{<}$0.01 & 62.5$ \rm{\pm}$0.4 & 33.9$ \rm{\pm}$0.9 & $ \rm{>}$1.4 & $ \rm{<}$0.01 & 5.0$ \rm{\pm}$0.1 & 5.6$ \rm{\pm}$0.2 & $ \rm{<}$-1.97 & $ \rm{<}$-2.1 & 7.01 & 45.1 & 0.01 \\ 
Ark347 & 0.77 & 0.05$ \rm{\pm}$0.04 & $ \rm{<}$0.06 & 70.0$ \rm{\pm}$0.7 & 1.29$ \rm{\pm}$0.02 & 0.06$ \rm{\pm}$0.04 & $ \rm{>}$39.76 & $ \rm{>}$12.75 & $ \rm{>}$1.0 & $\rm{>}$0.53 & 7.98 & 44.5 & -1.61 \\ 
NGC4151 & 1.93 & $ \rm{<}$0.02 & 31.5$ \rm{\pm}$0.2 & 60.0$ \rm{\pm}$0.1 & $ \rm{<}$0.01 & 1.0$ \rm{\pm}$0.004 & 8.02$ \rm{\pm}$0.03 & 11.0$ \rm{\pm}$0.01 & -1.301$ \rm{\pm}$0.002 & -1.620$ \rm{\pm}$0.002 & 7.42 & 44.1 & -1.44 \\ 
PG1211+143 & 0.53 & $ \rm{<}$0.01 & 60.0$ \rm{\pm}$0.2 & 32.1$ \rm{\pm}$0.2 & 0.66$ \rm{\pm}$0.03 & 0.8$ \rm{\pm}$0.1 & 5.0$ \rm{\pm}$0.1 & 5.7$ \rm{\pm}$0.2 & -1.6$ \rm{\pm}$0.02 & -1.85$ \rm{\pm}$0.02 & 7.65 & 44.7 & -1.08 \\ 
M106 & 3.29 & 0.24$ \rm{\pm}$0.02 & 39.9$ \rm{\pm}$0.3 & 45.0$ \rm{\pm}$0.04 & 1.35$ \rm{\pm}$0.01 & $ \rm{<}$0.01 & 35.0$ \rm{\pm}$0.02 & 9.0$ \rm{\pm}$0.01 & $ \rm{>}$1.0 & $\rm{>}$0.53 & 7.54 & 41.9 & -3.74 \\ 
NGC4388 & 1.65 & 0.9$ \rm{\pm}$0.02 & 86.0$ \rm{\pm}$1.5 & 27.8$ \rm{\pm}$0.8 & 1.0$ \rm{\pm}$0.01 & 1.04$ \rm{\pm}$0.04 & 15.0$ \rm{\pm}$0.1 & 8.9$ \rm{\pm}$0.1 & $ \rm{>}$1.0 & $\rm{>}$0.53 & 6.98 & 44.6 & -0.47 \\ 
NGC4507 & 2.44 & $ \rm{<}$0.01 & 66.1$ \rm{\pm}$0.1 & 30.3$ \rm{\pm}$0.1 & 1.02$ \rm{\pm}$0.005 & 1.45$ \rm{\pm}$0.03 & 35.0$ \rm{\pm}$0.1 & 9.0$ \rm{\pm}$0.1 & -0.124$ \rm{\pm}$0.003 & -0.560$ \rm{\pm}$0.003 & 7.87 & 44.7 & -1.23 \\ 
NGC4939 & 1.03 & $ \rm{<}$0.01 & $ \rm{>}$87.42 & 70.0$ \rm{\pm}$0.8 & $ \rm{<}$0.01 & 0.5$ \rm{\pm}$0.1 & 20.0$ \rm{\pm}$0.2 & 6.2$ \rm{\pm}$0.2 & $ \rm{>}$1.0 & $\rm{>}$0.53 & 7.91 & 43.7 & -2.32 \\ 
IISZ010 & 2.3 & $ \rm{<}$0.01 & 0.05$ \rm{\pm}$0.02 & $ \rm{>}$79.88 & 0.5$ \rm{\pm}$0.01 & $ \rm{<}$0.01 & 9.6$ \rm{\pm}$0.2 & 5.0$ \rm{\pm}$0.04 & 0.4$ \rm{\pm}$0.01 & -0.06$ \rm{\pm}$0.01 & 7.34 & 44.5 & -0.97 \\ 
MCG-03-34-064 & 4.83 & 0.65$ \rm{\pm}$0.01 & 79.9$ \rm{\pm}$0.1 & 18.2$ \rm{\pm}$0.4 & $ \rm{<}$0.01 & 1.11$ \rm{\pm}$0.04 & 2.23$ \rm{\pm}$0.02 & $ \rm{>}$12.99 & $ \rm{>}$1.0 & $\rm{>}$0.53 & 8.12 & 44.2 & -2.02 \\ 
MCG-06-30-015 & 0.47 & $ \rm{<}$0.01 & 56.0$ \rm{\pm}$0.1 & 37.7$ \rm{\pm}$0.2 & 1.41$ \rm{\pm}$0.01 & $ \rm{>}$1.49 & 35.0$ \rm{\pm}$0.1 & $ \rm{>}$12.96 & -1.0$ \rm{\pm}$0.004 & -1.38$ \rm{\pm}$0.004 & 6.94 & 43.7 & -1.34 \\ 
IC4329A & 0.8 & 0.18$ \rm{\pm}$0.01 & 66.3$ \rm{\pm}$0.1 & 26.0$ \rm{\pm}$0.2 & $ \rm{<}$0.01 & 0.5$ \rm{\pm}$0.02 & 2.79$ \rm{\pm}$0.01 & 11.0$ \rm{\pm}$0.04 & -1.301$ \rm{\pm}$0.004 & -1.620$ \rm{\pm}$0.004 & 7.51 & 44.8 & -0.85 \\ 
UM614 & 0.21 & $ \rm{<}$0.01 & 50.9$ \rm{\pm}$0.7 & 43.6$ \rm{\pm}$0.8 & $ \rm{<}$0.01 & 0.5$ \rm{\pm}$0.1 & 5.0$ \rm{\pm}$0.1 & 7.0$ \rm{\pm}$0.3 & -1.6$ \rm{\pm}$0.1 & -1.85$ \rm{\pm}$0.1 & 7.49 & 42.6 & -3.02 \\ 
Mrk279 & 0.8 & 0.4$ \rm{\pm}$0.02 & 78.1$ \rm{\pm}$0.1 & 13.9$ \rm{\pm}$0.2 & $ \rm{<}$0.01 & $ \rm{>}$1.47 & 2.23$ \rm{\pm}$0.02 & 9.0$ \rm{\pm}$0.04 & $ \rm{>}$1.0 & $\rm{>}$0.53 & 7.86 & 44.9 & -1.09 \\ 
PG1351+640 & 1.9 & $ \rm{<}$0.01 & 38.0$ \rm{\pm}$1.8 & $ \rm{>}$79.75 & $ \rm{<}$0.01 & 1.43$ \rm{\pm}$0.03 & 35.0$ \rm{\pm}$0.2 & $ \rm{<}$3.02 & $ \rm{>}$1.0 & $\rm{>}$0.53 & 8.07 & 44.0 & -2.17 \\ 
NGC5548 & 1.14 & 0.18$ \rm{\pm}$0.01 & 33.7$ \rm{\pm}$0.4 & 51.6$ \rm{\pm}$0.3 & $ \rm{<}$0.01 & $ \rm{<}$0.01 & 10.8$ \rm{\pm}$0.1 & 9.0$ \rm{\pm}$0.02 & $ \rm{>}$1.0 & $\rm{>}$0.53 & 7.79 & 44.7 & -1.15 \\ 
ESO511-G030 & 1.37 & $ \rm{<}$0.01 & 70.0$ \rm{\pm}$0.2 & 24.0$ \rm{\pm}$0.7 & 0.6$ \rm{\pm}$0.1 & $ \rm{<}$0.01 & $ \rm{<}$2.01 & $ \rm{<}$3.05 & $ \rm{<}$-1.99 & $ \rm{<}$-2.15 & 7.84 & 44.6 & -1.32 \\ 
PG1448+273 & 0.95 & $ \rm{<}$0.01 & $ \rm{<}$2.26 & 66.9$ \rm{\pm}$0.4 & 1.0$ \rm{\pm}$0.01 & $ \rm{<}$0.01 & 12.3$ \rm{\pm}$0.1 & 8.9$ \rm{\pm}$0.1 & -0.6$ \rm{\pm}$0.01 & -1.01$ \rm{\pm}$0.01 & 6.92 & 44.2 & -0.8 \\ 
IC4518W & 2.32 & 2.19$ \rm{\pm}$0.02 & 89.8$ \rm{\pm}$0.01 & $ \rm{<}$10.0 & 0.99$ \rm{\pm}$0.01 & 1.0$ \rm{\pm}$0.03 & 30.1$ \rm{\pm}$0.2 & 7.3$ \rm{\pm}$0.1 & $ \rm{>}$1.0 & $\rm{>}$0.53 & 7.71 & 44.1 & -1.71 \\ 
Mrk841 & 1.23 & $ \rm{<}$0.01 & 31.7$ \rm{\pm}$0.2 & 60.1$ \rm{\pm}$0.2 & 0.51$ \rm{\pm}$0.01 & 0.49$ \rm{\pm}$0.04 & 10.0$ \rm{\pm}$0.1 & $ \rm{>}$12.9 & -1.3$ \rm{\pm}$0.005 & -1.620$ \rm{\pm}$0.005 & 8.08 & 45.0 & -1.14 \\ 
Mrk1392 & 1.4 & $ \rm{<}$0.01 & $ \rm{<}$0.01 & $ \rm{>}$79.08 & 0.1$ \rm{\pm}$0.1 & 0.8$ \rm{\pm}$0.1 & 15.9$ \rm{\pm}$0.2 & $ \rm{>}$12.94 & $ \rm{>}$1.0 & $\rm{>}$0.53 & 8.3 & 44.7 & -1.68 \\ 
Mrk1393 & 1.02 & $ \rm{<}$0.01 & 33.8$ \rm{\pm}$0.3 & 59.9$ \rm{\pm}$0.2 & $ \rm{<}$0.01 & 1.0$ \rm{\pm}$0.01 & 11.1$ \rm{\pm}$0.1 & 11.0$ \rm{\pm}$0.03 & -1.3$ \rm{\pm}$0.01 & -1.62$ \rm{\pm}$0.01 & 8.61 & 44.8 & -1.92 \\ 
Mrk290 & 0.86 & $ \rm{<}$0.01 & 47.7$ \rm{\pm}$0.2 & 49.9$ \rm{\pm}$0.2 & $ \rm{<}$0.01 & 0.5$ \rm{\pm}$0.1 & 7.4$ \rm{\pm}$0.1 & 5.2$ \rm{\pm}$0.1 & -1.3$ \rm{\pm}$0.01 & -1.62$ \rm{\pm}$0.01 & 7.46 & 44.7 & -0.91 \\ 
ESO138-G001 & 1.84 & $ \rm{<}$0.02 & 11.3$ \rm{\pm}$0.6 & 72.3$ \rm{\pm}$0.4 & 1.26$ \rm{\pm}$0.01 & 0.5$ \rm{\pm}$0.01 & 20.0$ \rm{\pm}$0.1 & $ \rm{>}$12.98 & $ \rm{>}$1.0 & $\rm{>}$0.53 & 6.66 & 43.4 & -1.35 \\ 
ESO103-G35 & 0.54 & 1.11$ \rm{\pm}$0.03 & 87.2$ \rm{\pm}$1.9 & 23.8$ \rm{\pm}$0.6 & $ \rm{<}$0.01 & $ \rm{>}$1.37 & 3.7$ \rm{\pm}$0.1 & 8.8$ \rm{\pm}$0.2 & $ \rm{>}$1.0 & $\rm{>}$0.53 & 7.48 & 44.6 & -0.99 \\ 
Fairall51 & 3.22 & 0.34$ \rm{\pm}$0.01 & 15.5$ \rm{\pm}$2.1 & 50.0$ \rm{\pm}$0.2 & 1.002$ \rm{\pm}$0.004 & 0.06$ \rm{\pm}$0.01 & $ \rm{>}$39.96 & 11.0$ \rm{\pm}$0.03 & $ \rm{>}$1.0 & $\rm{>}$0.53 & 6.84 & 44.1 & -0.81 \\ 
ESO141-G055 & 1.13 & $ \rm{<}$0.01 & 70.1$ \rm{\pm}$0.1 & 20.4$ \rm{\pm}$0.1 & 0.5$ \rm{\pm}$0.01 & $ \rm{>}$1.49 & 3.3$ \rm{\pm}$0.03 & 9.0$ \rm{\pm}$0.03 & -0.63$ \rm{\pm}$0.01 & -1.04$ \rm{\pm}$0.01 & 7.5 & 45.3 & -0.28 \\ 
NGC6814 & 0.79 & 0.4$ \rm{\pm}$0.02 & 62.5$ \rm{\pm}$0.2 & 24.3$ \rm{\pm}$0.2 & $ \rm{>}$1.49 & 0.5$ \rm{\pm}$0.1 & $ \rm{>}$39.3 & $ \rm{>}$12.95 & $ \rm{>}$1.0 & $\rm{>}$0.53 & 7.08 & 43.5 & -1.73 \\ 
MCG+07-41-03 & 3.51 & $ \rm{<}$0.01 & $ \rm{>}$90.0 & $ \rm{<}$10.0 & $ \rm{<}$0.01 & $ \rm{>}$1.5 & 36.7$ \rm{\pm}$0.3 & 9.0$ \rm{\pm}$0.01 & 0.1$ \rm{\pm}$0.01 & -0.35$ \rm{\pm}$0.01 & 9.11 & 45.7 & -1.5 \\ 
IC5063 & 0.84 & 0.58$ \rm{\pm}$0.02 & 80.0$ \rm{\pm}$0.1 & 19.5$ \rm{\pm}$0.2 & 0.79$ \rm{\pm}$0.03 & 0.5$ \rm{\pm}$0.1 & 2.36$ \rm{\pm}$0.02 & $ \rm{>}$12.99 & $ \rm{>}$1.0 & $\rm{>}$0.53 & 7.63 & 44.2 & -1.52 \\ 
IIZw136 & 0.66 & 0.37$ \rm{\pm}$0.01 & 79.12$ \rm{\pm}$0.05 & 11.48$ \rm{\pm}$0.05 & $ \rm{<}$0.01 & 0.99$ \rm{\pm}$0.01 & 2.04$ \rm{\pm}$0.02 & 11.0$ \rm{\pm}$0.03 & -0.06$ \rm{\pm}$0.01 & -0.50$ \rm{\pm}$0.01 & 7.61 & 44.4 & -1.26 \\ 
NGC7213 & 7.02 & $ \rm{<}$0.01 & 28.6$ \rm{\pm}$0.4 & 60.0$ \rm{\pm}$0.1 & 0.5$ \rm{\pm}$0.01 & 0.64$ \rm{\pm}$0.02 & 10.0$ \rm{\pm}$0.02 & 7.0$ \rm{\pm}$0.03 & -1.301$ \rm{\pm}$0.003 & -1.62$ \rm{\pm}$0.003 & 7.71 & 43.3 & -2.5 \\ 
NGC7314 & 0.69 & $ \rm{<}$0.04 & 89.9$ \rm{\pm}$0.04 & $ \rm{<}$10.01 & 0.28$ \rm{\pm}$0.04 & 1.0$ \rm{\pm}$0.02 & 30.2$ \rm{\pm}$0.8 & 5.7$ \rm{\pm}$0.1 & -0.0$ \rm{\pm}$0.01 & -0.44$ \rm{\pm}$0.01 & 6.59 & 43.3 & -1.37 \\ 
PG2304+042 & 2.14 & $ \rm{<}$0.01 & 48.4$ \rm{\pm}$0.7 & 51.9$ \rm{\pm}$0.8 & 0.5$ \rm{\pm}$0.03 & $ \rm{<}$0.01 & 5.0$ \rm{\pm}$0.02 & 5.0$ \rm{\pm}$0.1 & $ \rm{<}$-1.99 & $ \rm{<}$-2.15 & $\dots$ & 44.3 & $\dots$ \\ 
PKS2356-61 & 0.19 & 1.24$ \rm{\pm}$0.03 & 44.4$ \rm{\pm}$1.2 & 49.4$ \rm{\pm}$1.0 & $ \rm{<}$0.02 & 0.5$ \rm{\pm}$0.1 & 12.4$ \rm{\pm}$0.3 & 7.0$ \rm{\pm}$0.1 & $ \rm{>}$1.0 & $\rm{>}$0.53 & 8.96 & 44.8 & -2.3 \\ 
\hline\hline
\end{tabular}
\end{center}
\caption{Resulting parameters when fitting the spectra of each object in the sample to [GoMar23]. The $\rm{\chi^2/dof}$ and $\rm{E(B-V)}$ are given in Cols.\,2\,\&\,3. We then include the model parameters (Cols.\,4-10): viewing angle ($i$), the half-opening angle of the torus ($\sigma$), the slope of the radial density distribution ($p$), the slope of the polar density distribution ($q$), the ratio between the outer and inner radius of the torus ($Y$), equatorial optical depth at ($\rm{\tau_{9.7}}$), maximum dust grain size ($P_{size}$), and average grain size ($\rm{<P>}$). It also includes general properties of the AGN in Cols.\,11-13 as the black-hole mass ($\rm{M_{BH}}$), bolometric luminosity ($\rm{L_{bol}}$) and accretion rate as the ratio between the bolometric luminosity and the Eddington luminosity ($\rm{L_{bol}/L_{Edd}}$). Note that these quantities are in log units. }
\label{tab:parresults}
\end{table*}

\clearpage

\begin{appendix}

\section{Spectral fitting evaluation procedure}\label{sec:stats}

We use the $\rm{\chi^2}$ statistics as the main mathematical support to infer if a model is able to reproduce an observed SED and to find the best-fit solution minimizing this statistic. The $\rm{\chi^2}$ is formally written as: 
\begin{equation}
\chi^2_{error} = \sum_{i} \frac {(data_{i} - model_{i})^2}{error_{i}^2} \label{eq:chierror}
\end{equation}
\noindent where $\rm{error_i}$ is the error bar associated with each spectral bin on the data. Note that this definition of $\rm{\chi^2}$ is called $\rm{\chi^2_{error}}$ (since it uses the error on the spectrum) in order to distinguish it from the definition below where the errors are not involved. Under this definition, the best result should have an associated $\rm{\chi^2}$ close to the degree of freedom (hereinafter \emph{dof}). Thus, the reduced $\rm{\chi^2}$ ($\rm{\chi_r^2=\chi^2/dof}$) should be close to unity. Lower values of $\rm{\chi_r^2}$ might indicate that the model has too many free parameters (i.e. the model shows a complexity that is not required by the data) or that the errors are overestimated. Large values could also indicate that the errors are underestimated or that the data show a complexity that cannot be recovered by the model. Indeed, \emph{Spitzer} spectra might show issues on the error bars due to a poor understanding of the uncertainties. To mitigate this issue, we also explore the following $\rm{\chi^2}$ definition: 
\begin{equation}
\chi^2_{model} = \sum_{i} \frac {(data_{i} - model_{i})^2}{model_{i}^2} \label{eq:chimodel}
\end{equation}
\noindent which has no dependence on the error bars of the data. In contrast with the previous definition of $\rm{\chi^2}$, $\rm{\chi_r^2}$ should approach zero to consider it as a good fit. 
We also use the Akaike information criterion (AIC) to compare the goodness of the fit of different models for a single AGN spectrum. This is a common method applied when different models (i.e. not nested) are compared \citep[e.g.][]{Emmanoulopoulos16,Osorio-Clavijo20,Esparza-Arredondo21,Martinez-Paredes21,Garcia-Bernete22}. This method allows an evaluation of the best fit by comparing the minimum $\rm{\chi^2_{r}}$ with any other model providing a statistically good fit to the data. In particular, we calculate the Bayes factor through the Akaike information criterion, $\rm{AIC_c}$. To this end, we use the eq.\,5 in \cite{Emmanoulopoulos16}:
\begin{equation}
    AIC_c = 2k - 2C_L + \chi^2 + \frac{2k(k+1)}{N-k-1}
\end{equation}
\noindent where $C_L$ is the constant likelihood of the true hypothetical model, $k$ is number of model parameters, and $N$ is the number of data points (where $\rm{dof = N - k - 1}$). We then calculate the difference between two different models, $\rm{\Delta[AIC_{c}]}$
\begin{equation}
    \Delta[AIC_{c}] = AIC_{c,2} - AIC_{c,1}
\end{equation}
\noindent where model 1 is the model providing the best statistic (i.e. the minimum $\rm{\chi^{2}}$). Finally, we estimate the evidence ratio, $\rm{\epsilon}$ 
\begin{equation}
    \epsilon = e^-\frac{\Delta[AIC_{c}]}{2}
\end{equation}
\noindent The evidence ratio or Bayes factor, is a measure of the relative likelihood of the best model versus another model. When the Bayes factor is $\rm{\leq 0.01}$, the best model is more likely to be the `correct' one because is at least 100 times better than the second model. If $\rm{> 0.01}$ we consider the second model as good as the first one.

In order to study if the fit improved by allowing to vary one of the parameters, we used the f-statistic test \citep[f-test][]{Barlow89}. Particularly, we use this test to explore if the dust grain size parameter provides a statistically better fit to the data compared to the model without this parameter. The likelihood ratio test (LRT) statistic or the related F-statistic is designed to choose between two models, where one model (the null model) is a simple or more parsimonious version of the other model \citep[][]{Protassov02}. The F-statistic is computed as:
\begin{equation}
    f = \frac{\chi_{1}^2-\chi_{2}^2} {dof_{1} - dof_{2}} 
\end{equation}
\noindent where $\rm{dof_{2}<dof_{1}}$ because the number of free parameters $k_1$ is larger than $k_2$ (with the same number of data points). We then compute the p-value associated with the F-statistic which is a measure of the probability that an observed difference could have occurred just by random chance. We consider that model 2 is needed when the p-value is $p<0.0001$. Note that this method cannot be applied to multiplicative models \citep[e.g. in our analysis of the need for foreground extinction][]{Protassov02} or for the inclusion of components affecting only a narrow range of the spectrum \citep[e.g. the need of emission lines][]{Orlandini12}. However, it is useful to test if allowing to vary one parameter of the model is needed by the data, which is the main purpose in our analysis.  

\section{The role of foreground extinction into the final fits}\label{sec:FGextinction}

\begin{figure*}[!ht]
\begin{center}
\includegraphics[width=0.49\columnwidth]{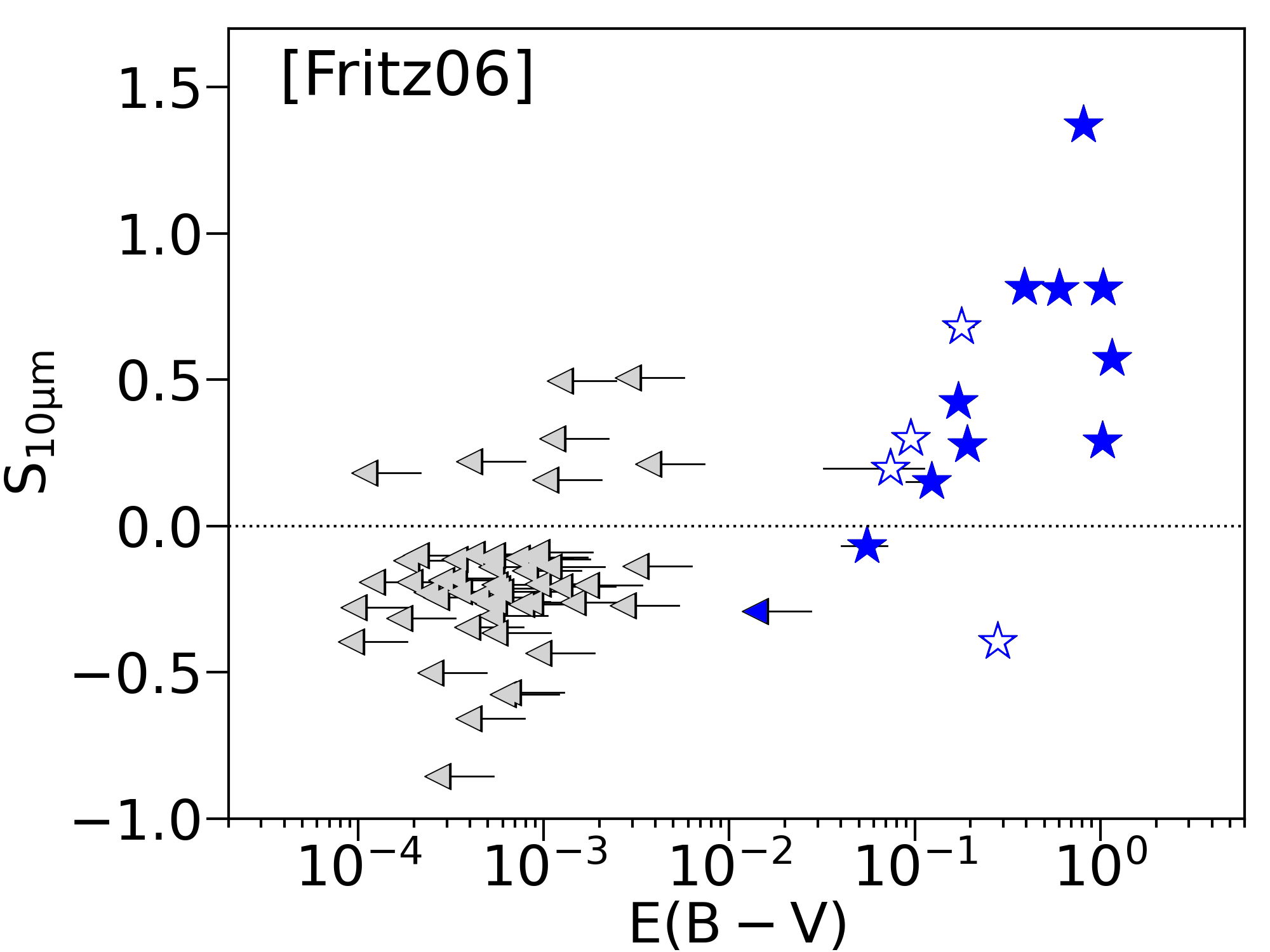}
\includegraphics[width=0.49\columnwidth]{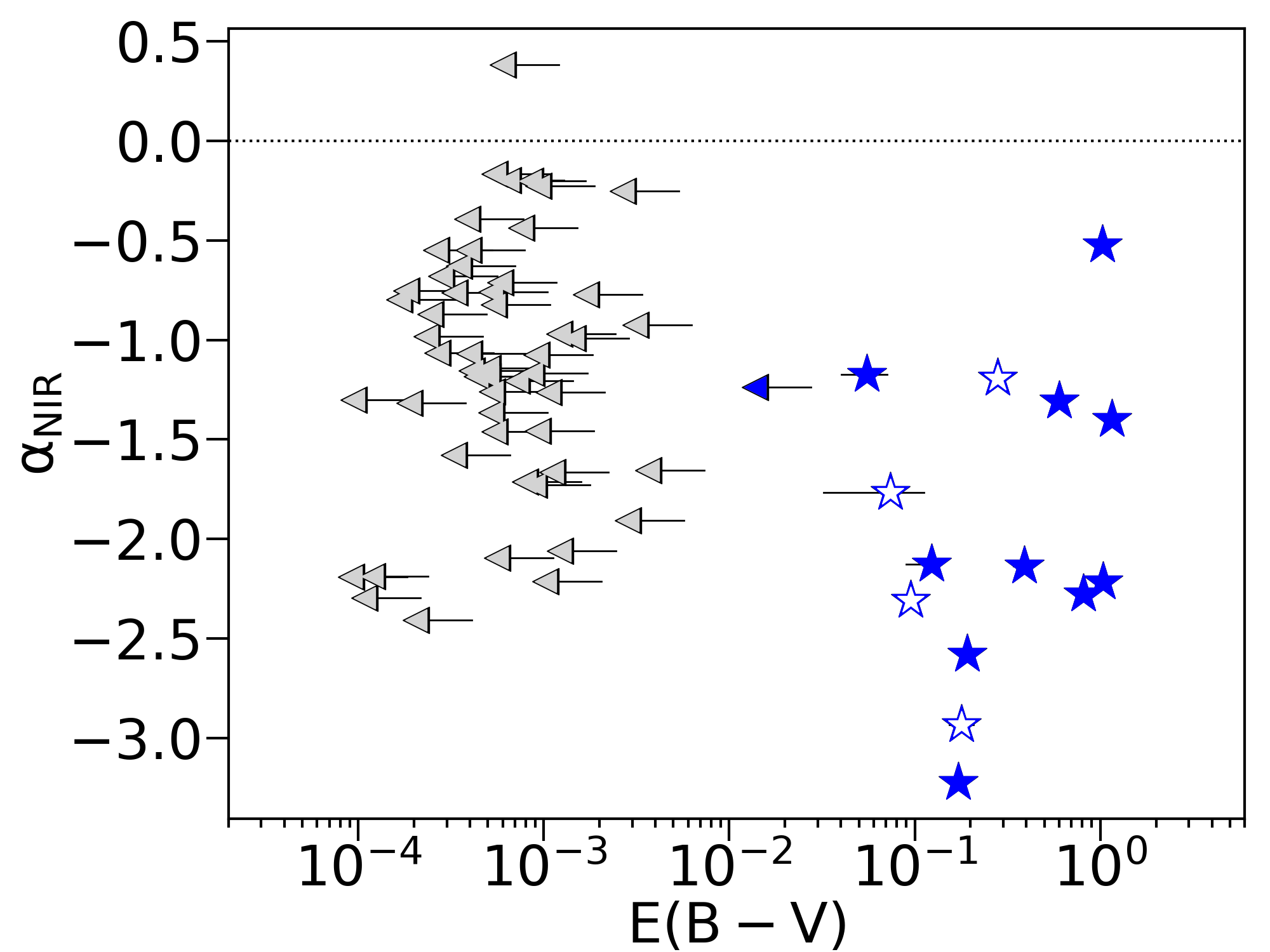}
\includegraphics[width=0.49\columnwidth]{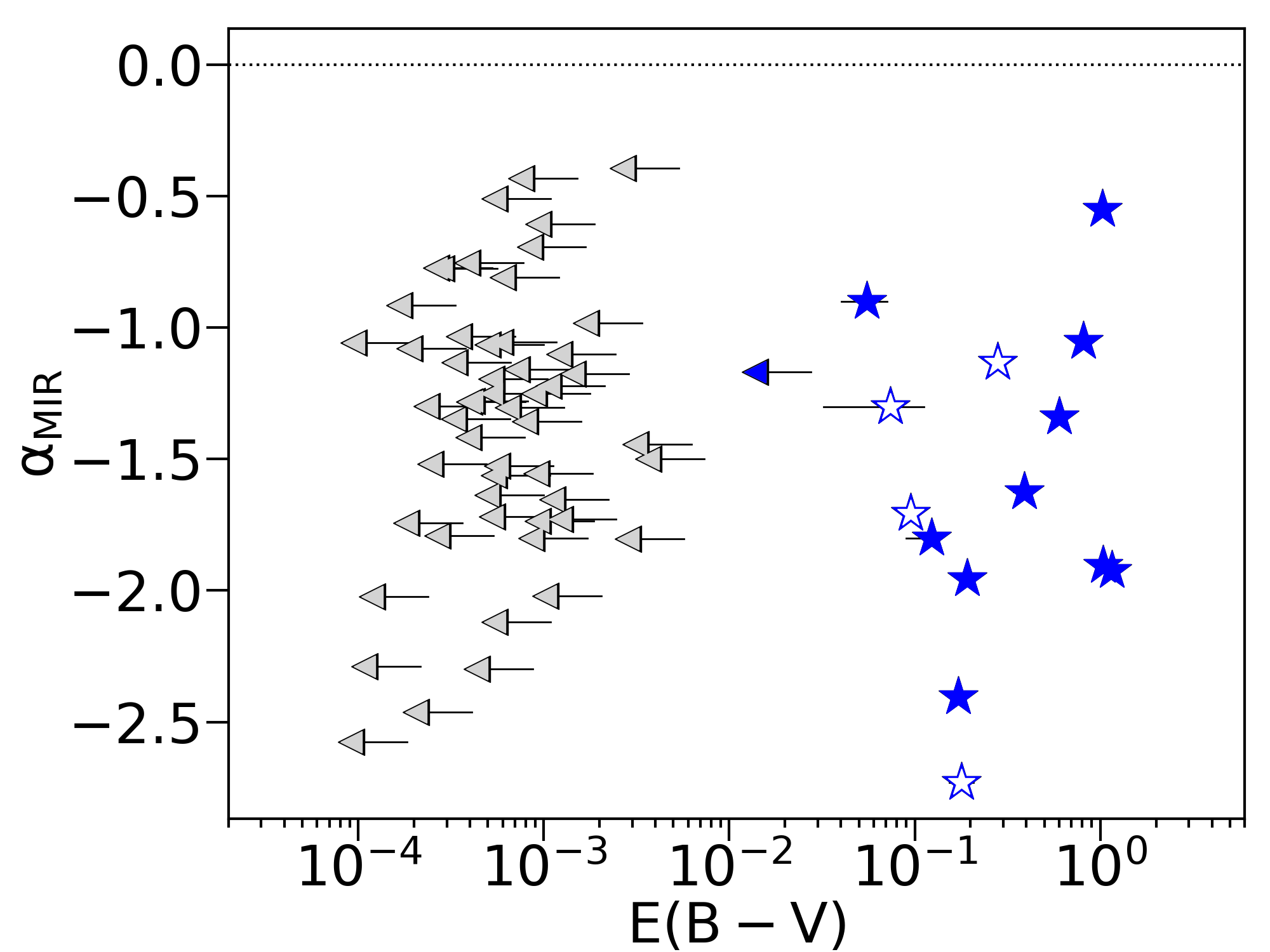}
\includegraphics[width=0.49\columnwidth]{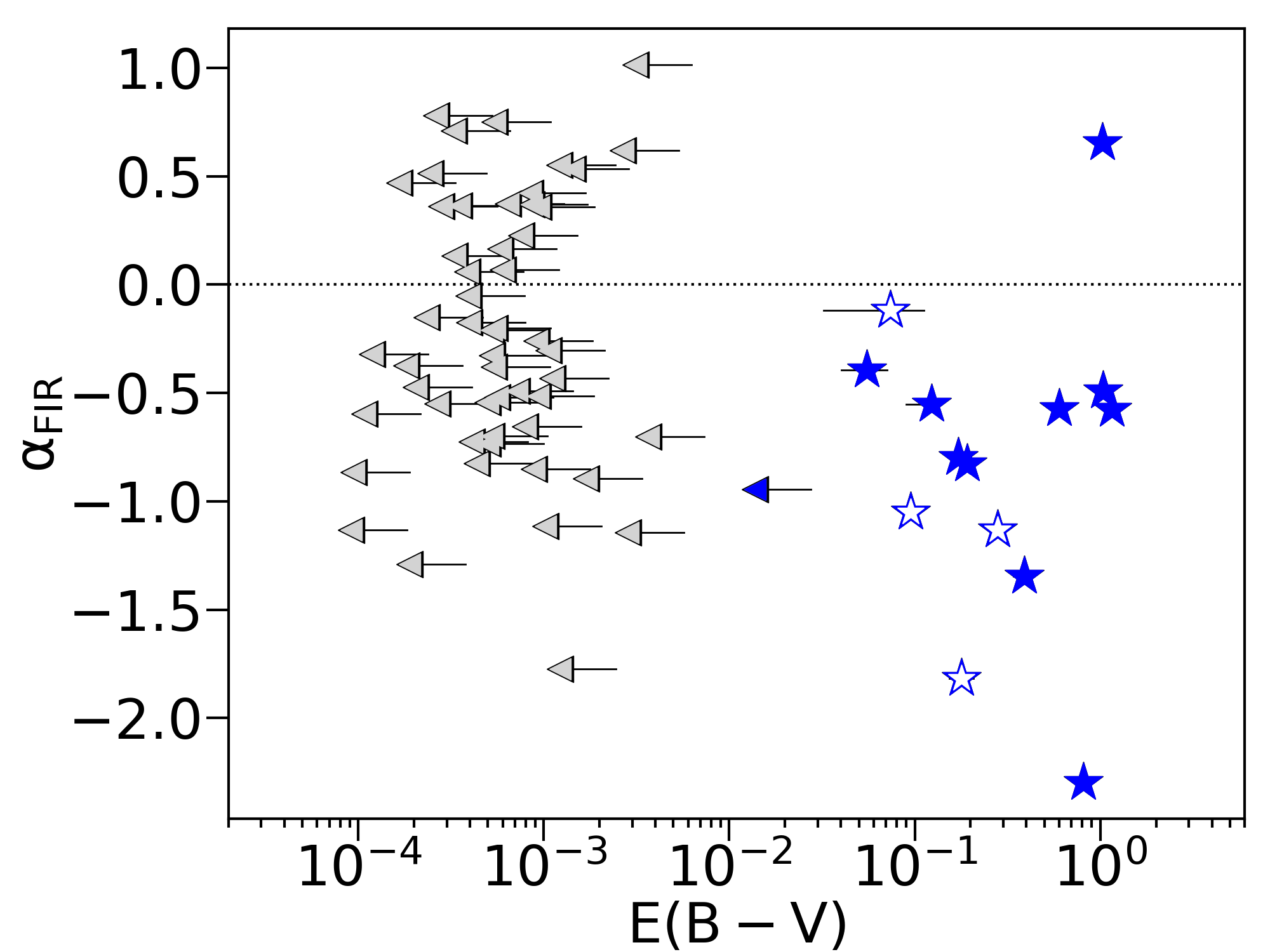}
\includegraphics[width=0.49\columnwidth]{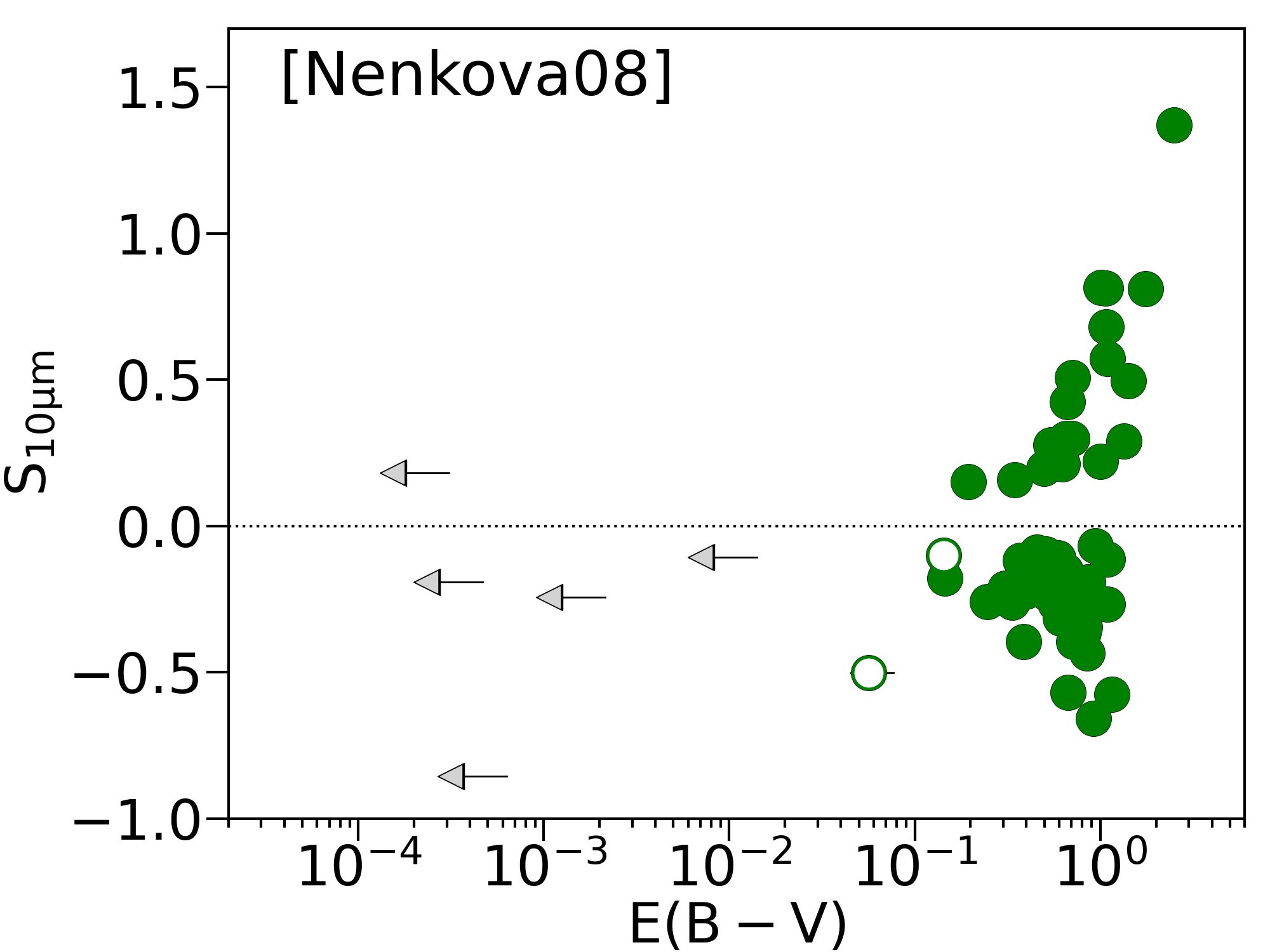}
\includegraphics[width=0.49\columnwidth]{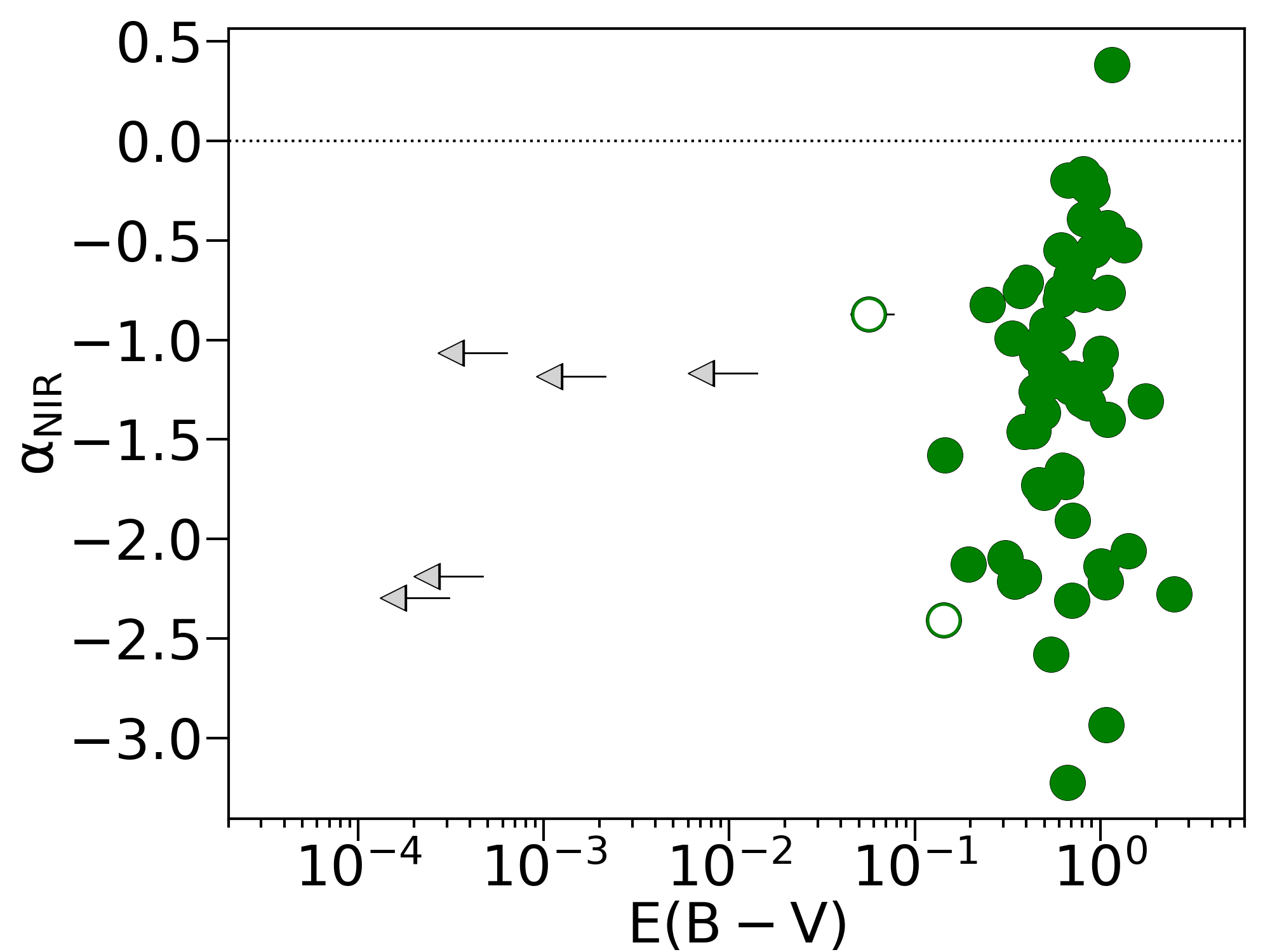}
\includegraphics[width=0.49\columnwidth]{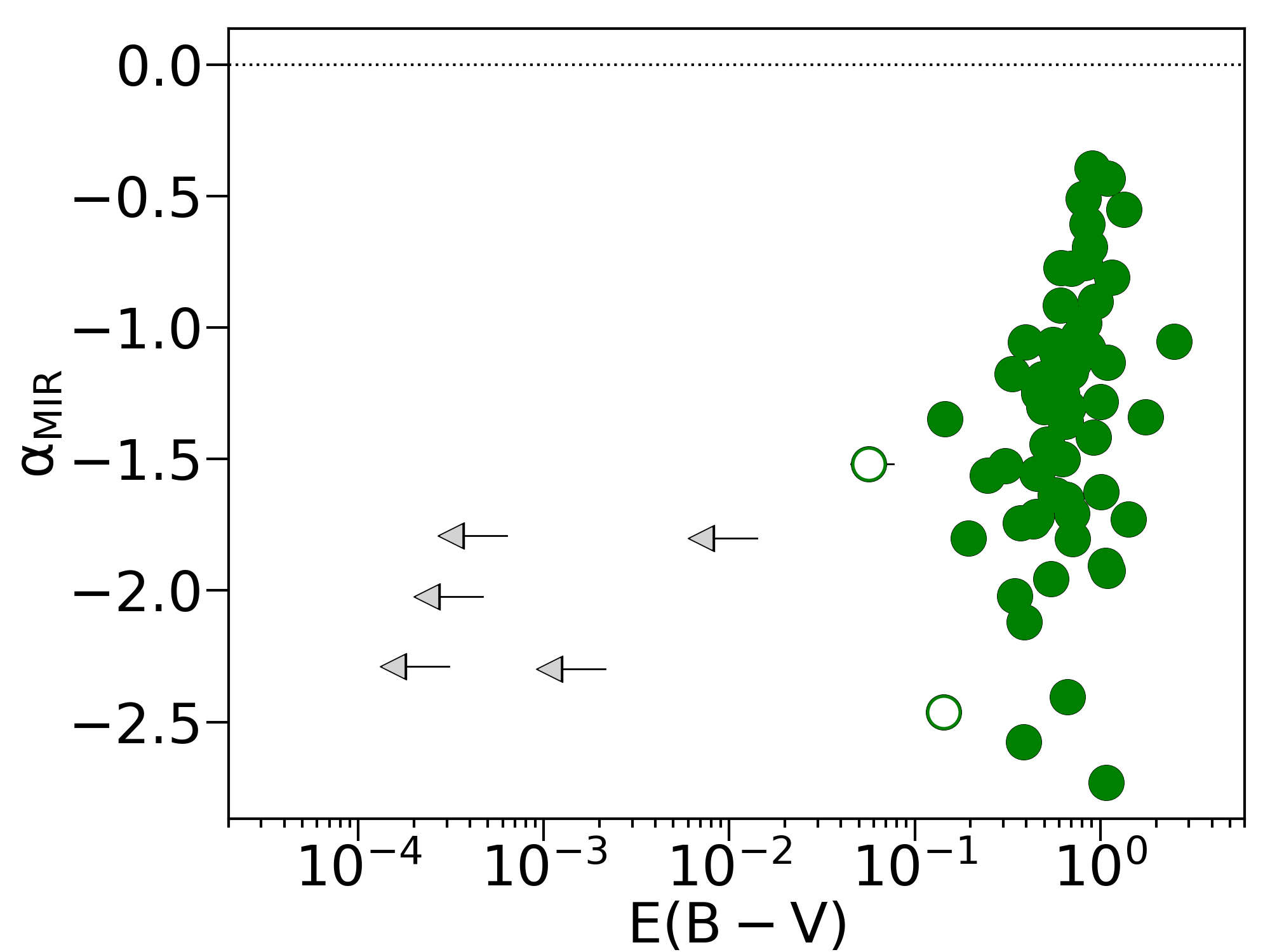}
\includegraphics[width=0.49\columnwidth]{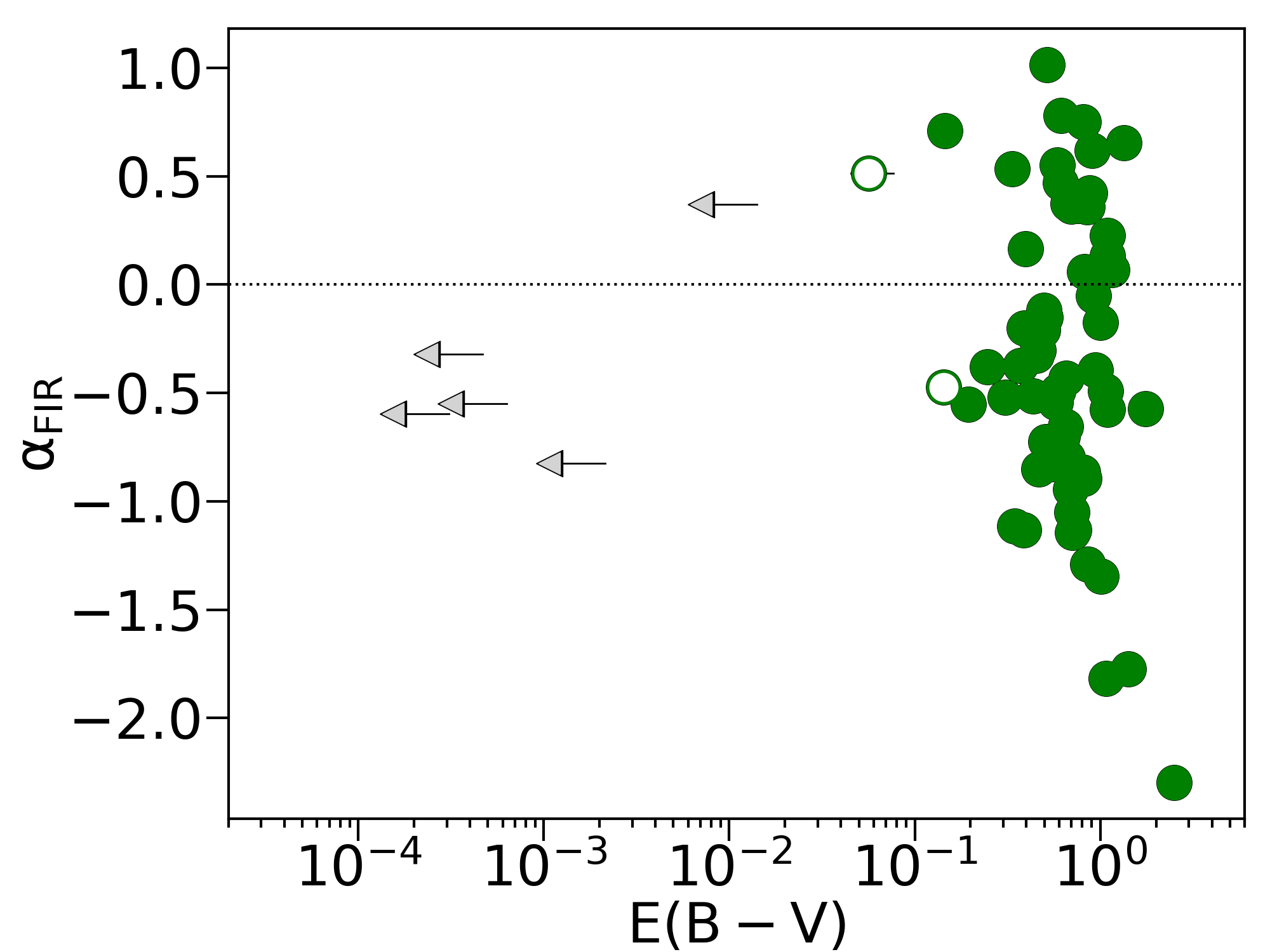}
\includegraphics[width=0.49\columnwidth]{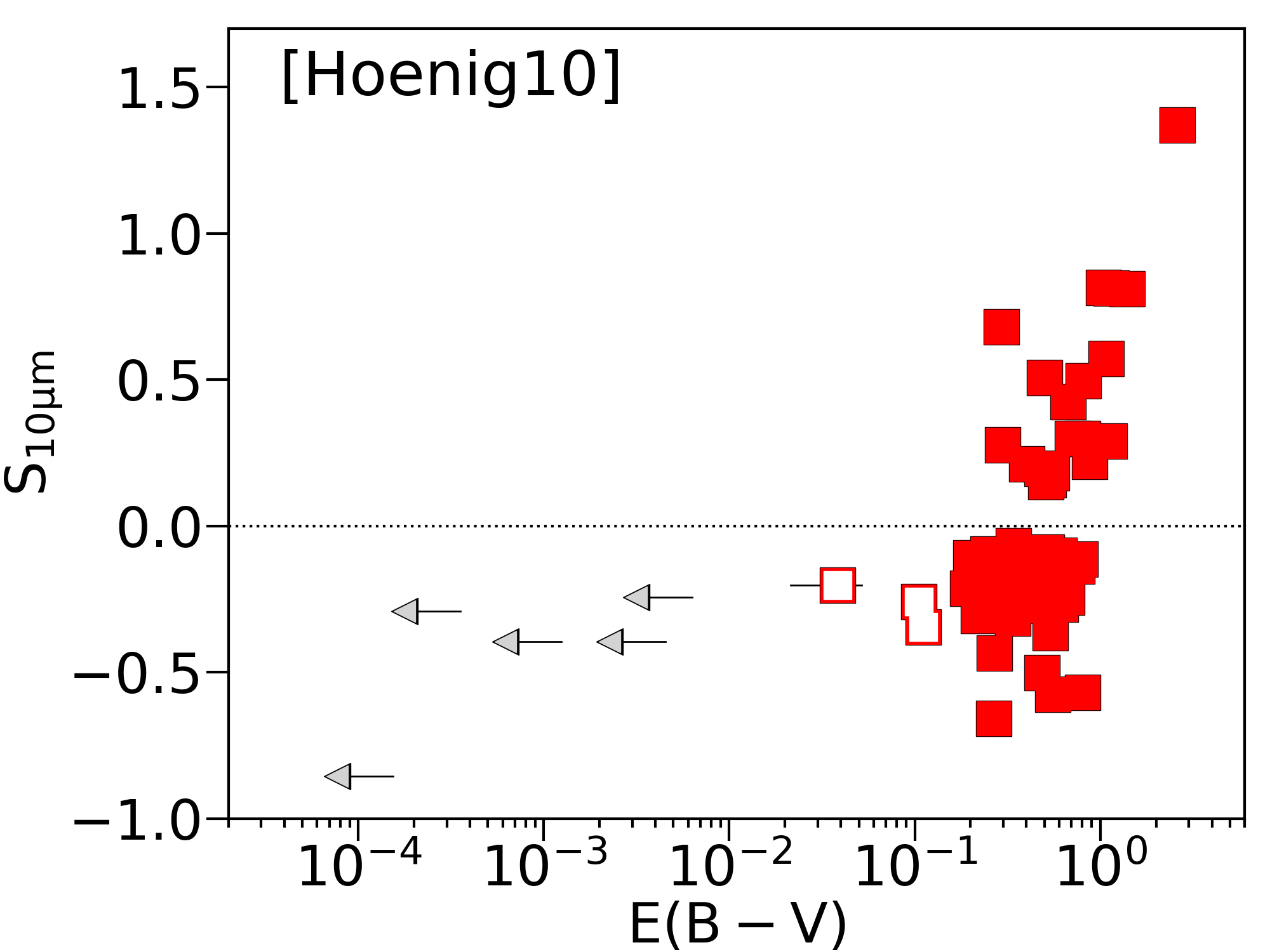}
\includegraphics[width=0.49\columnwidth]{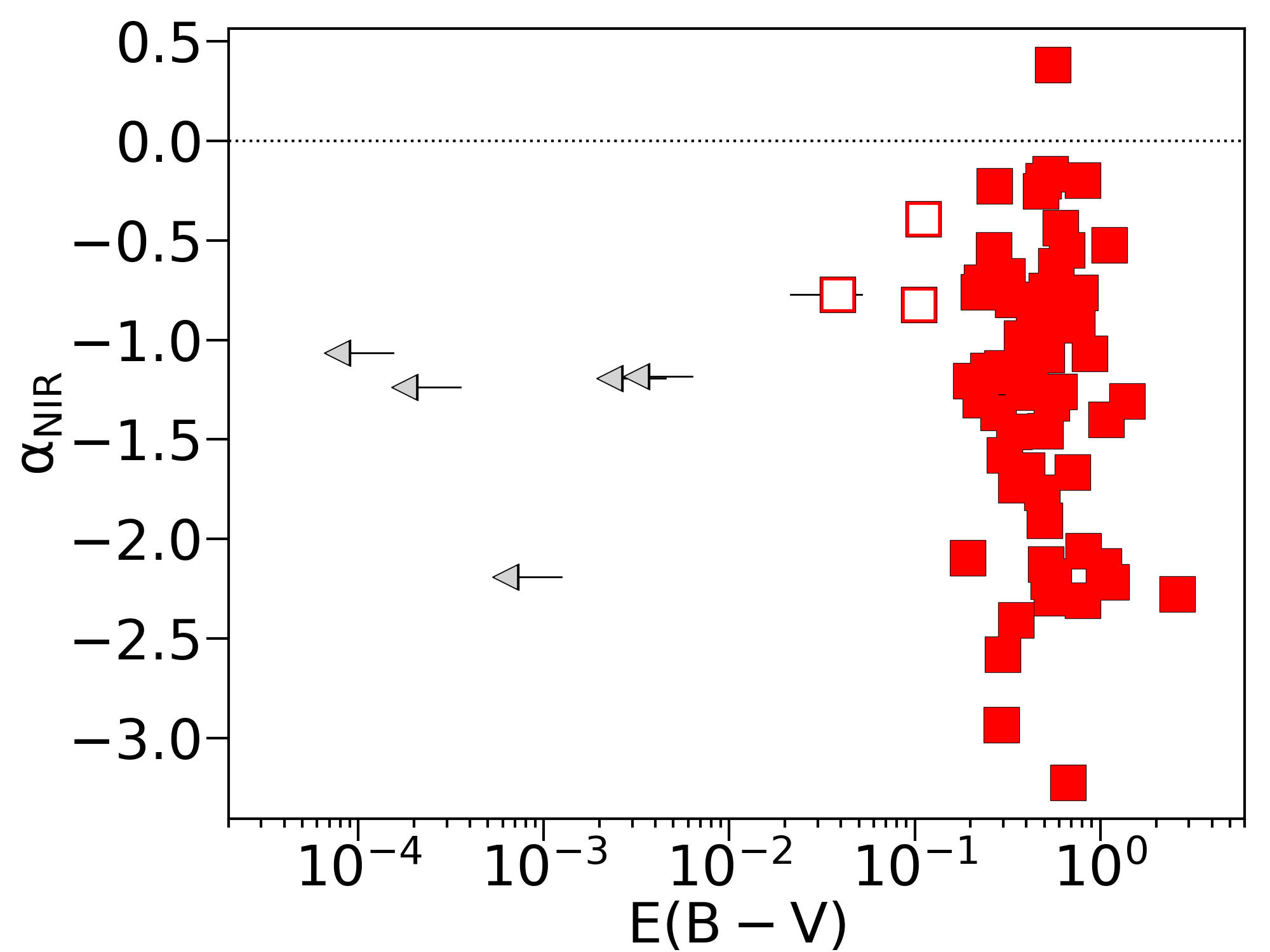}
\includegraphics[width=0.49\columnwidth]{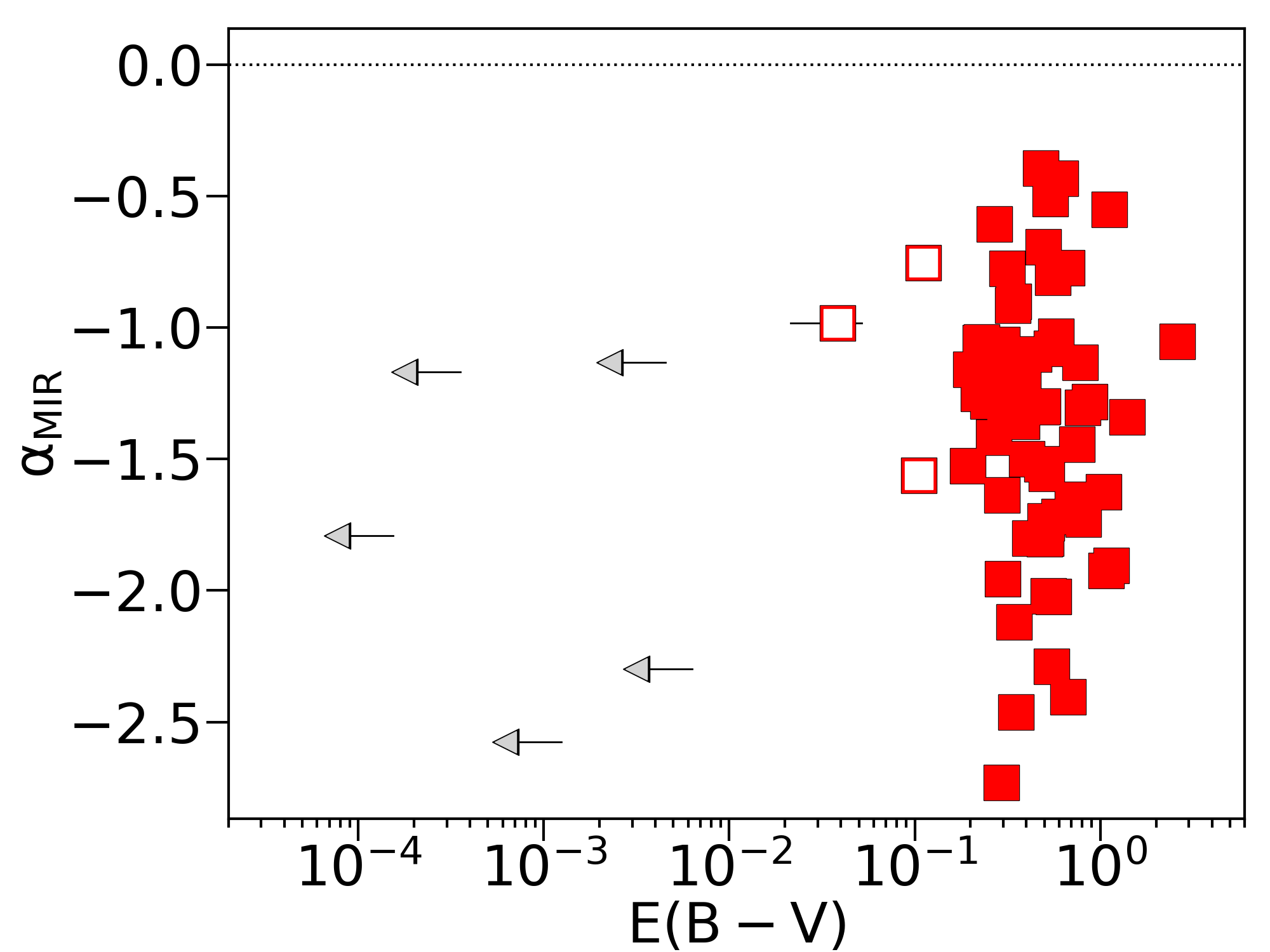}
\includegraphics[width=0.49\columnwidth]{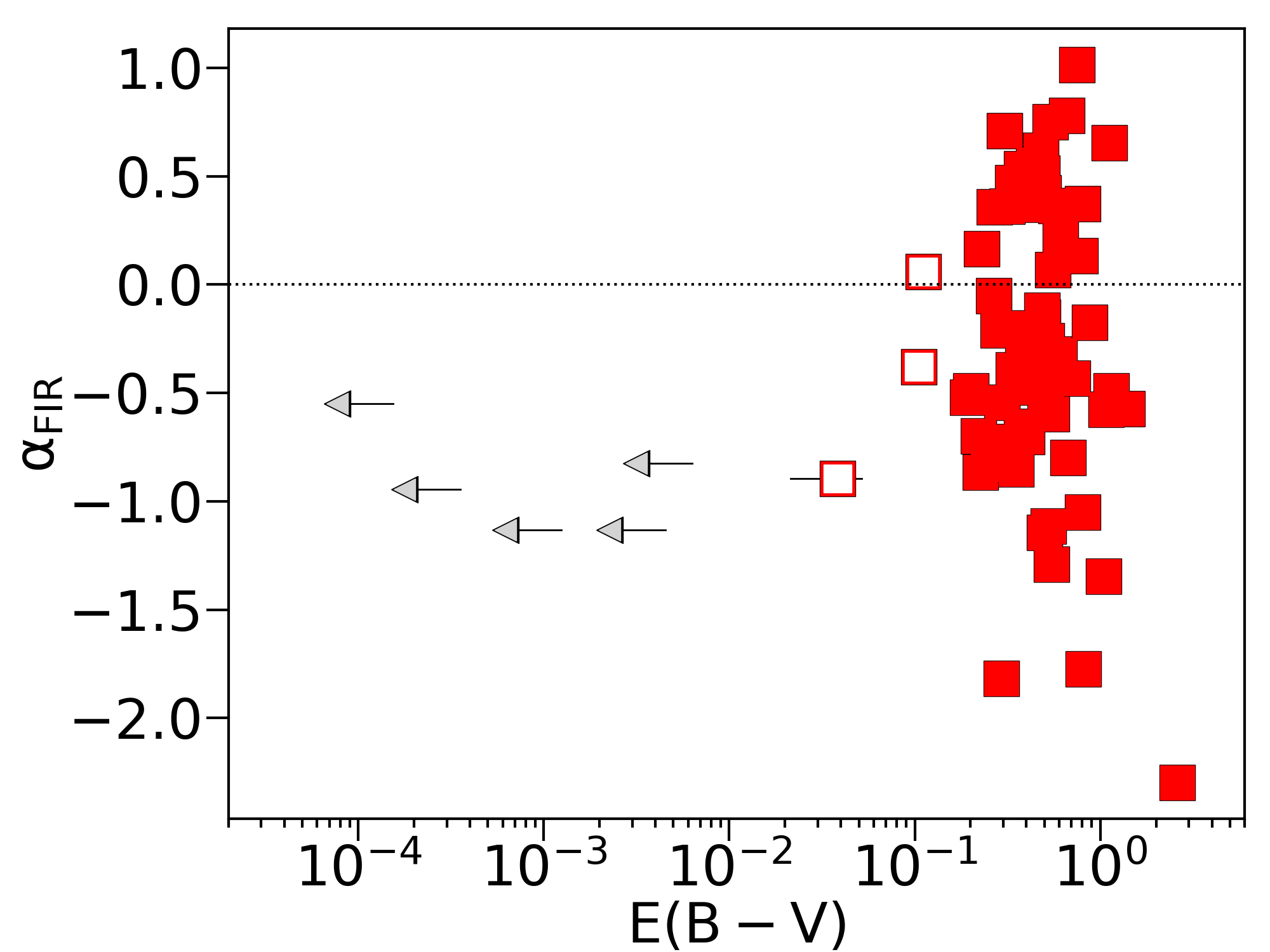}
\includegraphics[width=0.49\columnwidth]{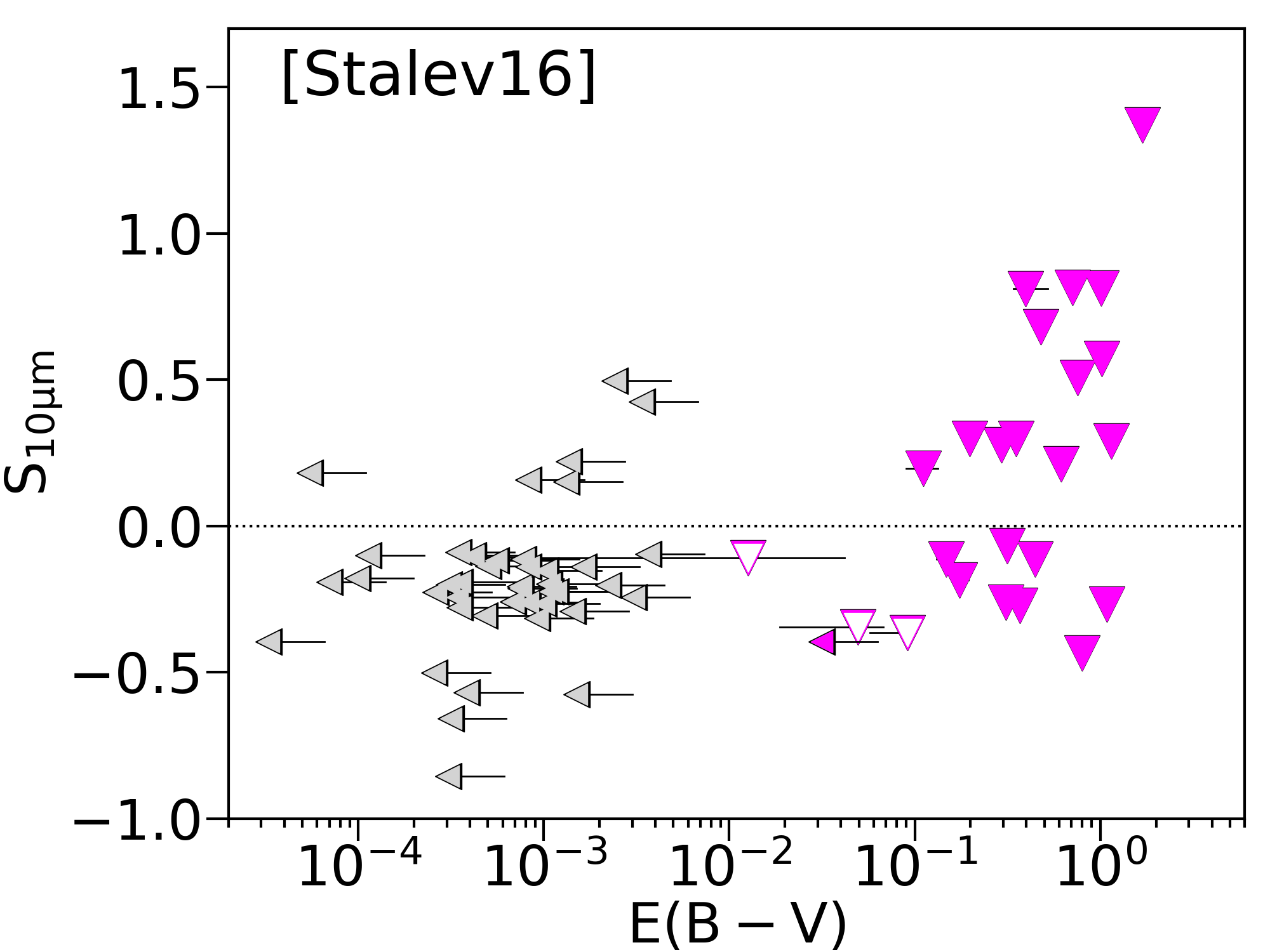}
\includegraphics[width=0.49\columnwidth]{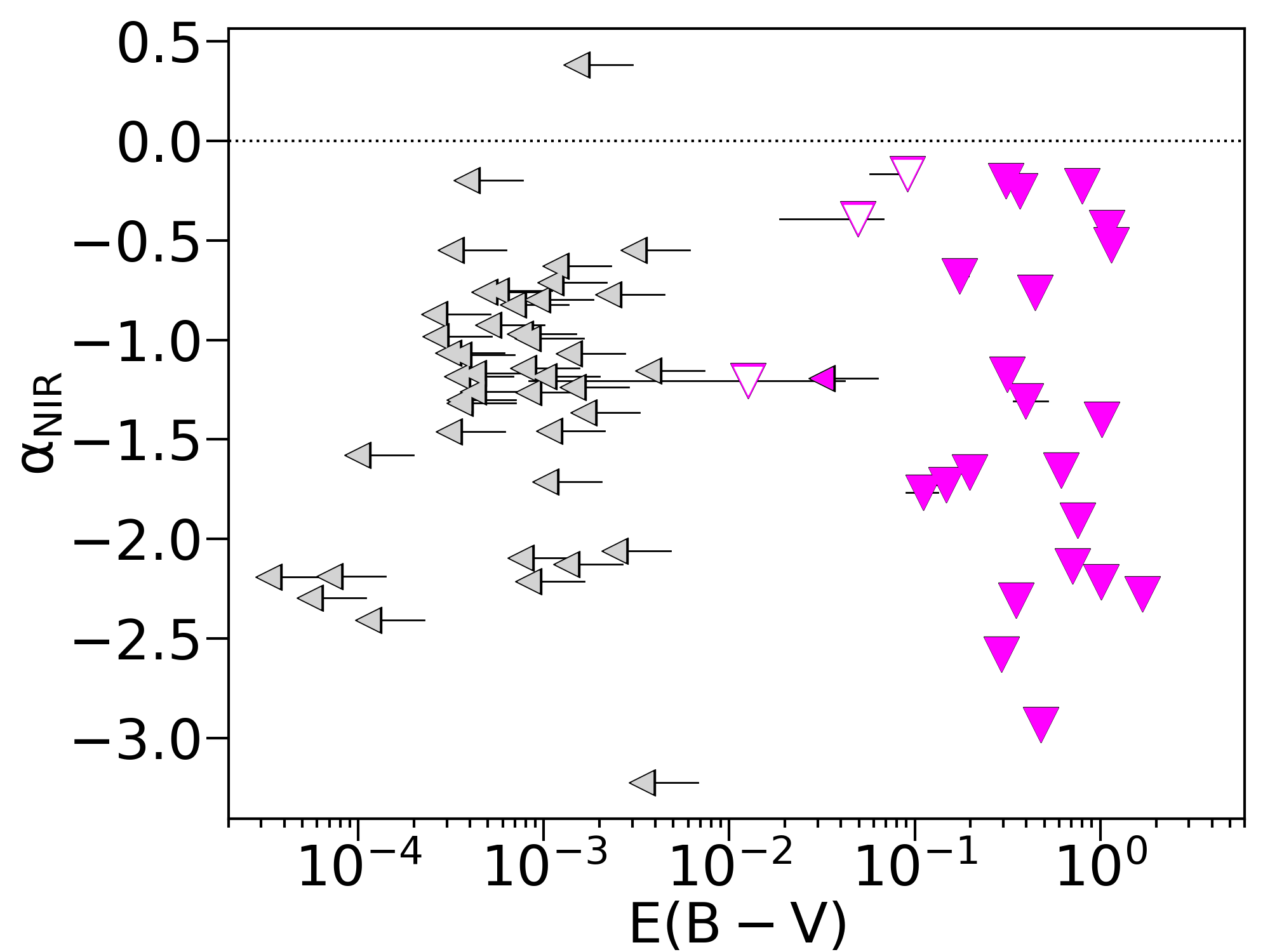}
\includegraphics[width=0.49\columnwidth]{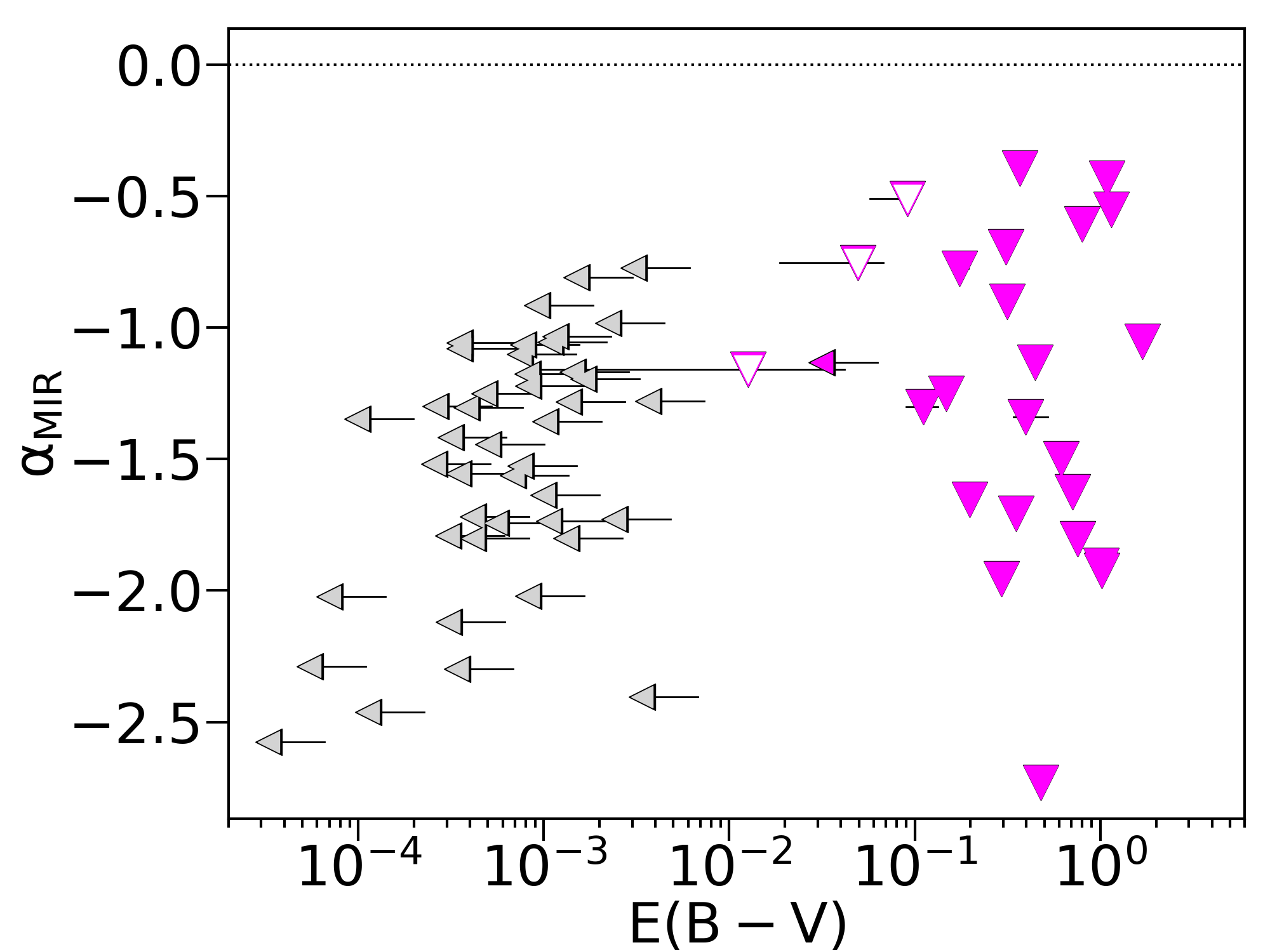}
\includegraphics[width=0.49\columnwidth]{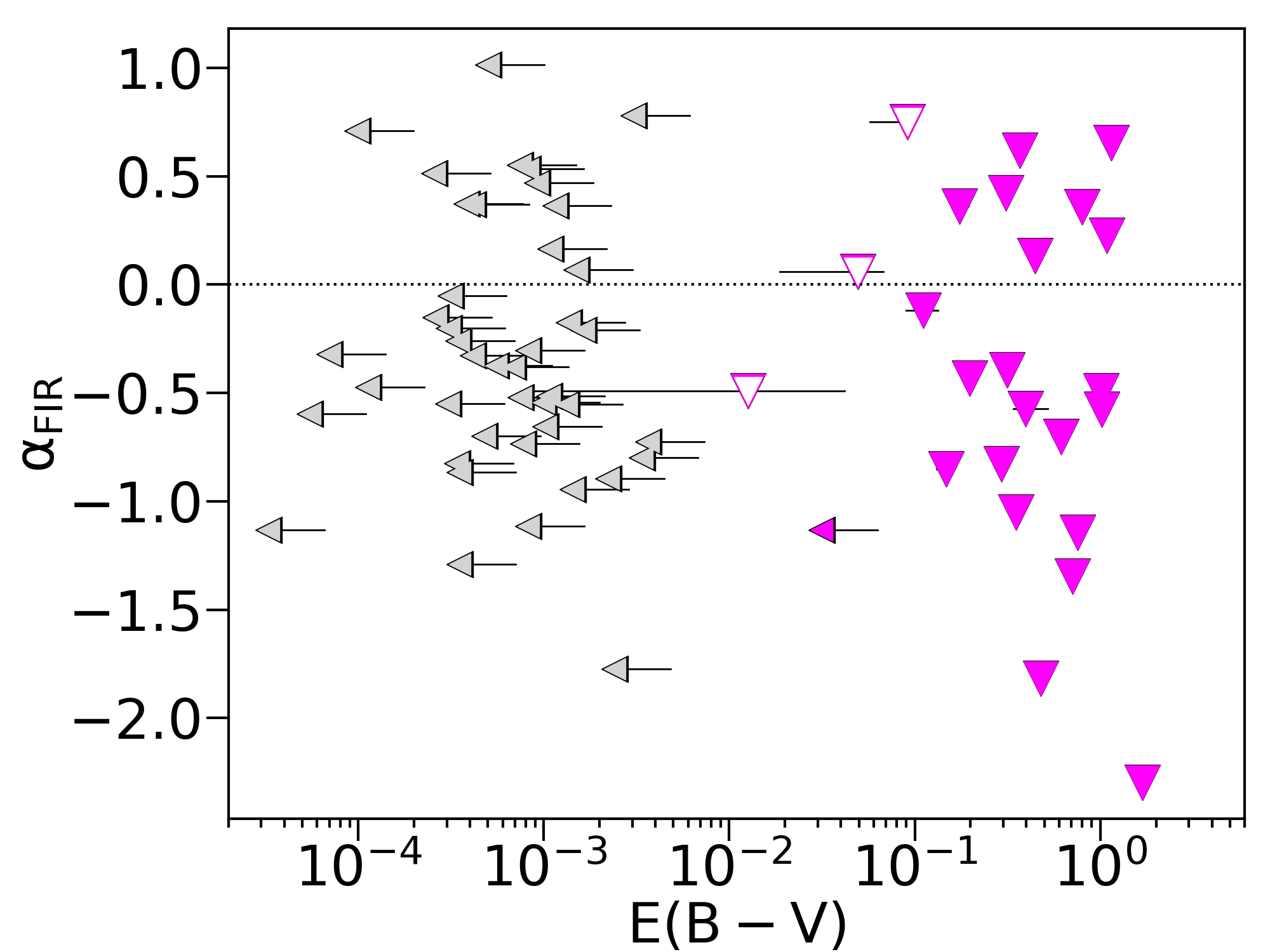}
\includegraphics[width=0.49\columnwidth]{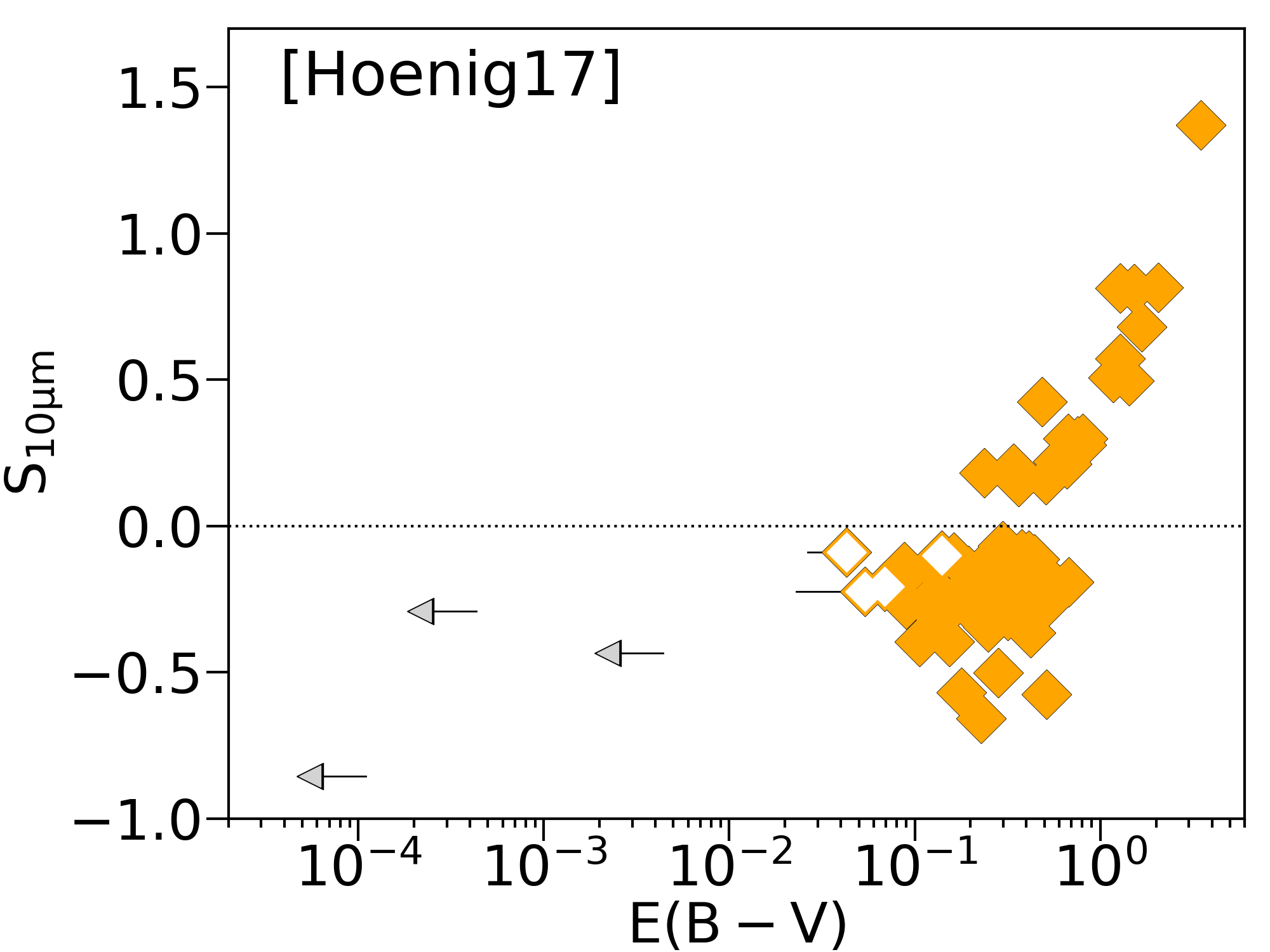}
\includegraphics[width=0.49\columnwidth]{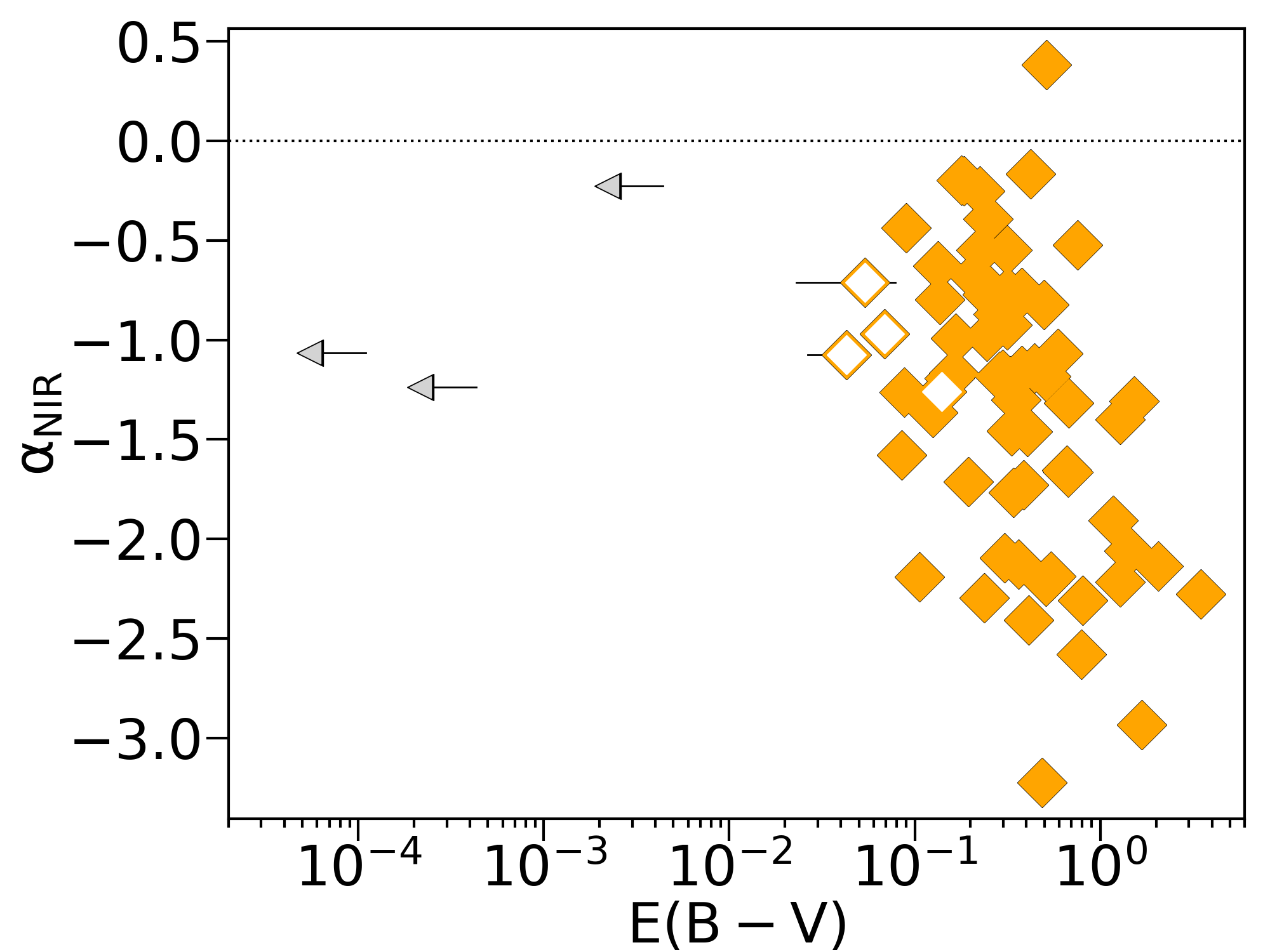}
\includegraphics[width=0.49\columnwidth]{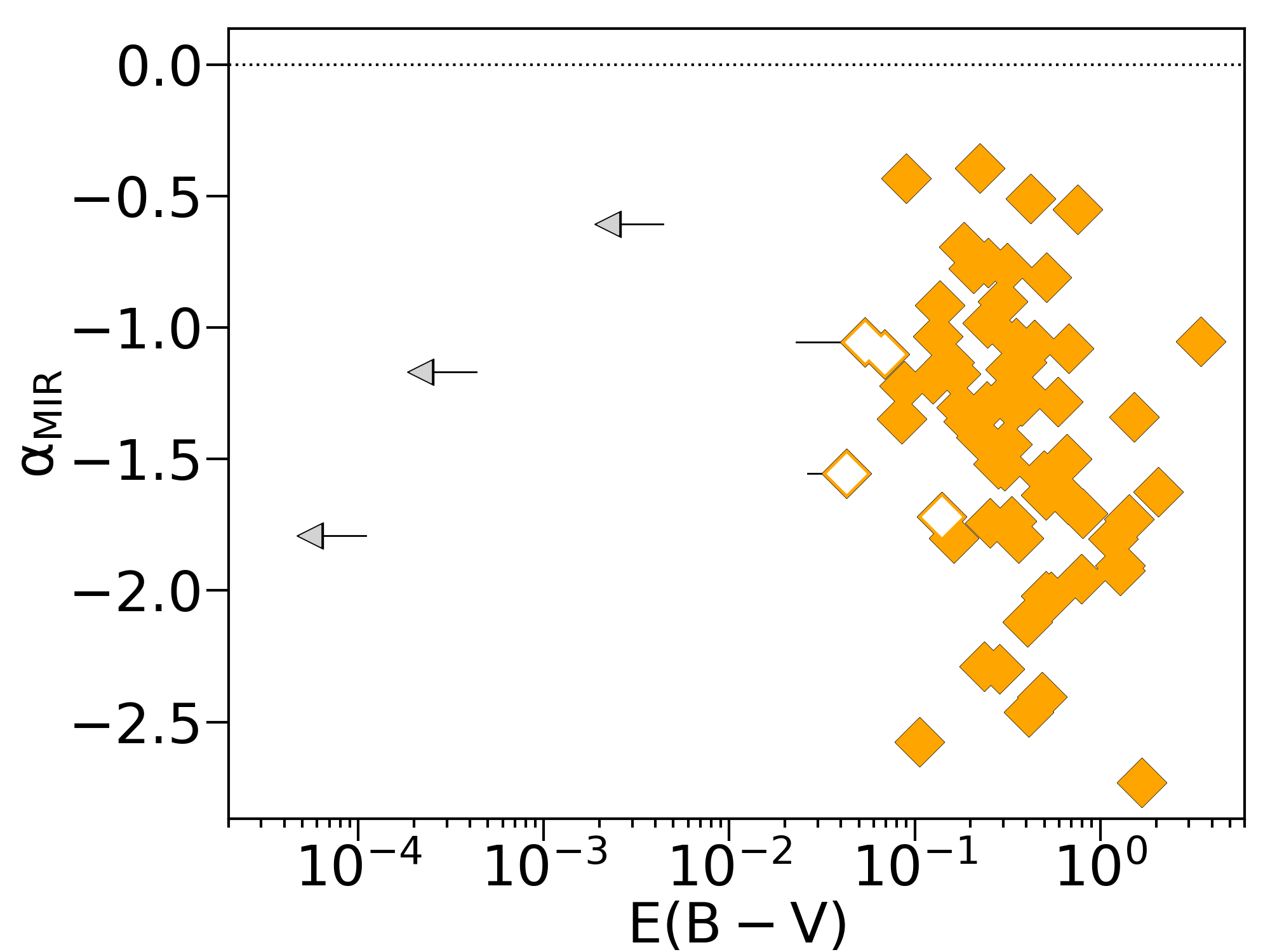}
\includegraphics[width=0.49\columnwidth]{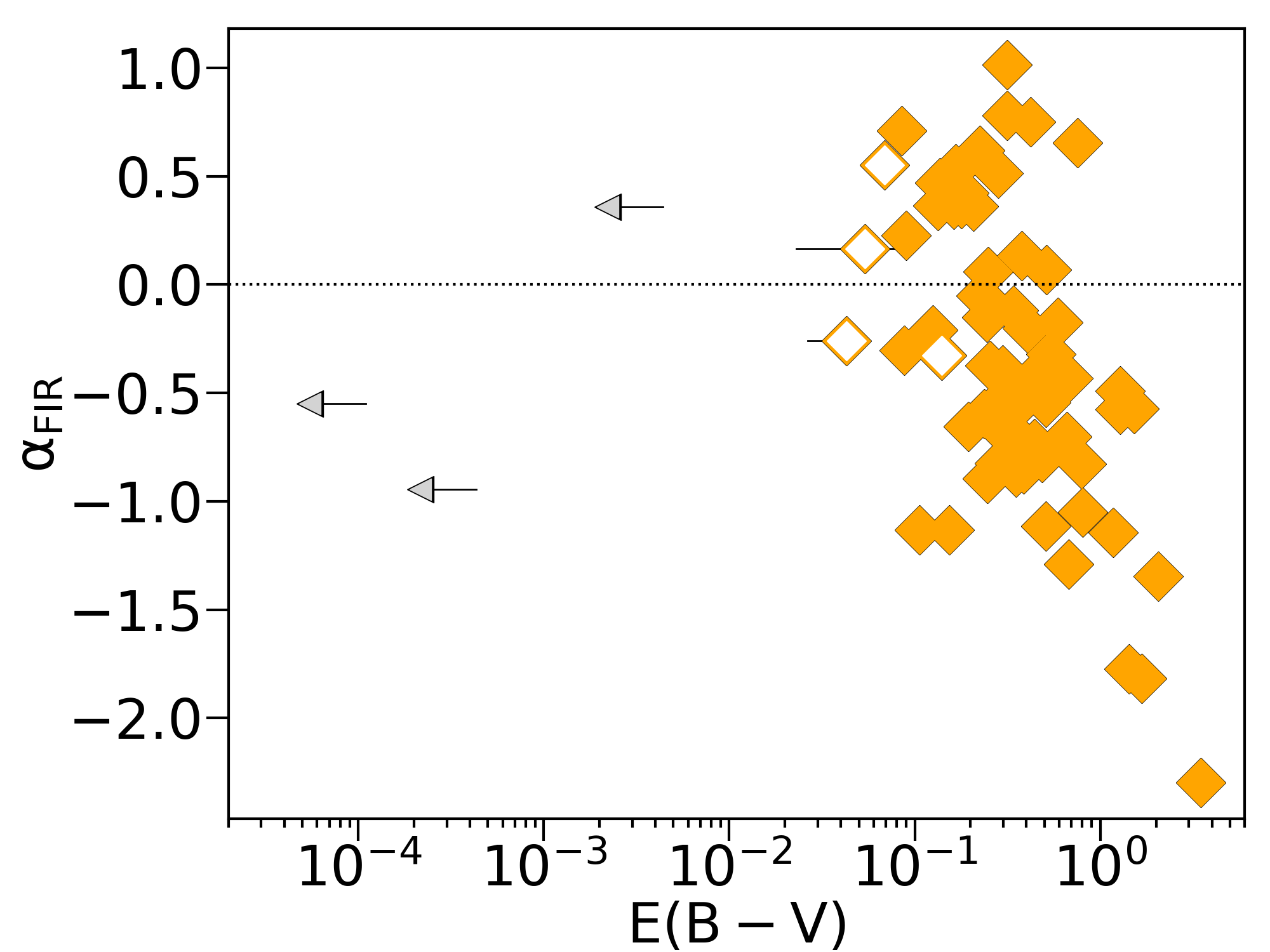}
\includegraphics[width=0.49\columnwidth]{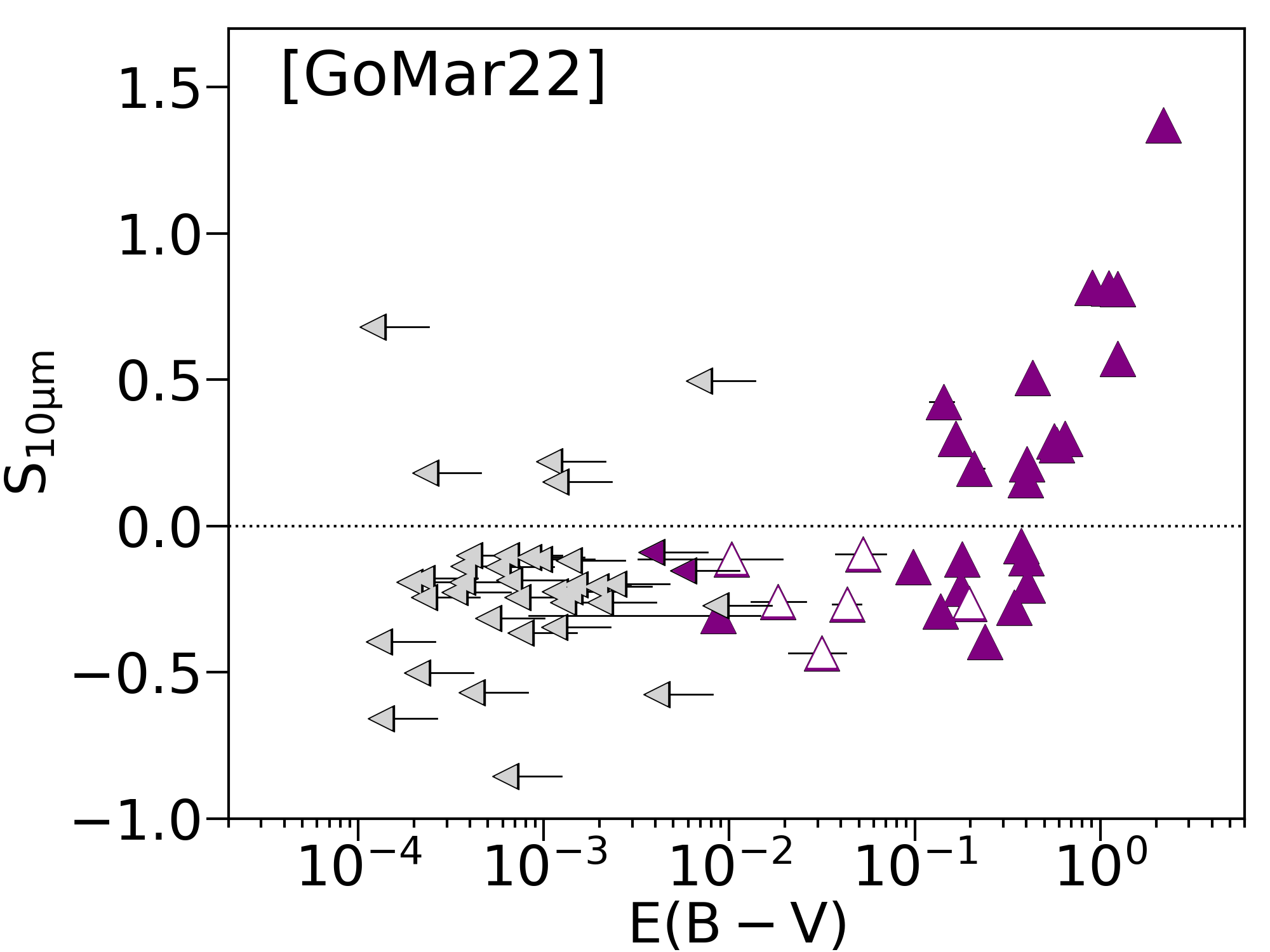}
\includegraphics[width=0.49\columnwidth]{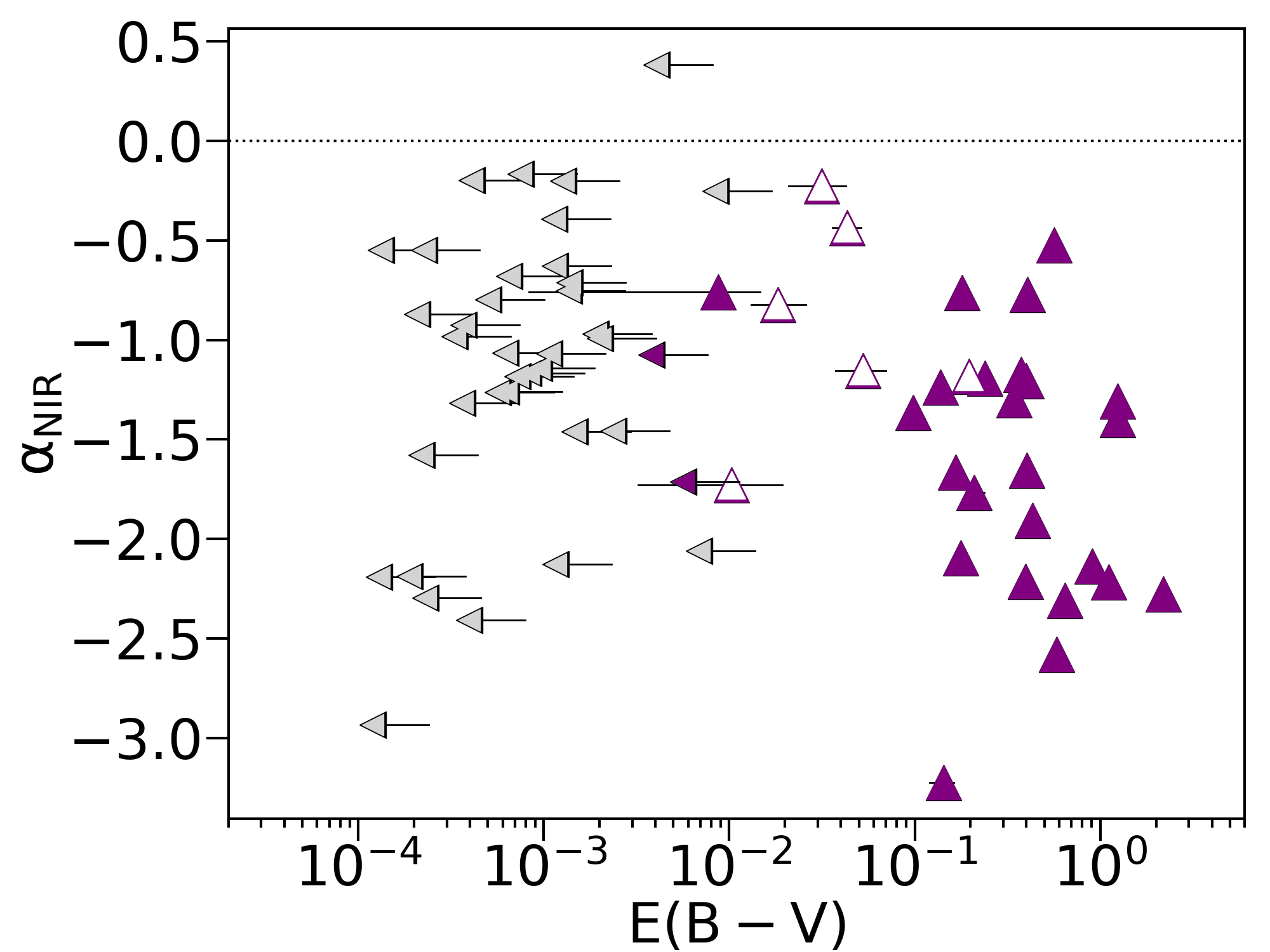}
\includegraphics[width=0.49\columnwidth]{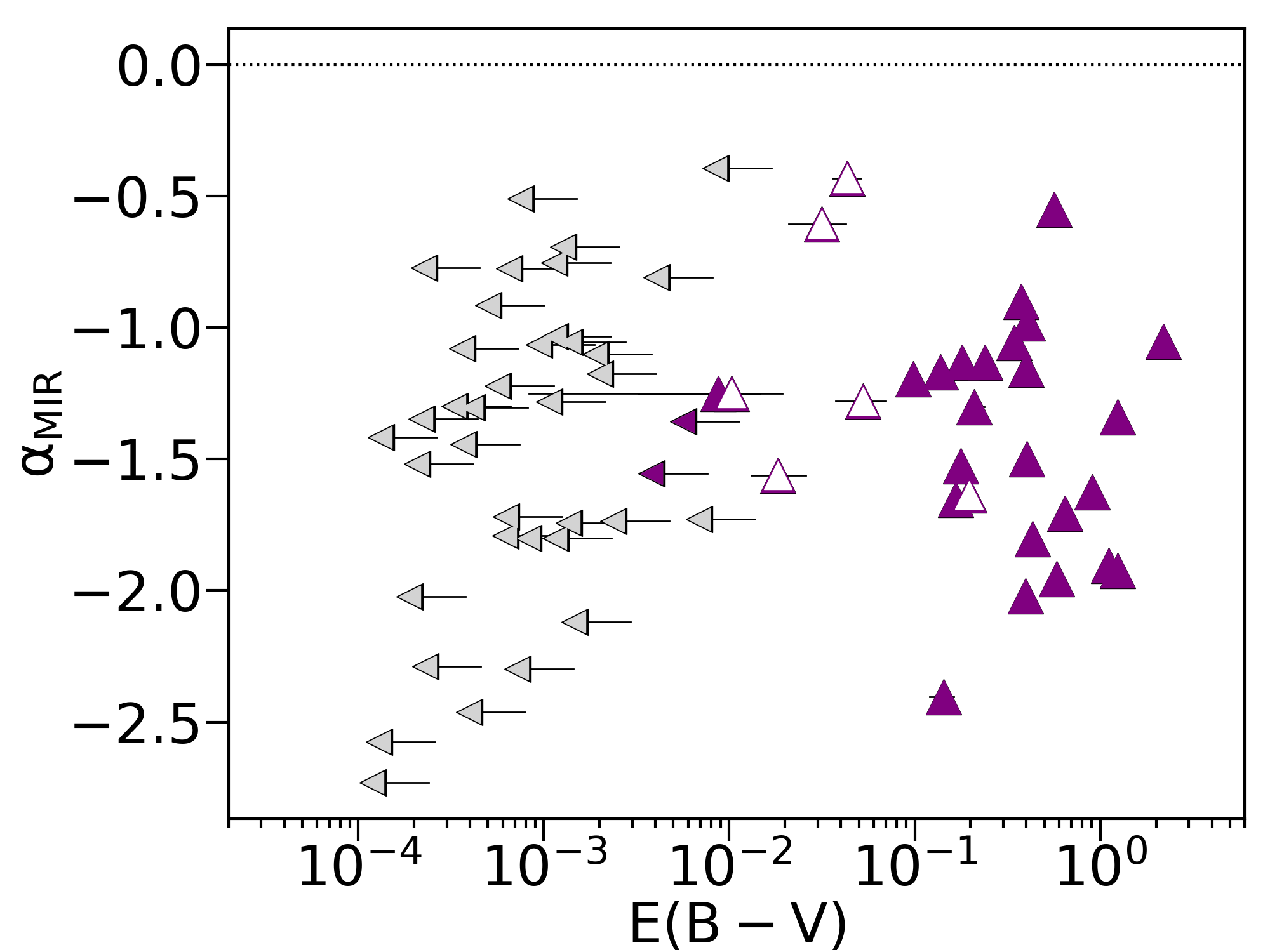}
\includegraphics[width=0.49\columnwidth]{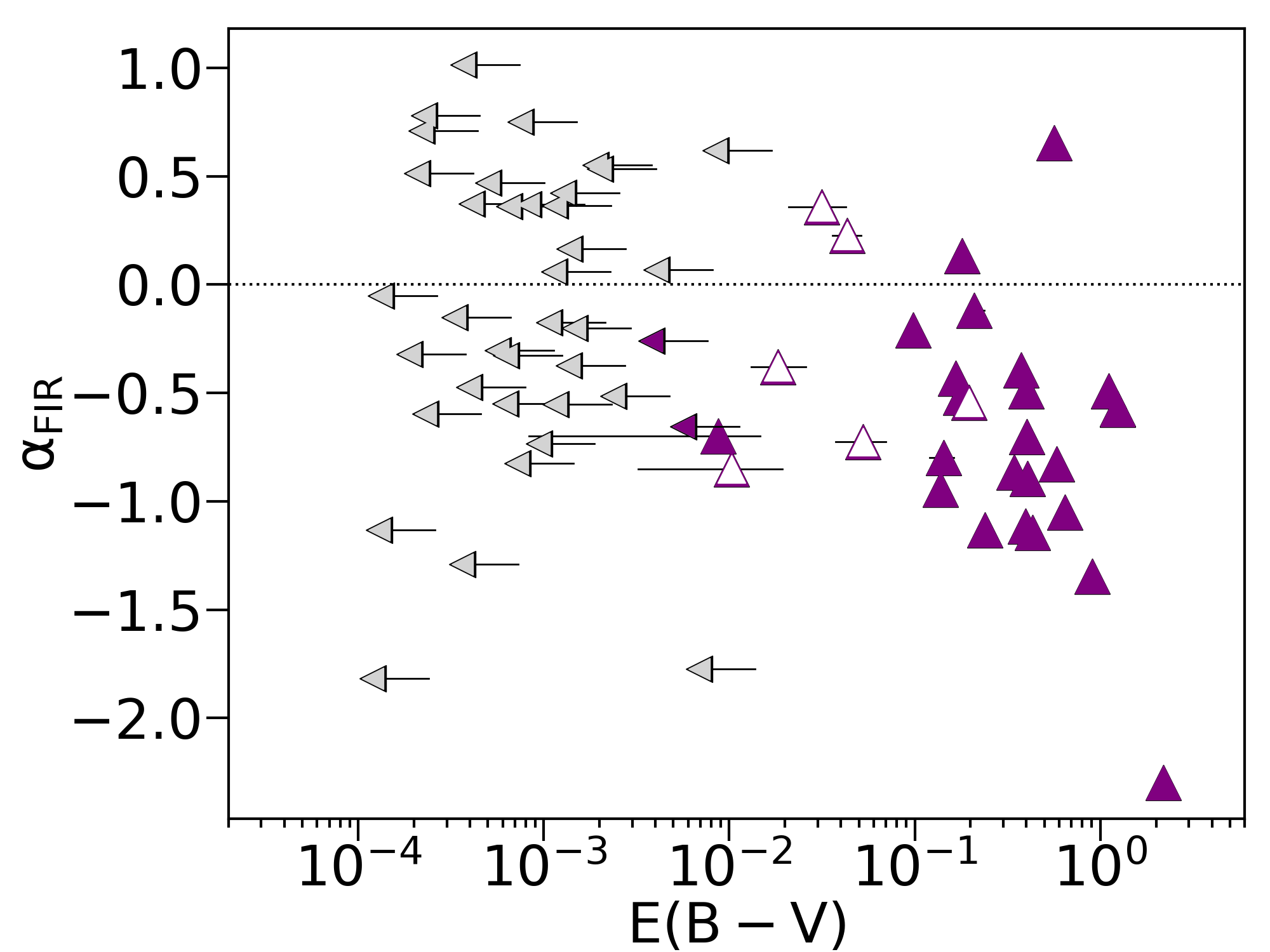}
\caption{Strength of the 9.7$\rm{\mu m}$ silicate feature (positive and negative values for emission and absorption features, respectively), $\rm{\alpha_{5.5-7.5\mu m}}$, $\rm{\alpha_{7.5-14 \mu m}}$, and $\rm{\alpha_{25-30\mu m}}$ versus E(B-V) foreground extinction for model fitting. The gray arrow indicates E(B-V) upper limits and empty symbols are objects where foreground extinction is constrained but its inclusion is not statistically needed by the data.}
\label{fig:FGextinction}
\end{center}
\end{figure*}

As discussed in Section\,\ref{sec:Results}, the inclusion of foreground extinction has a strong impact on the results for some of the available models. Here we further explore this by comparing the obtained foreground extinction and the spectral shape properties in our sample (see Fig.\,\ref{fig:FGextinction}). This plot clearly shows that [Nenkova08], [Hoenig10], and [Hoenig17] models statistically require the inclusion of foreground in a large percentage of objects and it is less required by [Fritz06], [Stalev16], and [GoMar23] (already quantified in Section\,\ref{sec:Results}). This is probably due to the inclusion of a smooth dust distribution for these latter models. When E(B-V) is constrained, the silicate strength is significantly correlated with E(B-V). Indeed, high absorption features require E(B-V)$\rm{\sim}$1, irrespective of the model used. This correlation is statistically significant ($\rm{r>0.5}$ and f-value$\rm{<10^{-4}}$) for all but [Fritz06] and [Stalev16] models (probably due to the low number of objects with constrained E(B-V)). Therefore, foreground extinction is required to reproduce deep silicate absorption features. Interestingly, among the infrared slopes used in this paper, we also found a significant anti-correlation between the far-infrared slope and E(B-V) for [Hoenig17] model ($\rm{r=0.6}$ and $\rm{f-value=2\times 10^{6}}$). Therefore, the inclusion of relatively large E(B-V) helps to reproduce steep (i.e. negative) FIR slopes with [Hoenig17] model, which indeed is one of the main issues of this model when compared with data (see Fig.\,\ref{fig:modelshapes}). 

Although E(B-V) roughly shows the same range of values irrespective of the model used, we also find significant differences when we compare the resulting E(B-V) found when used in combination with the tested models. [GoMar23] and [Stalev16] models show fully consistent E(B-V) values. This is expected since [Stalev16] is nested within [GoMar23] model. [Fritz06] model shows a factor of two lower values compared with [GoMar23] model while [Nenkova08], [Hoenig10], and [Hoenig17] systematically show at least a factor of two higher E(B-V) compared with [GoMar23] model and the disagreement is larger for lower E(B-V) values. 

\end{appendix}

\end{document}